\documentclass[aps,pre,reprint,superscriptaddress,footinbib,floatfix]{revtex4-1}

\usepackage{graphicx}
\usepackage{bm}
\usepackage{amssymb}
\usepackage{amsbsy}
\usepackage{amsmath}
\usepackage{mathtools}
\usepackage{xcolor}
\usepackage{stmaryrd}
\usepackage{mathrsfs}
\usepackage{tikz}
\usetikzlibrary{shapes}
\usepackage{cancel}
\usepackage[normalem]{ulem}
\makeatletter
\let\cat@comma@active\@empty
\makeatother

\makeatletter 
\gdef\@ptsize{2}% 12pt documents 
\let\@currsize\normalsize 
\makeatother 
\usepackage{multirow}
\usepackage{setspace}
\usepackage{enumitem}
\usepackage{verbatim}

\usepackage{amssymb}
\usepackage{amsbsy}
\usepackage[T1]{fontenc}
\usepackage{amsmath}
\usepackage{mathtools}

\usepackage{graphicx}
\usepackage{xcolor}
\usepackage{stmaryrd}
\usepackage{mathrsfs}
\usepackage{tikz}
\usetikzlibrary{shapes}
\usepackage{cancel}
\usepackage[normalem]{ulem}

\usepackage[colorinlistoftodos]{todonotes}
\usepackage{verbatim}
\usepackage{latexsym}
\usepackage{textcomp}
\usepackage{bm}
\usepackage[Euler]{upgreek}
\usepackage{longtable}
\usepackage{subfigure}

\usepackage[colorlinks,allcolors=cyan!70!black]{hyperref}

\def\tb{\textcolor{blue}}

\usepackage{epsfig}
\usepackage{dcolumn}

\usepackage{booktabs}

\newcommand{\qed}{\nobreak \ifvmode \relax \else
      \ifdim\lastskip<1.5em \hskip-\lastskip
      \hskip1.5em plus0em minus0.5em \fi \nobreak
      \vrule height0.75em width0.5em depth0.25em\fi}

\DeclareMathAlphabet\mathbfcal{OMS}{cmsy}{b}{n}
\DeclareMathAlphabet{\mathbfsf}{\encodingdefault}{\sfdefault}{bx}{sl}

\begin{document}
% % % %  Title and heading details
\title{Rouse model with fluctuating internal friction}
%\date{\today}
\author{R. Kailasham}
\affiliation{IITB-Monash Research Academy, Indian Institute of Technology Bombay, Mumbai, Maharashtra -  400076, India}
\affiliation{Department of Chemistry, Indian Institute of Technology Bombay, Mumbai, Maharashtra -  400076, India}
\affiliation{Department of Chemical Engineering, Monash University,
Melbourne, VIC 3800, Australia}
\author{Rajarshi Chakrabarti}
\email{rajarshi@chem.iitb.ac.in}
\affiliation{Department of Chemistry, Indian Institute of Technology Bombay, Mumbai, Maharashtra -  400076, India}
\author{J. Ravi Prakash}
\email{ravi.jagadeeshan@monash.edu}
\affiliation{Department of Chemical Engineering, Monash University,
Melbourne, VIC 3800, Australia}

\begin{abstract}
A coarse-grained bead-spring-dashpot chain model with the dashpots representing the presence of internal friction, is solved exactly numerically, for the case of chains with more than two beads. Using a decoupling procedure to remove the explicit coupling of a bead's velocity with that of its nearest neighbors, the governing set of stochastic differential equations are solved with Brownian dynamics simulations to obtain material functions in oscillatory and steady simple shear flow. Simulation results for the real and imaginary components of the complex viscosity have been compared with the results of previously derived semi-analytical approximations, and the difference in the predictions are seen to diminish with an increase in the number of beads in the chain. The inclusion of internal friction results in a non-monotonous variation of the viscosity with shear rate, with the occurrence of continuous shear-thickening following an initial shear-thinning regime. The onset of shear-thickening in the first normal stress coefficient is pushed to lower shear rates with an increase in the internal friction parameter. 
\end{abstract}

\maketitle

\section{\label{sec:intro} Introduction}

\begin{figure}[h]
\centering
\includegraphics[width=80mm]{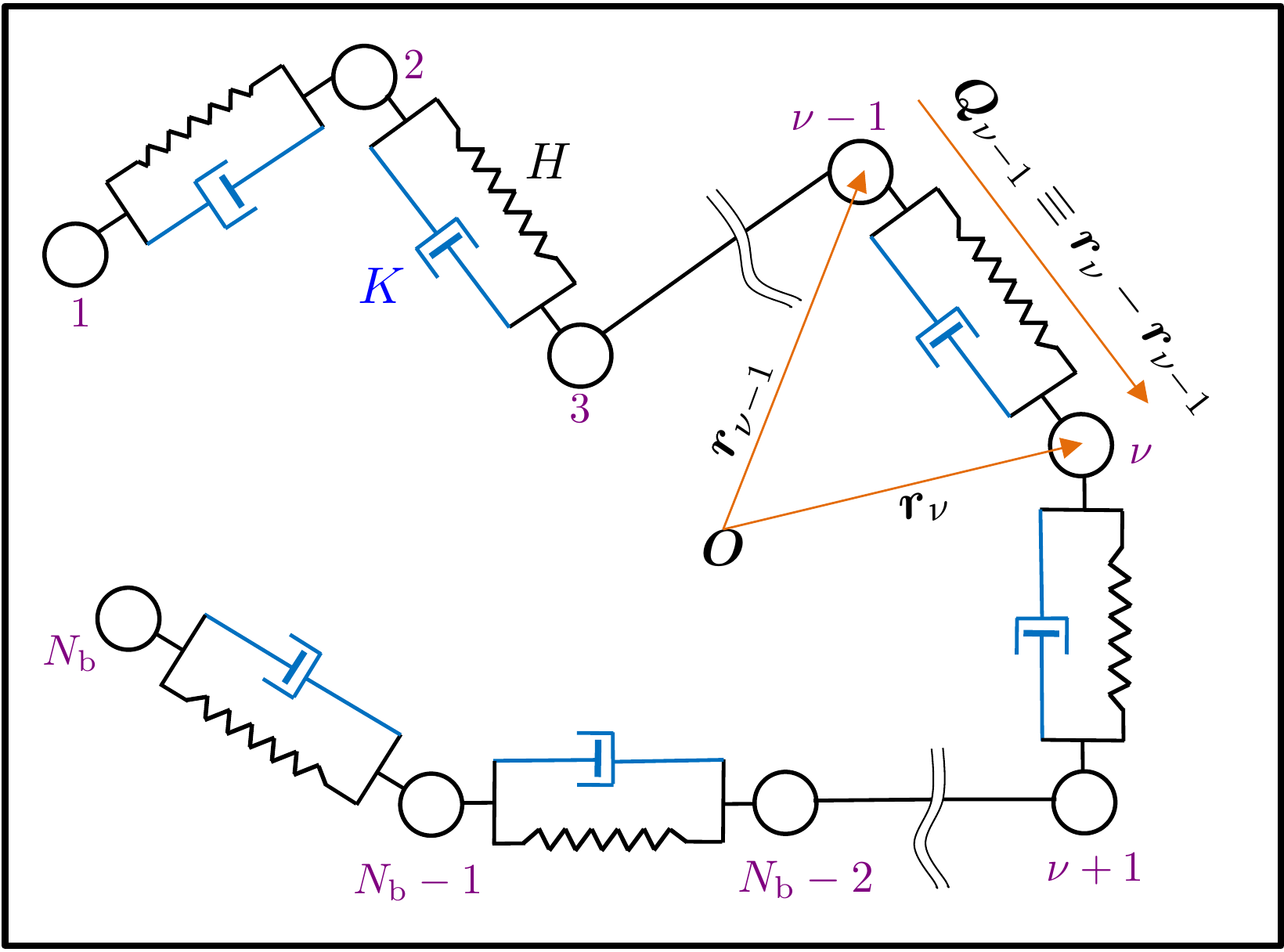}  
\caption{Representation of a polymer chain as a sequence of beads connected by spring-dashpots. The Hookean spring constant associated with each spring is $H$, and the damping coefficient of each dashpot is $K$.}
\label{fig:model}
\end{figure}

The presence of an additional mode of dissipation in polymer molecules arising from intramolecular interactions, denoted as internal friction or internal viscosity (IV)~\cite{kuhn1945bedeutung,Booij1970,degennes,Manke1985,Manke1988,Dasbach1992,Gerhardt1994}, has been invoked to reconcile the high values of dissipated work observed in force spectroscopic measurements on single molecules~\cite{Murayama2007,Alexander-Katz2009,Schulz2015,Kailasham2020}, the steepness of the probability distribution of polymer extensions in coil-stretch transitions observed in turbulent flow~\cite{dario2020}, and the dampened reconfiguration kinetics of biopolymers~\cite{Arbe2001,Poirier2002,Khatri20071825,Qiu20043398,Soranno201217800,Samanta2013,Samanta2014,Samanta2016,Sashi2016,Soranno07032017,Das2018,Mondal2020,Subramanian2020}. The discontinuous jump in the stress in polymer solutions upon the inception or cessation of flow~\cite{liang1993stress,Orr1996} has also been attributed to internal friction. Given its wide-ranging impact, a careful investigation of the consequences of the presence of internal friction is essential. As shown in Fig.~\ref{fig:model}, the bead-spring-chain model~\cite{Bird1987b,Ottinger1996}, widely used to describe the dynamics of flexible polymer chains, has been modified to include a dashpot in parallel with each spring to account for internal friction effects~\cite{Booij1970,Bird1987b,ravibook}. {The dashpot provides a restoring force proportional to the relative velocity between adjacent beads, and acts along the connector vector joining these beads.} The machinery for the solution of coarse-grained polymer models through Brownian dynamics (BD) simulations is well-established~\cite{Bird1987b,Ottinger1996}: the equation of motion for the connector vector velocities is combined with an equation of continuity in probability space to obtain a Fokker-Planck equation for the system, and the equivalent stochastic differential equation is integrated numerically. The inclusion of IV, however, results in a coupling of connector vector velocities and precludes a trivial application of the usual procedure for all but the simplest case of a dumbbell. By expanding the scope of an existing methodology for velocity-decoupling~\cite{Manke1988}, the exact set of governing stochastic differential equations for a bead-spring-dashpot chain with $N_{\mathrm{b}}$ beads, and its numerical solution using BD simulations is presented here. The thermodynamically consistent~\cite{Schieber1994} stress tensor expression for this model is derived, and material functions in simple shear and oscillatory shear flows have been calculated. {The decoupling procedure and the governing equations for bead-spring-chain dashpots derived in this work holds for arbitrary spring force laws. However, attention is restricted to the case of Hookean springs for the purposes of comparison with available approximate results in the literature~\cite{Manke1988,Dasbach1992,Schieber1993} where appropriate.}

{Given that the phenomenon of internal friction has both rheological and biophysical consequences, it is instructive to briefly discuss the chronology of research on this topic by both these communities, before presenting the results. In the rheology community, the expression for the internal viscosity force law in two dimensions was first introduced by \citet{kuhn1945bedeutung} in 1945. The three-dimensional analogue given by~\citet{Booij1970} in 1970 has since been used by many others~\cite{Manke1988,Dasbach1992,Schieber1993,Wedgewood1993,Kailasham2018} in the modeling and simulation of coarse-grained polymer models with an additional, solvent-viscosity-independent source of dissipation, and is also used in the present work. The effect of this additional mode of dissipation on conformational transitions in biomolecules was observed experimentally in the early 1990s~\cite{Ansari1992}, with the term ``internal friction'' predominantly used in the biophysics literature to refer to this phenomenon. A popularly used theoretical framework for the analysis of internal friction effects in experiments and simulations~\cite{Soranno201217800,ja211494h,Ameseder2018} on biopolymers is the Rouse model with internal friction (RIF), developed by McLeish and coworkers~\cite{Khatri2007rif,Khatri20071825}. The form of the internal friction force law in the RIF model is different from that used for the internal viscosity force in the rheology literature. We have recently shown [Ref.~\cite{ivfluc2021}, under review] that the force law in the RIF model is essentially a preaveraged treatment of the exact internal viscosity force law which accounts for fluctuations. We rationalize in Ref.~\cite{ivfluc2021} that the form of the internal viscosity force law used in the rheology literature may therefore be considered as ``fluctuating internal friction'', and that the same theoretical framework may consequently be applied for studying both rheological and biophysical phenomena. The present work considers only fluctuating internal friction, and the phrase ``internal friction'' or ``IV'' is henceforth used interchangeably and unambiguously.}

It is {worthwhile briefly surveying} the methods employed in the past, before turning our attention to the solution proposed in the present work. Booij and van Wiechen~\cite{Booij1970} used perturbation analysis to expand the configurational distribution function of a Hookean spring-dashpot in terms of the internal friction parameter, $\varphi=K/\zeta$, which is the ratio of the dashpot's damping coefficient, $K$, to the bead friction coefficient, $\zeta$, and predicted optical and rheological properties in the presence of steady shear flow. On the other hand, Williams and coworkers offered a semi-analytical approximate solution for the stress-jump~\cite{Manke1988} of bead-spring-dashpot-chains with an arbitrary number of beads, using a decoupling procedure which is discussed at length later in this paper. They also obtained predictions for the complex viscosity of such chains by writing the configurational distribution function as a series expansion in strain~\cite{Dasbach1992}. While the approach of \citet{Booij1970} is restricted to small values of the internal friction parameter, the solutions proposed by Williams and coworkers~\cite{Manke1988,Dasbach1992} are applicable only in the linear viscoelastic regime. The transient variation and steady-state values of viscometric functions of bead-spring-dashpot chains with arbitrary chain-length in shear flow have been predicted~\cite{Bazua1974,Manke1989} using the linearized rotational velocity~\cite{Cerf1957,cerf69,Peterlin1967} (LRV) approximation for the internal friction force. As mentioned previously, the form of the internal friction force proposed by Booij and van Wiechen~\cite{Booij1970} and Williams and coworkers~\cite{Manke1988,Dasbach1992} posits that the restoring force due to the dashpot is proportional to the relative velocity between adjacent beads projected along the tie-line joining these beads. The LRV approximation, however, assumes a more complicated dependence of the dashpot restoring force, and consequently results in erroneous predictions for the stress jump as indicated by~\citet{Manke1988}.  Furthermore, the LRV approximation predicts that the imaginary component of the complex viscosity, $\eta''$, vanishes at large values of internal friction parameter, $\varphi$, for all frequencies. ~\citet{Dasbach1992} showed this prediction to be in stark contrast to their semi-analytical approximation which predicts a limiting non-zero value for $\eta''$ as $\varphi\to\infty$. The LRV approximation was consequently established as being  incorrect~\cite{Booij1970,Manke1988,Dasbach1992} and its use was subsequently discarded. For the simplest case of a single-mode dumbbell-dashpot, it is straightforward to formulate the governing Fokker-Planck equation and obtain its equivalent stochastic differential equation. Both linear viscoelastic properties~\cite{Hua1995,Kailasham2018} and viscometric functions in steady-shear flow~\cite{Hua1996,Kailasham2018} have been calculated for this model using Brownian dynamics (BD) simulations, for arbitrary values of the internal friction parameter. The single-mode spring-dashpot model has also been solved using a Gaussian approximation for the distribution function~\cite{Schieber1993,Sureshkumar1995}. Upon comparison against exact BD simulation results, it was found that the Gaussian approximation (GA) offers accurate predictions of linear viscoelastic properties, but is unable to predict the shear-thickening of viscosity~\cite{Hua1995,Kailasham2018} predicted by the exact model. Furthermore, the predictions for the stress jump in the start up of shear flow,  obtained from BD simulations on the exact model, the Gaussian approximation, and the semi-analytical treatment of Manke and Williams agree with one another~\cite{Hua1995}. 

Fixman~\cite{Fixman1988} has shown that the effects of bond length and bond angle constraints in stiff polymer models may be sufficiently mimicked by a Rouse/Zimm-like chain with internal friction. A preaveraged form of the internal friction force was chosen for analytical tractability, and this simplified model yields predictions for equilibrium and linear viscoelastic properties,  such as the bond-vector correlations and storage and loss moduli, that are in reasonable agreement with that of stiff polymer chains. In the RIF model mention above, the standard continuum Rouse model is modified to include a rate-dependent dissipative force that resists changes in the curvature of the space-curve representing the polymer molecule. An expression for the autocorrelation of the end-to-end vector of the chain, and a closed-form expression for the {dynamic compliance} of the chain has been derived by them~\cite{Khatri2007rif}. The RIF model, and its variants~\cite{Cheng2013,Samanta2014,Samanta2016} which include excluded volume interactions and preaveraged hydrodynamic interactions, has been widely used to interpret the results of experiments and simulations on conformational transitions in biopolymers~\cite{Soranno201217800,ja211494h,Ameseder2018}, where the polymers do not experience a flow profile. To the best of our knowledge, an expression for the transient evolution of the mean-squared end-to-end vector of an RIF chain in flow is unknown, as are the viscometric functions. A thorough test of the accuracy of the preaveraging approximation, by comparing model predictions for observables at equilibrium and in flow, against exact BD simulations which account for fluctuations in internal friction, will be published as a separate study~\cite{ivfluc2021}.

There currently exists no methodology that is able to predict both linear viscoelastic properties and viscometric functions in shear flow for bead-spring-dashpot chains with arbitrary number of beads and magnitude of internal friction parameter. We address this deficiency by solving the bead-spring-dashpot chain problem exactly. We compare BD simulation results for the stress jump and complex viscosity against approximate analytical predictions given in Ref.~\citenum{Manke1988} and Ref.~\citenum{Dasbach1992}, respectively, and present steady-state results for viscometric functions in simple shear flow for the general case of $N_{\text{b}}>2$, for the first time.

\begin{figure*}[tbh]
    \centerline{
   {\includegraphics[width=120mm]{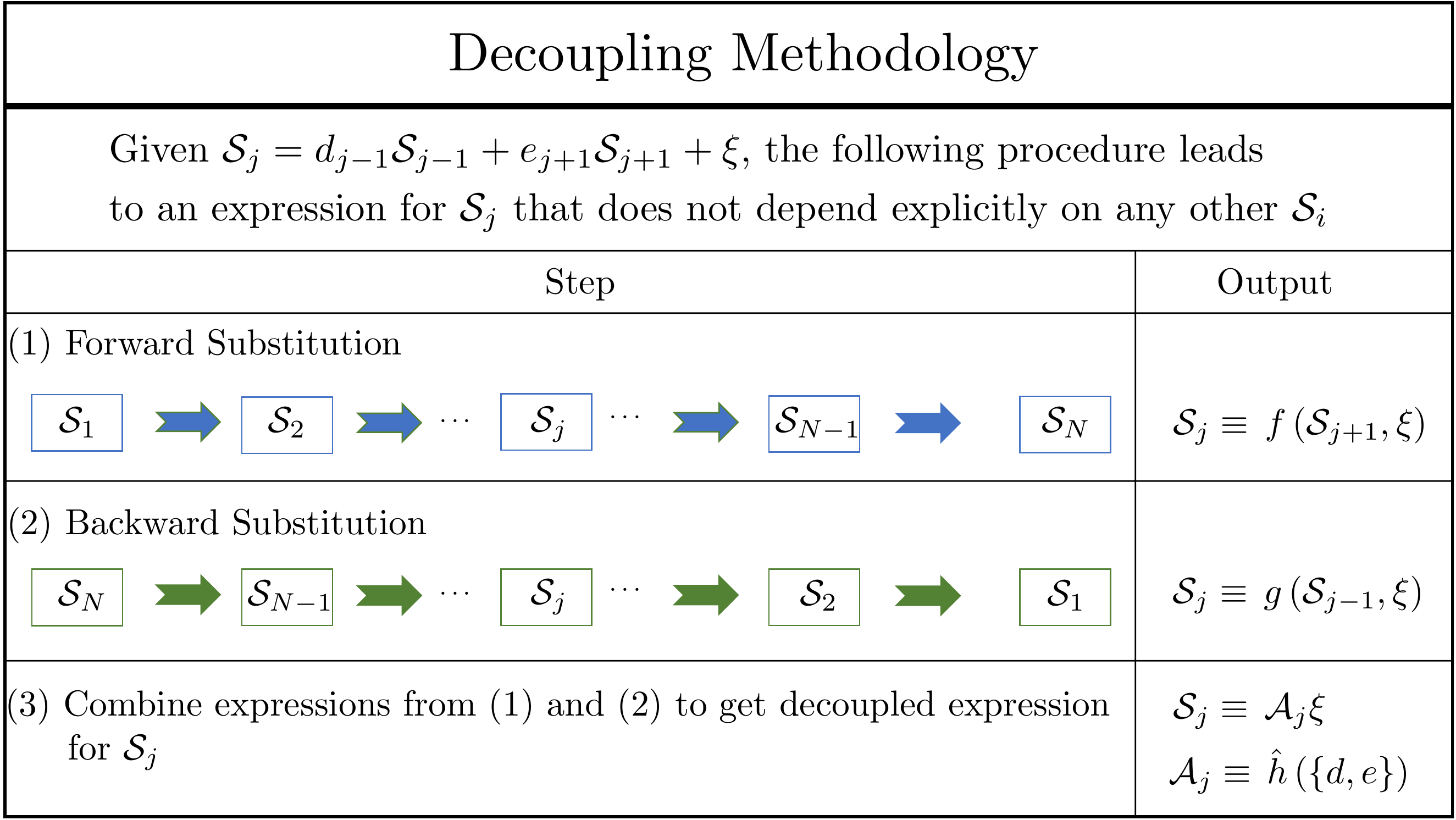}}     
    }
\caption{Schematic of three-step decoupling methodology introduced in Manke and Williams~\cite{Manke1988}.}
\label{fig:decoup_method}
\end{figure*}

A crucial step in our methodology is the decoupling of the connector vector velocities of neighboring beads. Stripping away all physical detail, the decoupling problem may be stated as follows: given a ``generating equation" for {some function ${\mathcal{S}}_j$ of $\{{X}_1,{X}_2,\dots,{X}_{N}\}$, which is of the form}
\begin{equation}
{\mathcal{S}}_j=d_{j-1}{\mathcal{S}}_{j-1}+e_{j+1}{\mathcal{S}}_{j+1}+\xi\left({X}_1,{X}_2,\dots,{X}_{N}\right)
\end{equation}
where $j\in[1,N]$, and $\left\{\xi,d_{j},e_{j}\right\}$ {are} some arbitrary functions of the ${X}_i$, is it possible to write an expression for ${\mathcal{S}}_j$ solely in terms of the ${X}_i$ that does not explicitly depend on any other ${\mathcal{S}}_i$? Manke and Williams~\cite{Manke1988} have proposed a three-step procedure for the solution of this problem. As the first step, the equation for ${\mathcal{S}}_{j}$ is successively substituted into ${\mathcal{S}}_{j+1}$, starting from $j=1$. At the end of this forward substitution step, an equation for ${\mathcal{S}}_{j}$ is obtained that only depends explicitly on ${\mathcal{S}}_{j+1}$ and ${X}_{i}$ with $1\leq\,i\leq\,j$. The second step is a backward substitution, where the equation for ${\mathcal{S}}_{j}$ is successively substituted into ${\mathcal{S}}_{j-1}$, starting from $j=N$. This results in an expression for ${\mathcal{S}}_{j}$ that only depends explicitly on ${\mathcal{S}}_{j-1}$ and ${X}_{i}$ with $j\leq\,i\leq\,N$. Finally, upon combining the results from the forward and backward substitution procedures, the decoupling procedure is completed, resulting in ${\mathcal{S}}_j=\mathcal{A}_j\xi\left({X}_1,{X}_2,\dots,{X}_{N}\right)$ where $\mathcal{A}_j$ is defined recursively in terms of $d$ and $e$. While the decoupling methodology developed by Manke and Williams has been adopted in the present work, we differ significantly in the generating equation which is subjected to the decoupling procedure. A schematic representation of the decoupling methodology is displayed in Fig.~\ref{fig:decoup_method}, {and the differences between the use of the methodology in the present work and that of Manke and Williams~\cite{Manke1988} is discussed briefly  in Sec.~\ref{sec:eqns}. In order to illustrate the use of the decoupling methodology, its application to a three-spring chain with internal friction is discussed in detail in Appendix~\ref{sec:app_a}. This simple case permits the demonstration of the essential aspects of the methodology, without excessive algebra.}

The rest of the paper is structured as follows. Sec.~\ref{sec:eqns} describes the bead-spring-dashpot chain model for a polymer, presents the governing stochastic differential equations and the stress tensor expression, and contains simulation details pertaining to the numerical integration of the governing equations. Sec.~\ref{sec:results}, which is a compilation of our results and the relevant discussion, is divided into three sections; Sec.~\ref{sec:code_valid} deals with code validation, Sec.~\ref{sec:osf} presents results for the complex viscosity calculated from oscillatory shear flow simulations, and Sec.~\ref{sec:steady_sr} contains results for steady shear viscometric functions,.
We conclude in Sec.~\ref{sec:conclusions}. Appendix~\ref{sec:app_a} contains the detailed steps showing the implementation of the decoupling algorithm represented in Fig.~\ref{fig:decoup_method} to a three-spring chain, and the derivation of the governing Fokker-Planck equation for the system. Additional details pertaining to the methodology developed in the present work and its implementation are provided in the Supplementary Material, and its various sections are referenced appropriately in the text below.

\section{\label{sec:eqns} Governing Equation and simulation details}

We consider $N_{\text{b}}$ massless beads, each of radius $a$, joined by $N\equiv\left(N_{\text{b}}-1\right)$ Hookean springs, with a dashpot in parallel with each spring, as shown in Fig.~\ref{fig:model}. The position of the $i^{\text{th}}$ bead is denoted as $\bm{r}_{i}$, and the connector vector joining adjacent beads is represented as $\bm{Q}_{i}\equiv\bm{r}_{i+1}-\bm{r}_{i}$. {The centre-of-mass of the chain is denoted by $\boldsymbol{r}_{\text{c}}\equiv\left(1/N_{\text{b}}\right)\sum_{\nu=1}^{N_{\text{b}}}\boldsymbol{r}_{\nu}$}. As will be seen shortly, in the absence of hydrodynamic interactions (HI), the inclusion of IV results in an explicit coupling of the connector vector velocities between nearest neighbors, and these velocities may be decoupled using the procedure suggested by Manke and Williams~\cite{Manke1988}. The simultaneous inclusion of fluctuating hydrodynamic interactions and internal friction, however, results in a one-to-all coupling of the connector vector velocities, which renders the problem intractable using the Manke and Williams approach~\cite{Manke1988}. It is noted, however, that in dumbbell models with internal friction, hydrodynamic interactions significantly magnify the stress jump and perceptibly affect the transient viscometric functions~\cite{Kailasham2018}. Coarse-grained polymer models which incorporate both fluctuating hydrodynamic interactions and internal friction are currently unsolved for the $N_{\text{b}}>2$ case. In this work, we restrict our attention to freely-draining bead-spring-dashpot chains. 

The chain, as shown in Fig.~\ref{fig:model}, is suspended in a Newtonian solvent of viscosity $\eta_{\text{s}}$ where the velocity $\bm{v}_{\text{f}}$ at any location $\bm{r}_{\text{f}}$ in the fluid is given by
$\bm{v}_{\text{f}}(\bm{r}_{\text{f}},t)\equiv\,\bm{v}_{0}+\boldsymbol{\kappa}(t)\cdot\bm{r}_{\text{f}}$, where $\bm{v}_{0}$ is a constant vector, and the transpose of the velocity gradient tensor is denoted as $\boldsymbol{\kappa}\equiv\left(\nabla\bm{v}_{\text{f}}\right)^{T}$. The chain is assumed to have completely equilibrated in momentum space, and its normalized configurational distribution function at any time $t$ is specified as $\Psi\equiv\Psi\left(\bm{r}_1,\bm{r}_2,...,\bm{r}_{{N_{\text{b}}}},t\right)=\left(1/\mathcal{Z}\right)\exp\left[-\phi/k_BT\right]$, where $\phi$ denotes the intramolecular potential energy stored in the springs joining the beads, $k_B$ is Boltzmann's constant, $T$ the absolute temperature, and $\mathcal{Z}=\int\exp\left[-\phi/k_BT\right]d\bm{Q}_1d\bm{Q}_2\dots\,d\bm{Q}_{N}$. {It is worth noting that the equilibrium configurational distribution function is unperturbed by the inclusion of internal friction}. The spring force in the $k^{\text{th}}$ connector vector is denoted as $\bm{F}^{\text{s}}_{k}=\partial \phi/\partial \bm{Q}_k$, with $\bm{F}^{\text{s}}_{k}=H\bm{Q}_k$ for a Hookean spring {where $H$ is the spring constant}. The expression for the IV force, $\bm{F}^{\text{IV}}_{k}$, in the $k^{\text{th}}$ connector vector may be written as 
\begin{equation}
\bm{F}^{\text{IV}}_{k}=K\left(\dfrac{\bm{Q}_k\bm{Q}_k}{Q_k^2}\right)\cdot\llbracket\dot{\bm{Q}}_k\rrbracket,
\end{equation}
{where $K$ is the damping coefficient of the dashpot}, and $\llbracket\dots\rrbracket$ denotes an average over momentum-space. Within the framework of polymer kinetic theory~\cite{Bird1987b}, the Fokker-Planck equation for the configurational distribution function is obtained by combining a force-balance on the beads (or connector vectors) with a continuity equation in probability space. The force-balance mandates that the sum of: (i) the internal friction force due to the dashpot, (ii) the restoring force from the Hookean spring, (iii) the random Brownian force arising from collisions with solvent molecules, and (iv) the hydrodynamic force which represents the solvent's resistance to the motion of the bead, equals zero. {It is convenient to work with connector vectors, rather than bead positions, for models with internal friction, and the following equations have been derived in Ref.~\citenum{ravibook} for the momentum-averaged velocity of the $k^{\text{th}}$ connector vector and the centre-of-mass, respectively,}
 \begin{align}\label{eq:qdot22}
 \llbracket\dot{\bm{Q}}_{k}\rrbracket&=\boldsymbol{\kappa}\cdot\bm{Q}_k-\dfrac{1}{\zeta}\sum^{N}_{l=1}\bm{{A}}_{kl}\cdot\Biggl(\uline{k_BT\dfrac{\partial \ln \Psi}{\partial \bm{Q}_l}}+\dashuline{\dfrac{\partial \phi}{\partial \bm{Q}_l}}\nonumber\\[5pt]
 &+K\dfrac{\bm{Q}_l\bm{Q}_l}{\bm{Q}^2_l}\cdot\llbracket\dot{\bm{Q}}_{l}\rrbracket\Biggr) \\
\llbracket\dot{\bm{r}}_{\text{c}}\rrbracket&=\bm{v}_{0}+\boldsymbol{\kappa}\cdot\bm{r}_{\text{c}}+\left(\dfrac{1}{N_{\text{b}}\zeta}\right)\sum^{N}_{k=1}\Biggl[k_BT\dfrac{\partial \ln\Psi}{\partial \bm{Q}_{k}}+\dfrac{\partial \phi}{\partial \bm{Q}_{k}} \nonumber\\[5pt]
& +K\dfrac{\bm{Q}_{k}\bm{Q}_{k}}{Q^2_{k}}\cdot\llbracket\dot{\bm{Q}}_{k}\rrbracket\Biggr]\end{align}
\noindent where $\zeta\coloneqq6\pi\eta_{\text{s}}a$ is the monomeric friction coefficient, and ${\bm{A}}_{kl}={A}_{kl}\bm{\delta}$ where ${A}_{kl}$ are the elements of the Rouse matrix, given by
 \begin{align}\label{eq:e_matrix_def}
A_{kl}= \left\{
\begin{array}{ll}
       2; &  k=l \\[15pt]
      -1; & |k-l|=1 \\[15pt]
       0 ; & \text{otherwise}.
\end{array} 
\right. 
\end{align}
As seen from Eq.~(\ref{eq:qdot22}), there is an explicit coupling between the velocity of the $k^{\text{th}}$ connector vector and its nearest neighbors which precludes a straightforward substitution into the equation of continuity for the configurational distribution function. This velocity-coupling may be removed by applying the decoupling scheme described in Fig.~\ref{fig:decoup_method}, to obtain the governing Fokker-Planck equation for the system, {as shown in Section~{SII} of the Supplementary Material. The crucial difference between the equation of motion for the connector vector used in the present work [Eq.~(\ref{eq:qdot22})], and that used by Manke and Williams can be understood by considering the solid and dashed underlined terms on the RHS of Eq.~(\ref{eq:qdot22}), which represent the Brownian and spring force contributions, respectively. Manke and Williams~\cite{Manke1988} were concerned only with the evaluation of the stress-jump, which occurs instantaneously upon the inception of flow. Since the Brownian and spring forces exactly balance each other at equilibrium, i.e., $k_BT\left(\partial \ln \Psi_{\text{eq}}\right/\partial \bm{Q}_{l})+ \left(\partial \phi/\partial\bm{Q}_l\right)=0$, Manke and Williams~\cite{Manke1988} ignore both these forces in their equation of motion. Here, however, we aim to find the exact governing equation that is valid both near and far away from equilibrium, and have consequently retained both the underlined terms in the force-balance equation. This is essentially the source of the difference between the two decoupling methodologies.}

Using $l_{H}\equiv\sqrt{k_BT/H}$ and $\lambda_{H}\equiv\zeta/4H$ as the length- and time-scales, respectively, and denoting dimensionless variables with an asterisk as superscript, the dimensionless form of the Fokker-Planck equation is given as
%\vspace{-20pt}
\begin{widetext}
\begin{align}\label{eq:fp_formal_dimless}
\dfrac{\partial \psi^{*}}{\partial t^{*}}&=-\sum_{j=1}^{N}\dfrac{\partial}{\partial \bm{Q}^{*}_j}\cdot\Biggl\{\Biggl[\boldsymbol{\kappa}^{*}\cdot\bm{Q}^{*}_j-\left(\dfrac{\varphi}{1+2\varphi}\right)\sum^{N}_{k=1}{\bm{U}}_{jk}\cdot\left(\boldsymbol{\kappa}^{*}\cdot\bm{Q}^{*}_k\right)-\dfrac{1}{4}\sum^{N}_{k=1}\widehat{\bm{A}}_{jk}\cdot\bm{F}^{*\text{s}}_{k}\\[5pt]
&-\dfrac{1}{4}\left(\dfrac{\varphi}{1+2\varphi}\right)\sum_{k=1}^{N}\dashuline{\dfrac{\partial}{\partial \bm{Q}^{*}_{k}}\cdot\bm{V}_{jk}^{T}}\Biggr]\psi^{*}\Biggr\}+\dfrac{1}{4}\sum^{N}_{j,k=1}\dfrac{\partial}{\partial \bm{Q}^{*}_j}\dfrac{\partial}{\partial \bm{Q}^{*}_{k}}:\left[\widehat{\bm{A}}_{jk}^{T}\psi^{*}\right]\nonumber
\end{align}
%\vspace{-5pt}
{where the dimensionless tensors $\bm{U}_{jk},\bm{V}_{jk},\text{and}\,\widehat{\boldsymbol{A}}_{jk}\equiv \boldsymbol{A}_{jk}-({\varphi}/{1+2\varphi})\boldsymbol{V}_{jk}$ are functions of the internal friction parameter, with definitions given in the Supplementary Material.} In the absence of internal friction, both $\bm{U}_{jk},\text{and}\,\bm{V}_{jk}$ reduce to $\bm{0}$, and $\widehat{\bm{A}}_{jk}$ becomes the Rouse matrix. {For the special case of a dumbbell ($N=1$), it may be shown (as detailed in Sec.~{SII}~{A} of the Supplementary Material) that Eq.~(\ref{eq:fp_formal_dimless}) is identical to the Fokker-Planck equations obtained in previous studies~\cite{Hua1995,Kailasham2018} on dumbbells with internal friction}. In order to simplify the notation, it is convenient to rewrite the Fokker-Planck equation in terms of collective coordinates. We define
\begin{equation}\label{eq:coll_coord}
\widetilde{\bm{Q}}^{*} \equiv \left[\bm{Q}^{*}_1,\,\bm{Q}^{*}_2,\,...,\,\bm{Q}^{*}_N\right]
\equiv\left[Q^{*1}_1,Q^{*2}_1,Q^{*3}_1,Q^{*1}_2,Q^{*2}_2,\,...,\,Q^{*3}_{N}\right]
\end{equation}
and write $\widetilde{Q}^{*}_{i}=Q^{*\beta}_{k}$, where $k=1,2,...,N$ and $\beta=1,2,3$ (represent Cartesian components in the $x,y,z$ directions, respectively), with $i$ related to $k$ and $\beta$ as $i=3\left(k-1\right)+\beta$. Similarly, $\widetilde{\bm{F}}^{*\text{s}}\equiv\left[\bm{F}^{*\text{s}}_{1},\,\bm{F}^{*\text{s}}_{2},\,...,\,\bm{F}^{*\text{s}}_{N}\right]$, and $\widetilde{\bm{v}}^{*}\equiv\left[\widehat{\bm{v}}^{*}_{1},\,\widehat{\bm{v}}^{*}_{2}\,...,\,\widehat{\bm{v}}^{*}_{N}\right]$, with $\widehat{\bm{v}}^{*}_{j}=\sum_{k=1}^{N}\left({\partial}/{\partial \bm{Q}^{*}_{k}}\right)\cdot\bm{V}_{jk}^{T}$. {The dimensionless $N\times N$ block diffusion matrix $\bm{\mathcal{D}}$ is a function of the internal friction parameter, whose elements are the tensors $\widehat{\bm{A}}_{jk}$}. The block matrix $\bm{\mathcal{K}}^{*}$ is defined such that its diagonal elements are given by the $3\times3$ matrix $\boldsymbol{\kappa}^{*}$, and its off-diagonal blocks are $\bm{0}$. Lastly, the block matrix $\bm{\mathcal{U}}$ consists of the tensors $\bm{U}_{jk}$. In terms of these collective variables, the stochastic differential equation equivalent to Eq.~(\ref{eq:fp_formal_dimless}), using It\^o's interpretation~\cite{Ottinger1996}, is given by
%\begin{widetext}
\begin{equation}\label{eq:sde_collective_dimless}
\begin{split}
d\widetilde{\bm{Q}}^{*}&=\Biggl[\bm{\mathcal{K}}^{*}\cdot\widetilde{\bm{Q}}^{*}-\left(\dfrac{\varphi}{1+2\varphi}\right){\bm{\mathcal{U}}}\cdot\left(\bm{\mathcal{K}}^{*}\cdot\widetilde{\bm{Q}}^{*}\right)-\dfrac{1}{4}\bm{\mathcal{D}}\tb{(\varphi)}\cdot\widetilde{\bm{F}}^{*\text{s}}-\dfrac{1}{4}\left(\dfrac{\varphi}{1+2\varphi}\right)\widetilde{\bm{v}}^{*}\Biggr]dt^{*}+{\dfrac{1}{\sqrt{2}}}\bm{\mathcal{B}}\tb{(\varphi)}\cdot d\bm{\widetilde{W}}^{*}
\end{split}
\end{equation}
where $\bm{\widetilde{W}}^{*}$ is a $3N-$dimensional Wiener process, and $\bm{\mathcal{B}}\cdot\bm{\mathcal{B}}^{T}=\bm{\mathcal{D}}$. The symmetricity and positive-definiteness of the diffusion matrix is established empirically in Section~{SII}~{C} of the Supplementary Material. The square-root of the diffusion matrix is found using Cholesky decomposition~\cite{press2007numerical}. {While a higher-order semi-implicit predictor-corrector algorithm for the solution of dumbbells with internal friction exists in the literature~\cite{Kailasham2018}, the complexity of the governing equations for the general case of $N_{\text{b}}>2$ precludes the construction of a higher order solver algorithm, and consequently,} Eq.~(\ref{eq:sde_collective_dimless}) is solved numerically using a simple explicit Euler method, as follows
\begin{equation}\label{eq:exp_eul_dimless}
\begin{split}
\widetilde{\bm{Q}}^{*}\left(t_{n+1}\right)=&\widetilde{\bm{Q}}^{*}_{n}+\Biggl[\bm{\mathcal{K}}^{*}\left(t_{n}\right)\cdot\widetilde{\bm{Q}}^{*}\left(t_{n}\right)-\left(\dfrac{\varphi}{1+2\varphi}\right){\bm{\mathcal{U}}}\left(t_{n}\right)\cdot\left(\bm{\mathcal{K}}^{*}\left(t_{n}\right)\cdot\widetilde{\bm{Q}}^{*}\left(t_{n}\right)\right)-\dfrac{1}{4}\bm{\mathcal{D}}\left(t_{n}\right)\cdot\widetilde{\bm{F}}^{*\,\text{s}}\left(t_{n}\right)\\[5pt]
&-\dashuline{\left(\dfrac{1}{4}\right)\left(\dfrac{\varphi}{1+2\varphi}\right)\widetilde{\bm{v}}^{*}\left(t_{n}\right)}\Biggr]\Delta t^{*}_{n}+ {\dfrac{1}{\sqrt{2}}}\Delta \bm{\widetilde{S}}^{*}_{n}
\end{split}
\end{equation}
where $\Delta \bm{\widetilde{S}}^{*}_{n}=\bm{\mathcal{B}}\cdot\Delta \bm{\widetilde{W}}^{*}_n$. 

{It is observed from Eq.~(\ref{eq:sde_collective_dimless}) that the inclusion of internal friction modifies both the drift and the diffusion (noise) components of the governing stochastic differential equation. Since the Hamiltonian of the system is unperturbed by internal friction, the equilibrium distribution function must be a solution to the Fokker-Planck equation [Eq.~(\ref{eq:fp_formal_dimless})] in the absence of the flow term, in order to satisfy the fluctuation-dissipation theorem.} {We have established (as detailed in Sec.~{SII}~{B} of the Supplementary Material) that the governing stochastic differential equation [Eq.~(\ref{eq:sde_collective_dimless})] satisfies the fluctuation-dissipation theorem by checking that the probability distribution of the end-to-end vector for a five-bead Rouse chain with internal friction at equilibrium, obtained by numerically integrating Eq.~(\ref{eq:exp_eul_dimless}) with $\bm{\mathcal{K}}^{*}=\bm{0}$, agrees with the analytical expression.}  

In order to relate the time-evolution of the connector vectors to macroscopically observable rheological properties, it is necessary to specify an appropriate stress tensor expression for the model discussed above. The formal, thermodynamically consistent stress tensor expression for free-draining models with internal friction may be obtained using the Giesekus expression~\cite{Schieber1994}, as follows
\begin{equation}\label{eq:giesekus_exp}
\begin{split}
{\boldsymbol{\tau}_{\text{p}}}&=\dfrac{n_{\text{p}}\zeta}{2}\left<\sum_{j=1}^{N}\sum_{k=1}^{N}{\mathscr{C}_{jk}}\bm{Q}_{j}\bm{Q}_{k}\right>_{(1)}=\dfrac{n_{\text{p}}\zeta}{2}\left[\dfrac{d}{dt}\left<\sum_{j,k}{\mathscr{C}_{jk}}\bm{Q}_{j}\bm{Q}_{k}\right>-\boldsymbol{\kappa}\cdot\left<\sum_{j,k}{\mathscr{C}_{jk}}\bm{Q}_{j}\bm{Q}_{k}\right>-\left<\sum_{j,k}{\mathscr{C}_{jk}}\bm{Q}_{j}\bm{Q}_{k}\right>\cdot\boldsymbol{\kappa}^{T}\right]
\end{split}
\end{equation}
where $\mathscr{C}_{jk}$ is the Kramers matrix~\cite{Bird1987b}. Upon simplification {following considerable algebra, as discussed in Sec.~{SIII} of the Supplementary Material,}
\begin{equation}\label{eq:formal_stress_tensor}
\begin{split}
{\boldsymbol{\tau}_{\text{p}}}&=n_{\text{p}}k_BT\left(N_{\text{b}}-1\right)\boldsymbol{\delta}-n_{\text{p}}\left<\sum_{k=1}^{N_{\text{b}}-1}\bm{Q}_k\bm{F}^{\text{c}}_{k}\right>
\end{split}
\end{equation}
which is formally similar to the Kramers expression~\cite{Bird1987b}, except that the force in the connector vector, $\bm{F}_{k}^{c}$, is redefined  to include contributions from both the spring and the dashpot (also noted in Refs.~[\citenum{Wedgewood1993,Schieber1994,Hua1996}] for Hookean dumbbells with IV), as follows, 
\begin{equation}\label{eq:force_def}
\begin{split}
\bm{F}^{\text{c}}_{k}&=\bm{F}^{\text{s}}_{k}+KC_k\bm{Q}_{k}
\end{split}
\end{equation}
where $C_k=\left({\bm{Q}_k\cdot\llbracket\dot{\bm{Q}}_{k}\rrbracket}/{Q^2_{k}}\right)$. Plugging Eq.~(\ref{eq:force_def}) into Eq.~(\ref{eq:formal_stress_tensor}) and using the closed-form expression for $C_k$ as {derived in the Supplementary Material}, the dimensionless stress tensor expression is obtained as 
\begin{equation}\label{eq:stress_tensor_dimless}
\begin{split}
\dfrac{\boldsymbol{\tau}_{\text{p}}}{n_{\text{p}}k_BT}&=\left(N_{\text{b}}-1\right)\boldsymbol{\delta}-\Biggl[\left<\sum_{k}\bm{Q}^{*}_k\bm{F}^{*\text{s}}_{k}\right>-\dfrac{1}{2}\left(\dfrac{\epsilon}{1+\epsilon}\right)\left<\sum_{k,l}\left(\bm{Q}^{*}_{k}\bm{F}^{*\text{s}}_{l}\right)\cdot\boldsymbol{\mu}_{kl}^{T}\right>\Biggr]-\dfrac{1}{2}\left(\dfrac{\epsilon}{1+\epsilon}\right)\Biggl[\left<\sum_{k}\boldsymbol{\mu}^{T}_{kk}\right>\\[15pt]
&+\left<\sum_{k,l}\bm{Q}^{*}_{k}\dotuline{\dfrac{\partial}{\partial \bm{Q}^{*}_l}\cdot\boldsymbol{\mu}_{kl}^{T}}\right>\Biggr]-\left(\dfrac{2\epsilon}{1+\epsilon}\right)\boldsymbol{\kappa}^{*}:\left<\sum_{k,l}\dfrac{\chi^{(k)}_{l}\bm{Q}^{*}_l\bm{Q}^{*}_l\bm{Q}^{*}_k\bm{Q}^{*}_k}{Q^{*}_lQ^{*}_k}\right>
\end{split}
\end{equation}
\end{widetext}
where $\epsilon=2\varphi$, and the definitions of $\boldsymbol{\mu}_{kl}\,\text{and}\,\chi^{(k)}_{l}$ are provided in {the Supplementary Material}.

The bead-spring-dashpot chain is subjected to steady simple shear flow and small amplitude oscillatory shear flow. The flow tensor, $\boldsymbol{\kappa}$, for simple shear flow has the following form
\begin{equation}\label{eq:kappa_shear_def}
\boldsymbol{\kappa} \equiv \boldsymbol{\kappa}^{*}\,\lambda^{-1}_{H}=   \dot{\gamma}
\begin{pmatrix}
0& 1 & {0}\\
0& 0& 0\\
{0} & 0 & 0 
\end{pmatrix}
\end{equation}
and is characterized by the following viscometric functions
\begin{align}\label{eq:visc_fun_def}
\begin{split}
\eta_{\text{p}}&=-\dfrac{\tau_{\text{p},xy}}{\dot{\gamma}}\\[5pt]
\Psi_1&=-\left[\dfrac{\tau_{\text{p},xx}-\tau_{\text{p},yy}}{\dot{\gamma}^2}\right]\\[5pt]
\Psi_2&=-\left[\dfrac{\tau_{\text{p},yy}-\tau_{\text{p},zz}}{\dot{\gamma}^2}\right]
\end{split}
\end{align}
where $\tau_{\text{p},xy}$ refers to the $xy$ element of the stress tensor, and $\eta_{\text{p}}$, $\Psi_1$, and $\Psi_2$ denote the shear viscosity, the first normal stress coefficient, and the second normal stress coefficient, respectively.
For small-amplitude oscillatory shear flow, we have
\begin{equation}\label{eq:osf_flow_def}
\boldsymbol{\kappa} = \dot{\gamma}_0\cos(\omega t)
\begin{pmatrix}
0& 1 & {0}\\
0& 0& 0\\
{0} & 0 & 0 
\end{pmatrix}
\end{equation}
The material functions relevant to this flow profile, $\eta'(\omega)$ and $\eta''(\omega)$, are given by
\begin{equation}\label{eq:fit_osf}
-\tau_{\text{p},xy}=\eta'(\omega)\dot{\gamma}_0\cos(\omega t)+\eta''(\omega)\dot{\gamma}_0\sin(\omega t)
\end{equation}
{The} variance reduction {algorithm proposed by~\citet{Wagner1997}} has been used in the evaluation of steady-shear viscometric functions at low shear rates ($\lambda_{H}\dot{\gamma}<1.0$), and for the calculation of oscillatory shear material functions at all frequencies reported in this work. {Additional details pertaining to the variance reduction algorithm employed in the present work, and examples illustrating the efficacy of this technique have been provided in Section~{SIV} Supplementary Material.}

Where appropriate, the viscometric functions have either been scaled by their respective Rouse chain values, $\eta^{\text{R}}$ and $\Psi^{\text{R}}_{1}$, in steady shear flow, given as~\cite{Bird1987b}
\begin{flalign}\label{eq:eta_rouse}
&\eta^{\text{R}}=n_{\text{p}}k_BT\lambda_{H}\left[\dfrac{N_{\text{b}}^2-1}{3}\right]
\end{flalign}
\begin{flalign}\label{eq:psi1_rouse}
&\Psi^{\text{R}}_{1}=2n_{\text{p}}k_BT\lambda^2_{H}\left[\dfrac{(N_{\text{b}}^2-1)(2N_{\text{b}}^2+7)}{45}\right]
\end{flalign}
or by the Rouse values of the real and imaginary portions of the complex viscosity given by~\cite{Bird1987b}
\begin{align}\label{eq:eta_prime_rouse}
{\left(\eta'\right)^{\text{R}}}&=n_{\text{p}}k_BT\sum_{j=1}^{N}\dfrac{\lambda_{j}}{1+\left(\lambda_{j}\omega\right)^2}
\end{align}
\begin{align}\label{eq:eta_dob_prime_rouse}
{\left(\eta''\right)^{\text{R}}}&=n_{\text{p}}k_BT\sum_{j=1}^{N}\dfrac{\lambda^2_{j}\omega}{1+\left(\lambda_{j}\omega\right)^2}
\end{align}
where $\lambda_{j}=2\lambda_{H}/a_{j}$, and $a_j=4\sin^2\left[j\pi/2N_{\text{b}}\right]$ are the eigenvalues of the Rouse matrix. Note that the dynamic viscosity, $\eta'$, has the solvent viscosity contribution subtracted off and the convention is followed throughout this paper. Shear rates and angular frequencies are scaled using $\lambda_{\text{p}}\equiv\left(\eta^{\text{R}}/n_{\text{p}}k_BT\right)$, which is the characteristic relaxation time defined using the Rouse viscosity. 

%%%%% simulation details %%%%%%%%%%%%%%%%%%%%%%%%%%%%%%%%%%%%
{The initial configurations of the chains are drawn from an ensemble of Rouse chain configurations, since it is known that the equilibrium configurational probability distribution is unaffected by internal friction. The timestep width of the integrator is chosen after testing for convergence, by checking that the simulation output is independent of the choice of $\Delta t^{*}$, as explained in detail in Section~{SII}~{D} of the Supplementary Material. For example, at low values of the dimensionless shear rate or frequency$(<1.0)$, $\Delta t^{*}=10^{-2}$ is found to suffice, while higher shear rates or frequency necessitate the use of smaller time steps in the range of $10^{-5}$ to $10^{-3}$. The ensemble size for oscillatory shear simulations is chosen in the range of $\mathcal{O}(10^3)$ to $\mathcal{O}(10^5)$ trajectories, with the lower limit corresponding to simulations at smaller frequencies. Components of the complex viscosity are extracted by fitting Eq.~(\ref{eq:fit_osf}) to three to five cycles of the observed stress, with the duration of a cycle given by $t^{*}_{\text{cycle}}=2\pi/\omega^{*}$. The ensemble size for steady-shear simulations is in the range of $\mathcal{O}(10^4)$ to $\mathcal{O}(10^5)$ trajectories.}
%%%%%%%%%%%%%%%%%%%%%%%%%%%%%%%%%%%%%%%%%%%%%%%%%%

The underlined \tb{divergence} terms in Eqs.~(\ref{eq:fp_formal_dimless}) and (\ref{eq:stress_tensor_dimless}) may be calculated by two routes: analytically, using recursive functions as explained in Sec.~{SV} of the Supplementary Material, or they can be calculated numerically. The connector vectors appearing in Eqs.~(\ref{eq:num_div_defn})\textendash(\ref{eq:num_grad_defn}) below are in their dimensionless form, with the asterisks omitted for the sake of clarity. The numerical route for the calculation of divergence is described below. Consider the general divergence, 
\begin{equation}\label{eq:num_div_defn}
\dfrac{\partial}{\partial \bm{Q}_{k}}\cdot\bm{\mathcal{G}}_{jk}=\sum_{\beta=1}^{3}\sum_{\gamma=1}^{3}\dfrac{\partial}{\partial {Q}^{\beta}_{k}}\left(\mathcal{G}^{\beta \gamma}_{jk}\right)\bm{e}_{\gamma}
\end{equation}
where $\beta$ and $\gamma$ run over the three Cartesian indices, $\bm{\mathcal{G}}_{jk}$ is a configuration-dependent tensor, and $\bm{e}_{\gamma}$ is a unit vector. The computation of the divergence requires the calculation of nine gradient terms, which are evaluated using the central-difference approximation. One such evaluation is shown here as an example: 
\begin{equation}\label{eq:num_grad_defn}
\begin{split}
\dfrac{\partial}{\partial {Q}^{1}_{k}}\left(\mathcal{G}^{12}_{jk}\right)&=\dfrac{1}{2\Delta_1}\Biggl[\mathcal{G}^{12}_{jk}\left(Q_k^{1}+\Delta_1,Q^{2}_k,Q^{3}_k\right)\\
&-\mathcal{G}^{12}_{jk}\left(Q_k^{1}-\Delta_1,Q^{2}_k,Q^{3}_k\right)\Biggr]
\end{split}
\end{equation}
where $\Delta_1$ is the spatial discretization width along one Cartesian direction, representing the infinitesimal change in $Q_k^{1}$. The error in the evaluation of the gradient using this approximation scales as $\mathcal{O}\left(\Delta^2_1\right)$. We have validated that the divergences calculated numerically agrees with that obtained using recursive functions, and have chosen the numerical route in view of its faster execution time that is largely invariant with chain length. In numerical computations, we set $\Delta_1=\Delta_2=\Delta_3=\Delta_{\text{d}}=10^{-5}$, unless noted otherwise. 

\section{\label{sec:results} Results}
\subsection{\label{sec:code_valid} Code Validation}
\begin{figure}[t]
\centering
\includegraphics[width=80mm]{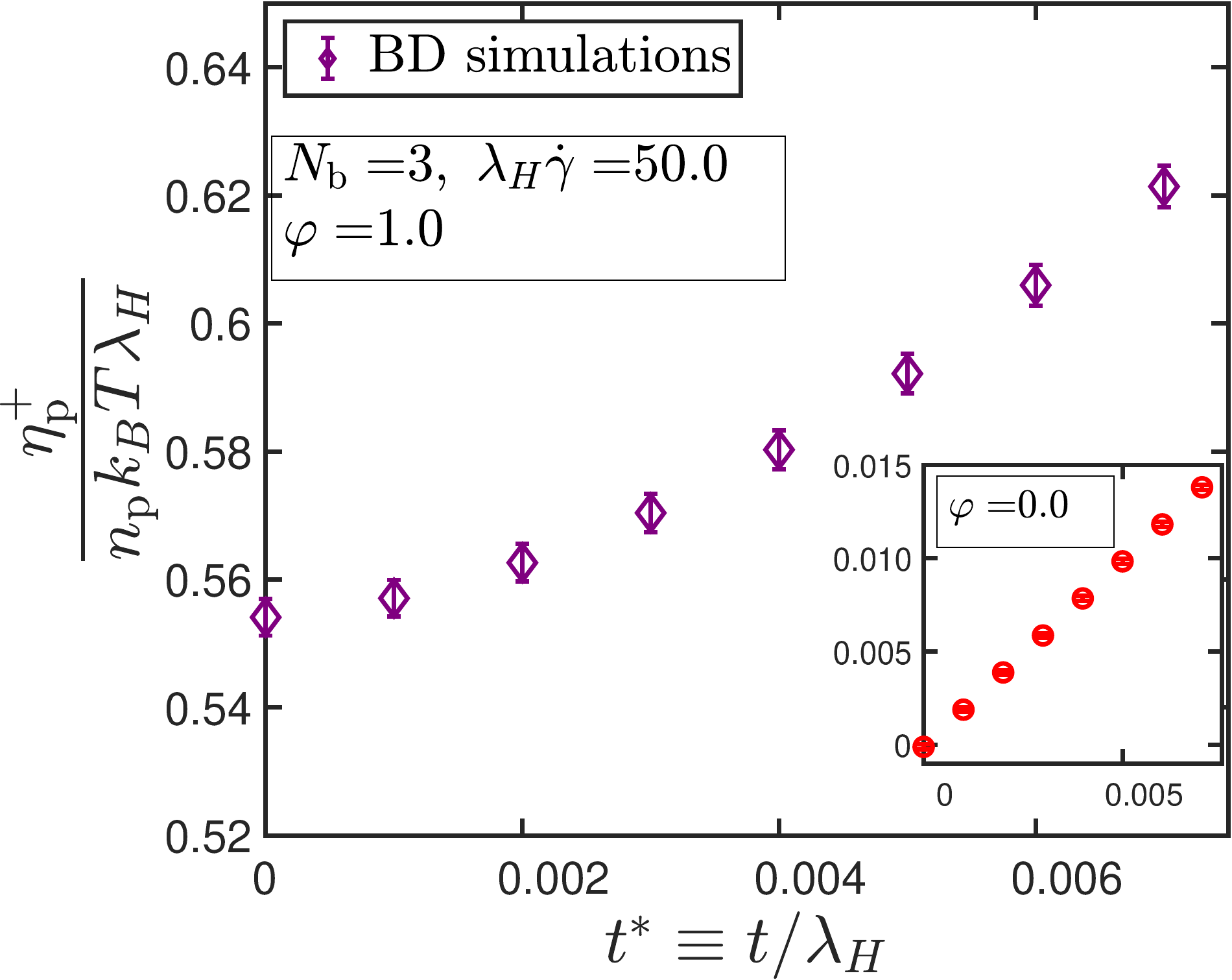}  
\caption{Time evolution of the shear viscosity of a three-bead Rouse chain with internal friction subjected to simple shear flow. The inset corresponds to the case without internal friction.}
\label{fig:sjump-proc}
\end{figure}

{In Fig.~\ref{fig:sjump-proc}, the transient viscosity of a three-bead Rouse chain with internal friction subjected to simple shear flow is plotted as a function of dimensionless time. An important rheological consequence of internal friction is the appearance of a discontinuous jump in viscosity at the inception of flow~\cite{Manke1988,Hua1995,Kailasham2018}. This phenomenon, called ``stress jump" is not predicted by bead-spring-chain models without internal friction, as evident from the inset. The numerical value of the stress jump is taken to be the shear viscosity at the startup $(t^{*}=0)$ of simple shear flow.}

\begin{figure}[t]
\centering
\includegraphics[width=80mm]{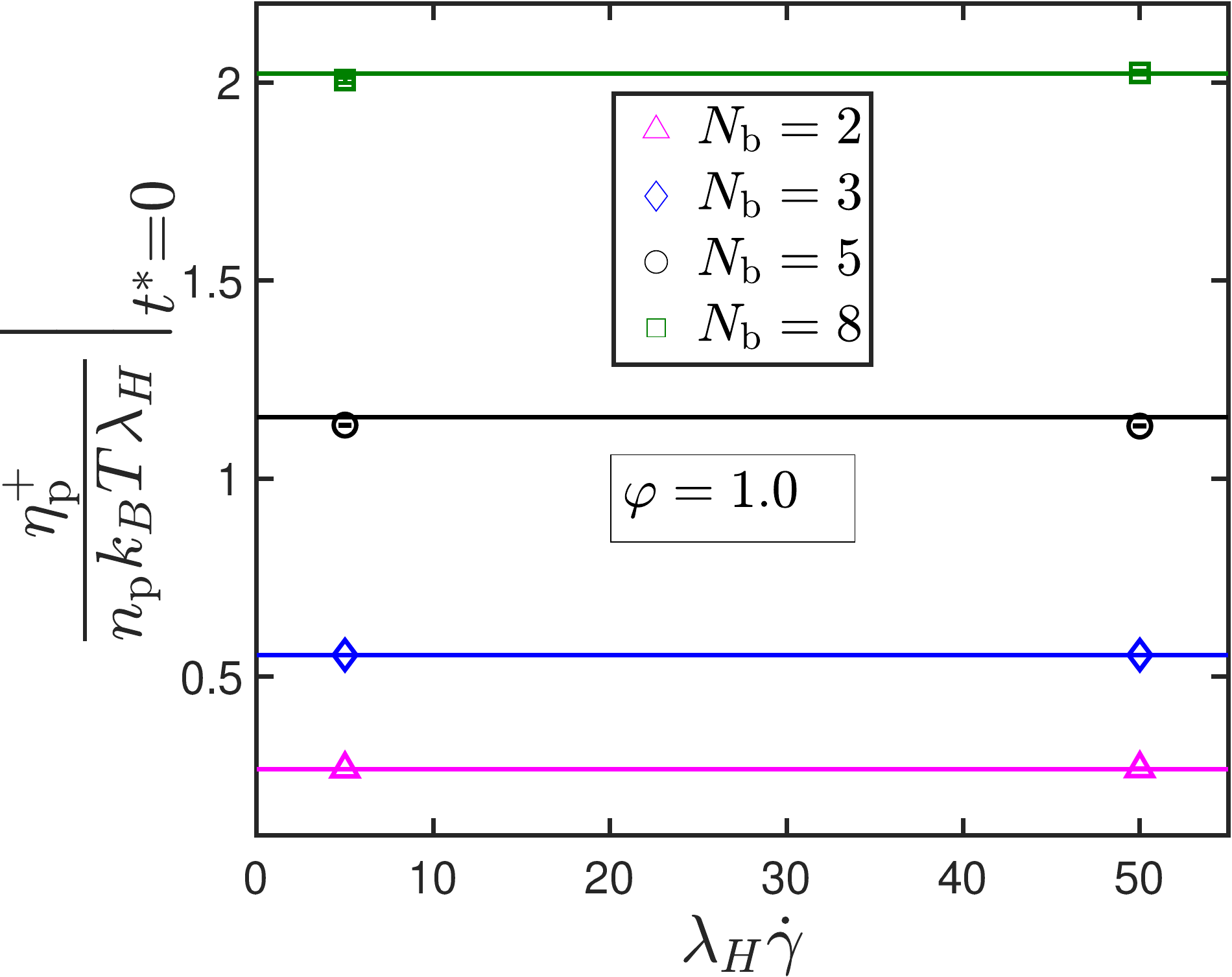}  
\caption{Stress jump as a function of dimensionless shear rate for various chain lengths and a fixed internal friction parameter of $\varphi=1.0$. Lines are approximate solutions by Manke and Williams~\cite{Manke1988}. Error bars are smaller than symbol size.}
\label{fig:sr_indep}
\end{figure}

{In Fig.~\ref{fig:sr_indep}, the stress-jump observed for different chain lengths is plotted as a function of the dimensionless shear rate}. It is observed that the stress jump is independent of the shear rate, in agreement with the theoretically expected trend~\cite{Gerhardt1994}. The horizontal lines in the figure represent the approximate analytical values for the stress jump evaluated by Manke and Williams~\cite{Manke1988}, and very good agreement is observed between the values estimated using the two approaches. 

\begin{figure}[t]
\begin{center}
\begin{tabular}{c}
\includegraphics[width=0.9\linewidth]{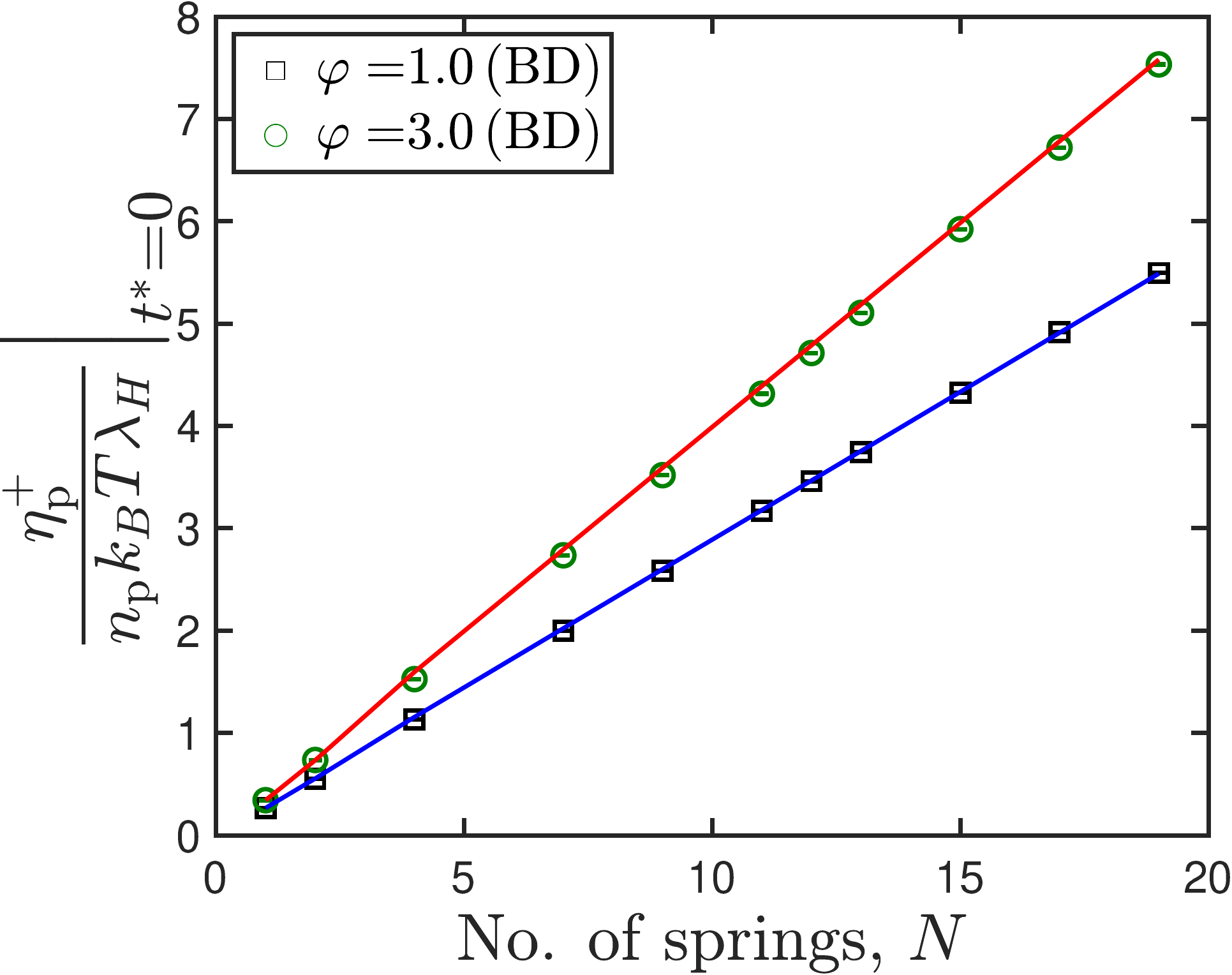}\\[5pt]
(a) \\[10pt]
\includegraphics[width=0.9\linewidth]{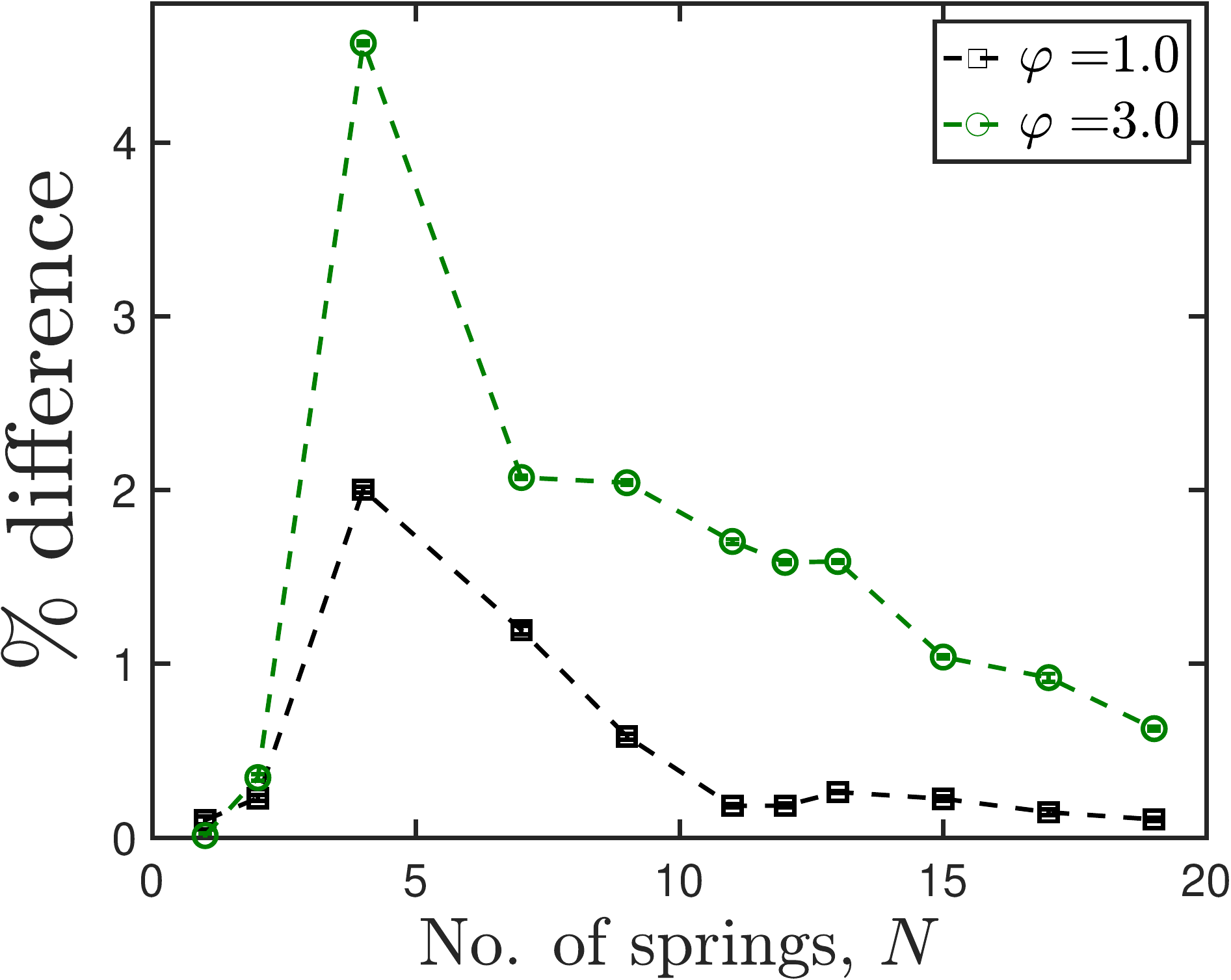}\\[5pt]
(b)  \\[5pt]
\end{tabular}
\end{center}
\caption{\small (a) Comparison of stress jump obtained using extrapolation of BD simulation data, and approximate solutions by Manke and Williams~\cite{Manke1988} indicated by solid lines. (b) Percentage difference between stress jump obtained using the two approaches, as a function of chain length. Dashed lines are drawn to guide the eye. Error bars are smaller than symbol size.}
\label{fig:mw_validation}
\end{figure}

In Fig.~\ref{fig:mw_validation}~(a), the stress jump evaluated from BD simulations for two different values of the internal friction parameter is plotted as a function of the number of springs in the chain. The semi-analytical approximation of Manke and Williams~\cite{Manke1988} is found to compare favourably against the exact simulation result. Furthermore, it is observed that the stress jump scales linearly with number of springs in the chain, with the slope of the line dependent on the internal friction parameter. It is instructive to first understand the simplifying {assumption} made in Manke and Williams~\cite{Manke1988} before interpreting the data in Fig.~\ref{fig:mw_validation}~(b) where the percentage difference between the analytical and simulation results is plotted as a function of chain length at a fixed value of the internal friction parameter. {Manke and Williams~\cite{Manke1988} assume that the terminal connector vectors and interior connector vectors contribute equally towards the stress jump.} This assumption is necessary only for chains with $N>2$, because there is no distinction between a terminal and interior connector vector for a dumbbell ($N=1$), and the two connector vectors for the $N=2$ case are shown by Manke and Williams~\cite{Manke1988} to contribute identically to the total stress jump. {It is likely that their assumption would most severely be tested in chains with fewer number of springs, where the terminal springs represent a larger fraction of the overall chain, and becomes progressively better with an increase in the number of springs.} The expected trend is clearly borne out by Fig.~\ref{fig:mw_validation}~(b), where the deviation between the exact simulation result and the analytical value first increases (beyond $N=1$) and later decreases with the number of springs in the chain.

\subsection{\label{sec:osf} Complex viscosity from oscillatory shear flow}

\begin{figure*}[t]
\begin{center}
\begin{tabular}{cc}
\includegraphics[width=0.5\linewidth]{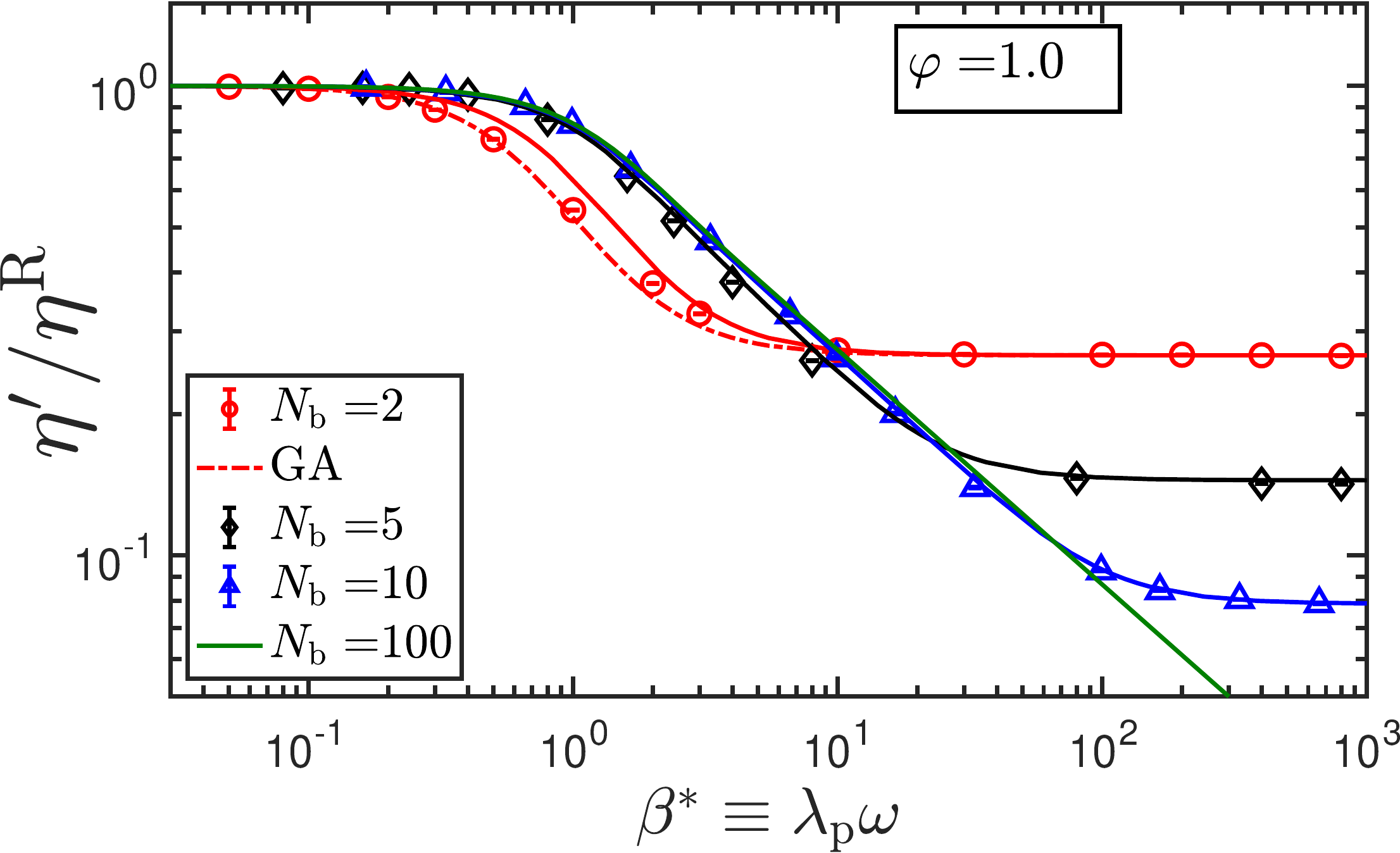}&
\includegraphics[width=0.5\linewidth]{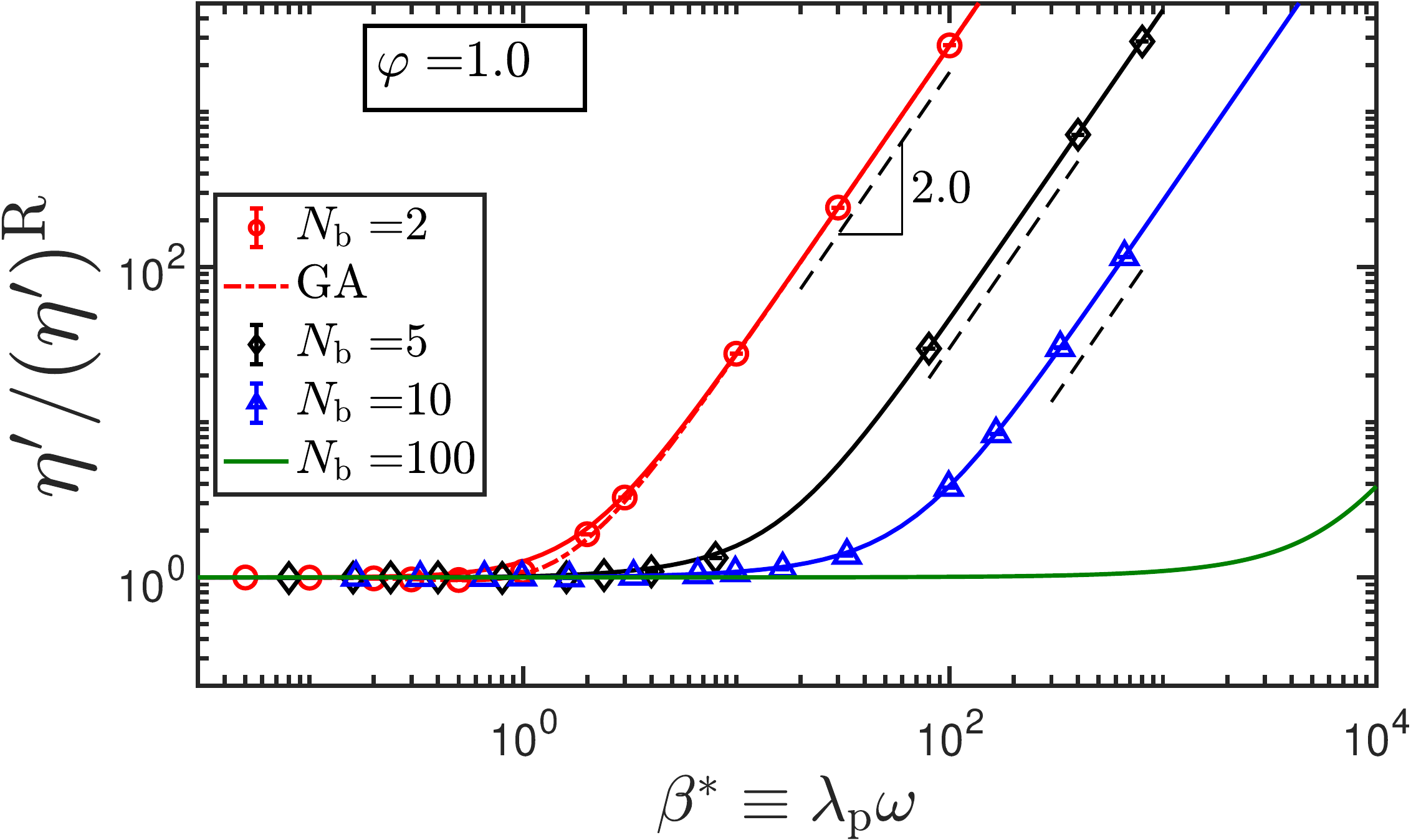}\\
(a)&(b)\\[5pt]
\includegraphics[width=0.5\linewidth]{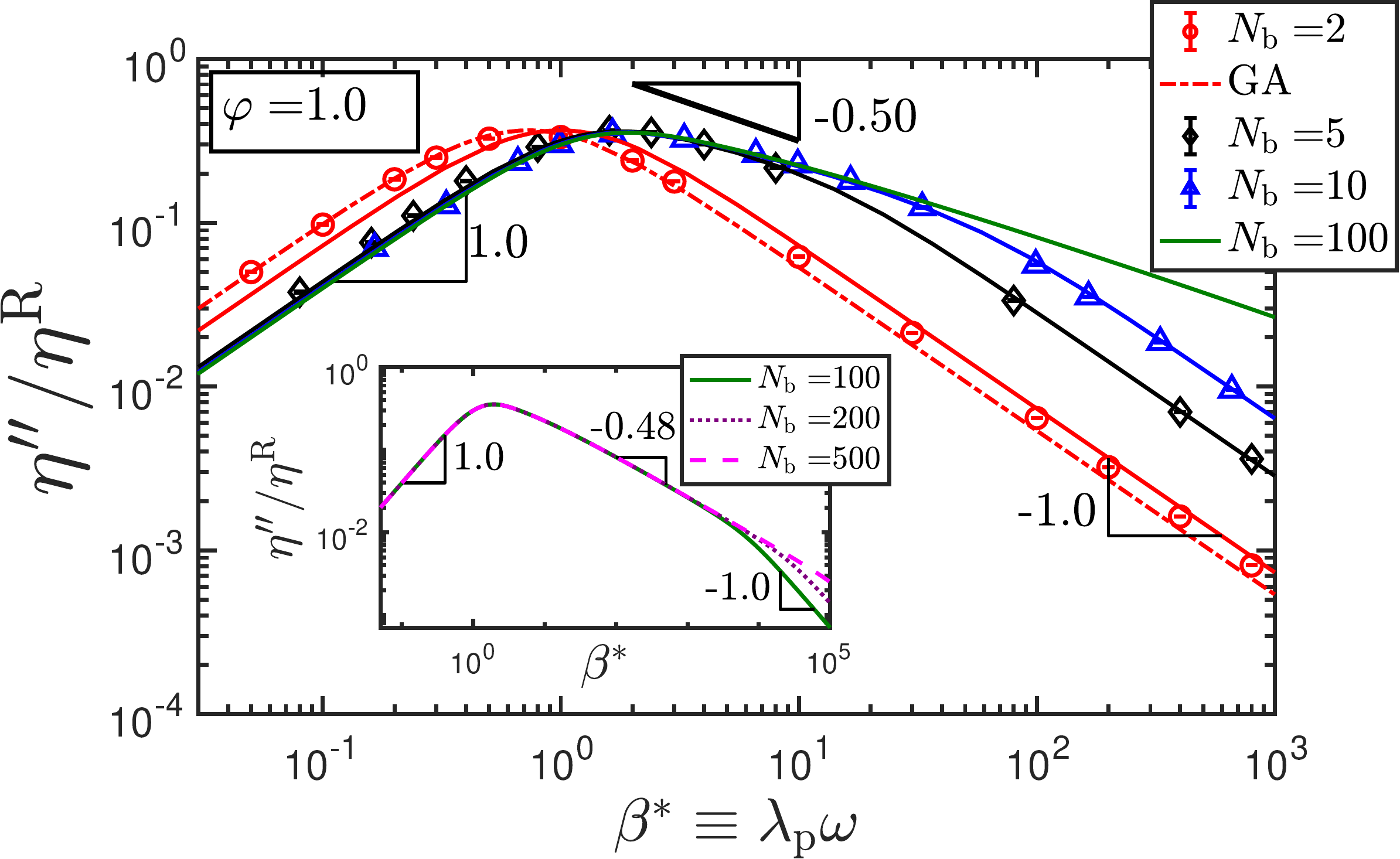}&
\includegraphics[width=0.5\linewidth]{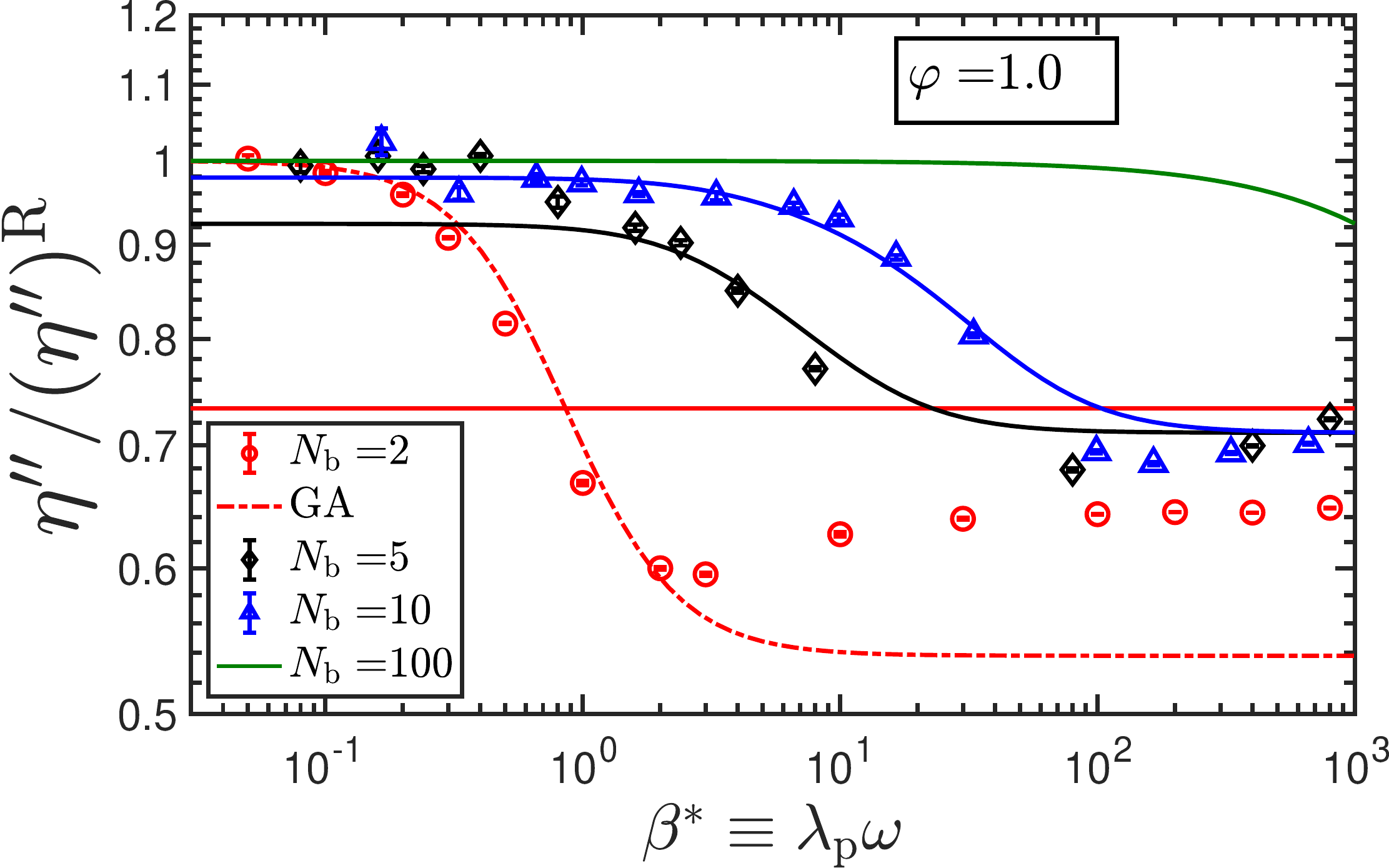}\\
(c)&(d)
\end{tabular}
\end{center}
\caption{\small Plots of the real ($\eta'$) and imaginary components ($\eta''$) of the complex viscosity, as a function of the scaled frequency, for a fixed value of the internal friction parameter and three different values for the number of beads in the chain. The solid lines are approximate solutions given in ~\citet{Dasbach1992}. The dash-dotted lines represent predictions obtained using the Gaussian approximation~\cite{Schieber1993}. {The lines in the inset of (c) are DMW results.} Error bars are smaller than symbol size.}
\label{fig:saos_fixed_iv}
\end{figure*}

In Fig.~\ref{fig:saos_fixed_iv}, the material functions in oscillatory shear flow are plotted for a fixed value of the internal friction parameter, and varying number of beads in the chain. The exact BD simulation results, indicated by symbols, are compared against the the approximate prediction given by \citet{Dasbach1992} (DMW), shown as solid lines. Schieber~\cite{Schieber1993} has obtained predictions for $\eta'$ and $\eta{''}$ for Hookean dumbbells ($N_{\text{b}}=2$) with internal viscosity, using a Gaussian approximation (GA), and these predictions have been shown using dash-dotted lines. The high-frequency-limiting value of $\eta'$ obtained by GA agrees with that derived by \citet{Dasbach1992}. Furthermore, while the functional form of $\eta''$ obtained by GA matches with the expression given by \citet{Dasbach1992}, they differ in the sense that the GA predicts a $\varphi$-dependent rescaling of the frequency which is absent in the latter work.

As seen from Fig.~\ref{fig:saos_fixed_iv}~(a), the inclusion of internal friction into the Rouse model introduces a qualitative change in the variation of the dynamic viscosity, $\eta'$, with the appearance of a plateau in the high-frequency regime, in contrast to the Rouse model where {$\eta\propto\omega^{-2}$} in the high-frequency limit. The numerical value of the plateau is equal to the stress jump, as seen from our simulations, which is in agreement with the theoretical prediction of~\citet{Gerhardt1994}. Since the stress jump scales linearly with the number of beads in the chain[Fig.~\ref{fig:mw_validation}~(a)], and the Rouse viscosity, $\eta^{\text{R}}$,  scales as $N^2$ [Eq.~(\ref{eq:eta_rouse})], the height of the high-frequency plateau decreases with an increase in the number of beads in the chain. The difference in the dynamic viscosity for the three different cases are less perceptible in the low frequency regime, where they are all seen to approach the respective Rouse viscosity. The GA prediction~\cite{Schieber1993} is seen to perform marginally better than the \citet{Dasbach1992} prediction at low frequencies. With the increase in the number of beads, the \citet{Dasbach1992} approximation compares satisfactorily against BD simulation results.

In Fig.~\ref{fig:saos_fixed_iv}~(b), the dynamic viscosity for chains with internal friction is scaled by its corresponding values for a Rouse chain and plotted as a function of scaled frequency. It is seen that the departure from Rouse prediction is pushed to higher values of the scaled frequency with an increase in the number of beads. Furthermore, since models with internal friction predict a saturation of the dynamic viscosity at high frequencies, and since the Rouse model prediction in the high frequency regime decays asymptotically as $\sim\omega^{-2}$ [Eq.~\ref{eq:eta_prime_rouse}], the scaled dynamic viscosity is expected to vary as $\sim\omega^{2}$ at high frequencies. This scaling is observed for all three cases examined in Fig.~\ref{fig:saos_fixed_iv}~(b). The long-chain ($N_{\text{b}}=100$) result predicted by the \citet{Dasbach1992} plotted on the same graph, further enunciates that for a fixed value of $\varphi$, the effect of internal friction decays with an increase in chain length. 

A similar weakening of internal friction effects has also been predicted by the RIF model~\cite{Khatri2007rif}, where the relaxation time of a mode $q$ is simply the sum of a mode-number-dependent Rouse contribution ($\tau^{\text{R}}_{q}\equiv\tau^{\text{R}}/q^2$), and an internal friction contribution ($\tau_{\text{int}}$) that is independent of mode-number. Here, $\tau^{\text{R}}=\left(N^2_{\text{b}}\zeta/\pi^2H\right)$ is the Rouse relaxation time, and $\tau_{\text{int}}=K/H$ is a characteristic timescale defined on the basis of the damping coefficient of the dashpot. The relative magnitude of the two timescales is then 
\begin{equation}
\dfrac{\tau_{\text{int}}}{\tau_{q}^{\text{R}}}=\left(\dfrac{\pi q}{N_{\text{b}}}\right)^2\varphi
\end{equation}
Two aspects are clear from the pre-averaged model predictions, for a fixed value of $\varphi$: Firstly, for a fixed chain length, the effects of internal friction are most pronounced at the higher mode numbers, i.e., at short time scales, and has the least impact on the global relaxation time, corresponding to the $q=1$ case. This aspect is qualitatively evident from Fig.~\ref{fig:saos_fixed_iv}~(b): at low frequencies, where long wavelength motions (low mode numbers) are perturbed, the dynamic viscosity for chains with internal friction is indistinguishable from the Rouse value. At higher frequencies, where short wavelength motions (large mode numbers) are probed, a clear departure from the Rouse value is observed, and one could consider that the deviation occurs at some critical mode number for a given chain length. Secondly, for a given mode number, the effect of internal friction diminishes with an increase in the chain length. This trend is also evident from Fig.~\ref{fig:saos_fixed_iv}~(b), where it is observed that the onset of deviation from the Rouse prediction is pushed to higher frequencies with an increase in chain length. Hagen and coworkers~\cite{Qiu20043398,Pabit2004,Hagen2010385} predict, based on experimental reconfiguration time measurements on proteins, that the effect of internal friction is most easily discernible in short molecules that fold on microsecond {(or faster)} timescales, and could scarcely be detected in longer molecules whose folding times are in the millisecond {(or slower)} range.

\begin{figure*}[t]
\begin{center}
\begin{tabular}{c c}
\includegraphics[width=3.5in,height=!]{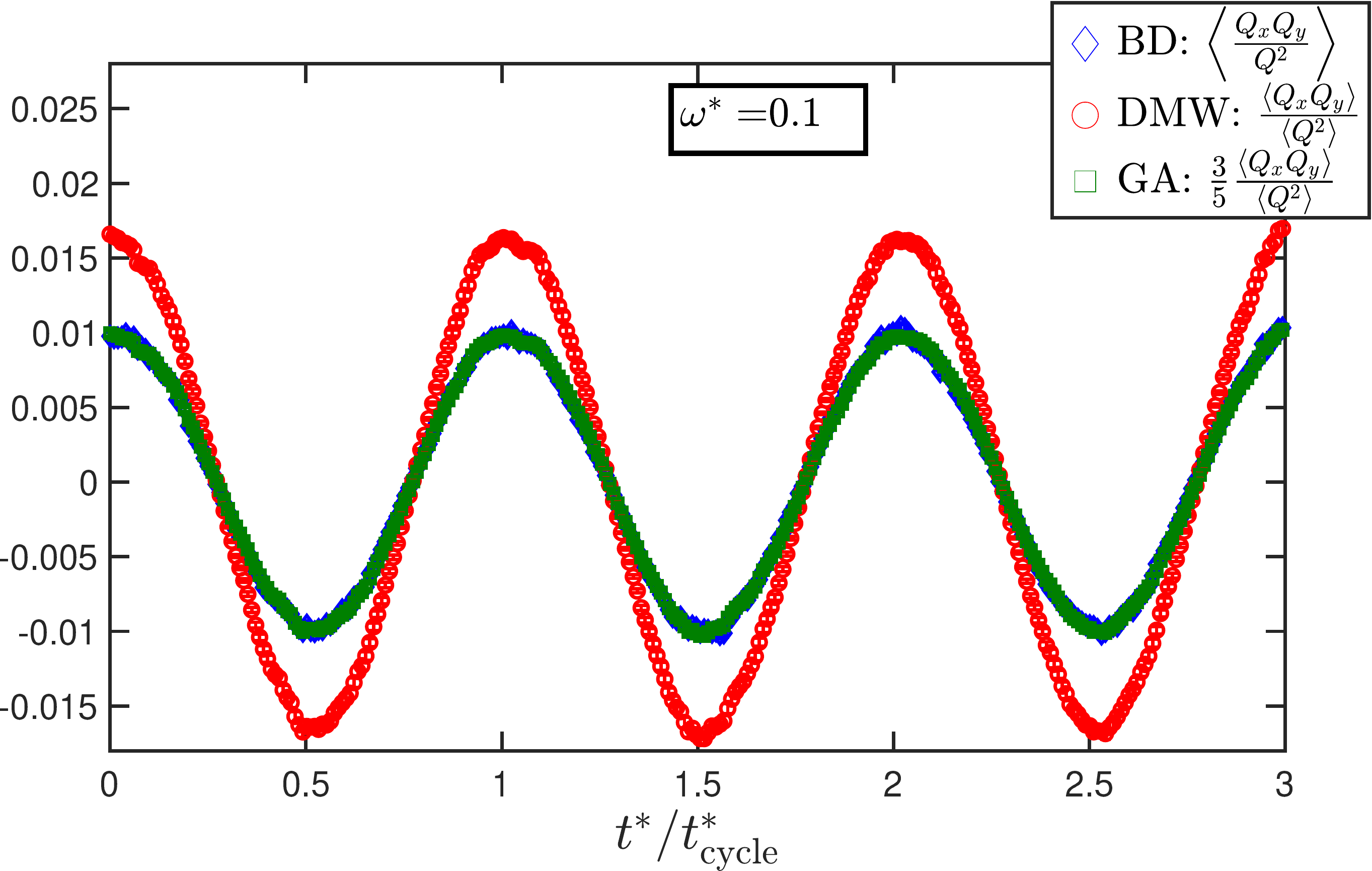}&
\includegraphics[width=3.5in,height=!]{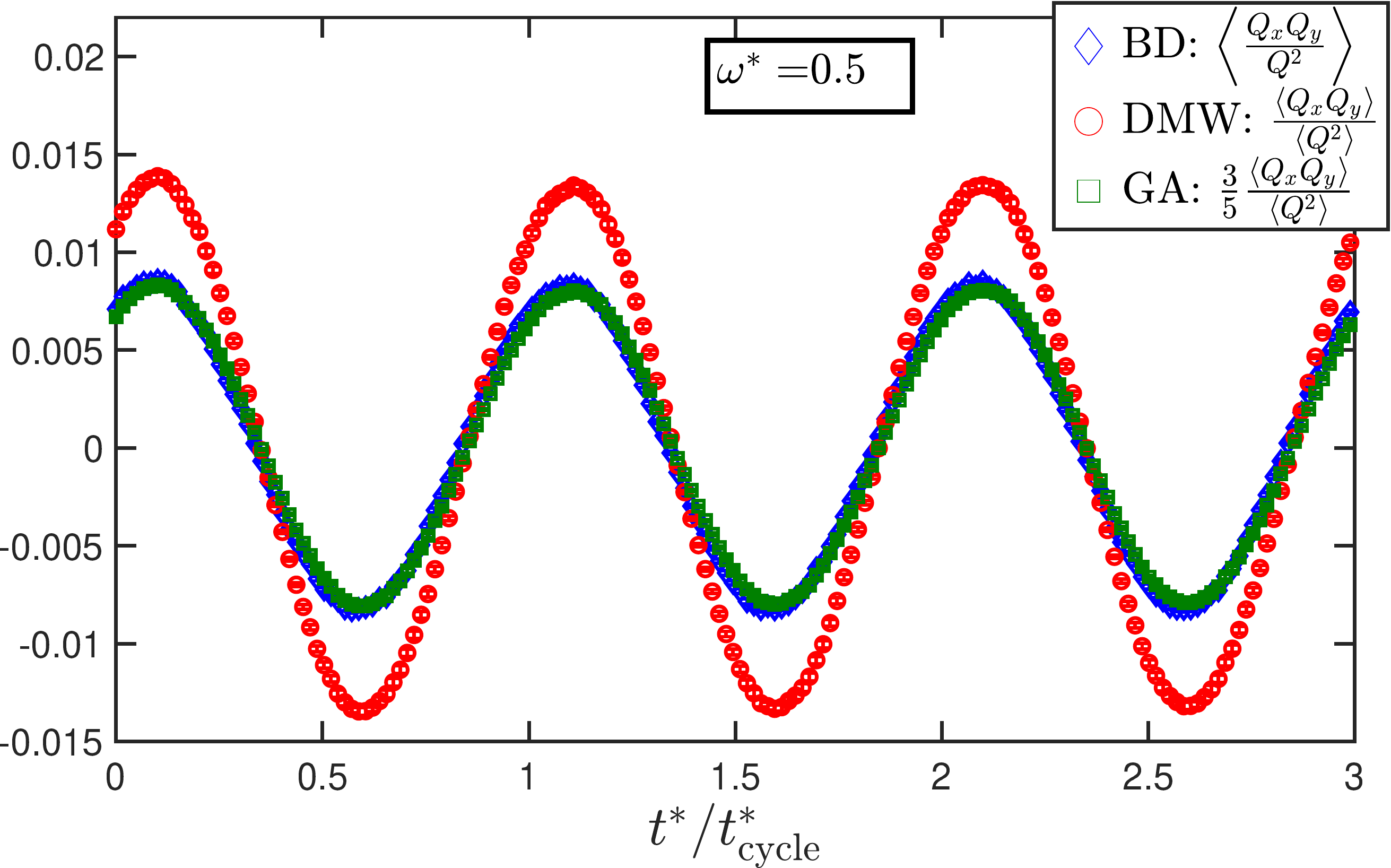}\\
(a) & (b)\\
\includegraphics[width=3.5in,height=!]{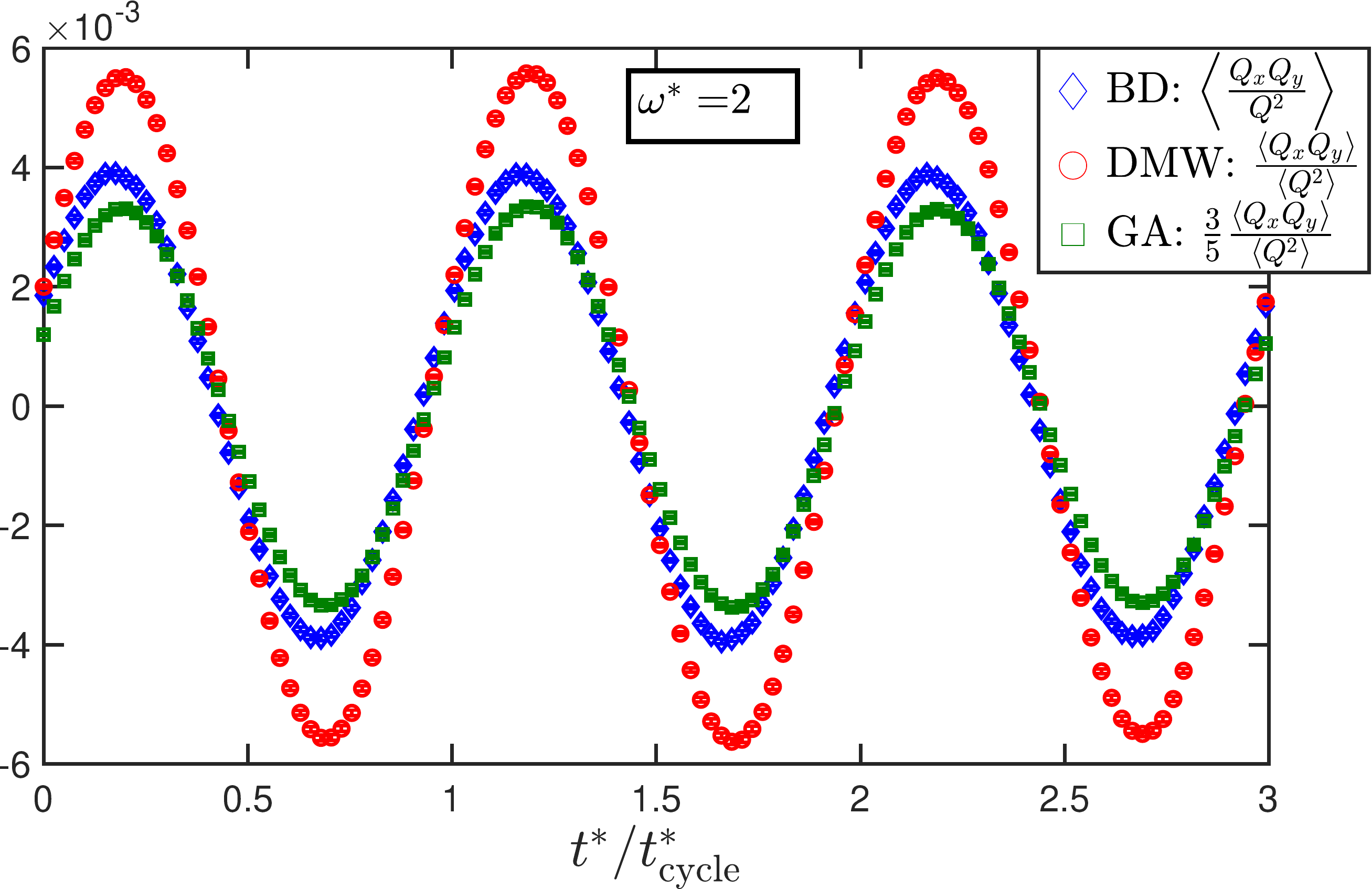}&
\includegraphics[width=3.5in,height=!]{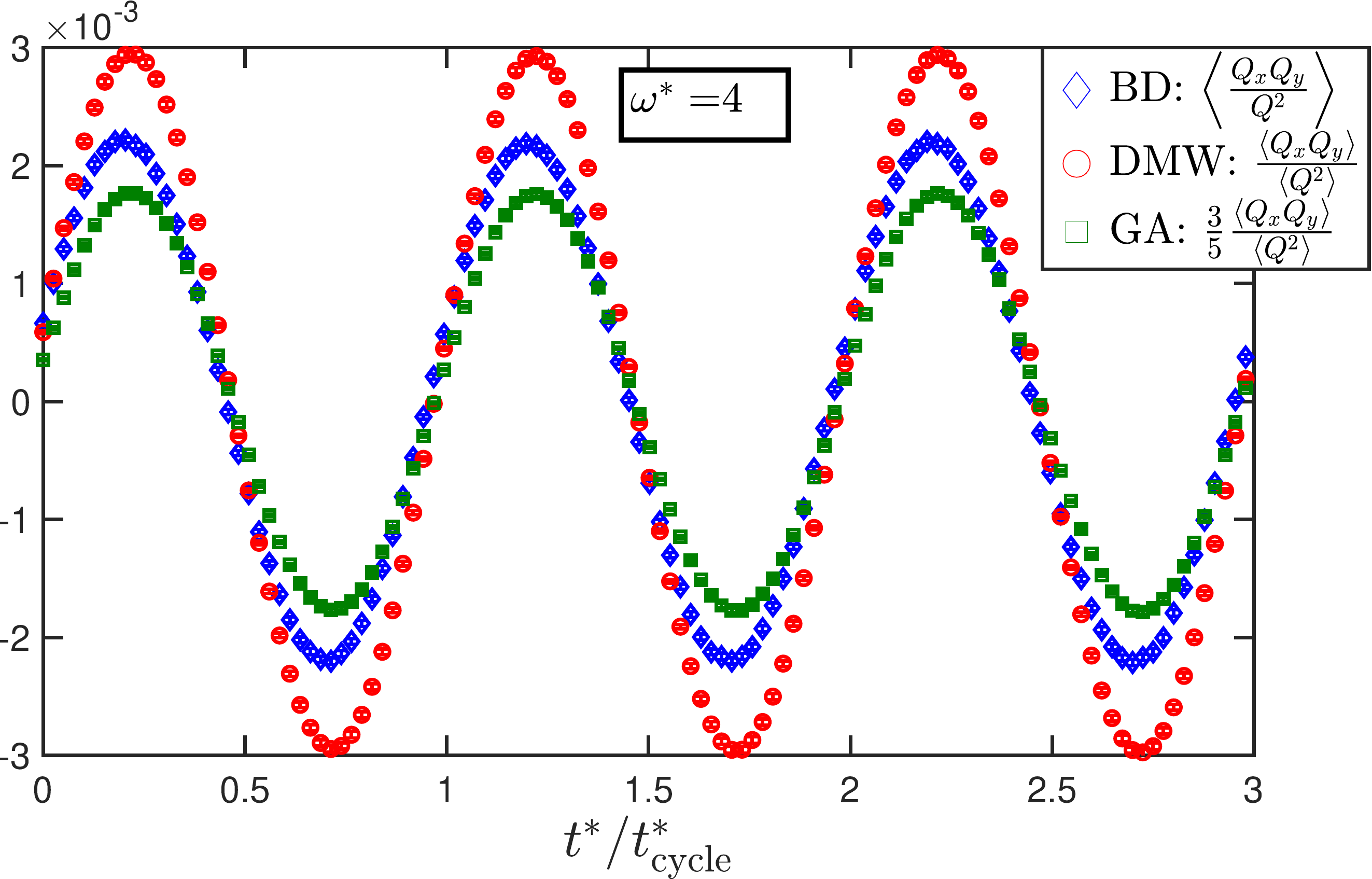}\\
(c) & (d)
\end{tabular}
\end{center}
\caption{The $xy$ component of the projection operator at four different values of the oscillatory frequency, $\omega^{*}$. The dimensionless cycle time is evaluated as $t^{*}_{\text{cycle}}=2\pi/\omega^{*}$. Blue symbols indicate the exact BD results, while the red and green data points denote the DMW and Gaussian approximations, respectively. Error bars are smaller than symbol size.}
\label{fig:proj_op_treatment}
\end{figure*}

In Fig.~\ref{fig:saos_fixed_iv}~(c), the imaginary component of the complex viscosity, $\eta''$, is normalized by the Rouse viscosity and plotted as a function of the scaled frequency. The {asymptotic (long chain)} Rouse scaling exponents at the low, intermediate and high frequency regimes are indicated in the figure. It is seen that inclusion of internal friction does not affect the Rouse scaling at low and high frequencies. {In the intermediate frequency regime, a power law region appears with an increase in the number of beads. The inset in Fig.~\ref{fig:saos_fixed_iv}~(c) shows the variation of $\eta''$ for three different bead numbers, $N_{\text{b}}=\{100,200,500\}$ and $\varphi=1.0$, as predicted by the DMW approximation. The portion of the curve in the frequency regime $\beta^{*}\in[80,300]$ is fitted using a power law of the form $y=cx^{n}$, in order to obtain a scaling exponent of $n=-0.48$. Upon fitting a power law to the Rouse chain results (without internal friction) for the above three bead numbers in the $\beta^{*}\in[80,300]$ regime, we obtain the same scaling exponent, within statistical error of the fit, which thereby hints at a finite size effect. We therefore conclude that the internal of internal friction does not affect the scaling of $\eta''$ in the intermediate frequency regime.} As observed in the case of $\eta'$, the accuracy of the \citet{Dasbach1992} prediction is seen to improve with an increase in the number of beads. Notably, for the two-bead case, the GA prediction for $\eta''$ is closer to the BD results at low frequencies, but a slight deviation is observed at values of the scaled frequency, $\beta^{*}>2$. {A more detailed examination of the predictions for the dumbbell case is presented below.}

In Fig.~\ref{fig:saos_fixed_iv}~(d), $\eta''$ is normalized by its corresponding value for a Rouse chain and plotted as a function of frequency. At the coarsest level of discretization ($N_{\text{b}}=2$), there is a striking, qualitative difference between the \citet{Dasbach1992} approximation and exact BD simulation results, in that the former predicts a frequency-independent response, while the latter exhibits a frequency-dependent variation which is also seen in models with higher number of beads. The GA prediction, however, captures the frequency dependence at the $N_{\text{b}}=2$ level, but is unable to account for the slight increase observed at $\beta^{*}>2$, and underestimates the magnitude of the high-frequency plateau. Furthermore, the low-frequency plateau for all the three values of the chain lengths ($N_{\text{b}}$) simulated is seen to approach unity, which is also the value predicted by the \citet{Dasbach1992} approximation in the long-chain ($N_{\text{b}}=100$) limit. Additionally, the onset of decrease in $\eta''/\left(\eta''\right)^{\text{R}}$ is pushed to higher frequencies as the number of beads in the chain is increased. {The appearance of a plateau in the high frequency region suggests that the frequency response of the elastic component of the complex viscosity of chains with internal friction is identical to that of chains without internal friction. The height of the high frequency plateau is not unity but a different constant that appears to be set by the internal friction parameter, and is independent of the chain length (except for the $N_{\text{b}}=2$ case which is discussed separately below)}. 

\begin{figure*}[t]
\begin{center}
\begin{tabular}{c c}
\includegraphics[width=3.5in,height=!]{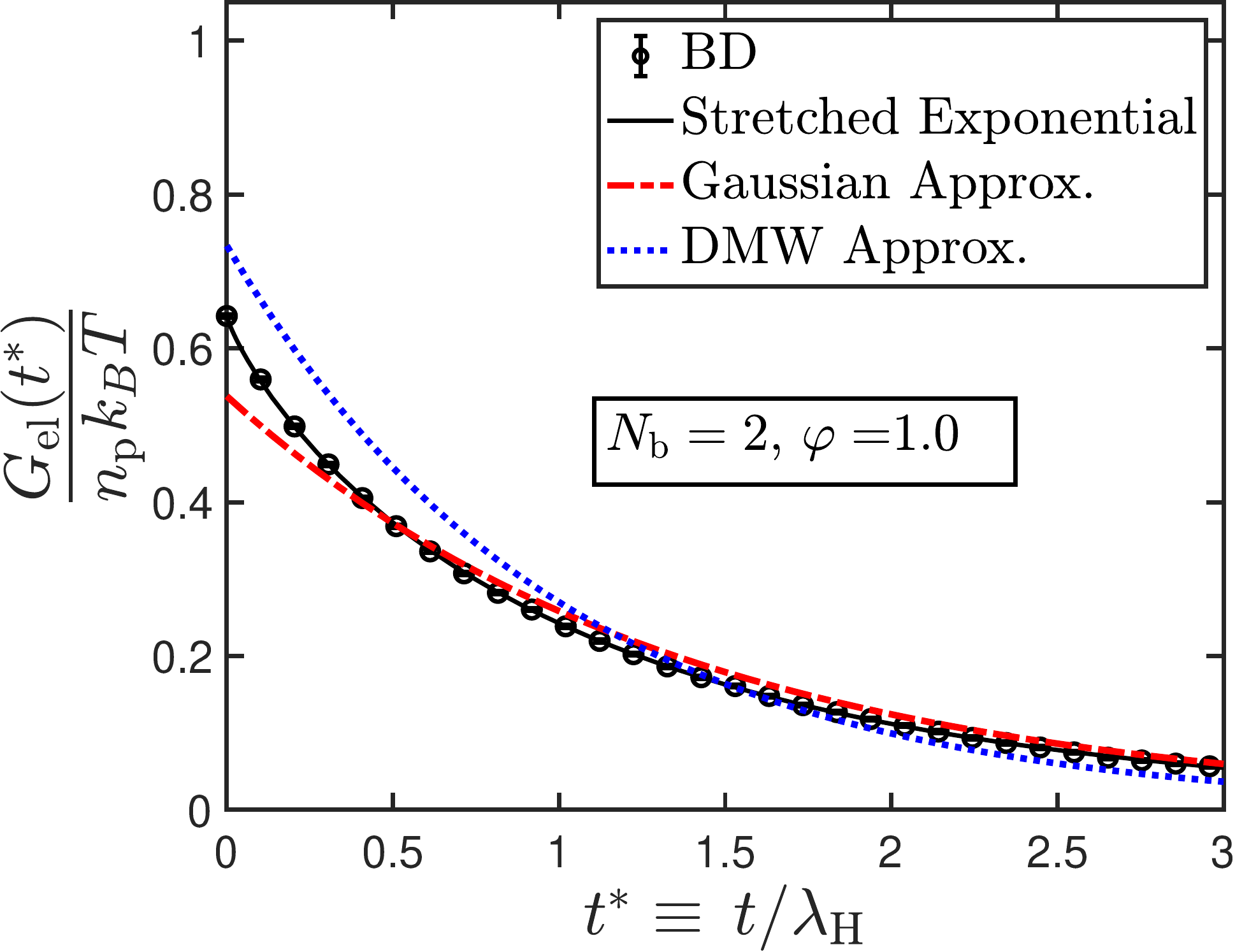}&
\includegraphics[width=3.5in,height=!]{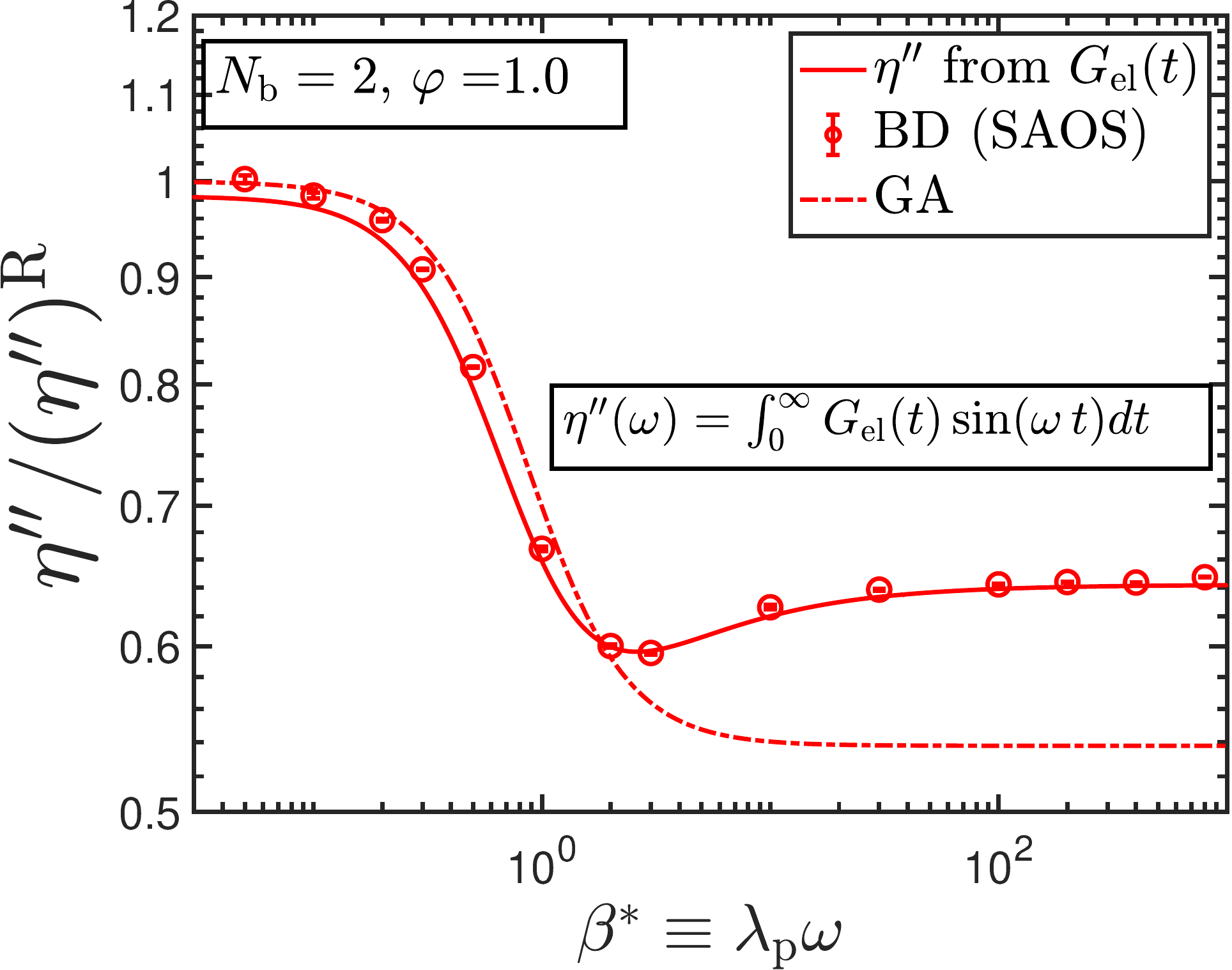}\\
(a)&(b)
\end{tabular}
\end{center}
\caption{(a) Elastic component of the stress relaxation modulus, and (b) the imaginary component of the complex viscosity. The stretched exponential has the following functional form given by Eq.~(\ref{eq:stretch_exp}), with $\tau_{\text{k}}=1.032,\,m=0.845$. Error bars are smaller than symbol size.}
\label{fig:gt_eta_dob}
\end{figure*}

{We present our investigations on the observed differences between the approximations and BD simulations at the dumbbell level in greater detail below. We reiterate that both BD simulations and the approximations (GA and DMW) use the Giesekus expression for the evaluation of the stress tensor, which may be written as follows~\cite{Hua1996} for Hookean dumbbells with internal friction
\begin{align}\label{eq:stress_tensor_db}
\boldsymbol{\tau}_{\text{p}}&=n_{\text{p}}k_BT\boldsymbol{\delta}-n_{\text{p}}H\left<\bm{Q}\bm{Q}\right>-\left(\dfrac{n_{\text{p}}K}{1+\epsilon}\right)\Biggl[\boldsymbol{\kappa}:\left<\dfrac{\bm{QQQQ}}{Q^2}\right>\nonumber\\[5pt]
&+\dfrac{6k_BT}{\zeta}\dashuline{\left<\dfrac{\bm{QQ}}{Q^2}\right>}-\dfrac{2H}{\zeta}\left<\bm{QQ}\right>\Biggr]
\end{align}
While BD simulations compute the dash-underlined term (the projection operator) exactly, the GA result in the linear viscoelastic limit is given by
\begin{equation}\label{eq:ga_proj_op}
\left<\dfrac{\bm{QQ}}{Q^2}\right>\approx\dfrac{2}{15}\boldsymbol{\delta}+\dfrac{3}{5}\dfrac{\left<\bm{QQ}\right>}{\left<Q^2\right>},
\end{equation}
and the DMW approximation replaces the underlined average of ratios by the ratio of averages, i.e.,
\begin{equation}\label{eq:dmw_proj_op}
\left<\dfrac{\bm{QQ}}{Q^2}\right>\approx\dfrac{\left<\bm{QQ}\right>}{\left<Q^2\right>},
\end{equation}
as a result of which the second and third terms within the square brackets in Eq.~(\ref{eq:stress_tensor_db}) cancel identically. These differing approaches for the approximation of the $\left<\bm{QQ}/Q^2\right>$ term manifests as the difference in the linear viscoelastic property prediction. The impact of the approximation on $\eta'$, however, is less significant than that on $\eta''$, and is most clearly revealed when scaled by the corresponding Rouse chain values for the components of the complex viscosity.}

{In Fig.~\ref{fig:proj_op_treatment} the $xy$ component of the projection operator, $\left<Q_{x}Q_{y}/Q^2\right>$, obtained from direct BD simulations is plotted at various frequencies and compared against the approximations given in Eqs.~(\ref{eq:ga_proj_op}) and~(\ref{eq:dmw_proj_op}). The $\left<\bm{QQ}\right>$ term on the RHS of Eq.~(\ref{eq:ga_proj_op}) ought strictly be evaluated using the analytical distribution function given in Ref., which would involve the solution of coupled integro-differential equations. We have chosen, rather, to use the value for $\left<\bm{QQ}\right>$ obtained from direct BD simulations, and believe that much insight may be gained into the problem despite this simplification.} 

{It is seen from the figure that the DMW approximation clearly deviates from the exact BD result at all frequencies considered. The Gaussian approximation, however, agrees remarkably at lower frequencies, before deviating from the exact result at $\omega^{*}>2$. We therefore believe this to be the reason for the failure of the GA at higher frequencies.}

{\citet{Hua1996} conclude, based on BD simulations on Hookean dumbbells with IV, that consequences of the GA and DMW approximations are more strongly observed in predictions for the stress relaxation modulus rather than those for the components of the complex viscosity. Before discussing our results for the stress relaxation modulus, $G(t)$, we note firstly that the complete expression for $G(t)$ contains a slowly decaying elastic component, $G_{\text{el}}(t)$,  and an instantaneously decaying singular component,
\begin{equation}
G(t)=G_{\text{el}}(t)+2\eta_{\text{v}}\delta(t)
\end{equation}
where $\eta_{\text{v}}$ equals the solvent viscosity for models without internal friction, and $\eta_{\text{v}}=\eta_{\text{s}}+\eta_{\text{jump}}$ for models with internal friction, where $\eta_{\text{jump}}$ denotes the instantaneous jump in viscosity at the inception of shear flow.
Since BD simulations are set up to begin at $t=0^{+}$, the singular component of the relaxation modulus is not calculated in these simulations~\cite{Hua1996}. In our simulations, $G_{\text{el}}(t)$ is calculated using the Green-Kubo relationship, which is based on the stress-autocorrelation of an ensemble of dumbbells at equilibrium~\cite{Kailasham2018}
\begin{equation}
{G^*_{\text{el}}(t^*)}=\dfrac{G_{\text{el}}(t^{*})}{n_{\text{p}}k_BT}=\left<\tau^{*}_{\text{p},xy}(t^*)\tau^{*}_{\text{p},xy}(0)\right>_{\text{eq}}
\end{equation}}

\begin{figure*}[t]
\begin{center}
\begin{tabular}{cc}
\includegraphics[width=0.5\linewidth]{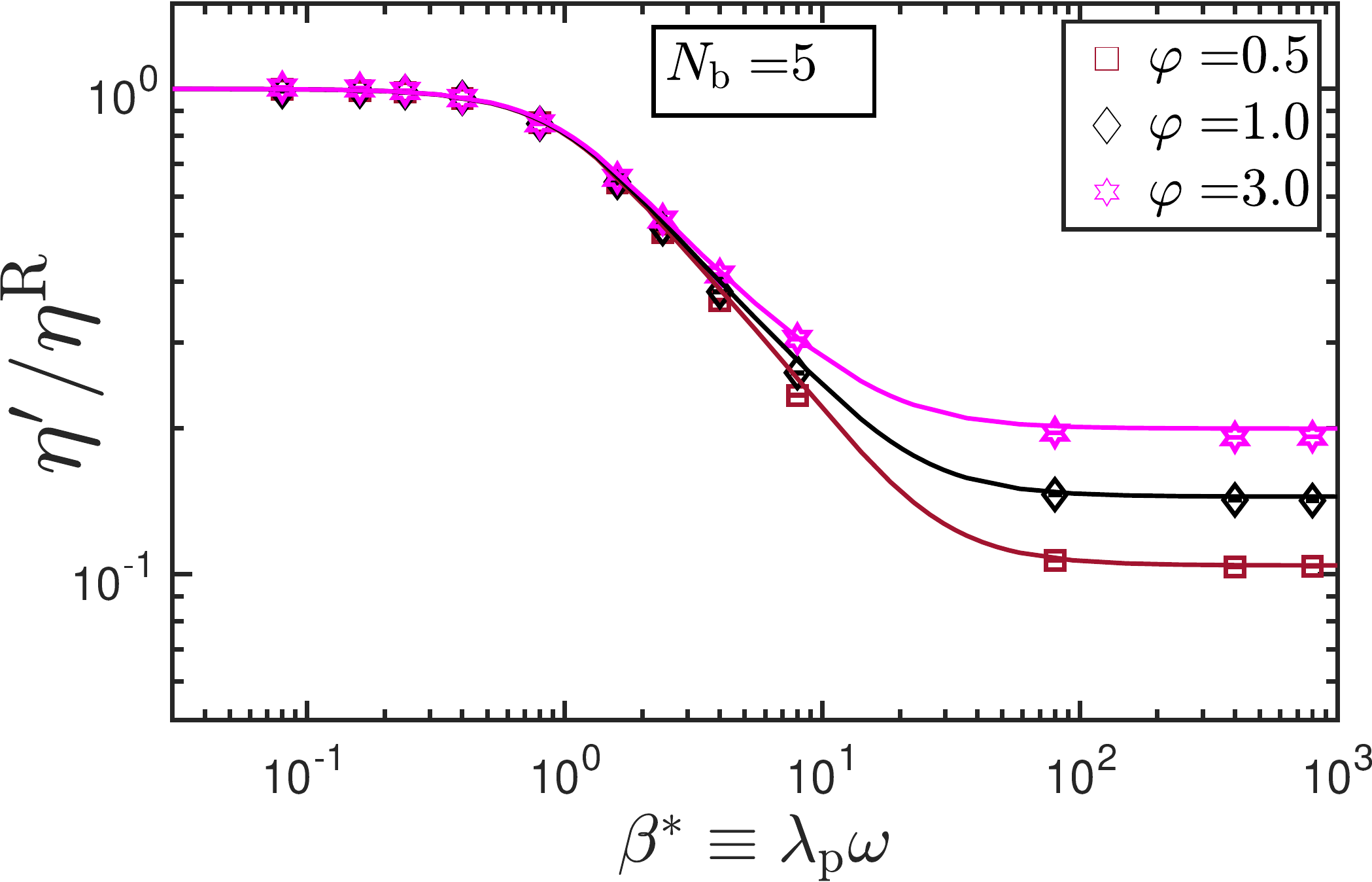}&
\includegraphics[width=0.5\linewidth]{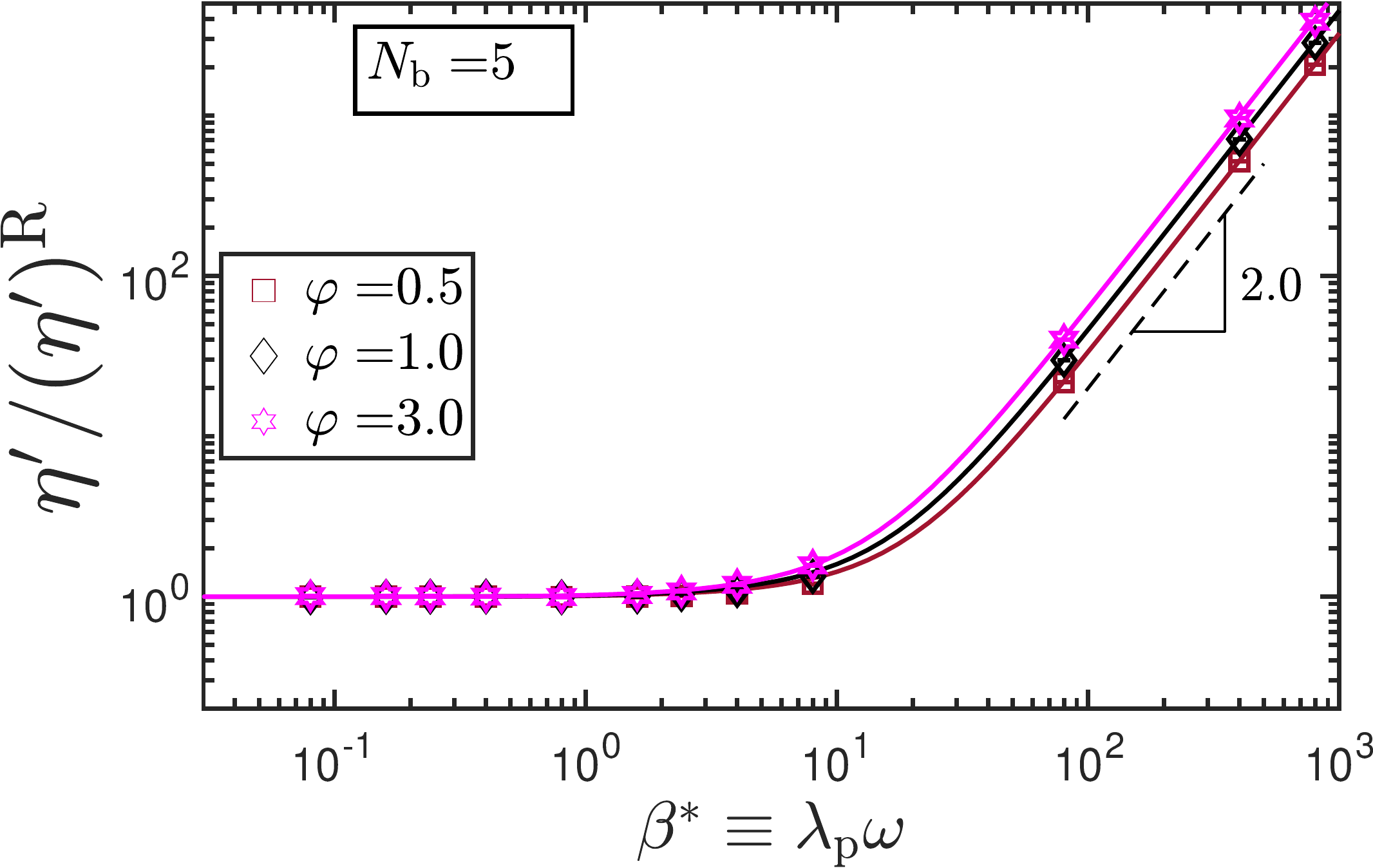}\\
(a)&(b)\\[5pt]
\includegraphics[width=0.5\linewidth]{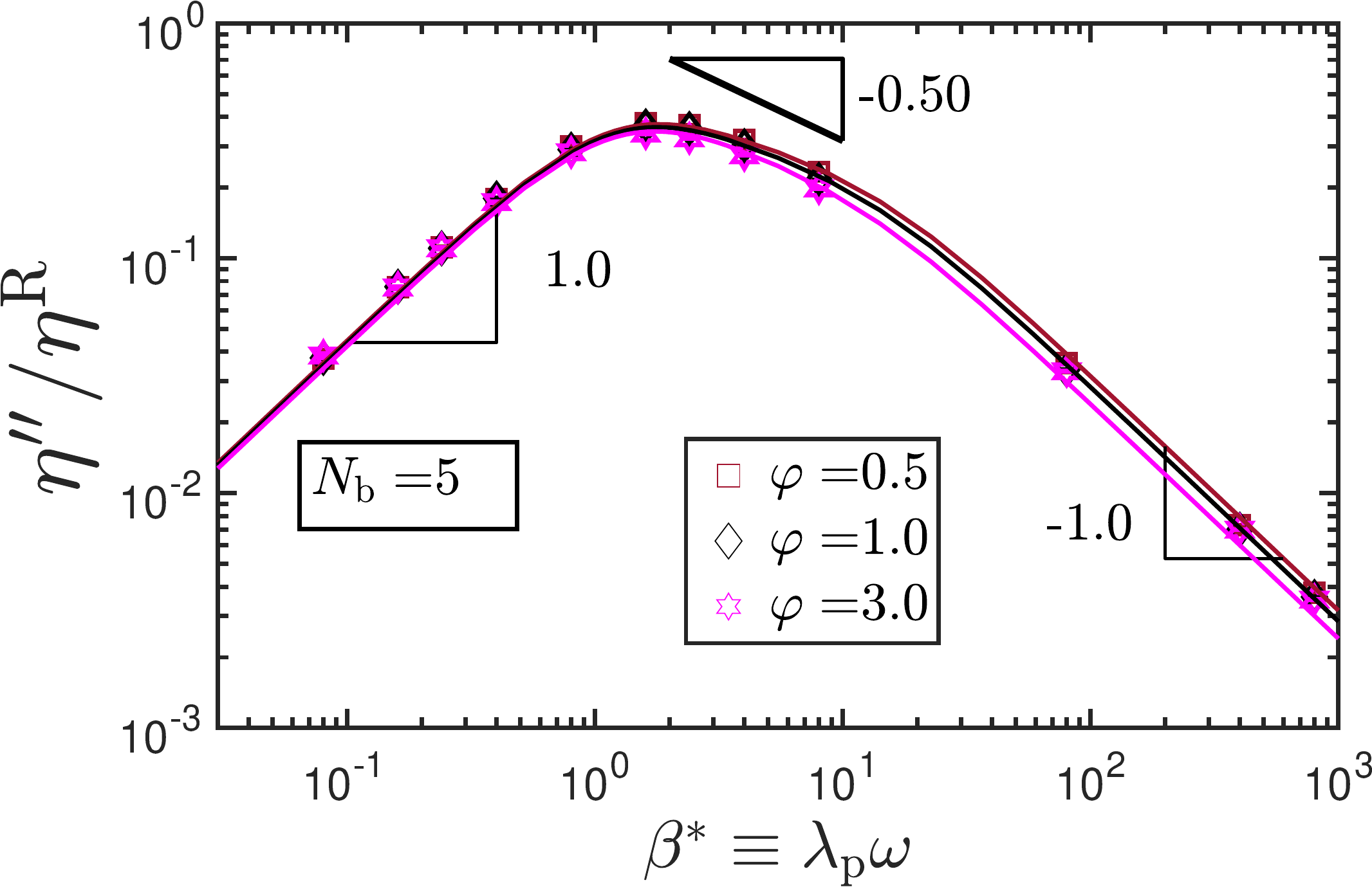}&
\includegraphics[width=0.5\linewidth]{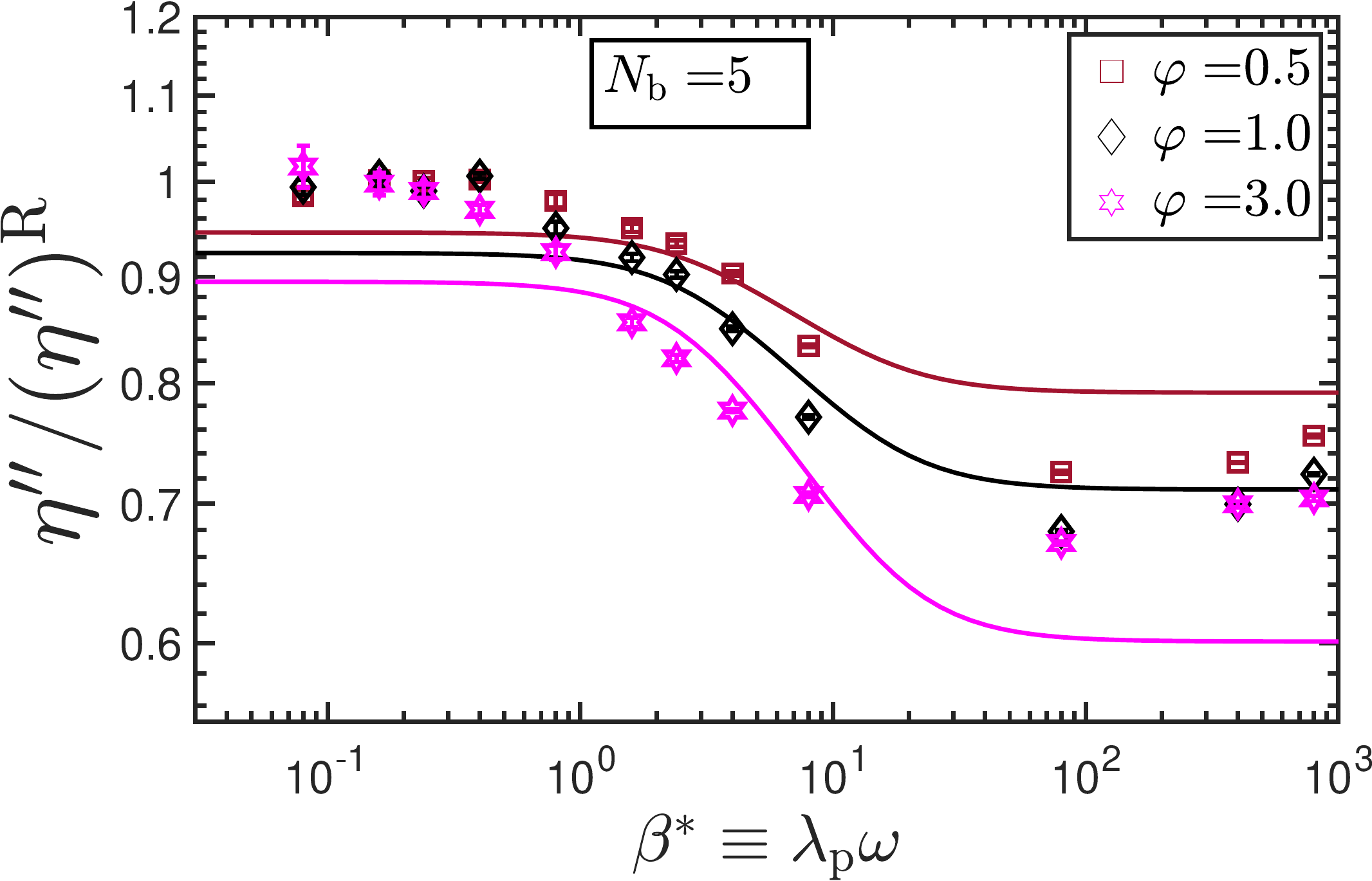}\\
(c)&(d)
\end{tabular}
\end{center}
\caption{\small Plots of the real ($\eta'$) and imaginary components ($\eta''$) of the complex viscosity, as a function of the scaled frequency, for a five-bead chain at three different values for the internal friction parameter. The lines are approximate solutions given in ~\citet{Dasbach1992}. Error bars are smaller than symbol size.}
\label{fig:saos_fixed_nb}
\end{figure*}

{In Fig.~\ref{fig:gt_eta_dob}~(a), the elastic component of the stress relaxation modulus for a Hookean dumbbell with internal friction is plotted as a function of dimensionless time. Both the Dasbach, Manke and Williams (DMW) approximation, and the Gaussian approximation (GA) are seen to agree with the exact BD simulation results at later times, with the deviation most strongly observed at early times. The BD simulation data is best fit by a stretched exponential of the form, 
\begin{equation}\label{eq:stretch_exp}
{G^*_{\text{el}}(t^*)}=\frac{G_{\text{el}}(0)}{n_{\text{p}}k_BT}\exp\left[(-t^*/\tau_{\text{k}})^{m}\right]
\end{equation}
where $\tau_{\text{k}}$ and $m$ are fitting parameters, and the fitted line is used for the calculation of $\eta^{''}$ as described below. The approximate theories, therefore, do not capture the linear viscoelastic properties correctly, as previously noted by~\citet{Hua1996}, who also observed that the accuracy of the approximations worsen as the IV parameter is increased.}

{The stretched exponential fitted to the stress relaxation modulus is subsequently used to evaluate $\eta''$
\begin{equation}\label{eq:eta_dob_prime_using_gt}
\eta^{''}(\omega)=\int_{0}^{\infty}G_{\mathrm{el}}(t)\sin(\omega\,t)dt
\end{equation}
As seen from Fig.~\ref{fig:gt_eta_dob}~(b), the imaginary component of the complex viscosity obtained from small amplitude oscillatory shear simulations, indicated by symbols, agrees well with the values obtained using Eq.~(\ref{eq:eta_dob_prime_using_gt}),
with a slight deviation at the lower frequencies. The scaled frequency at which the GA result deviates from BD simulation results is found to be $\beta^{*}\approx2$. It is possible to define a dimensionless timescale, $\tau_{\text{dev}}=1/\beta^{*}=0.5$, based on this frequency. This value of the timescale is remarkably close to the time at which the GA prediction for the stress relaxation modulus is seen to deviate from the exact BD simulation data. We have therefore established the correctness of the $\eta''$ predictions obtained from small amplitude oscillatory shear flow simulations, by comparison against the values obtained from the sine-transform of the stress relaxation modulus.}

In Fig.~\ref{fig:saos_fixed_nb}, the effect of the internal friction parameter on material functions in oscillatory shear flow is examined for a five-bead chain. The exact BD simulation results, indicated by symbols, are compared against the the approximate prediction given by \citet{Dasbach1992}, shown as lines. 

As seen from Fig.~\ref{fig:saos_fixed_nb}~(a), the height of the high-frequency plateau in the dynamic viscosity varies directly with the magnitude of the internal friction parameter. The low frequency, or long wavelength, response of the chain is unaffected by a variation in the internal friction parameter. In Fig.~\ref{fig:saos_fixed_nb}~(b), the dynamic viscosity normalized by its corresponding value for a Rouse chain and plotted as a function of the scaled frequency. This quantity is seen to increase as the square of the frequency, for the same reasons elaborated in connection with Fig.~\ref{fig:saos_fixed_iv}~(b).

In Fig.~\ref{fig:saos_fixed_nb}~(c), the imaginary component of the complex viscosity is scaled by the Rouse viscosity and plotted as a function of frequency. The effect of the variation in the internal friction parameter is almost negligible in the low frequency regime and is weak in the high frequency regime.

The difference between the predictions obtained using the \citet{Dasbach1992} approximation and the exact BD simulation results are most starkly visible in Fig.~\ref{fig:saos_fixed_nb}~(d), where the imaginary component of the complex viscosity is scaled by its corresponding value for a Rouse chain and plotted as a function of the scaled frequency. Firstly, the approximate model predicts a  low-frequency plateau that is dependent on the internal friction parameter. The simulation results, however, appear to converge on a low-frequency plateau value that is almost independent of the IV parameter. Secondly, the difference between the two predictions is seen to increase with the internal friction parameter {in the intermediate frequency regime, with the effect of the internal friction parameter becoming less perceptible in the high frequency regime.}

Thus we see from Figs.~\ref{fig:saos_fixed_iv} and~\ref{fig:saos_fixed_nb} that the effect of internal friction on the real (dissipative) component of the complex viscosity is more pronounced than that on the imaginary (elastic) component.

\begin{figure*}[t]
\begin{center}
\begin{tabular}{ccc}
\includegraphics[width=0.33\linewidth]{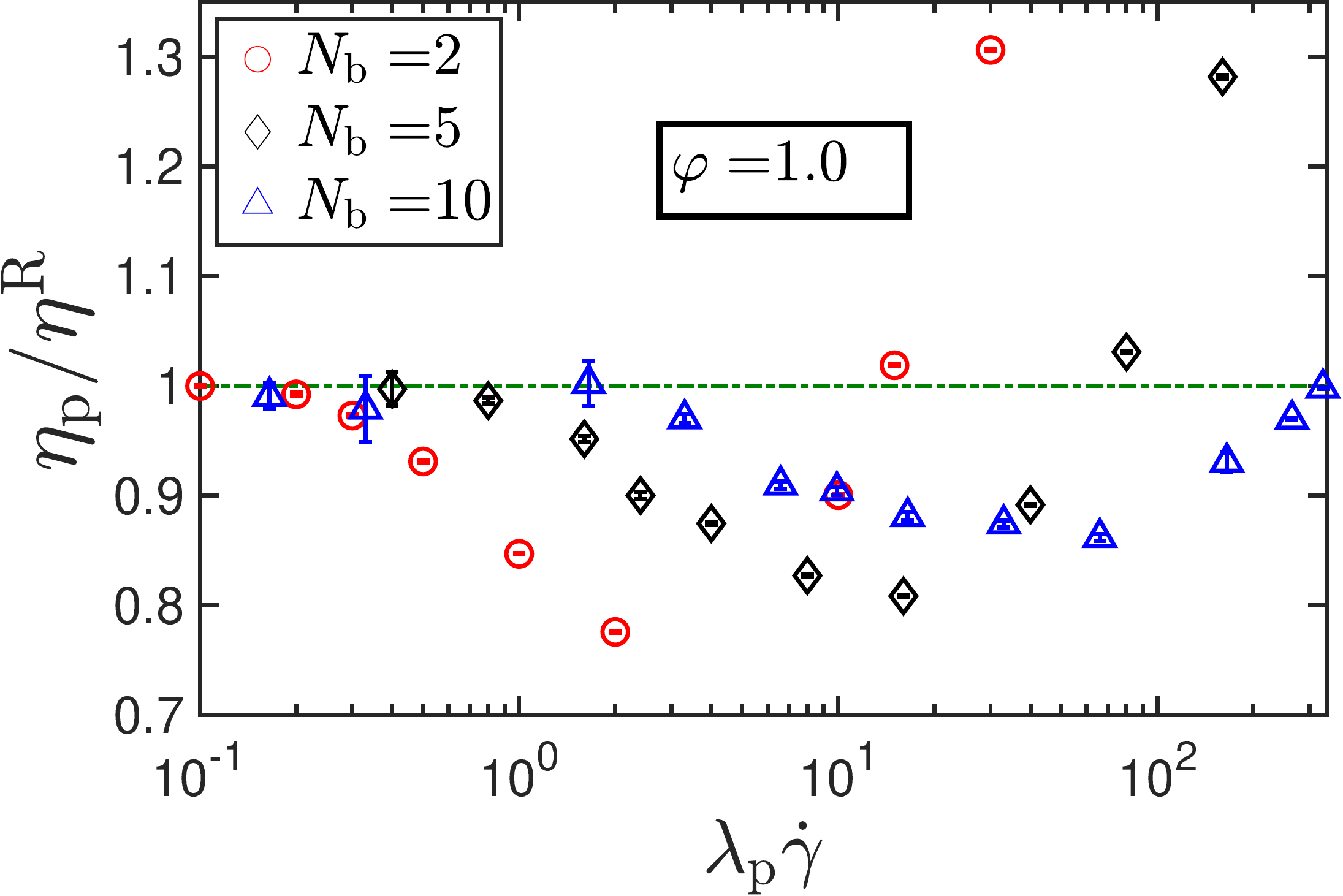}&
\includegraphics[width=0.33\linewidth]{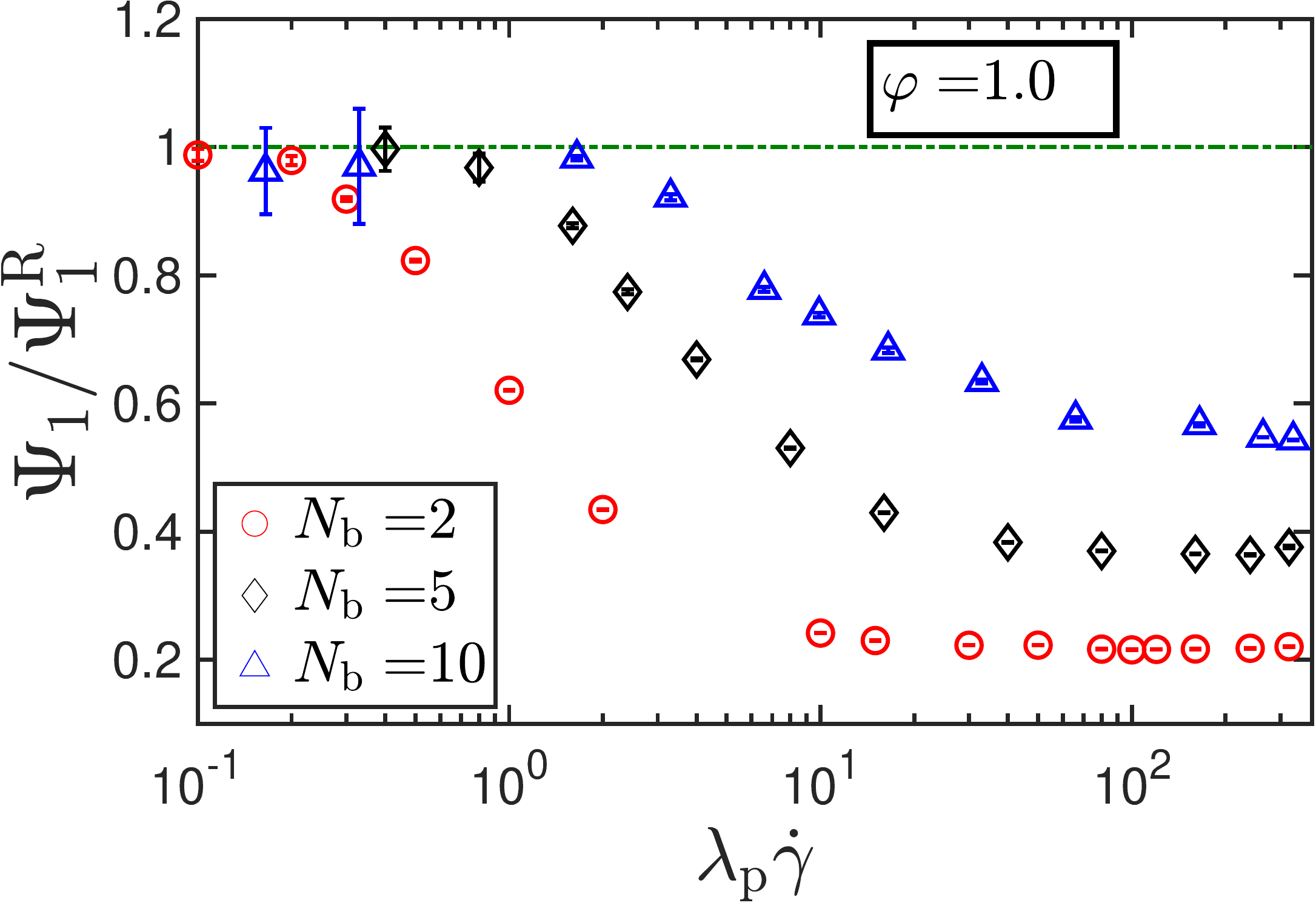}&
\includegraphics[width=0.33\linewidth]{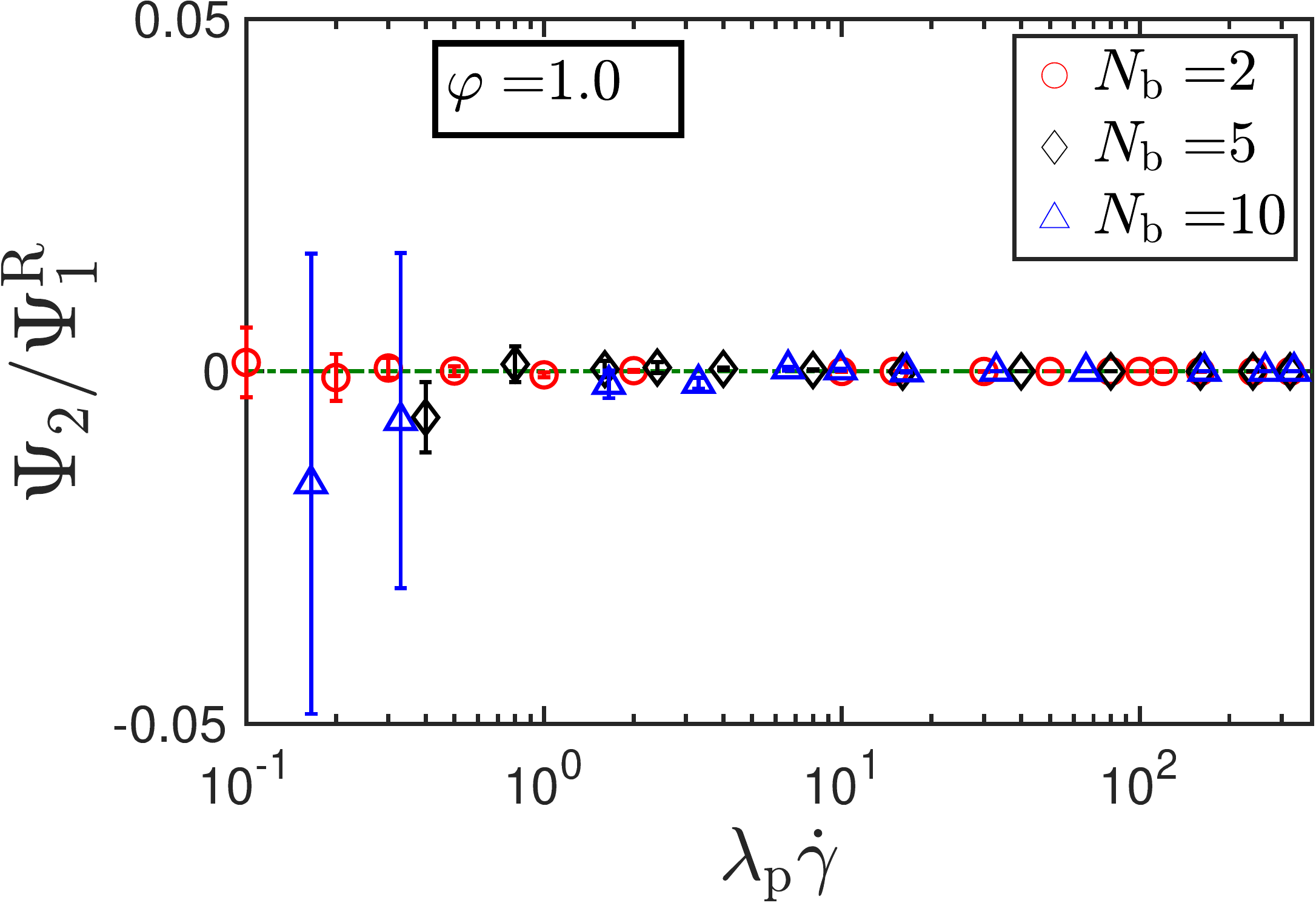}\\[5pt]
(a)&(b)&(c)\\[10pt]
\includegraphics[width=0.33\linewidth]{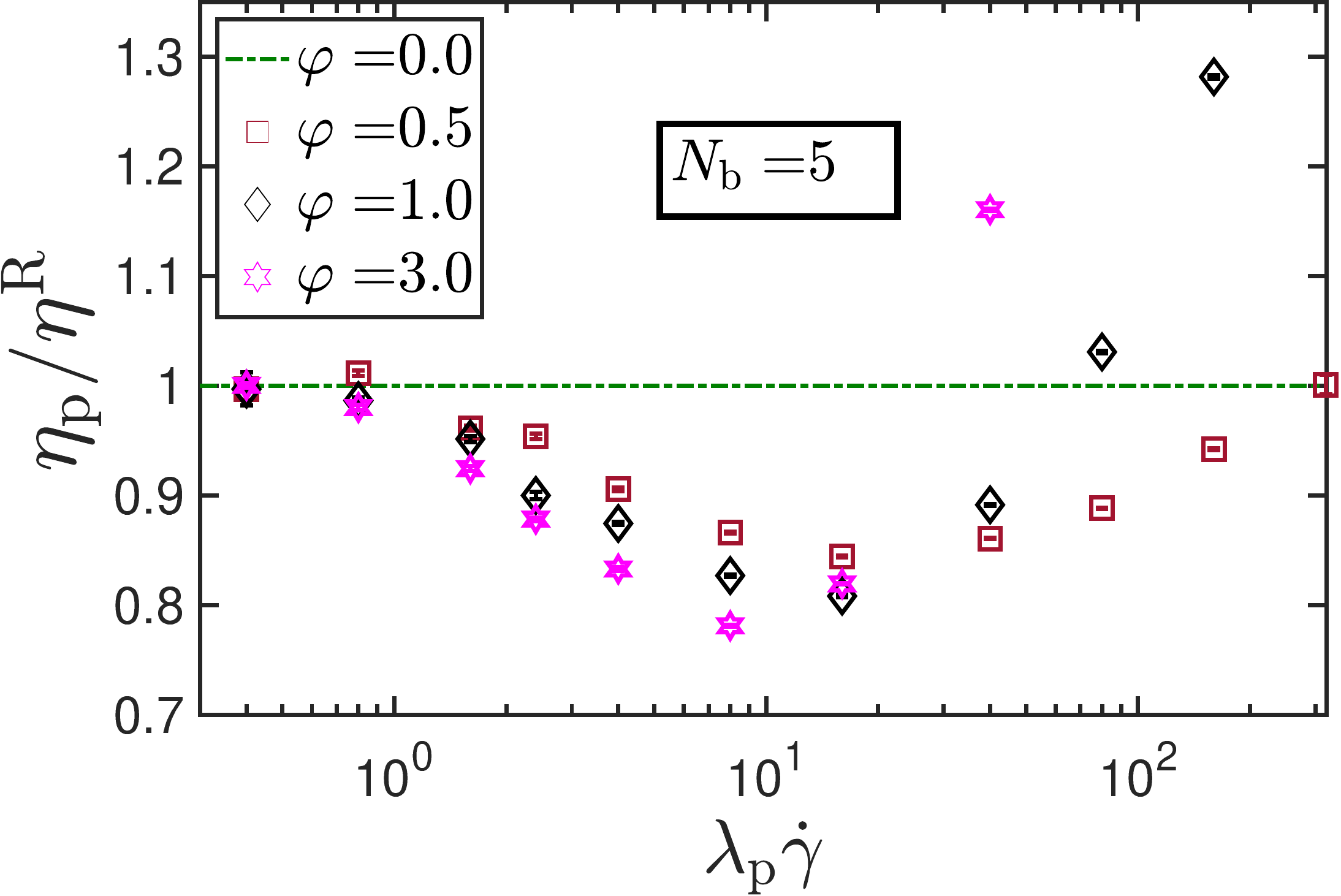}&
\includegraphics[width=0.33\linewidth]{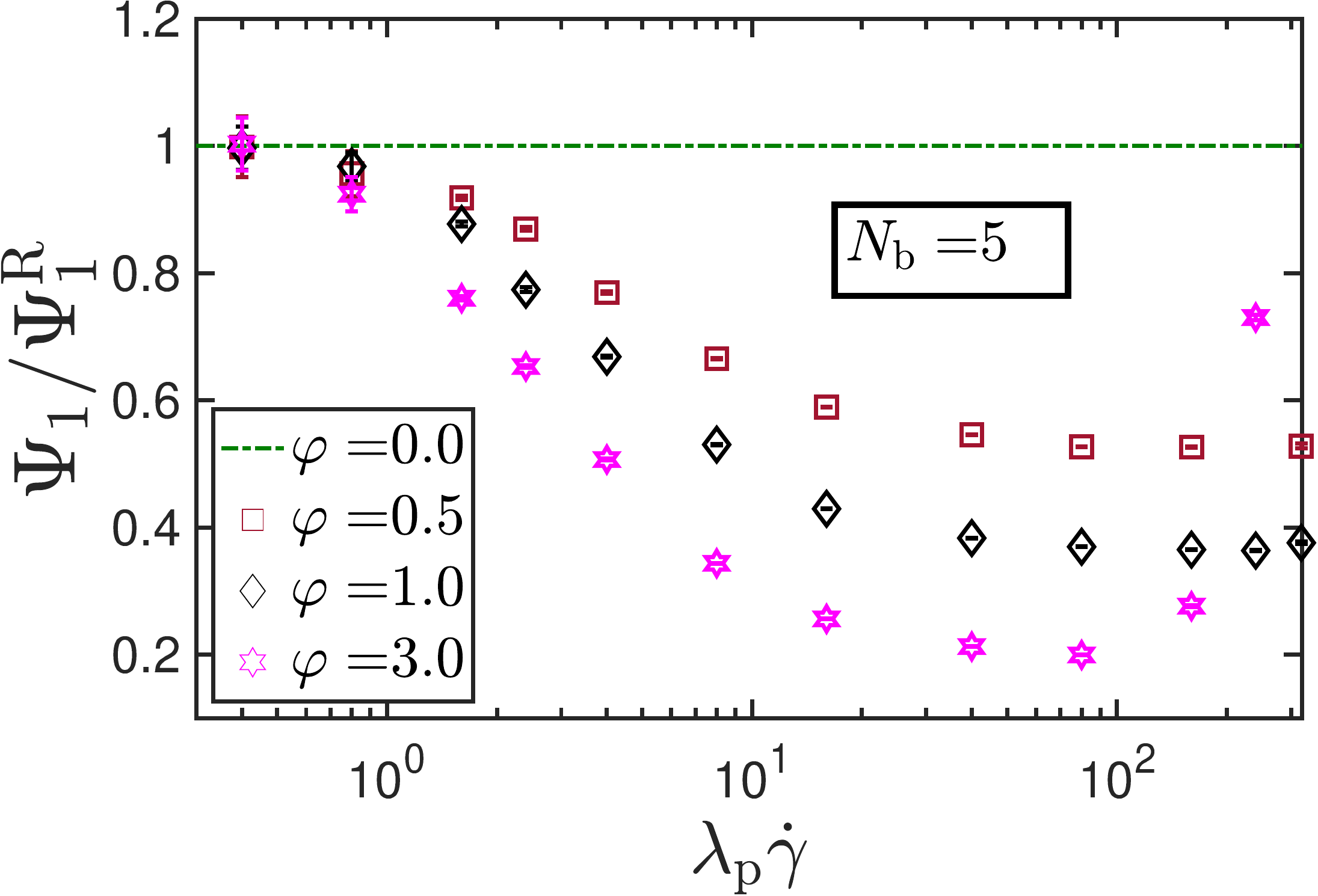}&
\includegraphics[width=0.33\linewidth]{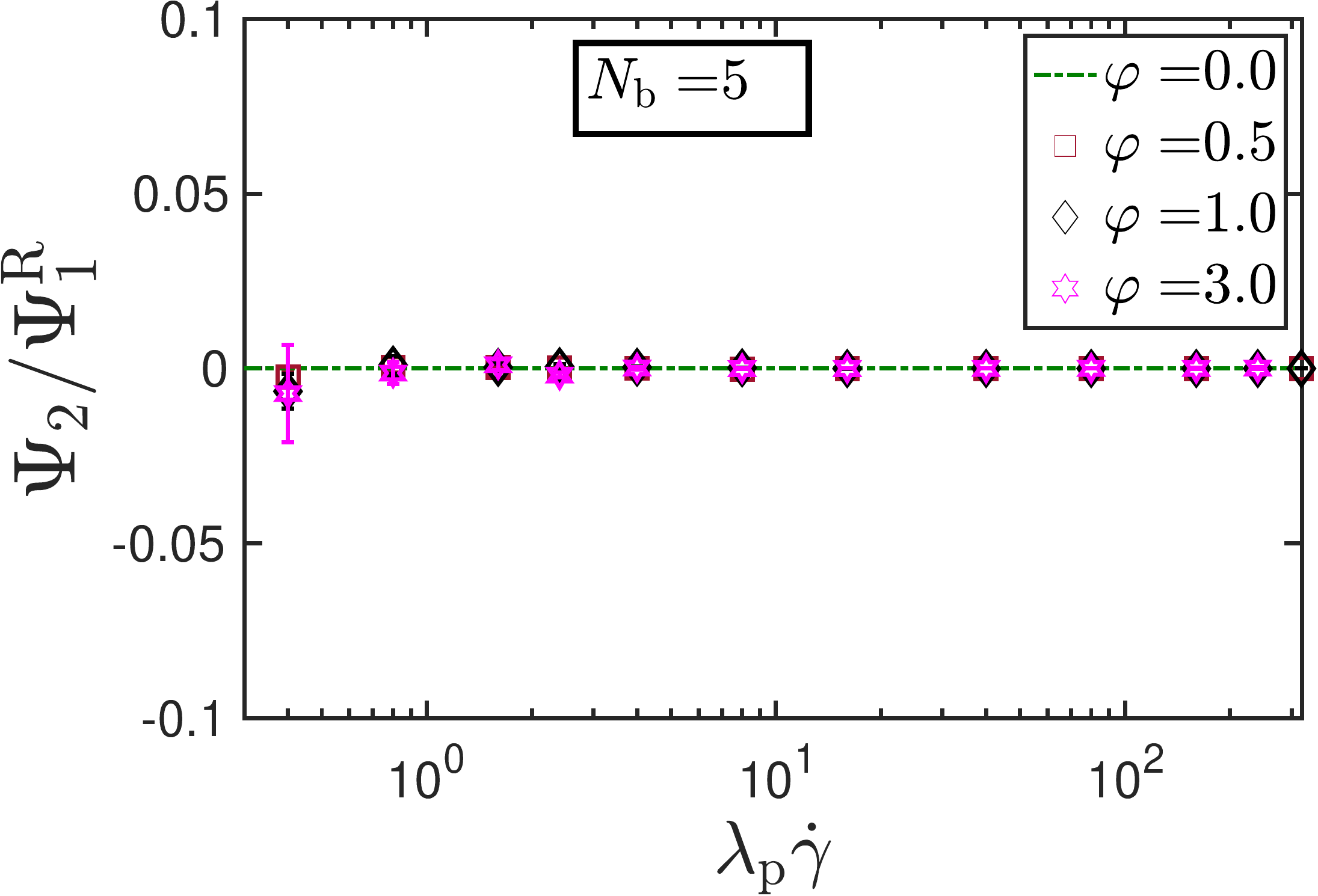}\\[5pt]
(d)&(e)&(f)
\end{tabular}
\end{center}
\caption{\small Steady-shear viscometric functions for bead-spring-dashpot chains with varying (a) number of beads in the chain and (b) values of the internal friction parameter. The horizontal lines in the figures indicate the material functions for a Rouse chain.}
\label{fig:steady_sr_plot}
\end{figure*}

A major motivation for the inclusion of internal friction in early theoretical models for polymeric solutions~\cite{Peterlin1967,PeterlinReinhold1967,Bazua1974} was to explain the high-frequency limiting value of the dynamic viscosity, $\eta'(\omega\to\infty)\equiv\eta'_{\infty}$, observed in experiments~\cite{Lamb1964,Philippoff1964,Massa1971}. An improvement to the Rouse/Zimm models was sought since they predict that the dynamic viscosity vanishes in the limit of high frequency, in contrast with experimental observations which in most instances indicate a positive limiting value~\cite{Lamb1964,Philippoff1964,Massa1971}. Models with internal friction, however, are able to successfully predict this plateau. There do exist systems, however, where the limiting value of the dynamic viscosity in the high frequency limit is \textit{negative}~\cite{Morris1988}. A detailed experimental investigation of the solvent molecule and polymer segment relaxation dynamics in such systems has been conducted by Lodge and coworkers~\cite{Morris1988,Lodge1993}. They conclude that such negative values of $\eta'_{\infty}$ cannot be explained within the existing polymer kinetic theory framework. Suggested modifications to the framework comprise the inclusion of an additional term in the stress tensor expression that accounts for coupling effects between the polymer molecules and the solvent~\cite{ByronBird1989}. It is not clear what factors determine when internal friction may be invoked to explain high frequency oscillatory shear data, and when additional physics needs to be considered. This is an important question that awaits theoretical and experimental investigation, but is beyond the scope of the present work.  

\subsection{\label{sec:steady_sr} Steady-shear viscometric functions}

In Fig.~\ref{fig:steady_sr_plot}, the steady-shear values of the material functions defined in Eq.~(\ref{eq:visc_fun_def}) are scaled by the corresponding values for a Rouse chain and plotted as a function of the characteristic shear rate. Schieber has shown~\cite{Schieber1993}, using the Gaussian approximation for dumbbells, that the zero-shear rate viscometric functions are unaffected by internal friction. The simulation data is found to concur with this prediction for all the three material functions. It is found that $\Psi_2$ is practically zero across the range of shear rates examined for all the cases.

As observed from Figs.~\ref{fig:steady_sr_plot}~(a) and (d), there is a striking similarity in the steady-shear variation of viscosity between Rouse chains with IV, and Rouse chains with hydrodynamic interactions~\cite{Zylka1991,Prabhakar2006}, in that there is shear-thinning followed by shear-thickening. For bead-spring-dashpot chains with a fixed number of beads, it is observed that the characteristic shear rate at which the minimum in the viscosity occurs is largely unaffected by the internal friction parameter. At shear rates larger than this critical value, the viscosity is found to increase with an increase in the IV parameter. For Rouse chains with hydrodynamic interactions, not only is the zero-shear-rate viscosity different from the free-draining case, the shear-dependence of viscosity is markedly dependent on the number of beads in the chain~\cite{Zylka1991}: for $N_{\text{b}}<6$, the Rouse viscosity is lower than the Zimm viscosity, and at large shear rates, where the effect of HI weakens, the viscosity values tend towards the Rouse value, and a shear-thinning is observed, following the Newtonian plateau at low shear rates. For $N_{\text{b}}\geq6$, however, the Rouse viscosity is greater than the Zimm viscosity, and at higher shear rates, the weakening of hydrodynamic interactions result in an upturn in the viscosity, causing it to approach the Rouse limit. An analogous explanation for the shear-thickening observed in Rouse chains with internal friction does not seem possible, since not only does the internal friction parameter result in a pronounced increase in shear thickening at high shear rates, but the viscosity is also seen to \textit{exceed} the Rouse value, clearly ruling out any weakening of the internal friction effects at high shear rates.

As seen from Fig.~\ref{fig:steady_sr_plot}~(b), the onset of shear-thinning in the first normal stress coefficient is pushed to higher shear rates, and the extent of shear-thinning reduced, with an increase in the number of beads at a fixed value of the internal friction parameter. {For an internal friction parameter value of $\varphi=1.0$, and the range of shear rates examined in the present work, there does not appear to be any pronounced shear-thickening for any of the three chain lengths.} BD simulations for Hookean dumbbells with internal friction by~\citet{Hua1995} show a similar plateauing in the first normal stress coefficient, as seen in the present work. There appears to be a suggestion, but no clear evidence of shear-thickening of $\Psi_1$, in their work~\cite{Hua1995}.

It is anticipated that a shear-thickening in $\Psi_1$ would be observed at higher values of the internal friction parameter, as evidenced by Fig.~\ref{fig:steady_sr_plot}~(e), where the effect of the internal friction parameter on the first normal stress coefficient is examined for a five-bead chain. At lower values of $\varphi$, there is no pronounced shear-thickening in $\Psi_1$, but a value of $\varphi=3.0$ results in the onset of a pronounced shear-thickening at $\lambda_{\text{p}}\dot{\gamma}\approx100.0$. Furthermore, this critical shear rate for the onset of shear-thickening in $\Psi_1$ is about an order-of-magnitude larger than that in the case of viscosity. 

It appears plausible that the shear-thickening in the viscosity and the first normal stress coefficient involves an interplay of internal friction and the number of beads in the chain.

A prevalent notion in the literature~\cite{Manke1986,Manke1989,Manke1991,Manke1993} is that the $\varphi\to\infty$ corresponds to the rigid-rod limit. This is supported by the following observation. The stress jump for rigid dumbbells with a Gaussian distribution of lengths is given by~\cite{Hua1996} $\eta_{\text{jump,rigid}}=0.4\,n_{\text{p}}k_BT\lambda_{H}$; while the stress jump for Hookean dumbbells with IV has the following form given by Manke and Williams~\cite{Manke1988}, $\eta_{\text{jump,IV}}=0.4\left[2\varphi/(1+2\varphi)\right]n_{\text{p}}k_BT\lambda_{H}$. Clearly, taking the $\varphi\to\infty$ limit for Hookean spring-dashpots gives the rigid-rod dumbbell result. There is a qualitative difference in the high-shear rate behavior of rigid-rod dumbbells and bead-rod chains, in that while the viscosity of rigid dumbbells shear-thins with an asymptotic exponent of $-1/3$, that of bead-rod-chains exhibits a high shear rate plateau~\cite{Petera1999,Pincus2020}. In flexible chains with internal friction, however, there is no qualitative difference between the dumbbell and chain results, with a pronounced increase in the shear-thickening of viscosity as $\varphi$ is increased. Furthermore, while the first normal stress coefficient for bead-rod chains shear-thins continuously~\cite{Petera1999}, chains with IV exhibit a slight shear-thickening at high shear rates, as discussed previously. 

A detailed comparison of the rheological properties of FENE dumbbells with IV and rigid dumbbells is given in Ref.~\citenum{Kailasham2018} where it is concluded that a combination of finite extensibility and a high value of the internal friction parameter ($\varphi\geq5$) is required to qualitatively mimic the steady-shear rheological response of rigid-rod models~\cite{Pincus2020}.

\section{\label{sec:conclusions} Conclusions}

The exact set of stochastic differential equations, and a thermodynamically consistent stress tensor expression for a Rouse chain with fluctuating internal friction has been derived. The BD simulation algorithm for the solution of these equations has been validated by comparison against approximate predictions available in the literature for the stress jump, and material functions in oscillatory and steady simple shear flows have been calculated. Semi-analytical predictions~\cite{Dasbach1992} for the dynamic viscosity are in near-quantitative agreement with the exact simulation results, with the accuracy improving with an increase in the number of beads in the chain. The difference between the predictions and the simulation results are more pronounced for the case of the imaginary component of the complex viscosity. The approximation by ~\citet{Dasbach1992} fails to capture the frequency dependence of $\eta''$ for the dumbbell case observed in exact BD simulations and predicted by the Gaussian approximation~\cite{Schieber1993}. 

The approach developed by Williams and coworkers~\cite{Manke1988,Dasbach1992}, however, is valid only in the linear viscoelastic regime, and cannot be used to obtain steady-shear viscometric predictions. The Gaussian approximation~\cite{Schieber1993} solution is only available for Hookean dumbbells with internal friction, and is unable to predict the shear-thickening in viscosity observed in exact Brownian dynamics simulations. {The present work therefore extends the applicability of the Manke and Williams approach~\cite{Manke1988,Dasbach1992} and permits the calculation of viscometric functions in the non-linear regime, and provides an exact solution to free-draining bead-spring-dashpot chains which was previously available for only the $N_{\text{b}}=2$ case~\cite{Hua1995}}.

Bead-spring-dashpot chains exhibit a non-monotonous variation in the viscosity with respect to the shear rate, with shear-thinning followed by shear-thickening. At a fixed value of the internal friction parameter, the shear-thickening effect is seen to weaken with an increase in the number of beads in the chain. Increasing the internal friction parameter at a fixed value of the number of beads in the chain leads to an increase in shear-thickening. The inclusion of internal friction results in a slight shear-thickening of the first normal stress coefficient, with the onset of thickening pushed to lower shear rates with an increase in the internal friction parameter.

The importance of hydrodynamic interactions in describing the dynamics of dilute polymer solutions is well-documented~\cite{Kailasham2018,Prakash2019}. However HI has not been considered in the present work, because its inclusion introduces an explicit coupling between all bead-pairs, and the procedure developed here is not applicable for the decoupling of the connector vector velocities. The solution of bead-spring-dashpot chains with hydrodynamic and excluded volume interactions is a subject for future study.

\begin{acknowledgments}
This work was supported by the MonARCH and MASSIVE computer clusters of Monash University, and the SpaceTime-2 computational facility of IIT Bombay. R. C. acknowledges SERB for funding (Project No. MTR/2020/000230 under MATRICS scheme). We also acknowledge the funding and general support received from the IITB-Monash Research Academy.
\end{acknowledgments}

%\vspace{-1cm}
\newpage

\appendix
\begin{widetext}
\section{\label{sec:app_a} Illustration of the decoupling methodology for a three-spring chain}

We define the quantity
\begin{align}\label{eq:ck_def}
C_k=\dfrac{\bm{Q}_k\cdot\llbracket\dot{\bm{Q}}_{k}\rrbracket}{Q^2_{k}};\quad k=1,2,\cdots,3
\end{align}

and take a dot-product on both sides of Eq.~(\ref{eq:qdot22}) with $\llbracket\dot{\bm{Q}}_{k}\rrbracket$ to obtain the following generating equation
\begin{align}\label{eq:ck_gen_eq_final}
C_k&=\left(\dfrac{\zeta}{\zeta+2K}\right)\left(\dfrac{\bm{Q}_k}{Q^2_{k}}\right)\cdot\left(\boldsymbol{\kappa}\cdot\bm{Q}_k\right)-{\left(\dfrac{k_BT}{\zeta+2K}\right)\left(\dfrac{\bm{Q}_k}{Q^2_{k}}\right)\cdot\left[-\dfrac{\partial \ln \Psi}{\partial \bm{Q}_{k-1}}+2\dfrac{\partial \ln \Psi}{\partial \bm{Q}_k}-\dfrac{\partial \ln \Psi}{\partial \bm{Q}_{k+1}}\right]}-\left(\dfrac{1}{\zeta+2K}\right)\left(\dfrac{\bm{Q}_k}{Q^2_{k}}\right)\cdot\Biggl[\nonumber\\[5pt]
&-\dfrac{\partial \phi}{\partial \bm{Q}_{k-1}}+2\dfrac{\partial \phi}{\partial \bm{Q}_k}-\dfrac{\partial \phi}{\partial \bm{Q}_{k+1}}\Biggr]+\left(\dfrac{K}{\zeta+2K}\right)\left[C_{k-1}L_{k-1}\left(\dfrac{{Q}_{k-1}}{Q_{k}}\right)+C_{k+1}L_{k}\left(\dfrac{{Q}_{k+1}}{Q_{k}}\right)\right]
 \end{align}
 where
 \begin{align}\label{eq:ldef_ex}
L_{k}\equiv\cos\theta_k=\dfrac{\boldsymbol{Q}_k\cdot\boldsymbol{Q}_{k+1}}{Q_kQ_{k+1}}
\end{align}
Eq.~(\ref{eq:ck_gen_eq_final}) is a coupled system of equations for the $C_k$, and it permits Eq.~(\ref{eq:qdot22}) to be written as 
 \begin{align}\label{eq:qdot_with_ck}
 \llbracket\dot{\bm{Q}}_{j}\rrbracket&=\boldsymbol{\kappa}\cdot\bm{Q}_j-\sum^{3}_{k=1}{\bm{A}}_{jk}\cdot\left(\dfrac{k_BT}{\zeta}\dfrac{\partial \ln \Psi}{\partial \bm{Q}_k}+\dfrac{1}{\zeta}\dfrac{\partial \phi}{\partial \bm{Q}_k}\right)-\varphi \sum^{3}_{k=1}{\bm{A}}_{jk}\cdot {\bm{Q}_k}C_k
 \end{align}
The objective, then, is to obtain a decoupled expression for $C_k$ that would yield a decoupled expression for $\llbracket\dot{\bm{Q}}_{j}\rrbracket$ which may then readily be substituted into the equation of continuity in probability space, to obtain the governing Fokker-Planck equation for the system. The expressions for $C_k$ for each of the three springs are as given below.
\begin{align}\label{eq:ck1}
C_1&=\left(\dfrac{\zeta}{\zeta+2K}\right)\left(\dfrac{\bm{Q}_1}{Q^2_{1}}\right)\cdot\left(\boldsymbol{\kappa}\cdot\bm{Q}_1\right)-{\left(\dfrac{k_BT}{\zeta+2K}\right)\left(\dfrac{\bm{Q}_1}{Q^2_{1}}\right)\cdot\uline{\left[2\dfrac{\partial \ln \Psi}{\partial \bm{Q}_1}-\dfrac{\partial \ln \Psi}{\partial \bm{Q}_{2}}\right]}}\nonumber\\[5pt]
&-\left(\dfrac{1}{\zeta+2K}\right)\left(\dfrac{\bm{Q}_1}{Q^2_{1}}\right)\cdot\uline{\Biggl[2\dfrac{\partial \phi}{\partial \bm{Q}_1}-\dfrac{\partial \phi}{\partial \bm{Q}_{2}}\Biggr]}+\left(\dfrac{K}{\zeta+2K}\right)\left[C_{2}L_{1}\left(\dfrac{{Q}_{2}}{Q_{1}}\right)\right]
\end{align}
\begin{align}\label{eq:ck2}
C_2&=\left(\dfrac{\zeta}{\zeta+2K}\right)\left(\dfrac{\bm{Q}_2}{Q^2_{2}}\right)\cdot\left(\boldsymbol{\kappa}\cdot\bm{Q}_2\right)-{\left(\dfrac{k_BT}{\zeta+2K}\right)\left(\dfrac{\bm{Q}_2}{Q^2_{2}}\right)\cdot\dashuline{\left[-\dfrac{\partial \ln \Psi}{\partial \bm{Q}_{1}}+2\dfrac{\partial \ln \Psi}{\partial \bm{Q}_2}-\dfrac{\partial \ln \Psi}{\partial \bm{Q}_{3}}\right]}}\nonumber\\[5pt]
&-\left(\dfrac{1}{\zeta+2K}\right)\left(\dfrac{\bm{Q}_2}{Q^2_{2}}\right)\cdot\dashuline{\Biggl[-\dfrac{\partial \phi}{\partial \bm{Q}_{1}}+2\dfrac{\partial \phi}{\partial \bm{Q}_2}-\dfrac{\partial \phi}{\partial \bm{Q}_{3}}\Biggr]}+\left(\dfrac{K}{\zeta+2K}\right)\left[C_{1}L_{1}\left(\dfrac{{Q}_{1}}{Q_{2}}\right)+C_{3}L_{2}\left(\dfrac{{Q}_{3}}{Q_{2}}\right)\right]
\end{align}
\begin{align}\label{eq:ck3}
C_3&=\left(\dfrac{\zeta}{\zeta+2K}\right)\left(\dfrac{\bm{Q}_3}{Q^2_{3}}\right)\cdot\left(\boldsymbol{\kappa}\cdot\bm{Q}_3\right)-{\left(\dfrac{k_BT}{\zeta+2K}\right)\left(\dfrac{\bm{Q}_3}{Q^2_{3}}\right)\cdot\dotuline{\left[-\dfrac{\partial \ln \Psi}{\partial \bm{Q}_{2}}+2\dfrac{\partial \ln \Psi}{\partial \bm{Q}_3}\right]}}\nonumber\\[5pt]
&-\left(\dfrac{1}{\zeta+2K}\right)\left(\dfrac{\bm{Q}_3}{Q^2_{3}}\right)\cdot\dotuline{\Biggl[-\dfrac{\partial \phi}{\partial \bm{Q}_{2}}+2\dfrac{\partial \phi}{\partial \bm{Q}_3}\Biggr]}+\left(\dfrac{K}{\zeta+2K}\right)\left[C_{2}L_{2}\left(\dfrac{{Q}_{2}}{Q_{3}}\right)\right]
\end{align}
It is clear from each pair of underlined terms in Eqs.~(\ref{eq:ck1}) - (\ref{eq:ck3}) that the structure of the Brownian force and the spring force terms are identical. In the forthcoming steps, therefore, only the spring force term shall be indicated for the sake of brevity. Once a general pattern has been identified, the final expression would contain both the Brownian and the spring force terms, multiplied by the appropriate prefactors.

We note from Eq.~(\ref{eq:ck_gen_eq_final}) that each value of $C_{k}$ depends on both its nearest neighbors, $C_{k-1}$ and $C_{k+1}$. The terminal connector vectors, however, represent a special case since $C_1$ would depend only on $C_2$, and $C_{N}$ would depend only on $C_{N-1}$, as evident from Eqs.~(\ref{eq:ck1}) - (\ref{eq:ck3}).

%\vspace{-0.75cm}

\subsection{\label{sec:fwd_sub} Forward substitution step}
In this step, the expression for $C_k$ is substituted into that for $C_{k+1}$, iteratively from $k=1$ to $k=2$, which removes all dependence of $C_{k}$ on $C_{k-1}$, and yields an expression for $C_{k}$ solely in terms of $C_{k+1}$.

Multiplying Eq.~(\ref{eq:ck1}) by $L_1(Q_1/Q_2)$, 
 \begin{align}\label{eq:c1_fwd_l_multip}
C_{1}L_{1}\left(\dfrac{{Q}_{1}}{Q_{2}}\right)&=L_1\left(\dfrac{\zeta}{\zeta+2K}\right)\left(\dfrac{1}{Q_{2}}\right)\left(\dfrac{\boldsymbol{Q}_1}{Q_{1}}\right)\cdot\left(\boldsymbol{\kappa}\cdot\boldsymbol{Q}_1\right)-\left(\dfrac{1}{\zeta+2K}\right)\left(\dfrac{L_1}{Q_{2}}\right)\left(\dfrac{\boldsymbol{Q}_1}{Q_{1}}\right)\cdot\left[2\dfrac{\partial \phi}{\partial \boldsymbol{Q}_1}-\dfrac{\partial \phi}{\partial \boldsymbol{Q}_{2}}\right]\\[5pt]
&+\left(\dfrac{K}{\zeta+2K}\right)C_2L^{2}_1
 \end{align}
and substituting into Eq.~(\ref{eq:ck2}), we obtain
  \begin{align}\label{eq:c2_fwd_intermed1}
&C_2=\left(\dfrac{K}{\zeta+2K}\right)C_{3}L_{2}\left(\dfrac{{Q}_{3}}{Q_{2}}\right)+\left(\dfrac{\zeta}{\zeta+2K}\right)\left(\dfrac{\boldsymbol{Q}_2}{Q^2_{2}}\right)\cdot\left(\boldsymbol{\kappa}\cdot\boldsymbol{Q}_2\right)+\left(\dfrac{1}{\zeta+2K}\right)\left(\dfrac{\boldsymbol{Q}_2}{Q^2_{2}}\right)\cdot\left[\dfrac{\partial \phi}{\partial \boldsymbol{Q}_{1}}\right]\nonumber\\[5pt]
&-2\left(\dfrac{1}{\zeta+2K}\right)\left(\dfrac{\boldsymbol{Q}_2}{Q^2_{2}}\right)\cdot\left[\dfrac{\partial \phi}{\partial \boldsymbol{Q}_2}\right]+\left(\dfrac{1}{\zeta+2K}\right)\left(\dfrac{\boldsymbol{Q}_2}{Q^2_{2}}\right)\cdot\left[\dfrac{\partial \phi}{\partial \boldsymbol{Q}_{3}}\right]+\dashuline{\left(\dfrac{K}{\zeta+2K}\right)^2C_2L^2_1}\nonumber\\[5pt]
&+L_1\left(\dfrac{K}{\zeta+2K}\right)\left(\dfrac{\zeta}{\zeta+2K}\right)\left(\dfrac{1}{Q_{2}}\right)\left(\dfrac{\boldsymbol{Q}_1}{Q_{1}}\right)\cdot\left(\boldsymbol{\kappa}\cdot\boldsymbol{Q}_1\right)\\[5pt]
&+\left(\dfrac{K}{\zeta+2K}\right)\Biggl\{\left(\dfrac{1}{\zeta+2K}\right)\left(\dfrac{L_1}{Q_{2}}\right)\left(\dfrac{\boldsymbol{Q}_1}{Q_{1}}\right)\cdot\left[\dfrac{\partial \phi}{\partial \boldsymbol{Q}_{2}}\right]-2\left(\dfrac{1}{\zeta+2K}\right)\left(\dfrac{L_1}{Q_{2}}\right)\left(\dfrac{\boldsymbol{Q}_1}{Q_{1}}\right)\cdot\left[\dfrac{\partial \phi}{\partial \boldsymbol{Q}_{1}}\right]\Biggr\}\nonumber
 \end{align}
 Defining
 \begin{equation}\label{eq:m2_def}
 M_2=\left(\dfrac{K}{\zeta+2K}\right)^2L^2_1
 \end{equation}
and grouping like terms together, the following expression for $C_2$ is obtained
 \begin{align}\label{eq:c2_fwd_fin}
C_2&=\left(\dfrac{1}{1-M_2}\right)\left(\dfrac{K}{\zeta+2K}\right)C_{3}L_{2}\left(\dfrac{{Q}_{3}}{Q_{2}}\right)\\[5pt]
&+\left(\dfrac{1}{1-M_2}\right)\left(\dfrac{\zeta}{\zeta+2K}\right)\left(\dfrac{1}{Q_2}\right)\left[\left(\dfrac{\boldsymbol{Q}_2}{Q_{2}}\right)\cdot\left(\boldsymbol{\kappa}\cdot\boldsymbol{Q}_2\right)+\left(\dfrac{K}{\zeta+2K}\right)L_1\left(\dfrac{\boldsymbol{Q}_1}{Q_{1}}\right)\cdot\left(\boldsymbol{\kappa}\cdot\boldsymbol{Q}_1\right)\right]\nonumber\\[5pt]
&+\left(\dfrac{1}{1-M_2}\right)\left(\dfrac{1}{\zeta+2K}\right)\left(\dfrac{\boldsymbol{Q}_2}{Q^2_{2}}\right)\cdot\left[\dfrac{\partial \phi}{\partial \boldsymbol{Q}_{3}}\right]\nonumber\\[5pt]
&-\left(\dfrac{1}{1-M_2}\right)\left(\dfrac{1}{\zeta+2K}\right){\left(\dfrac{1}{Q_{2}}\right)\left[-\left(\dfrac{K}{\zeta+2K}\right)L_1\left(\dfrac{\boldsymbol{Q}_1}{Q_1}\right)+2\left(\dfrac{\boldsymbol{Q}_2}{Q_2}\right)\right]}\cdot\left(\dfrac{\partial \phi}{\partial \boldsymbol{Q}_2}\right)\nonumber\\[10pt]
&-\left(\dfrac{1}{1-M_2}\right)\left(\dfrac{1}{\zeta+2K}\right){\left(\dfrac{1}{Q_{2}}\right)\left[2\left(\dfrac{K}{\zeta+2K}\right)L_1\left(\dfrac{\boldsymbol{Q}_1}{Q_1}\right)-\left(\dfrac{\boldsymbol{Q}_2}{Q_2}\right)\right]}\cdot\left(\dfrac{\partial \phi}{\partial \boldsymbol{Q}_1}\right)\nonumber
 \end{align}
Thus we see that the forward substitution procedure results in an expression for $C_2$ that depends only on $C_3$, which may be compared against Eq~(\ref{eq:ck2}) in which the expression for $C_2$ is coupled to both $C_1$ and $C_3$. 

As the next step, Eq.~(\ref{eq:c2_fwd_fin}) is multiplied by $L_2(Q_2/Q_3)$ and substituted into Eq.~(\ref{eq:ck3}), which results in
 \begin{align}\label{eq:c3_fwd_fin}
&C_3=\left(\dfrac{\zeta}{\zeta+2K}\right)\left(\dfrac{1}{Q_{3}}\right)\Biggl[\left(\dfrac{1}{1-M_3}\right)\left(\dfrac{\boldsymbol{Q}_3}{Q_{3}}\right)\cdot\left(\boldsymbol{\kappa}\cdot\boldsymbol{Q}_3\right)+\left(\dfrac{1}{1-M_3}\right)\left(\dfrac{1}{1-M_2}\right)\left(\dfrac{K}{\zeta+2K}\right)L_2\left(\dfrac{\boldsymbol{Q}_2}{Q_{2}}\right)\cdot\left(\boldsymbol{\kappa}\cdot\boldsymbol{Q}_2\right)\nonumber\\[5pt]
&+\left(\dfrac{1}{1-M_3}\right)\left(\dfrac{1}{1-M_2}\right)\left(\dfrac{K}{\zeta+2K}\right)^2L_2L_1\left(\dfrac{\boldsymbol{Q}_1}{Q_{1}}\right)\cdot\left(\boldsymbol{\kappa}\cdot\boldsymbol{Q}_1\right)\Biggr]\nonumber\\[5pt]
&-\left(\dfrac{1}{\zeta+2K}\right)\left(\dfrac{1}{Q_3}\right)\Biggl[-\left(\dfrac{1}{1-M_3}\right)\left(\dfrac{1}{1-M_2}\right)\left(\dfrac{K}{\zeta+2K}\right)L_2\left(\dfrac{\boldsymbol{Q}_2}{Q_{2}}\right)+2\left(\dfrac{1}{1-M_3}\right)\left(\dfrac{\boldsymbol{Q}_3}{Q_{3}}\right)\Biggr]\cdot\left(\dfrac{\partial \phi}{\partial \boldsymbol{Q}_{3}}\right)\nonumber\\[5pt]
&-\left(\dfrac{1}{\zeta+2K}\right)\left(\dfrac{1}{Q_{3}}\right)\Biggl[-\left(\dfrac{1}{1-M_3}\right)\left(\dfrac{1}{1-M_2}\right)\left(\dfrac{K}{\zeta+2K}\right)^2L_2L_1\left(\dfrac{\boldsymbol{Q}_1}{Q_1}\right)\nonumber\\[5pt]
&\qquad\qquad+2\left(\dfrac{1}{1-M_3}\right)\left(\dfrac{1}{1-M_2}\right)\left(\dfrac{K}{\zeta+2K}\right)L_2\left(\dfrac{\boldsymbol{Q}_2}{Q_2}\right)-\left(\dfrac{\boldsymbol{Q}_3}{Q_{3}}\right)\Biggr]\cdot\left(\dfrac{\partial \phi}{\partial \boldsymbol{Q}_2}\right)\nonumber\\[10pt]
&-\left(\dfrac{1}{\zeta+2K}\right)\left(\dfrac{1}{Q_{3}}\right)\Biggl[2\left(\dfrac{K}{\zeta+2K}\right)^2\left(\dfrac{1}{1-M_3}\right)\left(\dfrac{1}{1-M_2}\right)L_2L_1\left(\dfrac{\boldsymbol{Q}_1}{Q_1}\right)\\[5pt]
&\qquad\qquad\qquad\qquad-\left(\dfrac{K}{\zeta+2K}\right)\left(\dfrac{1}{1-M_3}\right)\left(\dfrac{1}{1-M_2}\right)L_2\left(\dfrac{\boldsymbol{Q}_2}{Q_2}\right)\Biggr]\cdot\left(\dfrac{\partial \phi}{\partial \boldsymbol{Q}_1}\right)\nonumber
 \end{align}
 where
 \begin{equation}\label{eq:m3_def}
 M_3=\left(\dfrac{K}{\zeta+2K}\right)^2\left(\dfrac{1}{1-M_2}\right)L^2_2
 \end{equation}
 
Since $C_3$ is the terminal connector vector in a three-spring chain, one obtains a completely decoupled expression for it by the application of the forward substitution procedure.
 
%\vspace{-0.5cm}
  
\subsection{\label{sec:bkwd_sub} Backward substitution step}

In this step, the expression for $C_{k}$ is substituted into that for $C_{k-1}$, iteratively from $k=3$ to $k=2$. Multiplying Eq.~(\ref{eq:ck3}) by $L_2(Q_3/Q_2)$,  
\begin{align}\label{eq:c3_bkwd_l_multip}
&C_3L_{2}\left(\dfrac{Q_3}{Q_{2}}\right)=\left(\dfrac{\zeta}{\zeta+2K}\right)\left(\dfrac{L_{2}}{Q_{2}}\right)\left(\dfrac{\boldsymbol{Q}_3}{Q_{3}}\right)\cdot\left(\boldsymbol{\kappa}\cdot\boldsymbol{Q}_3\right)+\left(\dfrac{K}{\zeta+2K}\right)\left[{C_{2}L^2_{2}}\right]\\[5pt]
&+\left(\dfrac{1}{\zeta+2K}\right)\left(\dfrac{L_{2}}{Q_{2}}\right)\left(\dfrac{\boldsymbol{Q}_3}{Q_{3}}\right)\cdot\left(\dfrac{\partial \phi}{\partial \boldsymbol{Q}_{2}}\right)-2\left(\dfrac{1}{\zeta+2K}\right)\left(\dfrac{L_{2}}{Q_{2}}\right)\left(\dfrac{\boldsymbol{Q}_3}{Q_{3}}\right)\cdot\left(\dfrac{\partial \phi}{\partial \boldsymbol{Q}_3}\right)\nonumber
 \end{align}
 and substituting into Eq.~(\ref{eq:ck2}), we obtain
 \begin{align}\label{eq:c2_bkwd_intermed1}
&C_{2}=\left(\dfrac{K}{\zeta+2K}\right)C_{1}L_{1}\left(\dfrac{{Q}_{1}}{Q_{2}}\right)+\left(\dfrac{\zeta}{\zeta+2K}\right)\left(\dfrac{\boldsymbol{Q}_{2}}{Q^2_{2}}\right)\cdot\left(\boldsymbol{\kappa}\cdot\boldsymbol{Q}_{2}\right)+\dashuline{\left(\dfrac{K}{\zeta+2K}\right)^2C_{2}L^2_{2}}\nonumber\\[5pt]
&+\left(\dfrac{1}{\zeta+2K}\right)\left(\dfrac{1}{Q_{2}}\right)\left(\dfrac{\boldsymbol{Q}_{2}}{Q_{2}}\right)\cdot\left(\dfrac{\partial \phi}{\partial \boldsymbol{Q}_{1}}\right)-2\left(\dfrac{1}{\zeta+2K}\right)\left(\dfrac{1}{Q_{2}}\right)\left(\dfrac{\boldsymbol{Q}_{2}}{Q_{2}}\right)\cdot\left(\dfrac{\partial \phi}{\partial \boldsymbol{Q}_{2}}\right)\nonumber\\[5pt]
&+\left(\dfrac{1}{\zeta+2K}\right)\left(\dfrac{1}{Q_{2}}\right)\left(\dfrac{\boldsymbol{Q}_{2}}{Q_{2}}\right)\cdot\left(\dfrac{\partial \phi}{\partial \boldsymbol{Q}_{3}}\right)+\left(\dfrac{\zeta}{\zeta+2K}\right)\left(\dfrac{K}{\zeta+2K}\right)L_{2}\left(\dfrac{1}{Q_{2}}\right)\left(\dfrac{\boldsymbol{Q}_3}{Q_{3}}\right)\cdot\left(\boldsymbol{\kappa}\cdot\boldsymbol{Q}_3\right)\\[5pt]
&+\left(\dfrac{1}{\zeta+2K}\right)\left(\dfrac{K}{\zeta+2K}\right)\left(\dfrac{1}{Q_{2}}\right)L_{2}\left(\dfrac{\boldsymbol{Q}_3}{Q_{3}}\right)\cdot\left(\dfrac{\partial \phi}{\partial \boldsymbol{Q}_{2}}\right)-2\left(\dfrac{1}{\zeta+2K}\right)\left(\dfrac{K}{\zeta+2K}\right)\left(\dfrac{1}{Q_{2}}\right)L_{2}\left(\dfrac{\boldsymbol{Q}_3}{Q_{3}}\right)\cdot\left(\dfrac{\partial \phi}{\partial \boldsymbol{Q}_3}\right)\nonumber
 \end{align}
  Defining
 \begin{equation}\label{eq:p_n_1_def}
 P_{2}={\left(\dfrac{K}{\zeta+2K}\right)^2L^2_{2}}
 \end{equation}
 and grouping like terms together, we have
 \begin{align}\label{eq:c2_bkwd_fin}
&C_{2}=\left(\dfrac{1}{1-P_{2}}\right)\left(\dfrac{K}{\zeta+2K}\right)C_{1}L_{1}\left(\dfrac{{Q}_{1}}{Q_{2}}\right)+\left(\dfrac{\zeta}{\zeta+2K}\right)\left(\dfrac{1}{Q_{2}}\right)\Biggl[\left(\dfrac{1}{1-P_{2}}\right)\left(\dfrac{\boldsymbol{Q}_{2}}{Q_{2}}\right)\cdot\left(\boldsymbol{\kappa}\cdot\boldsymbol{Q}_{2}\right)\nonumber\\[5pt]
&+\left(\dfrac{K}{\zeta+2K}\right)\left(\dfrac{1}{1-P_{2}}\right)L_{2}\left(\dfrac{\boldsymbol{Q}_3}{Q_{3}}\right)\cdot\left(\boldsymbol{\kappa}\cdot\boldsymbol{Q}_3\right)\Biggr]+\left(\dfrac{1}{\zeta+2K}\right)\left(\dfrac{1}{1-P_{2}}\right)\left(\dfrac{1}{Q_{2}}\right)\left(\dfrac{\boldsymbol{Q}_{2}}{Q_{2}}\right)\cdot\left(\dfrac{\partial \phi}{\partial \boldsymbol{Q}_{1}}\right)\nonumber\\[5pt]
&-\left(\dfrac{1}{\zeta+2K}\right)\left(\dfrac{1}{Q_{2}}\right)\Biggl[-\left(\dfrac{1}{1-P_{2}}\right)\left(\dfrac{K}{\zeta+2K}\right)L_{2}\left(\dfrac{\boldsymbol{Q}_3}{Q_{3}}\right)+2\left(\dfrac{1}{1-P_{2}}\right)\left(\dfrac{\boldsymbol{Q}_{2}}{Q_{2}}\right)\Biggr]\cdot\left(\dfrac{\partial \phi}{\partial \boldsymbol{Q}_{2}}\right)\nonumber\\[5pt]
&-\left(\dfrac{1}{\zeta+2K}\right)\left(\dfrac{1}{Q_{2}}\right)\Biggl[2\left(\dfrac{1}{1-P_{2}}\right)\left(\dfrac{K}{\zeta+2K}\right)L_{2}\left(\dfrac{\boldsymbol{Q}_3}{Q_{3}}\right)-\left(\dfrac{1}{1-P_{2}}\right)\left(\dfrac{\boldsymbol{Q}_{2}}{Q_{2}}\right)\Biggr]\cdot\left(\dfrac{\partial \phi}{\partial \boldsymbol{Q}_{3}}\right)
 \end{align}
 Thus we see that the backward substitution procedure results in an expression for $C_2$ that depends only on $C_1$, which may be compared against Eq~(\ref{eq:ck2}) in which the expression for $C_2$ is coupled to both $C_1$ and $C_3$. 
 Multiplying Eq.~(\ref{eq:c2_bkwd_fin}) by $L_1(Q_2/Q_1)$, 
  \begin{align}\label{eq:c2_bkwd_l_multip}
&C_{2}L_{1}\left(\dfrac{Q_{2}}{Q_{1}}\right)=\left(\dfrac{1}{1-P_{2}}\right)\left(\dfrac{K}{\zeta+2K}\right)C_{1}L^{2}_{1}\\[5pt]
&+\left(\dfrac{\zeta}{\zeta+2K}\right)\left(\dfrac{1}{Q_{1}}\right)\Biggl[\left(\dfrac{1}{1-P_{2}}\right)L_{1}\left(\dfrac{\boldsymbol{Q}_{2}}{Q_{2}}\right)\cdot\left(\boldsymbol{\kappa}\cdot\boldsymbol{Q}_{2}\right)+\left(\dfrac{K}{\zeta+2K}\right)\left(\dfrac{1}{1-P_{2}}\right)L_{1}L_{2}\left(\dfrac{\boldsymbol{Q}_3}{Q_{3}}\right)\cdot\left(\boldsymbol{\kappa}\cdot\boldsymbol{Q}_3\right)\Biggr]\nonumber\\[5pt]
&+\left(\dfrac{1}{\zeta+2K}\right)\left(\dfrac{1}{1-P_{2}}\right)L_{1}\left(\dfrac{1}{Q_{1}}\right)\left(\dfrac{\boldsymbol{Q}_{2}}{Q_{2}}\right)\cdot\left(\dfrac{\partial \phi}{\partial \boldsymbol{Q}_{1}}\right)\nonumber\\[5pt]
&-\left(\dfrac{1}{\zeta+2K}\right)\left(\dfrac{1}{Q_{1}}\right)\Biggl[-\left(\dfrac{1}{1-P_{2}}\right)\left(\dfrac{K}{\zeta+2K}\right)L_{1}L_{2}\left(\dfrac{\boldsymbol{Q}_3}{Q_{3}}\right)+2\left(\dfrac{1}{1-P_{2}}\right)L_{1}\left(\dfrac{\boldsymbol{Q}_{2}}{Q_{2}}\right)\Biggr]\cdot\left(\dfrac{\partial \phi}{\partial \boldsymbol{Q}_{2}}\right)\nonumber\\[5pt]
&-\left(\dfrac{1}{\zeta+2K}\right)\left(\dfrac{1}{Q_{1}}\right)\Biggl[2\left(\dfrac{1}{1-P_{2}}\right)\left(\dfrac{K}{\zeta+2K}\right)L_{1}L_{2}\left(\dfrac{\boldsymbol{Q}_3}{Q_{3}}\right)-\left(\dfrac{1}{1-P_{N-1}}\right)L_{1}\left(\dfrac{\boldsymbol{Q}_{2}}{Q_{2}}\right)\Biggr]\cdot\left(\dfrac{\partial \phi}{\partial \boldsymbol{Q}_{3}}\right)\nonumber
 \end{align} 
 and substituting into Eq.~(\ref{eq:ck1}), we obtain
  We thus have the following final expression for $C_{N-2}$,
 \begin{align}\label{eq:cN_2_bkwd_fin}
&C_{1}=\left(\dfrac{\zeta}{\zeta+2K}\right)\left(\dfrac{1}{Q_{1}}\right)\Biggl[\left(\dfrac{1}{1-P_{1}}\right)\left(\dfrac{\boldsymbol{Q}_{1}}{Q_{1}}\right)\cdot\left(\boldsymbol{\kappa}\cdot\boldsymbol{Q}_{1}\right)+\left(\dfrac{K}{\zeta+2K}\right)\left(\dfrac{1}{1-P_{1}}\right)\left(\dfrac{1}{1-P_{2}}\right)L_{1}\left(\dfrac{\boldsymbol{Q}_{2}}{Q_{2}}\right)\cdot\left(\boldsymbol{\kappa}\cdot\boldsymbol{Q}_{2}\right)\nonumber\\[5pt]
&+\left(\dfrac{K}{\zeta+2K}\right)^2\left(\dfrac{1}{1-P_{1}}\right)\left(\dfrac{1}{1-P_{2}}\right)L_{1}L_{2}\left(\dfrac{\boldsymbol{Q}_3}{Q_{3}}\right)\cdot\left(\boldsymbol{\kappa}\cdot\boldsymbol{Q}_3\right)\Biggr]\nonumber\\[5pt]
&-\left(\dfrac{1}{\zeta+2K}\right)\left(\dfrac{1}{Q_{1}}\right)\Biggl[-\left(\dfrac{K}{\zeta+2K}\right)\left(\dfrac{1}{1-P_{1}}\right)\left(\dfrac{1}{1-P_{2}}\right)L_{1}\left(\dfrac{\boldsymbol{Q}_{2}}{Q_{2}}\right)+2\left(\dfrac{1}{1-P_{1}}\right)\left(\dfrac{\boldsymbol{Q}_{1}}{Q_{1}}\right)\Biggr]\cdot\left(\dfrac{\partial \phi}{\partial \boldsymbol{Q}_{1}}\right)\nonumber\\[5pt]
&-\left(\dfrac{1}{\zeta+2K}\right)\left(\dfrac{1}{Q_{1}}\right)\Biggl[-\left(\dfrac{1}{1-P_{1}}\right)\left(\dfrac{1}{1-P_{2}}\right)\left(\dfrac{K}{\zeta+2K}\right)^2L_{1}L_{2}\left(\dfrac{\boldsymbol{Q}_3}{Q_{3}}\right)\nonumber\\[5pt]
&+2\left(\dfrac{1}{1-P_{1}}\right)\left(\dfrac{1}{1-P_{2}}\right)\left(\dfrac{K}{\zeta+2K}\right)L_{1}\left(\dfrac{\boldsymbol{Q}_{2}}{Q_{2}}\right)-\left(\dfrac{1}{1-P_{1}}\right)\left(\dfrac{\boldsymbol{Q}_{1}}{Q_{1}}\right)\Biggr]\cdot\left(\dfrac{\partial \phi}{\partial \boldsymbol{Q}_{2}}\right)\nonumber\\[5pt]
&-\left(\dfrac{1}{\zeta+2K}\right)\left(\dfrac{1}{Q_{1}}\right)\Biggl[2\left(\dfrac{1}{1-P_{1}}\right)\left(\dfrac{1}{1-P_{2}}\right)\left(\dfrac{K}{\zeta+2K}\right)^2L_{1}L_{2}\left(\dfrac{\boldsymbol{Q}_3}{Q_{3}}\right)\nonumber\\[5pt]
&-\left(\dfrac{1}{1-P_{1}}\right)\left(\dfrac{1}{1-P_{2}}\right)\left(\dfrac{K}{\zeta+2K}\right)L_{1}\left(\dfrac{\boldsymbol{Q}_{2}}{Q_{2}}\right)\Biggr]\cdot\left(\dfrac{\partial \phi}{\partial \boldsymbol{Q}_{3}}\right)
 \end{align}
 where
  \begin{equation}\label{eq:p_1_def}
 P_{1}=\left(\dfrac{1}{1-P_{2}}\right)\left(\dfrac{K}{\zeta+2K}\right)^2L^{2}_{1}
 \end{equation}
 
% \clearpage
% \newpage
 
 \subsection{\label{sec:comb}Combining output from forward and backward substitution steps to obtain decoupled expression}

Since $C_1$ and $C_3$ represent the terminal connector vectors in a three-spring chain, their decoupled expressions were obtained from a single iteration of the backward and forward steps, respectively. The decoupled expression for connector vectors not at the chain end [$C_2$ in the present case] is obtained by combining the output from the forward and backward substitution steps. For instance, the decoupled expression for $C_2$ may be obtained by substituting Eq.~(\ref{eq:c3_bkwd_l_multip}) into Eq.~(\ref{eq:c2_fwd_fin}). From the pattern observed in these decoupled expressions, a general expression for the decoupled $C_k$ may be written, by induction, to be
\begin{equation}\label{eq:decoup_ck}
\begin{split}
&C_k=\left(\dfrac{\zeta}{\zeta+2K}\right)\left(\dfrac{1}{Q_{k}}\right)\sum_{l=1}^{N}{\boldsymbol{\Lambda}}^{(k)}_{l}\cdot\left(\boldsymbol{\kappa}\cdot\boldsymbol{Q}_l\right)-\left(\dfrac{\zeta}{\zeta+2K}\right)\left(\dfrac{1}{Q_{k}}\right)\Biggl[\left(\dfrac{k_BT}{\zeta}\right)\sum_{l=1}^{N}{\boldsymbol{J}}^{(k)}_{l}\cdot\left(\dfrac{\partial \ln \Psi}{\partial \boldsymbol{Q}_{l}}\right)\\[5pt]
&+\left(\dfrac{1}{\zeta}\right)\sum_{l=1}^{N}{\boldsymbol{J}}^{(k)}_{l}\cdot\left(\dfrac{\partial \phi}{\partial \boldsymbol{Q}_{l}}\right)\Biggr]
\end{split}
\end{equation} 
The definitions of the vectors $\boldsymbol{\Lambda}^{(k)}_{l}$ and $\boldsymbol{J}^{(k)}_{l}$ for the $k=1$ and $k=3$ cases are readily apparent from comparison against Eqs.~(\ref{eq:c3_fwd_fin}) and~(\ref{eq:cN_2_bkwd_fin}), respectively. We provide below the definitions for the $k=2$ case
\begin{align}\label{eq:lam1_def}
\boldsymbol{\Lambda}^{(2)}_{1}=\left(\dfrac{1}{1-M_2-P_2}\right)\left(\dfrac{K}{\zeta+2K}\right)L_1\left(\dfrac{\boldsymbol{Q}_1}{Q_{1}}\right)
\end{align}
\begin{align}\label{eq:lam2_def}
\boldsymbol{\Lambda}^{(2)}_{2}=\left(\dfrac{1}{1-M_2-P_2}\right)\left(\dfrac{\boldsymbol{Q}_2}{Q_{2}}\right)
\end{align}
\begin{align}\label{eq:lam3_def}
\boldsymbol{\Lambda}^{(2)}_{3}=\left(\dfrac{1}{1-M_2-P_2}\right)\left(\dfrac{K}{\zeta+2K}\right)L_2\left(\dfrac{\boldsymbol{Q}_3}{Q_{3}}\right)
\end{align}
\begin{align}\label{eq:j1_def}
\boldsymbol{J}^{(2)}_{1}=\left(\dfrac{1}{1-M_2-P_2}\right)\Biggl[2\left(\dfrac{K}{\zeta+2K}\right)L_1\left(\dfrac{\boldsymbol{Q}_1}{Q_{1}}\right)-\left(\dfrac{\bm{Q}_{2}}{Q_2}\right)\Biggr]
\end{align}
\begin{align}\label{eq:j2_def}
\boldsymbol{J}^{(2)}_{2}=\left(\dfrac{1}{1-M_2-P_2}\right)\Biggl[2\left(\dfrac{\boldsymbol{Q}_2}{Q_{2}}\right)-\left(\dfrac{K}{\zeta+2K}\right)L_1\left(\dfrac{\boldsymbol{Q}_1}{Q_{1}}\right)-\left(\dfrac{K}{\zeta+2K}\right)L_{2}\left(\dfrac{\boldsymbol{Q}_3}{Q_{3}}\right)\Biggr]
\end{align}
\begin{align}\label{eq:j3_def}
\boldsymbol{J}^{(2)}_{3}=\left(\dfrac{1}{1-M_2-P_2}\right)\Biggl[2\left(\dfrac{K}{\zeta+2K}\right)L_2\left(\dfrac{\boldsymbol{Q}_3}{Q_{3}}\right)-\left(\dfrac{\boldsymbol{Q}_2}{Q_{2}}\right)\Biggr]
\end{align}
While the expressions for the vectors $\boldsymbol{\Lambda}^{(k)}_{l}$ and $\boldsymbol{J}^{(k)}_{l}$ have been calculated in this Appendix in a brute-force manner, to illustrate the working of the decoupling machinery, computer simulation of bead-spring-dashpot chains of arbitrary length ($N$) would require an algorithmic procedure for the construction of these terms, which has been described in detail in Sec.~{SII} of the Supplementary Material. Substituting Eq.~(\ref{eq:decoup_ck}) into Eq.~(\ref{eq:qdot_with_ck}) and simplifying results in the following general equation for the decoupled connector vector velocity
\begin{equation}\label{eq:subbing_ck}
\begin{split}
 \llbracket\dot{\boldsymbol{Q}}_{j}\rrbracket&=\boldsymbol{\kappa}\cdot\boldsymbol{Q}_j-\sum^{N}_{k=1}{\boldsymbol{A}}_{jk}\cdot\left(\dfrac{k_BT}{\zeta}\dfrac{\partial \ln \Psi}{\partial \boldsymbol{Q}_k}+\dfrac{1}{\zeta}\dfrac{\partial \phi}{\partial \boldsymbol{Q}_k}\right)-\left(\dfrac{\varphi}{1+2\varphi}\right)\sum^{N}_{k,l=1}{\boldsymbol{A}}_{jk}\cdot\boldsymbol{\alpha}_{kl}\cdot\left(\boldsymbol{\kappa}\cdot\boldsymbol{Q}_l\right)\\[5pt]
 &+\left(\dfrac{k_BT}{\zeta}\right)\left(\dfrac{\varphi}{1+2\varphi}\right)\sum^{N}_{k,l=1}{\boldsymbol{A}}_{jk}\cdot{\boldsymbol{\mu}_{kl}}\cdot\left(\dfrac{\partial \ln \Psi}{\partial \boldsymbol{Q}_{l}}\right)+\left(\dfrac{1}{\zeta}\right)\left(\dfrac{\varphi}{1+2\varphi}\right)\sum^{N}_{k,l=1}{\boldsymbol{A}}_{jk}\cdot\boldsymbol{\mu}_{kl}\cdot\left(\dfrac{\partial \phi}{\partial \boldsymbol{Q}_{l}}\right)
 \end{split}
\end{equation}
where the tensors
\begin{equation}
\begin{split}
\boldsymbol{\alpha}_{kl}&=\left(\dfrac{\boldsymbol{Q}_k}{Q_k}\right){\boldsymbol{\Lambda}}^{(k)}_{l}=\chi^{(k)}_l\left(\dfrac{\boldsymbol{Q}_k\boldsymbol{Q}_l}{Q_kQ_l}\right),\\[5pt]
\boldsymbol{\mu}_{kl}&={\left(\dfrac{\boldsymbol{Q}_k}{Q_k}\right){\boldsymbol{J}}^{(k)}_{l}}
\end{split}
\end{equation}
are used in the evaluation of the stress tensor expression as seen from Eq.~(\ref{eq:stress_tensor_dimless}). Additionally, the tensors $\bm{U}_{jk}$ and $\bm{V}_{jk}$ are defined as follows
\begin{equation}\label{eq:uv_def}
\begin{split}
\boldsymbol{U}_{jk}&=\sum_{l=1}^{N}\boldsymbol{A}_{jl}\cdot\boldsymbol{\alpha}_{lk};\quad
\boldsymbol{V}_{jk}=\sum_{l=1}^{N}\boldsymbol{A}_{jl}\cdot\boldsymbol{\mu}_{lk}
\end{split}
\end{equation}
and feature in the governing Fokker-Planck equation given by Eq.~(\ref{eq:fp_formal_dimless}).
The dimensionless Fokker-Planck equation for the three-spring chain is written in terms of collective coordinates (introduced in Eq.~(\ref{eq:coll_coord})) as 
\begin{align}\label{eq:fp_formal_dimless_collective}
\dfrac{\partial \psi^{*}}{\partial t^{*}}&=-\dfrac{\partial}{\partial \widetilde{\bm{Q}}^{*}}\cdot\Biggl\{\Biggl[\bm{\mathcal{K}}^{*}\cdot\widetilde{\bm{Q}}^{*}-\left(\dfrac{\varphi}{1+2\varphi}\right){\bm{\mathcal{U}}}\cdot\left(\bm{\mathcal{K}}^{*}\cdot\widetilde{\bm{Q}}^{*}\right)-\dfrac{1}{4}\bm{\mathcal{D}}(\varphi)\cdot\widetilde{\bm{F}}^{*\text{s}}-\dfrac{1}{4}\left(\dfrac{\varphi}{1+2\varphi}\right)\widetilde{\bm{v}}^{*}\Biggr]\psi^{*}\Biggr\}\nonumber\\[5pt]
&+\dfrac{1}{4}\dfrac{\partial}{\partial \widetilde{\bm{Q}}^{*}}\dfrac{\partial}{\partial \widetilde{\bm{Q}}^{*}}:\left[\bm{\mathcal{D}}(\varphi)\psi^{*}\right]\nonumber
\end{align}
where $\bm{\mathcal{D}}(\varphi)$ is an internal friction-dependent $3\times3$ block matrix with the following structure
\begin{equation}\label{eq:block_diff_def}
\bm{\mathcal{D}}(\varphi)=
\begin{pmatrix}
\widehat{\boldsymbol{A}}_{11}& \widehat{\boldsymbol{A}}_{12} & {\widehat{\boldsymbol{A}}_{13}}\\[5pt]
{\widehat{\boldsymbol{A}}_{21}}& {\widehat{\boldsymbol{A}}_{22}}& {\widehat{\boldsymbol{A}}_{23}}\\[5pt]
{\widehat{\boldsymbol{A}}_{31}} & {\widehat{\boldsymbol{A}}_{32}} & {\widehat{\boldsymbol{A}}_{33}}
\end{pmatrix};\quad \widehat{\boldsymbol{A}}_{jk}= \boldsymbol{A}_{jk}-\dfrac{\varphi}{1+2\varphi}\boldsymbol{V}_{jk}
\end{equation}

\end{widetext}

%\bibliographystyle{aipnum4-1}
%\bibliography{jor_iv_chain}

\end{document}

\beginsupplement
% % % %  Title and heading details
\title{Supplementary Material for: Rouse model with fluctuating internal friction}
%\date{\today}
\author{R. Kailasham}
\affiliation{IITB-Monash Research Academy, Indian Institute of Technology Bombay, Mumbai, Maharashtra -  400076, India}
\affiliation{Department of Chemistry, Indian Institute of Technology Bombay, Mumbai, Maharashtra -  400076, India}
\affiliation{Department of Chemical Engineering, Monash University,
Melbourne, VIC 3800, Australia}
\author{Rajarshi Chakrabarti}
\email{rajarshi@chem.iitb.ac.in}
\affiliation{Department of Chemistry, Indian Institute of Technology Bombay, Mumbai, Maharashtra -  400076, India}
\author{J. Ravi Prakash}
\email{ravi.jagadeeshan@monash.edu}
\affiliation{Department of Chemical Engineering, Monash University,
Melbourne, VIC 3800, Australia}

%\date{\today}

\maketitle

\section{\label{sec:intro} Introduction}

The main paper outlines a methodology for the derivation of the governing stochastic differential equations (SDE) for a free-draining bead-spring-dashpot chain model for a polymer with internal friction, and presents viscometric functions in steady simple shear and small amplitude oscillatory shear flows. The details corresponding to various aspects of the study have been presented here.

This document is structured as follows. The detailed steps for the implementation of the decoupling methodology described in the main text for the derivation of the Fokker-Planck equation for bead-spring-chains with internal friction is presented in Section~\ref{sec:decoup_method}. In Sec.~\ref{sec:db_compare}, it is shown that the governing equation for the simplest case of a dumbbell, derived using the methodology described in the present work, is identical to that derived in previous studies using an alternative approach. Sec.~\ref{sec:fdt_sat} discusses how the stochastic differential equation derived in the present work satisfies the fluctuation dissipation theorem. The positive definiteness and symmetricity of the diffusion tensor is presented in Sec.~\ref{sec:sqrt_calc}, and the topic of timestep convergence is covered in Sec.~\ref{sec:tstep_conv}. Sec.~\ref{sec:cpu_time} contains a discussion on the scaling of the computation time of the Brownian dynamics code as a function of the number of beads in the chain. The detailed steps for the derivation of the complete stress tensor expression is presented in Section~\ref{sec:stress_tens_derv}, followed by a brief discussion [Sec.~\ref{sec:var_red}] of the variance reduction algorithm used in the present work. Section~\ref{sec:app_b} presents a route for the conversion of finite continued fractions into ratios of recursive polynomial relations: Secs.~\ref{sec:fwd_poly} to ~\ref{sec:list_ident} present results for polynomial representations of continued fractions, and a list of tensor identities that is useful for the analytical calculation of the divergence terms appearing in the governing stochastic differential equation and the stress tensor expression. In Sec.~\ref{sec:ex_impl_fwd}, a detailed example of the use of recursive relations is provided. In Sec.~\ref{sec:num_div}, a comparison between the divergence calculated numerically and using recursive relations is presented. 

Summations are indicated explicitly, and the Einstein summation convention is not followed. 

\section{\label{sec:decoup_method} Implementation of decoupling methodology and derivation of governing Fokker-Planck equation}
Consider the quantity 
\begin{align}
C_k=\dfrac{\bm{Q}_k\cdot\llbracket\dot{\bm{Q}}_{k}\rrbracket}{Q^2_{k}};\quad k=1,2,\cdots,N
\end{align}
Upon taking a dot-product on both the sides of Eq.~(3) of the main text with $\bm{Q}_k/Q^2_{k}$, an equation for $C_k$ is obtained as
\begin{align}\label{eq:ck_gen_eq_intermed}
C_k&=\left(\dfrac{\bm{Q}_k}{Q^2_{k}}\right)\cdot\boldsymbol{\kappa}\cdot\bm{Q}_k-\left(\dfrac{k_BT}{\zeta}\right)\left(\dfrac{\bm{Q}_k}{Q^2_{k}}\right)\cdot\left[-\dfrac{\partial \ln \Psi}{\partial \bm{Q}_{k-1}}+2\dfrac{\partial \ln \Psi}{\partial \bm{Q}_k}-\dfrac{\partial \ln \Psi}{\partial \bm{Q}_{k+1}}\right]\nonumber\\[5pt]
&-\dfrac{1}{\zeta}\left(\dfrac{\bm{Q}_k}{Q^2_{k}}\right)\cdot\left[-\dfrac{\partial \phi}{\partial \bm{Q}_{k-1}}+2\dfrac{\partial \phi}{\partial \bm{Q}_k}-\dfrac{\partial \phi}{\partial \bm{Q}_{k+1}}\right]\\[5pt]
 &+\left(\dfrac{K}{\zeta}\right)C_{k-1}L_{k-1}\left(\dfrac{{Q}_{k-1}}{Q_{k}}\right)-\left(\dfrac{2K}{\zeta}\right)C_{k}+\left(\dfrac{K}{\zeta}\right)C_{k+1}L_{k}\left(\dfrac{{Q}_{k+1}}{Q_{k}}\right)\nonumber
 \end{align}
 where
 \begin{align}
L_{k}\equiv\cos\theta_k=\dfrac{\bm{Q}_k\cdot\bm{Q}_{k+1}}{Q_kQ_{k+1}}
\end{align}
Upon grouping together terms containing $C_k$ on the RHS and simplifying, the generating equation is obtained as
\begin{align}\label{eq:ck_gen_eq_final}
C_k&=\left(\dfrac{\zeta}{\zeta+2K}\right)\left(\dfrac{\bm{Q}_k}{Q^2_{k}}\right)\cdot\left(\boldsymbol{\kappa}\cdot\bm{Q}_k\right)-{\left(\dfrac{k_BT}{\zeta+2K}\right)\left(\dfrac{\bm{Q}_k}{Q^2_{k}}\right)\cdot\left[-\dfrac{\partial \ln \Psi}{\partial \bm{Q}_{k-1}}+2\dfrac{\partial \ln \Psi}{\partial \bm{Q}_k}-\dfrac{\partial \ln \Psi}{\partial \bm{Q}_{k+1}}\right]}\nonumber\\[5pt]
&-\left(\dfrac{1}{\zeta+2K}\right)\left(\dfrac{\bm{Q}_k}{Q^2_{k}}\right)\cdot\Biggl[-\dfrac{\partial \phi}{\partial \bm{Q}_{k-1}}+2\dfrac{\partial \phi}{\partial \bm{Q}_k}-\dfrac{\partial \phi}{\partial \bm{Q}_{k+1}}\Biggr]\\[5pt]
&+\left(\dfrac{K}{\zeta+2K}\right)\left[C_{k-1}L_{k-1}\left(\dfrac{{Q}_{k-1}}{Q_{k}}\right)+C_{k+1}L_{k}\left(\dfrac{{Q}_{k+1}}{Q_{k}}\right)\right]\nonumber
 \end{align}
Eq.~(\ref{eq:ck_gen_eq_final}) is then subjected to the forward and backward substitution schema discussed previously, to obtain a decoupled expression for $C_k$.  
In the forward substitution step, the equation for $C_{k}$ is substituted into that for $C_{k+1}$, starting with $k=1$. The general expression obtained at the end of this step can be shown, by induction, to be
\begin{align}\label{eq:fwd_sub}
&C_k\left(1-M_{k}\right)=\left(\dfrac{K}{\zeta+2K}\right)C_{k+1}L_{k}\left(\dfrac{{Q}_{k+1}}{Q_{k}}\right)+\left(\dfrac{\zeta}{\zeta+2K}\right)\left(\dfrac{1}{Q_{k}}\right)\sum_{l=1}^{k}\boldsymbol{\Gamma}^{(k)}_{l}\cdot\left(\boldsymbol{\kappa}\cdot\bm{Q}_l\right)\nonumber\\[10pt]
&+\left(1-\delta_{kN}\right)\left(\dfrac{k_BT}{\zeta+2K}\right)\left(\dfrac{1}{Q_{k}}\right)\left(\dfrac{\bm{Q}_k}{{Q}_k}\right)\cdot\left(\dfrac{\partial \ln \Psi}{\partial \bm{Q}_{k+1}}\right)\nonumber\\[5pt]
&+\left(1-\delta_{kN}\right)\left(\dfrac{1}{\zeta+2K}\right)\left(\dfrac{1}{Q_{k}}\right)\left(\dfrac{\bm{Q}_k}{{Q}_k}\right)\cdot\left(\dfrac{\partial \phi}{\partial \bm{Q}_{k+1}}\right)\\[10pt]
&-\left(\dfrac{k_BT}{\zeta+2K}\right)\left(\dfrac{1}{Q_{k}}\right)\sum_{l=1}^{k}\bm{E}^{(k)}_{l}\cdot\left(\dfrac{\partial \ln \Psi}{\partial \bm{Q}_{l}}\right)-\left(\dfrac{1}{\zeta+2K}\right)\left(\dfrac{1}{Q_{k}}\right)\sum_{l=1}^{k}\bm{E}^{(k)}_{l}\cdot\left(\dfrac{\partial \phi}{\partial \bm{Q}_{l}}\right)\nonumber
\end{align}
where  
\begin{align}\label{eq:m_def}
M_{k}=\left(\dfrac{K}{\zeta+2K}\right)^2\left(\dfrac{L^2_{k-1}}{1-M_{k-1}}\right); \quad \text {with} \quad M_{1}=0,
\end{align}
\begin{align}\label{eq:gamma_def}
\boldsymbol{\Gamma}^{(k)}_l=\left(\dfrac{K}{\zeta+2K}\right)^{k-l}\left[\prod_{i=l}^{k-1}\left(\dfrac{1}{1-M_i}\right)L_{i}\right]\left(\dfrac{\bm{Q}_l}{Q_{l}}\right),
\end{align}
and
\begin{equation}
\bm{E}^{(k)}_l=2\boldsymbol{\Gamma}^{(k)}_l-\boldsymbol{\Gamma}^{(k)}_{l-1}-\boldsymbol{\Gamma}^{(k)}_{l+1}
\end{equation}
Here, $\boldsymbol{\Gamma}^{(k)}_l$ is defined only for $0<l\leq k\leq N$, and is set to zero otherwise.
The backward substitution step involves plugging in the equation for $C_{k}$ into $C_{k-1}$, starting with $k=N$. The general expression at the end of this step can be shown, by induction, to be
\begin{align}\label{eq:bkwd_sub_intermed}
&C_kL_{k-1}\left(\dfrac{{Q}_{k}}{Q_{k-1}}\right)=\left(\dfrac{K}{\zeta+2K}\right)\left(\dfrac{1}{1-P_k}\right)C_{k-1}L^{2}_{k-1}+\left(\dfrac{\zeta}{\zeta+2K}\right)\left(\dfrac{1}{Q_{k-1}}\right)\sum_{l=k}^{N}\widetilde{\boldsymbol{\rho}}^{(k)}_{l}\cdot\left(\boldsymbol{\kappa}\cdot\bm{Q}_l\right)\nonumber\\[10pt]
&+\left(\dfrac{k_BT}{\zeta+2K}\right)\left(\dfrac{1}{1-P_k}\right)L_{k-1}\left(\dfrac{1}{Q_{k-1}}\right)\left(\dfrac{\bm{Q}_k}{{Q}_k}\right)\cdot\left(\dfrac{\partial \ln \Psi}{\partial \bm{Q}_{k-1}}\right)\nonumber\\[5pt]
&+\left(\dfrac{1}{\zeta+2K}\right)\left(\dfrac{1}{1-P_k}\right)L_{k-1}\left(\dfrac{1}{Q_{k-1}}\right)\left(\dfrac{\bm{Q}_k}{{Q}_k}\right)\cdot\left(\dfrac{\partial \phi}{\partial \bm{Q}_{k-1}}\right)\nonumber\\[10pt]
&-\left(\dfrac{k_BT}{\zeta+2K}\right)\left(\dfrac{1}{Q_{k-1}}\right)\sum_{l=k}^{N}\bm{\widetilde{G}}^{(k)}_{l}\cdot\left(\dfrac{\partial \ln \Psi}{\partial \bm{Q}_{l}}\right)-\left(\dfrac{1}{\zeta+2K}\right)\left(\dfrac{1}{Q_{k-1}}\right)\sum_{l=k}^{N}\bm{\widetilde{G}}^{(k)}_{l}\cdot\left(\dfrac{\partial \phi}{\partial \bm{Q}_{l}}\right)
\end{align}
where 
\begin{align}\label{eq:p_def}
P_{k}=\left(\dfrac{K}{\zeta+2K}\right)^2\left(\dfrac{L^{2}_{k}}{1-P_{k+1}}\right); \quad \text {with} \quad P_{N}=0,
\end{align}
and 
 \begin{align}\label{eq:rho_def}
\widetilde{\boldsymbol{\rho}}^{(k)}_{l}=\left(\dfrac{K}{\zeta+2K}\right)^{l-k}\left[\prod_{i=k}^{l}\left(\dfrac{1}{1-P_i}\right)L_{i-1}\right]\left(\dfrac{\bm{Q}_l}{Q_{l}}\right)
\end{align}
The vector $\bm{\widetilde{G}}^{(k)}_l$ appearing in Eq.~(\ref{eq:bkwd_sub_intermed}) is constructed using a slightly elaborate procedure. It is useful to first consider a block Rouse matrix, $\bm{R}$, of size $\Upsilon\times\Upsilon$, where $\Upsilon=\left(N-k\right)+1$, whose each element is a $3\times3$ matrix, and has the following structure, 
\begin{equation}\label{eq:rouse_block_def}
\bm{R} = 
\begin{pmatrix}
2\bm{\delta} & -\bm{\delta} & \bm{0} & \cdots & {} & {}\\
-\bm{\delta} & 2\bm{\delta} & -\bm{\delta} & \bm{0} & \cdots & {} \\
\bm{0} & -\bm{\delta} & 2\bm{\delta} & -\bm{\delta} &\cdots & {} \\
\vdots  & \vdots & \vdots & {}& {} & {} \\
\bm{0} & \bm{0}& \cdots & {} & -\bm{\delta}  & 2\bm{\delta}
\end{pmatrix}
\end{equation}
and 
define the intermediate quantity, 
\begin{align}\label{eq:ytilde_def}
\bm{\widetilde{Y}}^{(k)}_{s}=&\left(\dfrac{K}{\zeta+2K}\right)^{s-1}\left[\prod_{i=k}^{k+s-1}\left(\dfrac{1}{1-P_i}\right)L_{i-1}\right]\left(\dfrac{\bm{Q}_{s+k-1}}{Q_{s+k-1}}\right)
\end{align}
which is then used to populate a block matrix, $\widehat{\boldsymbol{\Theta}}^{(k)}$, of size $\Upsilon\times\Upsilon$ that has the following structure
\begin{equation}\label{eq:theta_hat_def}
\widehat{\boldsymbol{\Theta}}^{(k)} = 
\begin{pmatrix}
\bm{\widetilde{Y}}^{(k)}_{1} & \bm{\widetilde{Y}}^{(k)}_{1}  & \bm{0} & \cdots & {} & {}\\
\bm{\widetilde{Y}}^{(k)}_{2}   & \bm{\widetilde{Y}}^{(k)}_{2}   &\bm{\widetilde{Y}}^{(k)}_{2}   &\bm{0} & \cdots & {} \\
\bm{0} & \bm{\widetilde{Y}}^{(k)}_{3}  & \bm{\widetilde{Y}}^{(k)}_{3} & \bm{\widetilde{Y}}^{(k)}_{3}  &\cdots & {} \\
\vdots  & \vdots & \vdots & {}& {} & {} \\
\bm{0} & \bm{0}& \cdots & {} & \bm{\widetilde{Y}}^{(k)}_{\Upsilon} & \bm{\widetilde{Y}}^{(k)}_{\Upsilon} 
\end{pmatrix}
\end{equation}
We next consider the block matrix $\bm{Z}^{(k)}$ constructed from $\bm{R}$ and $\widehat{\boldsymbol{\Theta}}^{(k)}$, such that $\bm{Z}^{(k)}=\bm{R}\cdot\widehat{\boldsymbol{\Theta}}^{(k)}$. Now, $\bm{\widetilde{G}}^{(k)}_{k+m}=\bm{Z}^{(k)}_{m+1,m+1}$, which is the $\left(m+1\right)^{\text{th}}$ diagonal element of $\bm{Z}^{(k)}$. A change of variable, $k\to\left(k+1\right)$, in Eq.~(\ref{eq:bkwd_sub_intermed}) permits us to write
\begin{align}\label{eq:bkwd_sub}
&C_{k+1}L_{k}\left(\dfrac{{Q}_{k+1}}{Q_{k}}\right)=\left(\dfrac{K}{\zeta+2K}\right)\left(\dfrac{1}{1-P_{k+1}}\right)C_{k}L^{2}_{k}+\left(\dfrac{\zeta}{\zeta+2K}\right)\left(\dfrac{1}{Q_{k}}\right)\sum_{l=k+1}^{N}\widetilde{\boldsymbol{\rho}}^{(k+1)}_{l}\cdot\left(\boldsymbol{\kappa}\cdot\bm{Q}_l\right)\nonumber\\[10pt]
&+\left(1-\delta_{kN}\right)\left(\dfrac{1}{1-P_{k+1}}\right)L_{k}\left(\dfrac{1}{Q_{k}}\right)\Biggl[\left(\dfrac{k_BT}{\zeta+2K}\right)\left(\dfrac{\bm{Q}_{k+1}}{{Q}_{k+1}}\right)\cdot\left(\dfrac{\partial \ln \Psi}{\partial \bm{Q}_{k}}\right)\nonumber\\[5pt]
&+\left(\dfrac{1}{\zeta+2K}\right)\left(\dfrac{\bm{Q}_{k+1}}{{Q}_{k+1}}\right)\cdot\left(\dfrac{\partial \phi}{\partial \bm{Q}_{k}}\right)\Biggr]-\left(\dfrac{k_BT}{\zeta+2K}\right)\left(\dfrac{1}{Q_{k}}\right)\sum_{l=k+1}^{N}\bm{\widetilde{G}}^{(k+1)}_{l}\cdot\left(\dfrac{\partial \ln \Psi}{\partial \bm{Q}_{k}}\right)\\[10pt]
&-\left(\dfrac{1}{\zeta+2K}\right)\left(\dfrac{1}{Q_{k}}\right)\sum_{l=k+1}^{N}\bm{\widetilde{G}}^{(k+1)}_{l}\cdot\left(\dfrac{\partial \phi}{\partial \bm{Q}_{k}}\right)\nonumber
\end{align}
Lastly, by inserting the backward substitution result, Eq.~(\ref{eq:bkwd_sub}), into the equation obtained from forward substitution, Eq.~(\ref{eq:fwd_sub}), the decoupled expression is obtained as 
\begin{equation}\label{eq:decoup_init}
\begin{split}
&C_k\left(1-M_k-P_k\right)=\left(\dfrac{\zeta}{\zeta+2K}\right)\left(\dfrac{1}{Q_{k}}\right)\sum_{l=1}^{N}\widehat{\boldsymbol{\Lambda}}^{(k)}_{l}\cdot\left(\boldsymbol{\kappa}\cdot\bm{Q}_l\right)\\[5pt]
&-\left(\dfrac{1}{Q_{k}}\right)\Biggl[\left(\dfrac{k_BT}{\zeta+2K}\right)\sum_{l=1}^{N}\widehat{\bm{J}}^{(k)}_{l}\cdot\left(\dfrac{\partial \ln \Psi}{\partial \bm{Q}_{l}}\right)+\left(\dfrac{1}{\zeta+2K}\right)\sum_{l=1}^{N}\widehat{\bm{J}}^{(k)}_{l}\cdot\left(\dfrac{\partial \phi}{\partial \bm{Q}_{l}}\right)\Biggr]
\end{split}
\end{equation}
with
 \begin{align}\label{eq:flow_coeffdef}
\widehat{\boldsymbol{\Lambda}}^{(k)}_{l}= \left\{
\begin{array}{ll}
       \boldsymbol{\Gamma}^{(k)}_l; &  l<k\\[15pt]
       \left(\dfrac{\bm{Q}_k}{{Q}_k}\right); & l=k\\[15pt]
       \left(\dfrac{K}{\zeta+2K}\right)\widetilde{\boldsymbol{\rho}}^{(k+1)}_{l} ; & l>k
\end{array} 
\right. 
\end{align}
and
 \begin{align}\label{eq:stiff_coeffdef}
\widehat{\bm{J}}^{(k)}_{l}= \left\{
\begin{array}{ll}
       \bm{E}^{(k)}_l; &  l<k\\[15pt]
       \bm{E}^{(k)}_l - \left(1-\delta_{kN}\right)\left(\dfrac{1}{1-P_{k+1}}\right)\left(\dfrac{K}{\zeta+2K}\right)L_{k}\left(\dfrac{\bm{Q}_{k+1}}{{Q}_{k+1}}\right); & l=k\\[15pt]
        \left(1-\delta_{kN}\right)\left[\left(\dfrac{K}{\zeta+2K}\right)\widetilde{\bm{G}}^{(k+1)}_{l}-\left(\dfrac{\bm{Q}_{k}}{{Q}_{k}}\right)\right] ; & l=k+1\\[15pt]
       \left(\dfrac{K}{\zeta+2K}\right)\widetilde{\bm{G}}^{(k+1)}_{l}; & l>\left(k+1\right)
\end{array} 
\right. 
\end{align}
The procedure for the construction of $\bm{\widetilde{G}}^{(k+1)}_{l}$ which appears in Eq.~(\ref{eq:stiff_coeffdef}) is fairly similar to that described in Eqs.~(\ref{eq:rouse_block_def})\textendash(\ref{eq:theta_hat_def}) for the construction of $\bm{\widetilde{G}}^{(k)}_l$, with the only caveat that the size of the block matrices, $\bm{R}$ and the $\widehat{\boldsymbol{\Theta}}^{(k+1)}$, remain $\Upsilon\times\Upsilon$, where $\Upsilon=\left(N-k\right)+1$. This procedure for the construction of $\boldsymbol{\widetilde{G}}^{(k+1)}_{l}$ allows for an easier mathematical description. The implementation of this calculation in the computer code, however, follows the formula:
\begin{equation}\label{eq:gdef_simple}
{\boldsymbol{\widetilde{G}}}^{(k+1)}_{k+1+m}=2{\boldsymbol{\widetilde{Y}}}^{(k+1)}_{m+1}-{\boldsymbol{\widetilde{Y}}}^{(k+1)}_{m}-{\boldsymbol{\widetilde{Y}}}^{(k+1)}_{m+2};\quad 1\leq\,k\leq\,N   
\end{equation}
While Eq.~(\ref{eq:gdef_simple}) appears to be relatively simple in comparison to the detailed procedure described above, its implementation requires a cascade of ``if-else'' statements to ensure a valid value for each term on its RHS. For example, we see from Eq.~(\ref{eq:ytilde_def}) that ${\boldsymbol{\widetilde{Y}}}^{(k+1)}_{m}$ has a non-zero value only if: (a) $m\geq 1$, and (b) $1\leq\left(m+k\right)\leq N$, and is set to zero otherwise.
Defining
\begin{equation}\label{eq:cosmetic}
\begin{split}
{\boldsymbol{\Lambda}}^{(k)}_{l}&=\left(\dfrac{1}{1-M_k-P_k}\right)\widehat{\boldsymbol{\Lambda}}^{(k)}_{l}\\[5pt]
{\bm{J}}^{(k)}_{l}&=\left(\dfrac{1}{1-M_k-P_k}\right)\widehat{\bm{J}}^{(k)}_{l}
\end{split}
\end{equation}
Eq.~(\ref{eq:decoup_init}) may be rewritten as
\begin{equation}\label{eq:decoup}
\begin{split}
&C_k=\left(\dfrac{1}{1+2\varphi}\right)\left(\dfrac{1}{Q_{k}}\right)\sum_{l=1}^{N}{\boldsymbol{\Lambda}}^{(k)}_{l}\cdot\left(\boldsymbol{\kappa}\cdot\bm{Q}_l\right)-\left(\dfrac{1}{1+2\varphi}\right)\left(\dfrac{1}{Q_{k}}\right)\Biggl[\left(\dfrac{k_BT}{\zeta}\right)\sum_{l=1}^{N}{\bm{J}}^{(k)}_{l}\cdot\left(\dfrac{\partial \ln \Psi}{\partial \bm{Q}_{l}}\right)\\[5pt]
&+\left(\dfrac{1}{\zeta}\right)\sum_{l=1}^{N}{\bm{J}}^{(k)}_{l}\cdot\left(\dfrac{\partial \phi}{\partial \bm{Q}_{l}}\right)\Biggr]
\end{split}
\end{equation}
to give the decoupled expression for $C_k$.
Noting that the equation for the momentum-averaged velocity of the $j^{\text{th}}$ connector vector in a freely-draining chain is given by
 \begin{align}\label{eq:qdot_with_ck}
 \llbracket\dot{\bm{Q}}_{j}\rrbracket&=\boldsymbol{\kappa}\cdot\bm{Q}_j-\sum^{N}_{k=1}{\bm{A}}_{jk}\cdot\left(\dfrac{k_BT}{\zeta}\dfrac{\partial \ln \Psi}{\partial \bm{Q}_k}+\dfrac{1}{\zeta}\dfrac{\partial \phi}{\partial \bm{Q}_k}\right)-\varphi \sum^{N}_{k=1}{\bm{A}}_{jk}\cdot {\bm{Q}_k}C_k
 \end{align}
 where $\bm{A}_{jk}=A_{jk}\boldsymbol{\delta}$ with $A_{jk}$ denoting the elements of the Rouse matrix~\citep{Bird1987b}, substituting the expression for $C_k$ from Eq.~(\ref{eq:decoup}) into Eq.~(\ref{eq:qdot_with_ck}) gives,  
\begin{align}\label{eq:subbing_ck}
 &\llbracket\dot{\bm{Q}}_{j}\rrbracket=\boldsymbol{\kappa}\cdot\bm{Q}_j-\sum^{N}_{k=1}{\bm{A}}_{jk}\cdot\left(\dfrac{k_BT}{\zeta}\dfrac{\partial \ln \Psi}{\partial \bm{Q}_k}+\dfrac{1}{\zeta}\dfrac{\partial \phi}{\partial \bm{Q}_k}\right)-\left(\dfrac{\varphi}{1+2\varphi}\right)\sum^{N}_{k,l=1}{\bm{A}}_{jk}\cdot\dashuline{\left(\dfrac{\bm{Q}_k}{Q_k}\right){\boldsymbol{\Lambda}}^{(k)}_{l}}\cdot\left(\boldsymbol{\kappa}\cdot\bm{Q}_l\right)\nonumber\\[5pt]
 &+\left(\dfrac{k_BT}{\zeta}\right)\left(\dfrac{\varphi}{1+2\varphi}\right)\sum^{N}_{k,l=1}{\bm{A}}_{jk}\cdot\dotuline{\left(\dfrac{\bm{Q}_k}{Q_k}\right){\bm{J}}^{(k)}_{l}}\cdot\left(\dfrac{\partial \ln \Psi}{\partial \bm{Q}_{l}}\right)\nonumber\\[5pt]
 &+\left(\dfrac{1}{\zeta}\right)\left(\dfrac{\varphi}{1+2\varphi}\right)\sum^{N}_{k,l=1}{\bm{A}}_{jk}\cdot\left(\dfrac{\bm{Q}_k}{Q_k}\right){\bm{J}}^{(k)}_{l}\cdot\left(\dfrac{\partial \phi}{\partial \bm{Q}_{l}}\right)
\end{align}
The dashed and dotted underlined terms in Eq.~(\ref{eq:subbing_ck}) are denoted by the tensors $\boldsymbol{\alpha}_{kl}$ and $\boldsymbol{\mu}_{kl}$, respectively. The following simplifications in notation are introduced before proceeding to the next step: 
\begin{equation}\label{eq:uv_def}
\begin{split}
\boldsymbol{\alpha}_{kl}&=\left(\dfrac{\bm{Q}_k}{Q_k}\right){\boldsymbol{\Lambda}}^{(k)}_{l};\quad
\boldsymbol{\mu}_{kl}={\left(\dfrac{\bm{Q}_k}{Q_k}\right){\bm{J}}^{(k)}_{l}}\\[5pt]
\bm{U}_{jl}&=\sum_{k=1}^{N}\bm{A}_{jk}\cdot\boldsymbol{\alpha}_{kl};\quad
\bm{V}_{jl}=\sum_{k=1}^{N}\bm{A}_{jk}\cdot\boldsymbol{\mu}_{kl}
\end{split}
\end{equation}
Using Eqs.~(\ref{eq:flow_coeffdef}) and ~(\ref{eq:cosmetic}), an alternate definition for $\boldsymbol{\alpha}_{kl}$ that is more convenient for the construction of the stress tensor expression, may be constructed as,
\begin{align}\label{eq:alpha_def}
\boldsymbol{\alpha}_{kl}=\chi^{(k)}_l\left(\dfrac{\bm{Q}_k\bm{Q}_l}{Q_kQ_l}\right)
\end{align}
where
 \begin{align}\label{eq:chi_def}
\chi^{(k)}_{l}= \left\{
\begin{array}{ll}
       \left(\dfrac{1}{1-M_k-P_k}\right)\left(\dfrac{K}{\zeta+2K}\right)^{k-l}\left[\prod_{i=l}^{k-1}\left(\dfrac{1}{1-M_i}\right)L_{i}\right]; &  l<k\\[15pt]
       \left(\dfrac{1}{1-M_k-P_k}\right); & l=k\\[15pt]
       \left(\dfrac{1}{1-M_k-P_k}\right)\left(\dfrac{K}{\zeta+2K}\right)^{l-k}\left[\prod_{i=k+1}^{l}\left(\dfrac{1}{1-P_i}\right)L_{i-1}\right]; & l>k
\end{array} 
\right. 
\end{align}
The equation for $ \llbracket\dot{\bm{Q}}_{j}\rrbracket$ then becomes
\begin{equation}\label{eq:subbing_ck_simp1}
\begin{split}
\llbracket\dot{\bm{Q}}_{j}\rrbracket&=\boldsymbol{\kappa}\cdot\bm{Q}_j-\sum^{N}_{k=1}{\bm{A}}_{jk}\cdot\left(\dfrac{k_BT}{\zeta}\dfrac{\partial \ln \Psi}{\partial \bm{Q}_k}+\dfrac{1}{\zeta}\dfrac{\partial \phi}{\partial \bm{Q}_k}\right)-\left(\dfrac{\varphi}{1+2\varphi}\right)\Biggl[\sum^{N}_{l=1}{\bm{U}}_{jl}\cdot\left(\boldsymbol{\kappa}\cdot\bm{Q}_l\right)\\[5pt]
 &-\left(\dfrac{k_BT}{\zeta}\right)\sum^{N}_{l=1}{\bm{V}}_{jl}\cdot\left(\dfrac{\partial \ln \Psi}{\partial \bm{Q}_{l}}\right)+\left(\dfrac{1}{\zeta}\right)\sum^{N}_{l=1}{\bm{V}}_{jl}\cdot\left(\dfrac{\partial \phi}{\partial \bm{Q}_{l}}\right)\Biggr]
\end{split}
\end{equation}
which is simplified to
\begin{equation}\label{eq:subbing_ck_simp3}
\begin{split}
 \llbracket\dot{\bm{Q}}_{j}\rrbracket&=\Biggl[\boldsymbol{\kappa}\cdot\bm{Q}_j-\left(\dfrac{\varphi}{1+2\varphi}\right)\sum^{N}_{k=1}{\bm{U}}_{jk}\cdot\left(\boldsymbol{\kappa}\cdot\bm{Q}_k\right)\Biggr]-\dfrac{1}{\zeta}\sum^{N}_{k=1}\left(\bm{A}_{jk}-\dfrac{\varphi}{1+2\varphi}\bm{V}_{jk}\right)\cdot\left(\dfrac{\partial \phi}{\partial \bm{Q}_{k}}\right)\\[5pt]
&-\dfrac{k_BT}{\zeta}\sum^{N}_{k=1}\left(\bm{A}_{jk}-\dfrac{\varphi}{1+2\varphi}\bm{V}_{jk}\right)\cdot\left(\dfrac{\partial \ln \Psi}{\partial \bm{Q}_{k}}\right)
\end{split}
\end{equation}   
As the next step, the expression for $ \llbracket\dot{\bm{Q}}_{j}\rrbracket$ will be substituted into the equation of continuity, recognizing that the homogeneous flow profile allows one to write the continuity equation solely in terms of the relative coordinates, $\bm{Q}_j$. This means that the distribution function $\Psi\left(\bm{r}_c,\bm{Q}_1,\bm{Q}_2,...\bm{Q}_N\right)$ can be replaced by $\psi\left(\bm{Q}_1,\bm{Q}_2,...\bm{Q}_N\right)$, and we have
 \begin{equation}\label{eq:cont_eq_fp}
 \begin{split}
 &\dfrac{\partial \psi}{\partial t}=-\sum_{j=1}^{N}\dfrac{\partial}{\partial \bm{Q}_j}\cdot\left\{\llbracket\dot{\bm{Q}}_{j}\rrbracket\psi\right\}=-\sum_{j=1}^{N}\dfrac{\partial}{\partial \bm{Q}_j}\cdot\Biggl\{\Biggl[\boldsymbol{\kappa}\cdot\bm{Q}_j-\left(\dfrac{\varphi}{1+2\varphi}\right)\sum^{N}_{k=1}{\bm{U}}_{jk}\cdot\left(\boldsymbol{\kappa}\cdot\bm{Q}_k\right)\\
 &-\dfrac{1}{\zeta}\sum^{N}_{k=1}\left(\bm{A}_{jk}-\dfrac{\varphi}{1+2\varphi}\bm{V}_{jk}\right)\cdot\left(\dfrac{\partial \phi}{\partial \bm{Q}_{k}}\right)\Biggr]\psi\Biggr\}+\dfrac{k_BT}{\zeta}\sum^{N}_{j,k=1}\dfrac{\partial}{\partial \bm{Q}_j}\cdot\left(\bm{A}_{jk}-\dfrac{\varphi}{1+2\varphi}\bm{V}_{jk}\right)\cdot\dfrac{\partial \psi}{\partial \bm{Q}_{k}}
 \end{split}
  \end{equation}
The last term on the RHS of Eq.~(\ref{eq:cont_eq_fp}) must be processed further, in order to render the Fokker-Planck equation amenable to the It\^o interpretation~\cite{Ottinger1996}. Invoking the identity 
\begin{align}
\dfrac{\partial}{\partial \bm{Q}_j}\cdot\boldsymbol{\Xi}_{jl}\cdot\dfrac{\partial h}{\partial \bm{Q}_l}&=\dfrac{\partial}{\partial \bm{Q}_j}\dfrac{\partial}{\partial \bm{Q}_l}:\left[\boldsymbol{\Xi}_{jl}^{T}h\right]-\dfrac{\partial}{\partial \bm{Q}_j}\cdot\left[h\dfrac{\partial}{\partial \bm{Q}_l}\cdot\boldsymbol{\Xi}_{jl}^{T}\right],
\end{align}
where $\boldsymbol{\Xi}_{jl}$ is an arbitrary tensor, and $h$ is a scalar, we may write
\begin{equation}\label{eq:dot_use}
\begin{split}
\dfrac{\partial}{\partial \bm{Q}_j}\cdot\left(\bm{A}_{jk}-\dfrac{\varphi}{1+2\varphi}\bm{V}_{jk}\right)\cdot\dfrac{\partial \psi}{\partial \bm{Q}_{k}}&=\dfrac{\partial}{\partial \bm{Q}_j}\dfrac{\partial}{\partial \bm{Q}_{k}}:\Biggl[\Biggl(\bm{A}_{jk}-\dfrac{\varphi}{1+2\varphi}\bm{V}_{jk}\Biggr)^{T}\psi\Biggr]\\[5pt]
&-\dfrac{\partial}{\partial \bm{Q}_j}\cdot\Biggl\{\psi \dfrac{\partial}{\partial \bm{Q}_{k}}\cdot\Biggl(\bm{A}_{jk}-\dfrac{\varphi}{1+2\varphi}\bm{V}_{jk}\Biggr)^{T}\Biggr\}
\end{split}
\end{equation}
Recognizing that the Rouse matrix, $\bm{A}_{jk}$ is composed only of constant values independent of the chain connector vectors, we have $\left(\partial/\partial \bm{Q}_k\right)\cdot\bm{A}^{T}_{jk}=0$, and can simplify Eq.~(\ref{eq:dot_use}) to the form
\begin{align*}
\dfrac{\partial}{\partial \bm{Q}_j}\cdot\left(\bm{A}_{jk}-\dfrac{\varphi}{1+2\varphi}\bm{V}_{jk}\right)\cdot\dfrac{\partial \psi}{\partial \bm{Q}_{k}}&=\dfrac{\partial}{\partial \bm{Q}_j}\dfrac{\partial}{\partial \bm{Q}_{k}}:\left[\left(\bm{A}_{jk}-\dfrac{\varphi}{1+2\varphi}\bm{V}_{jk}\right)^{T}\psi\right]\nonumber\\[5pt]
&+\left(\dfrac{\varphi}{1+2\varphi}\right)\dfrac{\partial}{\partial \bm{Q}_j}\cdot\left\{\psi \dfrac{\partial}{\partial \bm{Q}_{k}}\cdot\bm{V}_{jk}^{T}\right\},
\end{align*}
which, upon substitution into Eq.~(\ref{eq:cont_eq_fp}), results in
\begin{align}\label{eq:fp_formal}
\dfrac{\partial \psi}{\partial t}&=-\sum_{j=1}^{N}\dfrac{\partial}{\partial \bm{Q}_j}\cdot\Biggl\{\Biggl[\boldsymbol{\kappa}\cdot\bm{Q}_j-\left(\dfrac{\varphi}{1+2\varphi}\right)\sum^{N}_{k=1}{\bm{U}}_{jk}\cdot\left(\boldsymbol{\kappa}\cdot\bm{Q}_k\right)\nonumber\\[5pt]
&-\dfrac{1}{\zeta}\sum^{N}_{k=1}\left(\bm{A}_{jk}-\dfrac{\varphi}{1+2\varphi}\bm{V}_{jk}\right)\cdot\left(\dfrac{\partial \phi}{\partial \bm{Q}_{k}}\right)-\left(\dfrac{k_BT}{\zeta}\right)\left(\dfrac{\varphi}{1+2\varphi}\right)\sum^{N}_{k=1}{\dfrac{\partial}{\partial \bm{Q}_{k}}\cdot\bm{V}_{jk}^{T}}\Biggr]\psi\Biggr\}\\[5pt]
 &+\dfrac{k_BT}{\zeta}\sum^{N}_{j,k=1}\dfrac{\partial}{\partial \bm{Q}_j}\dfrac{\partial}{\partial \bm{Q}_{k}}:\left[\left(\bm{A}_{jk}-\dfrac{\varphi}{1+2\varphi}\bm{V}_{jk}\right)^{T}\psi\right]\nonumber
\end{align}
Defining
\begin{equation}\label{eq:a_hat}
\widehat{\bm{A}}_{jk}=\bm{A}_{jk}-\dfrac{\varphi}{1+2\varphi}\bm{V}_{jk}
\end{equation}
and non-dimensionalizing Eq.~(\ref{eq:fp_formal}) using the length- and time-scales, $l_{H}$ and $\lambda_{H}$, the dimensionless Fokker-Planck equation may be written as shown in Eq.~(6) of the main text. 

\subsection{\label{sec:db_compare} Comparison of dumbbell equations to prior work}

Due to the absence of a coupling between connector vector velocities for the simple case of a dumbbell with internal viscosity, the governing the governing Fokker-Planck equation can be derived, using a procedure different from that outlined in the present document, as shown in Refs.~\citenum{Hua1995,Kailasham2018}, to be:
\begin{equation}\label{eq:db_prior}
\begin{split}
\dfrac{\partial \psi}{\partial t}&=-\dfrac{\partial}{\partial \bm{Q}}\cdot\Biggl\{\Biggl[\left(\boldsymbol{\delta}-\dfrac{2\varphi}{2\varphi+1}\dfrac{\bm{Q}\bm{Q}}{Q^2}\right)\cdot\left(\boldsymbol{\kappa}\cdot\bm{Q}-\dfrac{2}{\zeta}\dfrac{\partial \phi}{\partial \bm{Q}}\right)-\left(\dfrac{4k_BT}{\zeta}\right)\left(\dfrac{2\varphi}{2\varphi+1}\right)\dfrac{\bm{Q}}{Q^2}\Biggr]\psi\Biggr\}\\[10pt]
&+\left(\dfrac{k_BT}{\zeta}\right)\dfrac{\partial}{\partial \bm{Q}}\dfrac{\partial}{\partial \bm{Q}}:\left[2\left(\boldsymbol{\delta}-\dfrac{2\varphi}{2\varphi+1}\dfrac{\bm{Q}\bm{Q}}{Q^2}\right)\psi\right]
\end{split}
\end{equation}
The force term $(\partial \phi/\partial \bm{Q})$ may be of the Hookean or FENE type, and does not affect the development presented below.

Setting $N=1$ in Eq.~(\ref{eq:fp_formal}), and denoting $\bm{Q}_1$ simply as $\bm{Q}$, we get
\begin{align}\label{eq:fp_dumbbell}
\dfrac{\partial \psi}{\partial t}&=-\dfrac{\partial}{\partial \bm{Q}}\cdot\Biggl\{\Biggl[\boldsymbol{\kappa}\cdot\bm{Q}-\left(\dfrac{\varphi}{1+2\varphi}\right){\bm{U}}_{11}\cdot\left(\boldsymbol{\kappa}\cdot\bm{Q}\right)\\[5pt]
 &-\dfrac{1}{\zeta}\left(\bm{A}_{11}-\dfrac{\varphi}{1+2\varphi}\bm{V}_{11}\right)\cdot\left(\dfrac{\partial \phi}{\partial \bm{Q}}\right)-\left(\dfrac{k_BT}{\zeta}\right)\left(\dfrac{\varphi}{1+2\varphi}\right){\dfrac{\partial}{\partial \bm{Q}}\cdot\bm{V}_{11}^{T}}\Biggr]\psi\Biggr\}\nonumber\\[5pt]
 &+\dfrac{k_BT}{\zeta}\dfrac{\partial}{\partial \bm{Q}}\dfrac{\partial}{\partial \bm{Q}}:\left[\left(\bm{A}_{11}-\dfrac{\varphi}{1+2\varphi}\bm{V}_{11}\right)^{T}\psi\right]\nonumber
\end{align}
Recognizing that $M_1=P_1=0$ by definition for a dumbbell, and using the methodology described in Sec.~\ref{sec:decoup_method} above, we can write
\begin{equation}\label{eq:simpli_db}
\begin{split}
\bm{A}_{11}&=2\boldsymbol{\delta}\\[10pt]
\bm{U}_{11}&=\bm{A}_{11}\cdot\left(\dfrac{\bm{Q}}{Q}\right)\boldsymbol{\Lambda}^{(1)}_1=2\left(\dfrac{\bm{Q}\bm{Q}}{Q^2}\right)\\[10pt]
\bm{V}_{11}&=\bm{A}_{11}\cdot\left(\dfrac{\bm{Q}}{Q}\right)\boldsymbol{J}^{(1)}_1=4\left(\dfrac{\bm{Q}\bm{Q}}{Q^2}\right)\\[10pt]
\left(\bm{A}_{11}-\dfrac{\varphi}{1+2\varphi}\bm{V}_{11}\right)&=\left(\bm{A}_{11}-\dfrac{\varphi}{1+2\varphi}\bm{V}_{11}\right)^{T}=2\left(\boldsymbol{\delta}-\dfrac{2\varphi}{2\varphi+1}\dfrac{\bm{Q}\bm{Q}}{Q^2}\right)\\[10pt]
\dfrac{\partial}{\partial \bm{Q}}\cdot\bm{V}_{11}^{T}&=\dfrac{\partial}{\partial \bm{Q}}\cdot\bm{V}_{11}=4\dfrac{\partial}{\partial \bm{Q}}\cdot\left(\dfrac{\bm{Q}\bm{Q}}{Q^2}\right)=8\left(\dfrac{\bm{Q}}{Q^2}\right)
\end{split}
\end{equation}
Substituting the results from Eq.~(\ref{eq:simpli_db}) into Eq.~(\ref{eq:fp_dumbbell}), 
\begin{align}\label{eq:fp_dumbbell_2}
\dfrac{\partial \psi}{\partial t}&=-\dfrac{\partial}{\partial \bm{Q}}\cdot\Biggl\{\Biggl[\boldsymbol{\kappa}\cdot\bm{Q}-\dfrac{2\varphi}{1+2\varphi}\dfrac{\bm{Q}\bm{Q}}{Q^2}\cdot\left(\boldsymbol{\kappa}\cdot\bm{Q}\right)\\[5pt]
 &-\dfrac{2}{\zeta}\left(\boldsymbol{\delta}-\dfrac{2\varphi}{2\varphi+1}\dfrac{\bm{Q}\bm{Q}}{Q^2}\right)\cdot\left(\dfrac{\partial \phi}{\partial \bm{Q}}\right)-\left(\dfrac{4k_BT}{\zeta}\right)\left(\dfrac{2\varphi}{1+2\varphi}\right)\left(\dfrac{\bm{Q}}{Q^2}\right)\Biggr]\psi\Biggr\}\nonumber\\[5pt]
 &+\dfrac{2k_BT}{\zeta}\dfrac{\partial}{\partial \bm{Q}}\dfrac{\partial}{\partial \bm{Q}}:\left[\left(\boldsymbol{\delta}-\dfrac{2\varphi}{2\varphi+1}\dfrac{\bm{Q}\bm{Q}}{Q^2}\right)\psi\right]\nonumber,
\end{align}
which, upon, re-arrangement, yields
\begin{align}\label{eq:fp_db_fin}
\dfrac{\partial \psi}{\partial t}&=-\dfrac{\partial}{\partial \bm{Q}}\cdot\Biggl\{\Biggl[\left(\boldsymbol{\delta}-\dfrac{2\varphi}{2\varphi+1}\dfrac{\bm{Q}\bm{Q}}{Q^2}\right)\cdot\left(\boldsymbol{\kappa}\cdot\bm{Q}-\dfrac{2}{\zeta}\dfrac{\partial \phi}{\partial \bm{Q}}\right)-\left(\dfrac{4k_BT}{\zeta}\right)\left(\dfrac{2\varphi}{2\varphi+1}\right)\dfrac{\bm{Q}}{Q^2}\Biggr]\psi\Biggr\}\nonumber\\[10pt]
&+\left(\dfrac{k_BT}{\zeta}\right)\dfrac{\partial}{\partial \bm{Q}}\dfrac{\partial}{\partial \bm{Q}}:\left[2\left(\boldsymbol{\delta}-\dfrac{2\varphi}{2\varphi+1}\dfrac{\bm{Q}\bm{Q}}{Q^2}\right)\psi\right]
\end{align}
By comparing Eqs.~(\ref{eq:fp_db_fin}) and (\ref{eq:db_prior}), we conclude that the procedure outlined in the present work yields the same Fokker-Planck equation for the dumbbell as was obtained using an alternate approach. As stated in Sec.~{II} of the main text, we use the following notation to denote the diffusion tensor
\begin{equation}\label{eq:diff_def}
\bm{D}(\varphi)=\widehat{\boldsymbol{A}}_{11}=\left(\bm{A}_{11}-\dfrac{\varphi}{1+2\varphi}\bm{V}_{11}\right)=2\left(\boldsymbol{\delta}-\dfrac{2\varphi}{2\varphi+1}\dfrac{\bm{Q}\bm{Q}}{Q^2}\right)
\end{equation}
which is clearly dependent on the internal friction parameter. While the functional dependence of $\bm{D}$ on $\varphi$ is easily discerned for a dumbbell, one needs to go through the iterative procedure described in Sec.~\ref{sec:decoup_method} to know the exact form of $\bm{D}$ in chains with greater than one spring.

The dimensionless form of the Fokker-Planck equation is then given by
\begin{align}\label{eq:fp_db_dimless}
\dfrac{\partial \psi^{*}}{\partial t^{*}}&=-\dfrac{\partial}{\partial \bm{Q}^{*}}\cdot\Biggl\{\Biggl[\left(\boldsymbol{\delta}-\dfrac{2\varphi}{2\varphi+1}\dfrac{\bm{Q}^{*}\bm{Q}^{*}}{Q^{*2}}\right)\cdot\left(\boldsymbol{\kappa}^{*}\cdot\bm{Q}^{*}-\dfrac{1}{2}\dfrac{\partial \phi^{*}}{\partial \bm{Q}^{*}}\right)\nonumber\\[10pt]
&-\left(\dfrac{2\varphi}{2\varphi+1}\right)\dfrac{\bm{Q}^{*}}{Q^{*2}}\Biggr]\psi^{*}\Biggr\}+\dfrac{1}{2}\Biggl\{\dfrac{1}{2}\dfrac{\partial}{\partial \bm{Q}^{*}}\dfrac{\partial}{\partial \bm{Q}^{*}}:\left[\boldsymbol{D}(\varphi)\psi^{*}\right]\Biggr\}
\end{align}

\begin{figure}[t]
\centering
\includegraphics[width=0.79\linewidth]{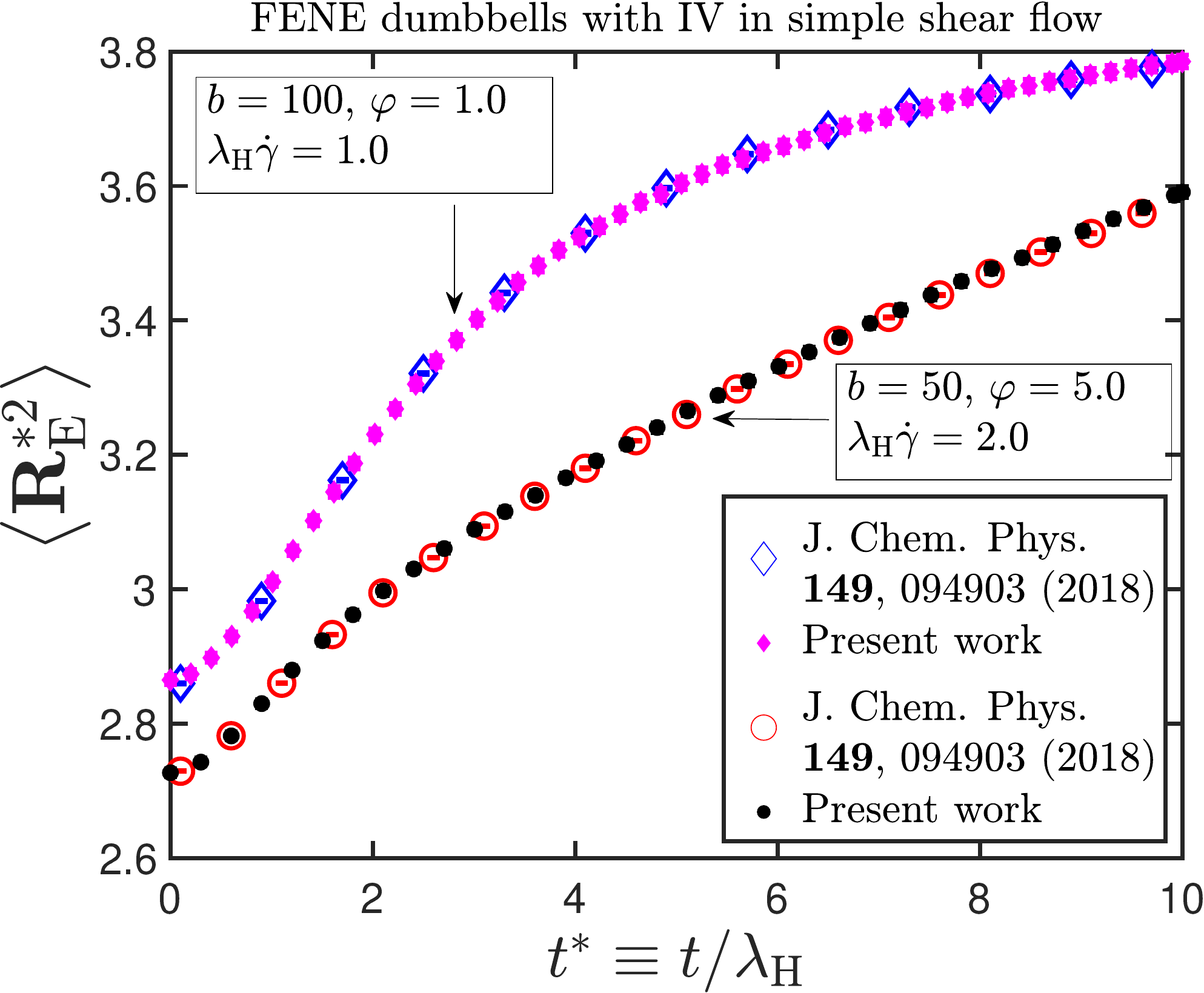}  
\caption{Transient evolution of the dimensionless mean-squared end-to-end distance of freely-draining FENE dumbbells with internal friction subjected to steady shear flow, obtained using two different approaches: the solid symbols denote BD simulation results obtained using the methodology described in the present work, using the explicit Euler scheme, while the hollow symbols represent results obtained by simulating the equations derived in Ref.~\citenum{Kailasham2018} using a higher order semi-implicit predictor-corrector algorithm.}
\label{fig:db_sc_compare}
\end{figure}

The stochastic differential equation corresponding to Eq.~(\ref{eq:fp_db_dimless}), may be written, using It\^{o}'s interpretation, as
\begin{equation}\label{eq:sde_gov}
\begin{split}
d\bm{Q}^{*}&=\Biggl[\left(\boldsymbol{\delta}-\dfrac{\epsilon}{2\epsilon+1}\dfrac{\bm{Q}^{*}\bm{Q}^{*}}{Q^{*2}}\right)\cdot\left(\boldsymbol{\kappa}^{*}\cdot\bm{Q}^{*}-\dfrac{1}{2}\dfrac{\partial \phi^{*}}{\partial \bm{Q}^{*}}\right)-\left(\dfrac{\epsilon}{\epsilon+1}\right)\dfrac{\bm{Q}^{*}}{Q^{*2}}\Biggr]dt^{*}\\[5pt]
&+\dfrac{1}{\sqrt{2}}\bm{B}(\epsilon)\cdot\,d\bm{W}_{t}
\end{split}
\end{equation}
where $\epsilon=2\varphi$, and $\bm{B}(\epsilon)$ is the square-root of the diffusion tensor, defined as $\bm{B}\cdot\bm{B}^{T}=\bm{D}$. It is worth noting that $\bm{B}$ is not unique for a given $\bm{D}$. For the special form of $\bm{D}$ given by Eq.~(\ref{eq:diff_def}), an analytical expression for the square root may be written using the method prescribed in Refs.~[\citenum{Ottinger1996,Prabhakar2002}], 
\begin{equation}\label{eq:sqrt_analytical}
\bm{B}=\sqrt{2}\left[\boldsymbol{\delta}-\left(1-\sqrt{\dfrac{1}{\epsilon+1}}\right)\dfrac{\bm{Q}^{*}\bm{Q}^{*}}{Q^{*2}}\right]
\end{equation}
Upon substitution of Eq.~(\ref{eq:sqrt_analytical}) into Eq.~(\ref{eq:sde_gov}), the governing stochastic differential equation given in Ref.~\citenum{Hua1995} is obtained, and it is seen that an IV dependent prefactor multiplies the Wiener process. We, on the other hand, propose that the square-root may be calculated using the Cholesky decomposition algorithm, in order to accommodate the general case where the square root of the block diffusion tensor may not be known analytically. This approach for the calculation of the square root of the diffusion tensor does not affect the simulation estimates, as evident from Fig.~\ref{fig:db_sc_compare} and discussed below.

In Fig.~\ref{fig:db_sc_compare}, a comparison of the mean-squared end-to-end distance of FENE dumbbells, obtained from BD simulations with the square-root of the diffusion tensor calculated as in Eq.~(\ref{eq:sqrt_analytical}), and that obtained using the procedure developed in the present work is plotted, and an excellent agreement is observed. 

%The excellent agreement between the two results suggests that the Brownian force term used in the present work also satisfies the fluctuation dissipation theorem.

\subsection{\label{sec:fdt_sat}Satisfaction of fluctuation dissipation theorem}

In Brownian dynamics simulations, the requirement of satisfying the fluctuation dissipation theorem (FDT) is enforced by requiring that the square-root of the diffusion tensor is the multiplicative prefactor to the Wiener noise term, and we have followed this procedure in our work. 

\begin{figure}[t]
\centering
\includegraphics[width=0.79\linewidth]{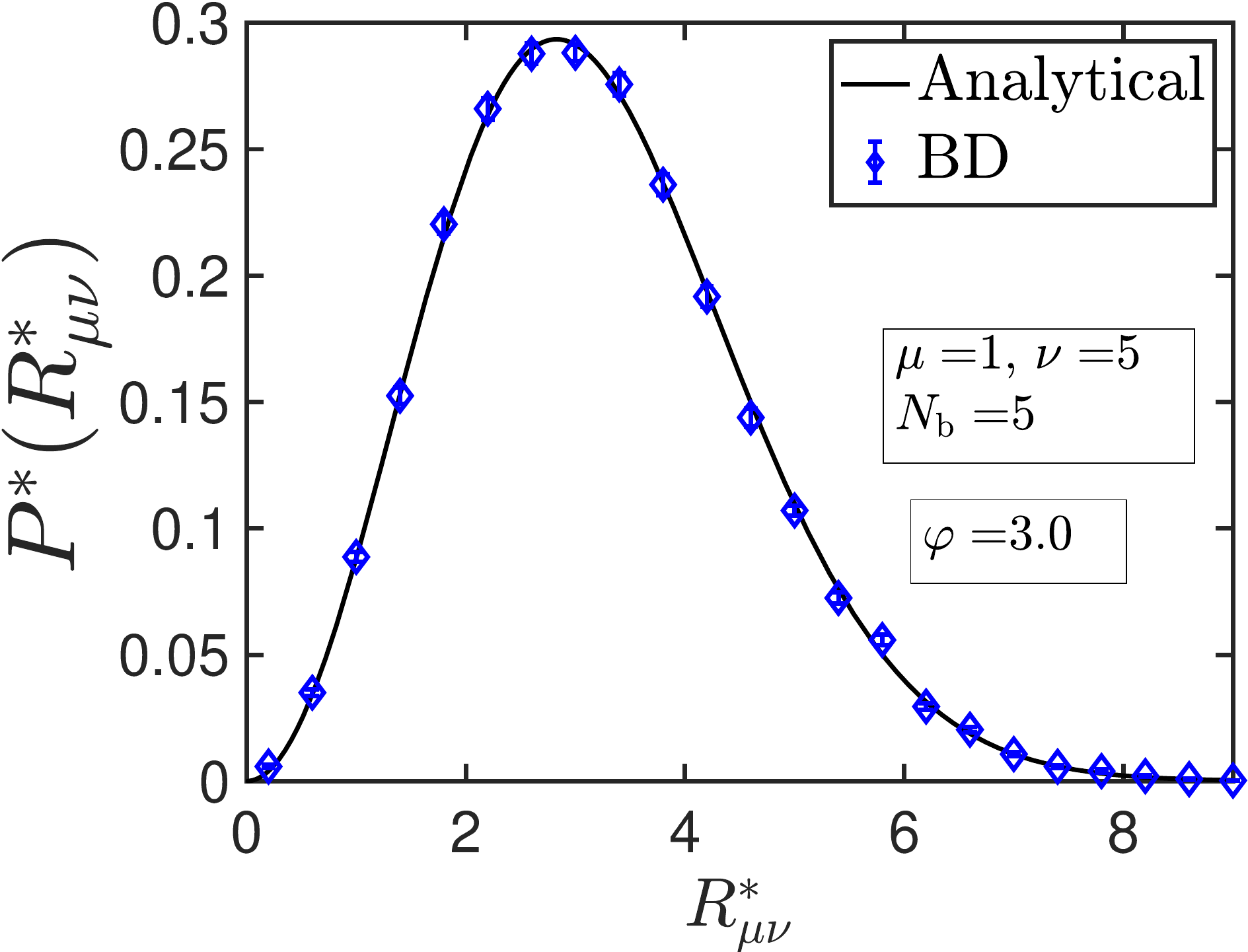}  
\caption{Probability distribution of the lengths of the end-to-end vector of a five-bead chain with internal friction. Symbols are BD simulation results, and the line represents the analytical result given by Eq.~(\ref{eq:prob_dist_eqn}).}
\label{fig:fdt_pdf}
\end{figure}

It is known that the probability distribution of the length of the interbead connector vector~\cite{doi-edwards,Bird1987b}, $R^{*}_{\mu\nu}\equiv\,|\bm{R^{*}}_{\mu\nu}|=|\bm{r}^{*}_{\nu}-\bm{r}^{*}_{\mu}|$,
\begin{equation}\label{eq:prob_dist_eqn}
P^{*}\left(R^{*}_{\mu\nu}\right)=4\pi\,R^{*2}_{\mu\nu}\left[\dfrac{1}{2\pi|\nu-\mu|}\right]^{3/2}\exp\left[-\dfrac{R^{*2}_{\mu\nu}}{2|\nu-\mu|}\right],
\end{equation}
is unaffected by the inclusion of internal friction. In Fig.~\ref{fig:fdt_pdf}, the probability distribution of the length of the end-to-end vector [$R^{*}_{1\,N_{\text{b}}}$] for a five-bead chain with internal friction is plotted. The good agreement between the simulation and analytical results indicate that the stochastic differential equation derived in the present work satisfies FDT.
The use of the probability distribution of chain lengths at equilibrium for testing the correctness of the governing equations was employed in our previous work~\cite{Kailasham2018} on dumbbells with fluctuating internal friction and hydrodynamic interactions. Our stochastic differential equation differed from the one quoted in a prior study~\cite{Hua19961473}, in the definition of one term, $g_2$. Upon comparing the probability distributions of the spring lengths obtained by simulating the two equations, our equation was found to yield the correct distribution, while the other equation deviated significantly from the analytical distribution and the difference between the two equations was put down to a typographical error in the other study~\cite{Hua19961473}.

\subsection{\label{sec:sqrt_calc} Symmetricity and positive-definiteness of the diffusion tensor}

A pre-requisite to the use of the Cholesky decomposition method is that the matrix be positive-definite~\cite{press2007numerical}. We are not able to prove analytically that the diffusion tensor, $\bm{\mathcal{D}}$, appearing in Eq.~(6) of the main text is positive-definite. However, we checked for a hundred different random initial configurations of a forty-five spring chain that the eigenvalues of the diffusion matrix are real and positive.

\begin{figure}[t]
    \centerline{
   {\includegraphics[width=120mm]{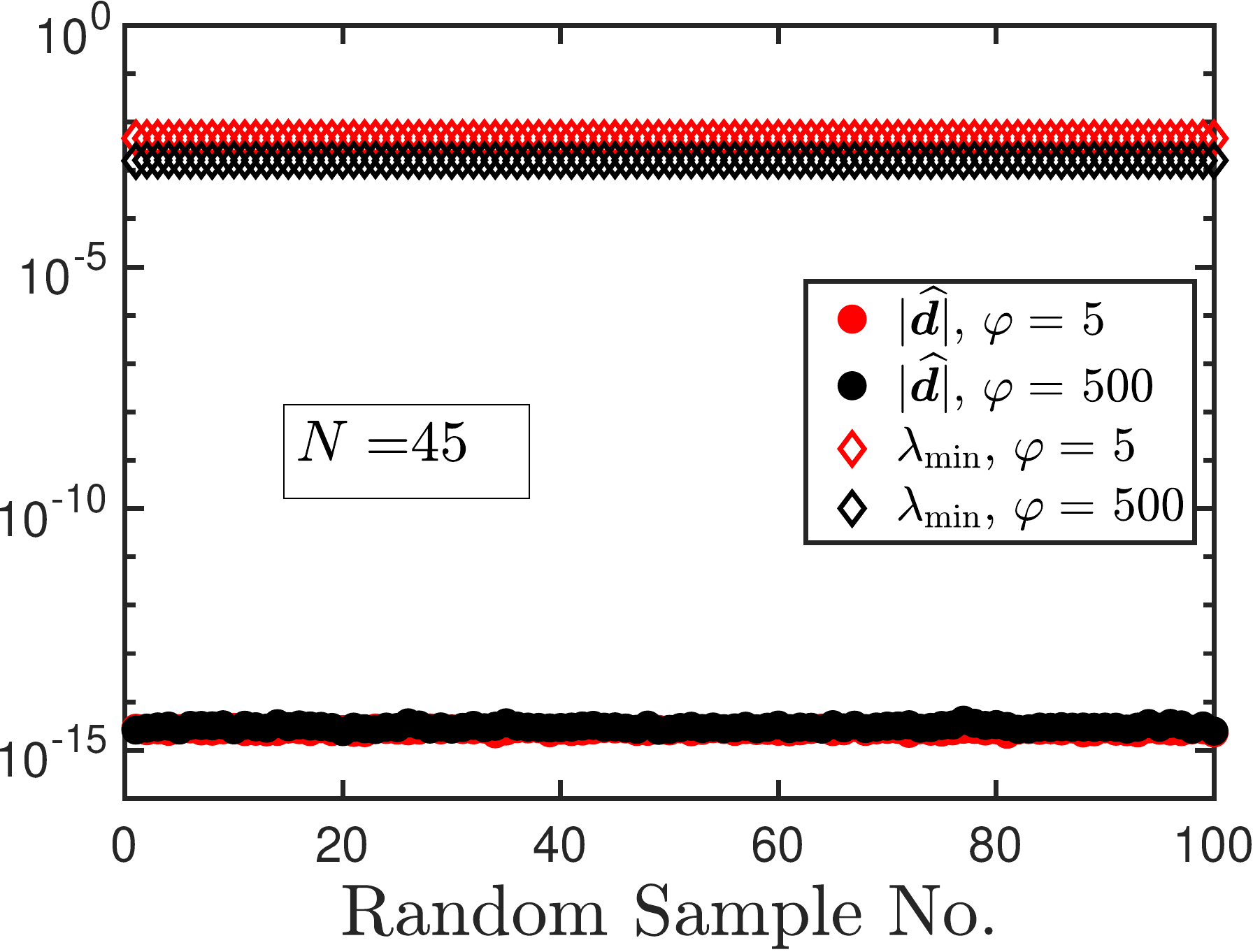}}     
    }
\caption{Test for symmetricity, and smallest eigenvalue of the diffusion tensor, for a hundred randomly chosen initial values of the chain configuration for a forty-five spring chain.}
\label{fig:eig_val_sym_check}
\end{figure}

In Fig.~\ref{fig:eig_val_sym_check}, the smallest eigenvalue for each sample configuration, and difference between the diffusion matrix and its transpose, are plotted for two different values of the internal friction parameter.  The difference is computed as follows: firstly, the diffusion matrix and its transpose are subtracted, to generate a $3N\times3N$ matrix. This matrix is then unwrapped to give an array, $\bm{\widehat{d}}$, of $9N^2$ elements. Finally, the $2-$norm of this array is computed, and taken to be a numerical measure of the difference between the diffusion matrix and its transpose, as
\begin{equation}
\begin{split}
||\bm{\mathcal{D}}^{T}-\bm{\mathcal{D}}||&\equiv|\bm{\widehat{d}}|=\sqrt{\left(\widehat{d}_{1}\right)^2+\left(\widehat{d}_{2}\right)^2+\dots+\left(\widehat{d}_{9N^2}\right)^2}
\end{split}
\end{equation}
The difference computed in this manner is $\mathcal{O}(10^{-14})$, meaning that the diffusion matrix may be considered symmetric for all practical purposes.

\subsection{\label{sec:tstep_conv}Establishing timestep convergence}

The governing stochastic differential equations (SDE) for the simplest case of a free-draining dumbbell ($N_{\text{b}}>2$) have already been derived by Refs.~\citenum{Hua1995} and~\citenum{Kailasham2018} . While the SDEs in the two references are identical, the former employs an explicit Euler solver, while the latter uses a higher order semi-implicit predictor-corrector scheme. For the general case of $N_{\text{b}}>2$, however, the complexity of the set of governing stochastic differential equations makes the construction of a higher-order solver algorithm quite difficult, and we have therefore used the explicit Euler method.

In Fig.~\ref{fig:db_sc_compare}, the time-evolution of the dimensionless mean-squared end-to-end distance of FENE dumbbells with internal friction subjected to simple shear flow, obtained using two different approaches, is presented. The excellent agreement between the results obtained using the higher order solver (shown as hollow symbols) and the explicit Euler scheme employed in the present work (indicated by solid symbols) establishes the reliability of our solution methodology. For the $\{b=100,\varphi=1.0,\lambda_{H}\dot{\gamma}=1.0\}$ case, the higher-order solver algorithm uses three different timestep widths, $\Delta t^{*}=\{5\times10^{-2},2\times10^{-2},1\times10^{-2}\}$, followed by timestep width extrapolation to the $\Delta t^{*}\to 0$ limit. The corresponding set of timestep widths for the $\{b=50,\varphi=5.0,\lambda_{H}\dot{\gamma}=2.0\}$ is 
$\Delta t^{*}=\{1\times10^{-3},5\times10^{-4},2\times10^{-4}\}$. The explicit Euler results presented in Fig.~\ref{fig:db_sc_compare} for the $b=50$ and $b=100$ cases have been obtained using a timestep width of $10^{-4}$, and $10^{-5}$, respectively, and have been found to be statistically indistinguishable from results obtained using timestep widths that are an order of magnitude smaller.

\begin{figure*}[t]
\begin{center}
\begin{tabular}{c c}
\includegraphics[width=3.in,height=!]{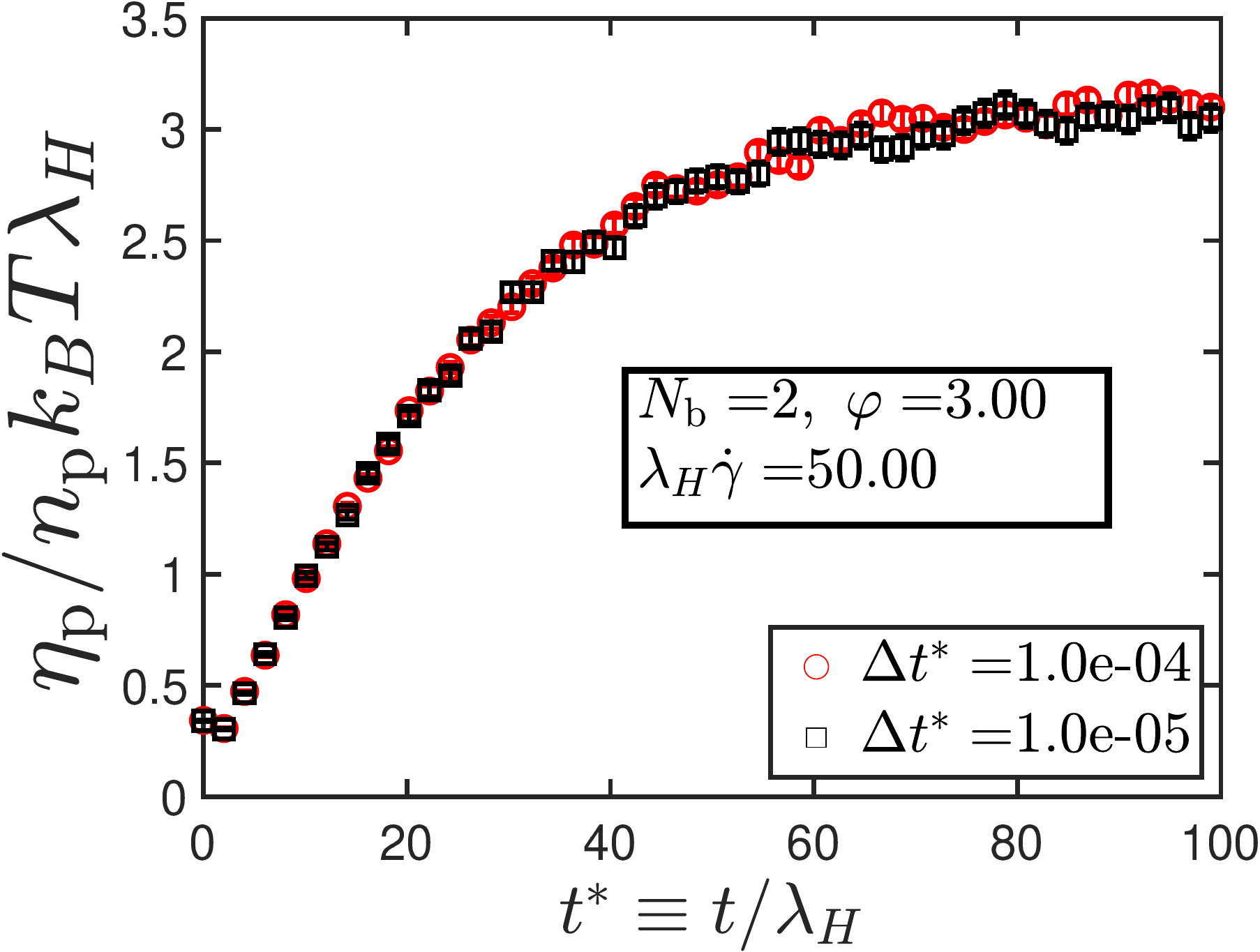}&
\includegraphics[width=3.in,height=!]{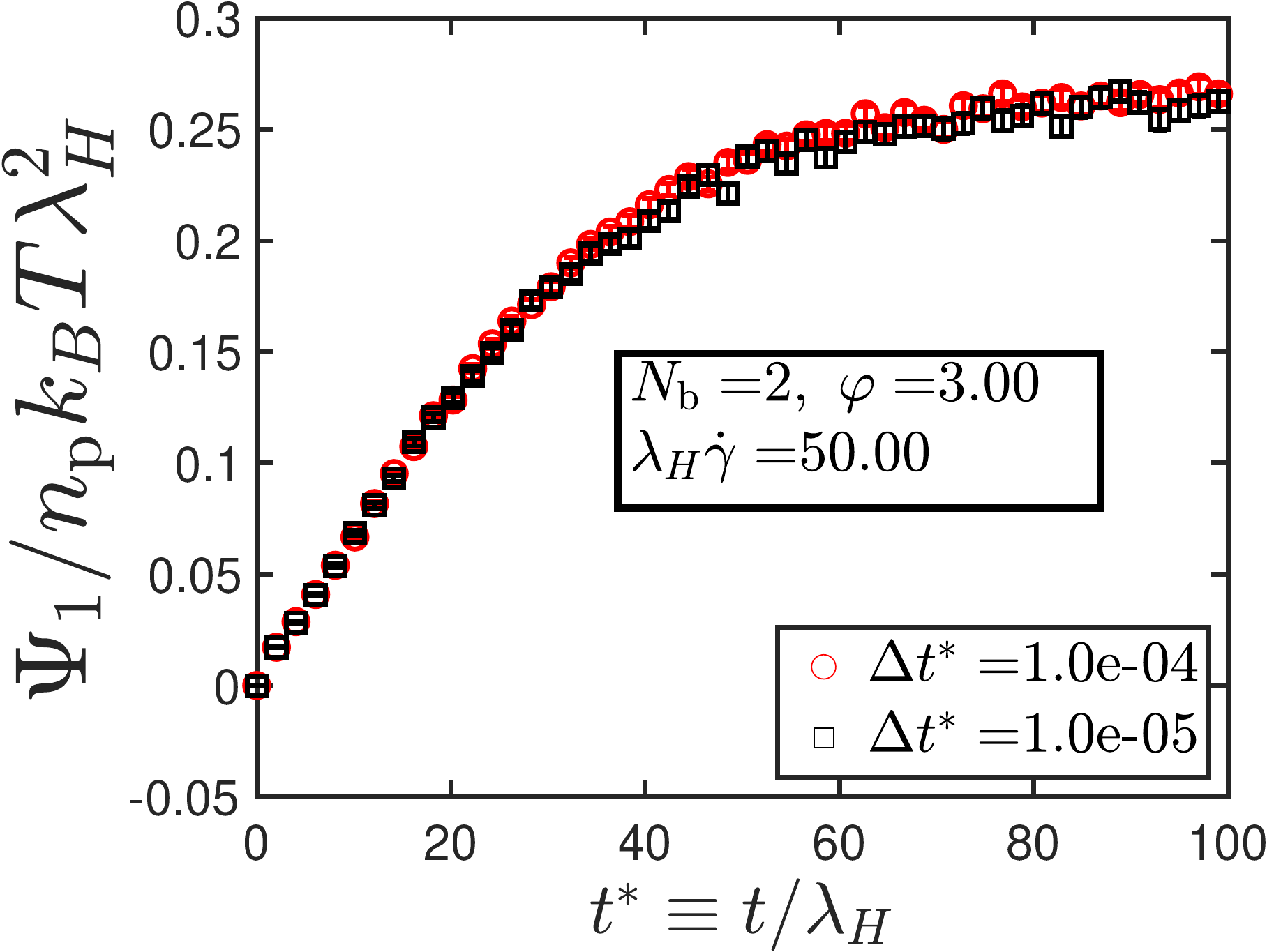}\\
(a)&(b)\\
\includegraphics[width=3.in,height=!]{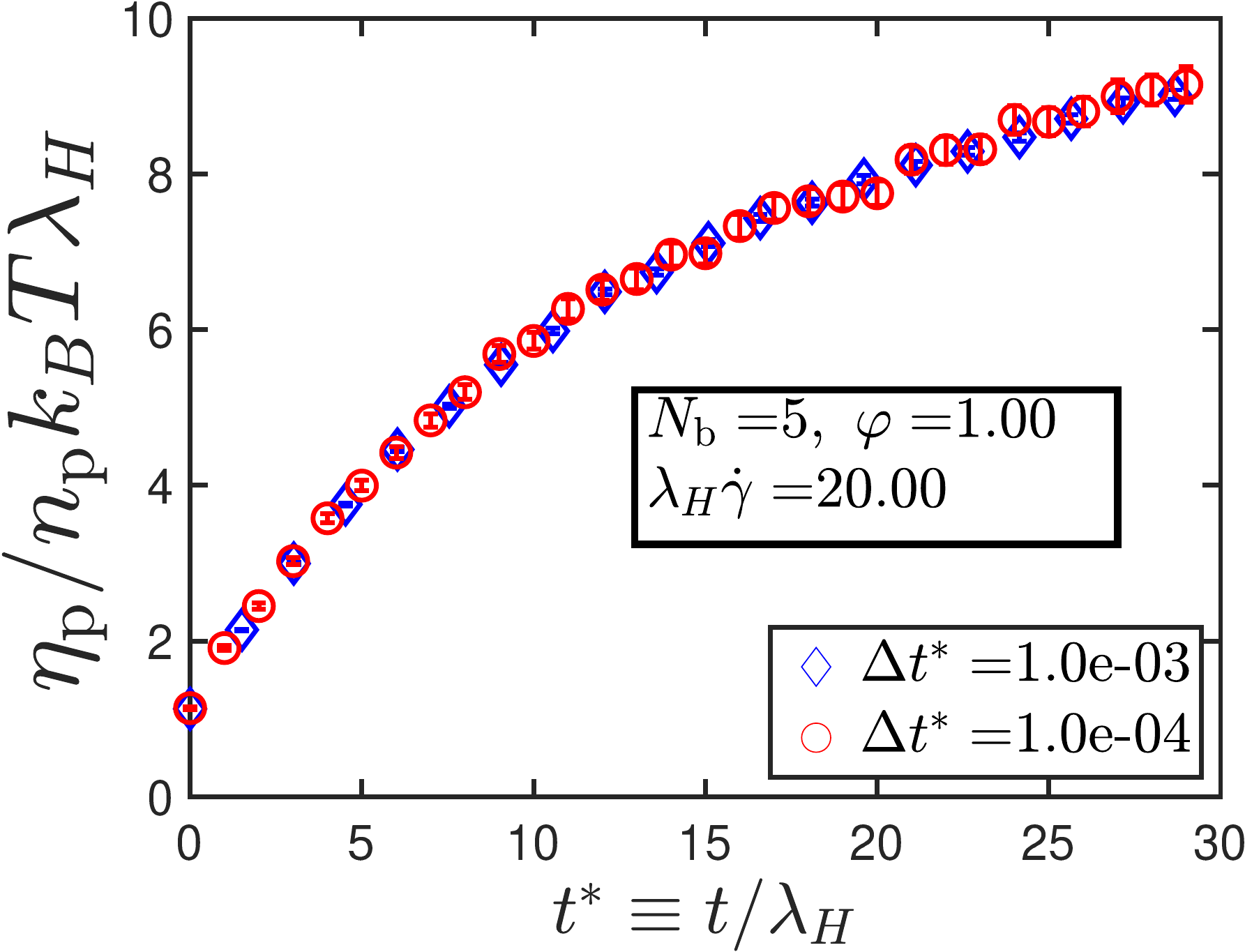}&
\includegraphics[width=3.in,height=!]{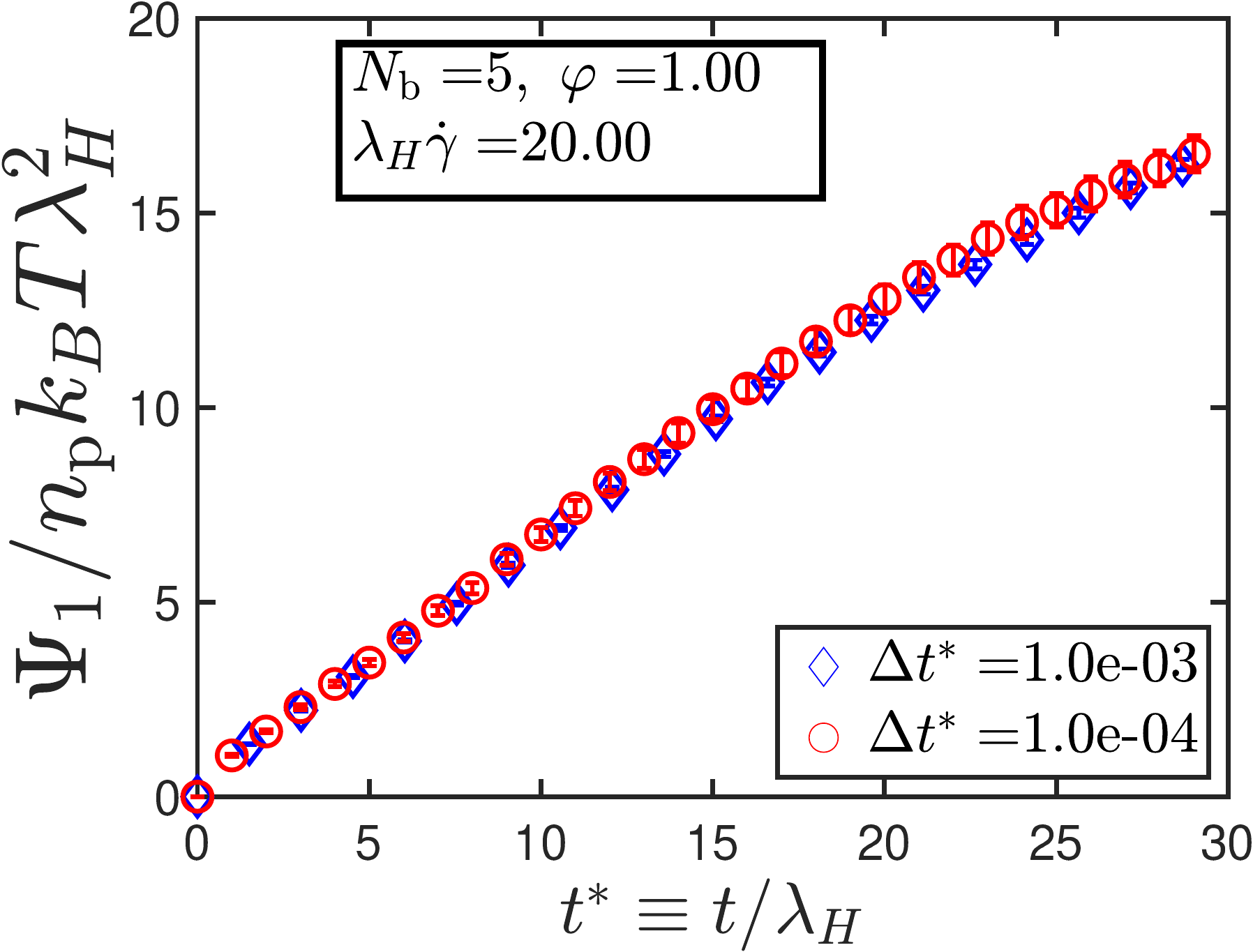}\\
(c)&(d)
\end{tabular}
\end{center}
\caption{Establishing timestep convergence by plotting simulation results at two different values of the timestep width. The transient evolution of the shear viscosity and the first normal stress coefficient are plotted for a dumbbell ((a) and (b)) and a five-bead Rouse chain ((c) and (d)) with internal friction.}
\label{fig:tstep_equivalence_figs}
\end{figure*}

In Fig.~\ref{fig:tstep_equivalence_figs}, the transient shear viscosity and the first normal stress coefficient for a dumbbell and five-bead chain with internal friction, obtained by integrating the governing equation using two different values of the timestep width, is presented. The excellent agreement between the results for different values of $\Delta t^{*}$ establishes timestep convergence for our simulations. Similar checks have also been performed for the viscometric functions in small amplitude oscillatory shear flow in order to pick the appropriate values for the timestep.

\subsection{\label{sec:cpu_time} CPU time scaling}

\begin{figure}[h]
    \centerline{
   {\includegraphics[width=120mm]{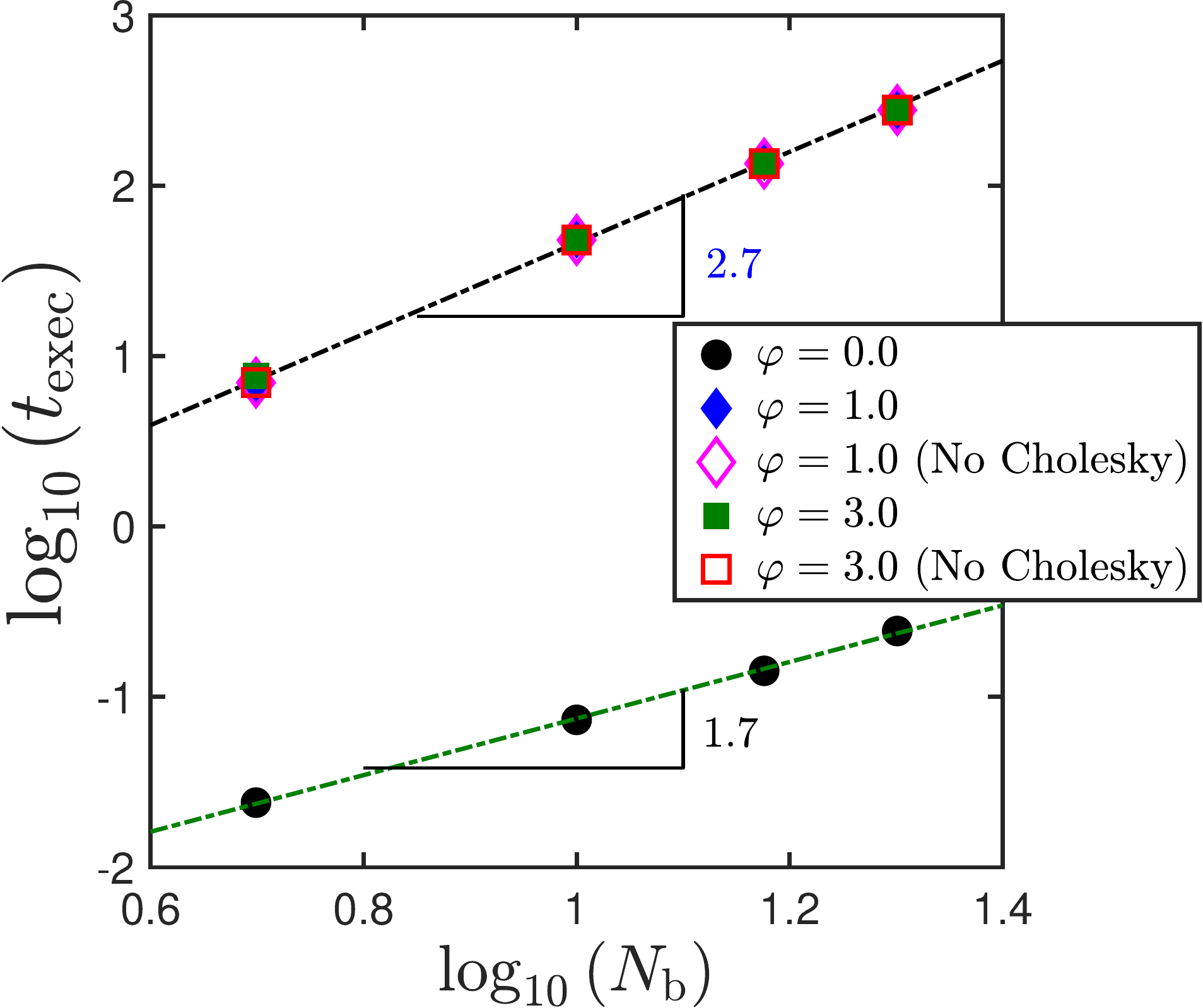}}     
    }
\caption{Scaling of simulation time as a function of number of beads in the chain. Execution time is calculated as an average over one hundred trajectories at each value of $N_{\text{b}}$. Each trajectory is 100 dimensionless times ($\lambda_{{H}}$) long. Simulations performed on MonARCH, the HPC hosted at Monash University, with all the runs executed on the same type of processor [16 core Xeon-E5-2667-v3 @ 3.20GHz servers with 100550MB usable memory]. A step size of $\Delta t^*=10^{-2}$ is used for all the simulation runs.}
\label{fig:scaling}
\end{figure}

In Fig.~\ref{fig:scaling}, the scaling of the execution time as a function of the chain length is presented for a code written using the simple explicit Euler method for solving the SDE. The time taken for simulation of simple Rouse chains scales with an exponent of $1.7$ with respect to the chain length, whereas chains with internal friction scale with an exponent of $2.7$. Moreover, running simulations on chains with internal friction after dropping the noise term from the governing equation does not seem to significantly affect the execution time, indicating that the calculation of the square root of the diffusion tensor, using Cholesky decomposition, represents only a minor portion of the total workload. Furthermore, the execution time and scaling for chains with internal friction is practically unaffected by the value of the internal friction parameter.
The code is written in a way that the simulation of chains without internal friction involves neither the construction of the $\bm{\mathcal{U}}$ and $\bm{\mathcal{D}}$ matrices, nor the evaluation of the divergence terms or the square-root of the diffusion tensor. In fact, the Rouse case is simulated exactly as given in Eq.~(4.4) of Ref.~\citenum{Ottinger1996}, which is a significant simplification over the case with internal friction turned on.

In absolute numbers, the execution time for one trajectory of a ten-bead Rouse chain is $0.07$ seconds, whereas that for a ten-bead Rouse chain with internal friction ($\varphi=1.0$) is $49.6$ seconds, representing an increase that is nearly three-orders in magnitude. 

%%%%%%%%%%%%

%%%%%%%%%%%%%%%%%%%%%%
\section{\label{sec:stress_tens_derv} Stress tensor derivation}

The Giesekus expression for the stress tensor is written as follows
\begin{equation}\label{eq:giesekus_exp_start_point}
\begin{split}
{\boldsymbol{\tau}_{\text{p}}}=\dfrac{n_{\text{p}}\zeta}{2}\left<\sum_{u=1}^{N}\sum_{v=1}^{N}{\mathscr{C}_{uv}}\boldsymbol{Q}_{u}\boldsymbol{Q}_{v}\right>_{(1)}=\dfrac{n_{\text{p}}\zeta}{2}\Biggl[&\dashuline{\dfrac{d}{dt}\left<\sum_{u,v}{\mathscr{C}_{uv}}\boldsymbol{Q}_{u}\boldsymbol{Q}_{v}\right>}-\boldsymbol{\kappa}\cdot\left<\sum_{u,v}{\mathscr{C}_{uv}}\boldsymbol{Q}_{u}\boldsymbol{Q}_{v}\right>\\[5pt]
&-\left<\sum_{u,v}{\mathscr{C}_{uv}}\boldsymbol{Q}_{u}\boldsymbol{Q}_{v}\right>\cdot\boldsymbol{\kappa}^{T}\Biggr]
\end{split}
\end{equation}
where $\mathscr{C}_{uv}$ represents elements of the symmetric Kramers matrix~\citep{Bird1987b}, which is inverse to the Rouse matrix, i.e.,
\begin{equation}\label{eq:rouse_kram_inv}
\sum_{j}A_{kj}\mathscr{C}_{ju}=\delta_{ku}
\end{equation}
We identify $\widetilde{\boldsymbol{B}}\equiv\sum_{u,v=1}^{N}{\mathscr{C}_{uv}}\boldsymbol{Q}_{u}\boldsymbol{Q}_{v}$ in Eq.~(\ref{eq:giesekus_exp_start_point}) and aim to find an expression for $\dfrac{d}{dt}\left<\widetilde{\boldsymbol{B}}\right>$, which is also referred to as the equation of change~\citep{Bird1987b}.

We start with the continuity equation for a bead-spring-dashpot chain in a homogeneous flow profile, which may be written as
\begin{equation}\label{eq:cont_eq_psi}
\dfrac{\partial \psi}{\partial t}=-\sum_{j=1}^{N}\dfrac{\partial}{\partial \boldsymbol{Q}_j}\cdot\left\{\llbracket\dot{\boldsymbol{Q}}_{j}\rrbracket\psi\right\}
\end{equation} 
The normalized distribution function for the internal coordinates ($\psi$) may be used instead of the distribution function based on all the bead positions ($\Psi$), due to the homogeneous flow profile being considered in the derivation of the stress tensor expression.
The equation of change for the second-order tensor $\widetilde{\boldsymbol{B}}$ may be found by multiplying both sides of Equation~(\ref{eq:cont_eq_psi}) and integrating over all space as follows
\begin{equation}\label{eq:multip_both}
\begin{split}
\int\widetilde{\boldsymbol{B}}\dfrac{\partial \psi}{\partial t}d\boldsymbol{Q}=-\int\left(\sum_{j=1}^{N}\dfrac{\partial}{\partial \boldsymbol{Q}_j}\cdot\left\{\llbracket\dot{\boldsymbol{Q}}_{j}\rrbracket\psi\right\}\right)\widetilde{\boldsymbol{B}}d\boldsymbol{Q}
\end{split}
\end{equation}
where $d\boldsymbol{Q}\equiv\,d\boldsymbol{Q}_1d\boldsymbol{Q}_2\cdots\,d\boldsymbol{Q}_{N}$
On the LHS of Eq.~(\ref{eq:multip_both}), since $\widetilde{\boldsymbol{B}}$ does not depend explicitly on $t$, we may write
\begin{equation}\label{eq:time_indep}
\int\widetilde{\boldsymbol{B}}\dfrac{\partial \psi}{\partial t}d\boldsymbol{Q}=\dfrac{\partial}{\partial t}\int\widetilde{\boldsymbol{B}}\psi\,d\boldsymbol{Q}=\dfrac{d}{dt}\left<\widetilde{\boldsymbol{B}}\right>
\end{equation}
We represent
\begin{equation}
\begin{split}
\boldsymbol{Q}_{j}=\sum_{\alpha}Q^{\alpha}_{j}\boldsymbol{e}_{\alpha};\,\,
\llbracket\dot{\boldsymbol{Q}}_{j}\rrbracket=\sum_{\beta}\llbracket\dot{Q}^{\beta}_{k}\rrbracket\,\boldsymbol{e}_{\beta};\,\,
\widetilde{\boldsymbol{B}}=\sum_{m,n}B^{mn}\boldsymbol{e}_{m}\boldsymbol{e}_{n}
\end{split}
\end{equation}
and the term within the summation on the RHS of Eq.~(\ref{eq:multip_both}) may be written as
\begin{equation}
\begin{split}
\dfrac{\partial}{\partial \boldsymbol{Q}_j}\cdot\left\{\llbracket\dot{\boldsymbol{Q}}_{j}\rrbracket\psi\right\}\widetilde{\boldsymbol{B}}&=\sum_{m,n,\alpha}\left[B^{mn}\dfrac{\partial \left(\llbracket\dot{Q}^{\alpha}_{j}\rrbracket\psi\right)}{\partial {Q}^{\alpha}_j}\right]\boldsymbol{e}_{m}\boldsymbol{e}_{n}\\[5pt]
&=\sum_{m,n,\alpha}\left[\dfrac{\partial \left(B^{mn}\llbracket\dot{Q}^{\alpha}_{j}\rrbracket\psi\right)}{\partial {Q}^{\alpha}_j}-\llbracket\dot{Q}^{\alpha}_{j}\rrbracket\psi\dfrac{\partial \left(B^{mn}\right)}{\partial {Q}^{\alpha}_j}\right]\boldsymbol{e}_{m}\boldsymbol{e}_{n}\\[5pt]
&=\dfrac{\partial}{\partial \boldsymbol{Q}_j}\cdot\left\{\llbracket\dot{\boldsymbol{Q}}_{j}\rrbracket\psi\widetilde{\boldsymbol{B}}\right\}-\psi\llbracket\dot{\boldsymbol{Q}}_{j}\rrbracket\cdot\dfrac{\partial \widetilde{\boldsymbol{B}}}{\partial \boldsymbol{Q}_{j}},
\end{split}
\end{equation}
leading to
\begin{equation}
\begin{split}
-\sum_{j=1}^{N}\left(\dfrac{\partial}{\partial \boldsymbol{Q}_j}\cdot\left\{\llbracket\dot{\boldsymbol{Q}}_{j}\rrbracket\psi\right\}\right)\widetilde{\boldsymbol{B}}
&=-\sum_{j=1}^{N}\dfrac{\partial}{\partial \boldsymbol{Q}_j}\cdot\left\{\llbracket\dot{\boldsymbol{Q}}_{j}\rrbracket\psi\widetilde{\boldsymbol{B}}\right\}+\sum_{j=1}^{N}\psi\llbracket\dot{\boldsymbol{Q}}_{j}\rrbracket\cdot\dfrac{\partial \widetilde{\boldsymbol{B}}}{\partial \boldsymbol{Q}_{j}},\
\end{split}
\end{equation}
and finally
\begin{equation}\label{eq:simp_integr}
\begin{split}
-\int\left(\sum_{j=1}^{N}\dfrac{\partial}{\partial \boldsymbol{Q}_j}\cdot\left\{\llbracket\dot{\boldsymbol{Q}}_{j}\rrbracket\psi\right\}\right)\widetilde{\boldsymbol{B}}d\boldsymbol{Q}&=-\int\left(\sum_{j=1}^{N}\dfrac{\partial}{\partial \boldsymbol{Q}_j}\cdot\left\{\llbracket\dot{\boldsymbol{Q}}_{j}\rrbracket\psi\widetilde{\boldsymbol{B}}\right\}\right)d\boldsymbol{Q}\\[5pt]
&+\int\left(\sum_{j=1}^{N}\psi\llbracket\dot{\boldsymbol{Q}}_{j}\rrbracket\cdot\dfrac{\partial \widetilde{\boldsymbol{B}}}{\partial \boldsymbol{Q}_{j}}\right)d\boldsymbol{Q}
\end{split}
\end{equation}
The first term on the RHS of Eq.~(\ref{eq:simp_integr}) vanishes, due to the Gauss divergence theorem~\citep{Bird1987b}, since the configurational distribution function is expected to vanish on a surface that is infinitely large. As a result, 
\begin{equation}\label{eq:simp1}
-\int\left(\sum_{j=1}^{N}\dfrac{\partial}{\partial \boldsymbol{Q}_j}\cdot\left\{\llbracket\dot{\boldsymbol{Q}}_{j}\rrbracket\psi\right\}\right)\widetilde{\boldsymbol{B}}d\boldsymbol{Q}=\int\left(\sum_{j=1}^{N}\psi\llbracket\dot{\boldsymbol{Q}}_{j}\rrbracket\cdot\dfrac{\partial \widetilde{\boldsymbol{B}}}{\partial \boldsymbol{Q}_{j}}\right)d\boldsymbol{Q}=\left<\sum_{j=1}^{N}\left(\llbracket\dot{\boldsymbol{Q}}_{j}\rrbracket\cdot\dfrac{\partial \widetilde{\boldsymbol{B}}}{\partial \boldsymbol{Q}_{j}}\right)\right>
\end{equation}
From Eqs.~(\ref{eq:multip_both}), \ref{eq:time_indep}, and (\ref{eq:simp1}), we thus have the following simplified equation of change for $\widetilde{\boldsymbol{B}}$,
\begin{equation}
\dfrac{d}{dt}\left<\widetilde{\boldsymbol{B}}\right>=\left<\sum_{j=1}^{N}\left(\llbracket\dot{\boldsymbol{Q}}_{j}\rrbracket\cdot\dfrac{\partial \widetilde{\boldsymbol{B}}}{\partial \boldsymbol{Q}_{j}}\right)\right>
\end{equation}
For the specific choice of $\widetilde{\boldsymbol{B}}$ considered in this appendix, we have
\begin{equation}
\begin{split}
\dfrac{\partial \widetilde{\boldsymbol{B}}}{\partial \boldsymbol{Q}_{j}}&=\sum_{u,v}\mathscr{C}_{uv}\dfrac{\partial}{\partial \boldsymbol{Q}_{j}}\left(\boldsymbol{Q}_{u}\boldsymbol{Q}_{v}\right)\\[5pt]
&=\sum_{u,v}\mathscr{C}_{uv}\sum_{\alpha,\beta,\gamma}\left[Q^{\beta}_{u}\dfrac{\partial Q^{\gamma}_{v}}{\partial Q^{\alpha}_{j}}+Q^{\gamma}_{v}\dfrac{\partial Q^{\beta}_{u}}{\partial Q^{\alpha}_{j}}\right]\boldsymbol{e}_{\alpha}\boldsymbol{e}_{\beta}\boldsymbol{e}_{\gamma},
\end{split}
\end{equation}
and obtain
\begin{equation}\label{eq:b_part}
\dfrac{\partial \widetilde{\boldsymbol{B}}}{\partial \boldsymbol{Q}_{j}}=\sum_{u}\sum_{\alpha,\beta}\mathscr{C}_{uj}Q^{\beta}_{u}\boldsymbol{e}_{\alpha}\boldsymbol{e}_{\beta}\boldsymbol{e}_{\alpha}+\sum_{v}\sum_{\alpha,\gamma}\mathscr{C}_{jv}Q^{\gamma}_{v}\boldsymbol{e}_{\alpha}\boldsymbol{e}_{\alpha}\boldsymbol{e}_{\gamma}
\end{equation}
The equation of motion for the connector vector velocity is given by Eq.~(\ref{eq:qdot_with_ck}). The following definition holds for~\citep{Sunthar2005} arbitrary spring-force laws, with the Hookean stiffness of the spring denoted by $H$ and maximum stretched length of the spring given by $Q_0$
\begin{equation}\label{eq:gen_spforce_def}
\begin{split}
\boldsymbol{F}_{k}^{\text{s}}\equiv\dfrac{\partial \phi}{\partial \boldsymbol{Q}_k}=H\boldsymbol{Q}_{k}f\left(Q_{k}/Q_0\right)
\end{split}
\end{equation}
The functional form of the scalar entity, $f\left(Q_k/Q_0\right)$ depends on the spring type used, and is unity for Hookean springs. The following notation is introduced $f_{k}\equiv\,f\left(Q_k/Q_0\right)$ for convenience.
The underlined term in Eq.~(\ref{eq:giesekus_exp_start_point}) may then be processed as 
\begin{equation}\label{eq:process1}
\begin{split}
\dfrac{d}{dt}\left<\sum_{u,v}{\mathscr{C}_{uv}}\boldsymbol{Q}_{u}\boldsymbol{Q}_{v}\right>\equiv\dfrac{d}{dt}\left<\widetilde{\boldsymbol{B}}\right>=\left<\sum_{j=1}^{N}\left(\llbracket\dot{\boldsymbol{Q}}_{j}\rrbracket\cdot\dfrac{\partial \widetilde{\boldsymbol{B}}}{\partial \boldsymbol{Q}_{j}}\right)\right>
\end{split}
\end{equation}
Substituting Eq.~(\ref{eq:qdot_with_ck})in Eq.~(\ref{eq:process1}), we obtain
\begin{equation}\label{eq:process2}
\begin{split}
\dfrac{d}{dt}\left<\widetilde{\boldsymbol{B}}\right>&=\left<\sum_{j}\left[\boldsymbol{\kappa}\cdot\boldsymbol{Q}_{j}\right]\cdot\dfrac{\partial \widetilde{\boldsymbol{B}}}{\partial \boldsymbol{Q}_{j}}\right>-\dfrac{k_BT}{\zeta}\left<\sum_{j,k}\left(A_{jk}\dfrac{\partial \ln \psi}{\partial \boldsymbol{Q}_k}\right)\cdot\dfrac{\partial \widetilde{\boldsymbol{B}}}{\partial \boldsymbol{Q}_{j}}\right>\\[5pt]
&-\dfrac{H}{\zeta}\left<\sum_{j,k}A_{jk}f_k\boldsymbol{Q}_{k}\cdot\dfrac{\partial \widetilde{\boldsymbol{B}}}{\partial \boldsymbol{Q}_{j}}\right>-\dfrac{K}{\zeta}\left<\sum_{j,k}A_{jk}C_{k}\boldsymbol{Q}_{k}\cdot\dfrac{\partial \widetilde{\boldsymbol{B}}}{\partial \boldsymbol{Q}_{j}}\right>
\end{split}
\end{equation}
The four terms on the RHS of Eq.~(\ref{eq:process2}) are identified respectively as the flow, Brownian force, spring force, and internal friction force contributions. These terms are processed sequentially, as shown below.
Firstly, we recognize that
\begin{equation}\label{eq:kappa_dot}
\begin{split}
\boldsymbol{\kappa}\cdot\boldsymbol{Q}_{j}&=\left[\sum_{m,n}{\kappa}^{mn}\boldsymbol{e}_m\boldsymbol{e}_n\right]\cdot\left[\sum_{\alpha}Q^{\alpha}_{j}\boldsymbol{e}_{\alpha}\right]=\sum_{m,n}\kappa^{mn}Q_{j}^{n}\boldsymbol{e}_m
\end{split}
\end{equation}
Using Eq.~(\ref{eq:kappa_dot}) and (\ref{eq:b_part}), we may write
\begin{equation}
\begin{split}
\left<\sum_{j}\left[\boldsymbol{\kappa}\cdot\boldsymbol{Q}_{j}\right]\cdot\dfrac{\partial \widetilde{\boldsymbol{B}}}{\partial \boldsymbol{Q}_{j}}\right>&=\Biggl<\sum_{j,u}\sum_{\alpha,\beta,n}\left(\mathscr{C}_{uj}Q^{\beta}_{u}Q_{j}^{n}\right)\kappa^{\alpha n}\boldsymbol{e}_{\beta}\boldsymbol{e}_{\alpha}\Biggr>+\Biggl<\sum_{j,v}\sum_{\alpha,\beta,n}\kappa^{\alpha n}\left(\mathscr{C}_{jv}Q_{j}^{n}Q^{\gamma}_{v}\right)\boldsymbol{e}_{\alpha}\boldsymbol{e}_{\gamma}\Biggr>
\end{split}
\end{equation}
Recognizing that
\begin{equation}\label{eq:kappa_prod1}
\begin{split}
\boldsymbol{\kappa}\cdot(\boldsymbol{Q}_{u}\boldsymbol{Q}_{v})&=\sum_{m,n,r,s}\kappa^{mn}Q^{r}_{u}Q^{s}_{v}\left[(\boldsymbol{e}_m\boldsymbol{e}_n)\cdot(\boldsymbol{e}_{r}\boldsymbol{e}_{s})\right]\\[5pt]
&=\sum_{m,r,s}\kappa^{mr}Q^{r}_{u}Q^{s}_{v}\boldsymbol{e}_m\boldsymbol{e}_{s},
\end{split}
\end{equation}
and
\begin{equation}\label{eq:kappa_prod2}
\begin{split}
(\boldsymbol{Q}_{u}\boldsymbol{Q}_{v})\cdot\boldsymbol{\kappa}^{T}&=\sum_{m,n,r,s}\left(Q^{r}_{u}Q^{s}_{v}\boldsymbol{e}_{r}\boldsymbol{e}_{s}\right)\cdot\left(\kappa^{nm}\boldsymbol{e}_{m}\boldsymbol{e}_{n}\right)\\[5pt]
&=\sum_{m,n,r}Q^{r}_{u}Q^{m}_{v}\kappa^{nm}\boldsymbol{e}_{r}\boldsymbol{e}_{n},
\end{split}
\end{equation}
we may write
\begin{equation}\label{eq:flow_term_final}
\begin{split}
\left<\sum_{j}\left[\boldsymbol{\kappa}\cdot\boldsymbol{Q}_{j}\right]\cdot\dfrac{\partial \widetilde{\boldsymbol{B}}}{\partial \boldsymbol{Q}_{j}}\right>&=\left<\sum_{j,u}\mathscr{C}_{uj}\boldsymbol{Q}_{u}\boldsymbol{Q}_{j}\right>\cdot\boldsymbol{\kappa}^{T}+\boldsymbol{\kappa}\cdot\left<\sum_{j,v}\mathscr{C}_{jv}\boldsymbol{Q}_{j}\boldsymbol{Q}_{v}\right>\\[10pt]
&=\left<\sum_{u,v}\mathscr{C}_{uv}\boldsymbol{Q}_{u}\boldsymbol{Q}_{v}\right>\cdot\boldsymbol{\kappa}^{T}+\boldsymbol{\kappa}\cdot\left<\sum_{u,v}\mathscr{C}_{uv}\boldsymbol{Q}_{u}\boldsymbol{Q}_{v}\right>
\end{split}
\end{equation}
Next, the Brownian force term may be processed as 
\begin{equation}\label{eq:brown1}
\begin{split}
&\Biggl<\sum_{j,k}\left(A_{jk}\dfrac{\partial \ln \psi}{\partial \boldsymbol{Q}_k}\right)\cdot\dfrac{\partial \widetilde{\boldsymbol{B}}}{\partial \boldsymbol{Q}_{j}}\Biggr>\\[5pt]
&=\Biggl<\sum_{j,k}\sum_{s}\left(A_{jk}\dfrac{\partial \ln\psi}{\partial Q^{s}_k}\boldsymbol{e}_{s}\right)\cdot\Biggl[\sum_{u}\sum_{\alpha,\beta}\mathscr{C}_{uj}Q^{\beta}_{u}\boldsymbol{e}_{\alpha}\boldsymbol{e}_{\beta}\boldsymbol{e}_{\alpha}+\sum_{v}\sum_{\alpha,\gamma}\mathscr{C}_{jv}Q^{\gamma}_{v}\boldsymbol{e}_{\alpha}\boldsymbol{e}_{\alpha}\boldsymbol{e}_{\gamma}\Biggr]\Biggr>\\[5pt]
&=\uline{\Biggl<\Biggl[\sum_{j,k,u}\,\,\sum_{\alpha,\beta}A_{jk}\dfrac{\partial \ln\psi}{\partial Q^{\alpha}_k}\mathscr{C}_{uj}Q^{\beta}_{u}\boldsymbol{e}_{\beta}\boldsymbol{e}_{\alpha}\Biggr]\Biggr>}+\dotuline{\Biggl<\Biggl[\sum_{j,k,v}\,\,\sum_{\alpha,\gamma}A_{jk}\dfrac{\partial \ln\psi}{\partial Q^{\alpha}_k}\mathscr{C}_{jv}Q^{\gamma}_{v}\boldsymbol{e}_{\alpha}\boldsymbol{e}_{\gamma}\Biggr]\Biggr>}
\end{split}
\end{equation}
The solid-underlined term in Eq.~(\ref{eq:brown1}) may be simplified as shown below
\begin{equation}\label{eq:brown_underlined}
\begin{split}
\Biggl<\Biggl[\sum_{j,k,u}\,\,\sum_{\alpha,\beta}A_{jk}\dfrac{\partial \ln\psi}{\partial Q^{\alpha}_k}\mathscr{C}_{uj}Q^{\beta}_{u}\boldsymbol{e}_{\beta}\boldsymbol{e}_{\alpha}\Biggr]\Biggr>&=\Biggl<\sum_{k,u}\sum_{\alpha,\beta}\left(\sum_{j}A_{kj}\mathscr{C}_{ju}\right)\dfrac{\partial \ln\psi}{\partial Q^{\alpha}_k}Q^{\beta}_{u}\boldsymbol{e}_{\beta}\boldsymbol{e}_{\alpha}\Biggr>\\[5pt]
&=\sum_{k}\sum_{\alpha,\beta}\left[\int\left(\dfrac{1}{\psi}\right)\left(\dfrac{\partial \psi}{\partial Q_k^{\alpha}}\right)Q^{\beta}_{k}\psi\,d\boldsymbol{Q}\right]\boldsymbol{e}_{\beta}\boldsymbol{e}_{\alpha}
\end{split}
\end{equation}
The integral in Eq.~(\ref{eq:brown_underlined}) may be simplified as 
\begin{equation}\label{eq:brown2}
\begin{split}
\int\left(\dfrac{\partial \psi}{\partial Q_k^{\alpha}}\right)Q^{\beta}_{k}\,d\boldsymbol{Q}&=\int\left[\dfrac{\partial}{\partial Q^{\alpha}_{k}}\left(Q^{\beta}_{k}\psi\right)-\psi\dfrac{\partial Q^{\beta}_{k}}{\partial Q^{\alpha}_k}\right]d\boldsymbol{Q}\\[5pt]
&=\dashuline{\int\dfrac{\partial}{\partial Q^{\alpha}_{k}}\left(Q^{\beta}_{k}\psi\right)d\boldsymbol{Q}}-\delta^{\alpha\beta}\int\psi\,d\boldsymbol{Q}
\end{split}
\end{equation}
The underlined term in Eq.~(\ref{eq:brown2}) vanishes due to the Gauss divergence theorem, and the integral in the second term is unity due to the normalization condition, and we have
\begin{equation}\label{eq:brown3}
\begin{split}
\int\left(\dfrac{\partial \psi}{\partial Q_k^{\alpha}}\right)Q^{\beta}_{k}\,d\boldsymbol{Q}
&=-\delta^{\alpha\beta}
\end{split}
\end{equation}
Combining Eq.~(\ref{eq:brown_underlined}) and ~(\ref{eq:brown3}), we may show that
\begin{equation}\label{eq:ul1}
\begin{split}
\Biggl<\Biggl[\sum_{j,k,u}\,\,\sum_{\alpha,\beta}A_{jk}\dfrac{\partial \ln\psi}{\partial Q^{\alpha}_k}\mathscr{C}_{uj}Q^{\beta}_{u}\boldsymbol{e}_{\beta}\boldsymbol{e}_{\alpha}\Biggr]\Biggr>=-\sum_{k=1}^{N}\sum_{\alpha,\beta}\delta^{\alpha\beta}\boldsymbol{e}_{\beta}\boldsymbol{e}_{\alpha}=-N\boldsymbol{\delta}
\end{split}
\end{equation}
Proceeding identically to the steps outlined in Eq.~(\ref{eq:brown_underlined}) to (\ref{eq:brown3}), the dot-underlined term on the RHS of Eq.~(\ref{eq:brown1}) also evaluates to be
\begin{equation}\label{eq:ul2}
\begin{split}
\Biggl<\Biggl[\sum_{j,k,v}\,\,\sum_{\alpha,\gamma}A_{jk}\dfrac{\partial \ln\psi}{\partial Q^{\alpha}_k}\mathscr{C}_{jv}Q^{\gamma}_{v}\boldsymbol{e}_{\alpha}\boldsymbol{e}_{\gamma}\Biggr]\Biggr>=-N\boldsymbol{\delta}
\end{split}
\end{equation}
From Equations~(\ref{eq:brown1}), (\ref{eq:ul1}) and (\ref{eq:ul2}), we may write
\begin{equation}\label{eq:brown_term_final}
\Biggl<\sum_{j,k}\left(A_{jk}\dfrac{\partial \ln \psi}{\partial \boldsymbol{Q}_k}\right)\cdot\dfrac{\partial \widetilde{\boldsymbol{B}}}{\partial \boldsymbol{Q}_{j}}\Biggr>=-2N\boldsymbol{\delta}
\end{equation}
The last two terms on the RHS of Equation~(\ref{eq:process2}) may be processed as follows
\begin{equation}\label{eq:force1}
\begin{split}
&\dfrac{H}{\zeta}\left<\sum_{j,k}A_{jk}f_k\boldsymbol{Q}_{k}\cdot\dfrac{\partial \widetilde{\boldsymbol{B}}}{\partial \boldsymbol{Q}_{j}}\right>+\dfrac{K}{\zeta}\left<\sum_{j,k}A_{jk}C_{k}\boldsymbol{Q}_{k}\cdot\dfrac{\partial \widetilde{\boldsymbol{B}}}{\partial \boldsymbol{Q}_{j}}\right>\\[5pt]
&=\dfrac{1}{\zeta}\Biggl<\sum_{j,k}A_{jk}\left[Hf_k+KC_k\right]\boldsymbol{Q}_{k}\cdot\dfrac{\partial \widetilde{\boldsymbol{B}}}{\partial \boldsymbol{Q}_{j}}\Biggr>
\end{split}
\end{equation}
Substituting Eq.~\ref{eq:b_part} into Eq.~(\ref{eq:force1}), we write
\begin{align}\label{eq:force2}
&\Biggl<\sum_{j,k}A_{jk}\left[Hf_k+KC_k\right]\boldsymbol{Q}_{k}\cdot\dfrac{\partial \widetilde{\boldsymbol{B}}}{\partial \boldsymbol{Q}_{j}}\Biggr>\nonumber\\[5pt]
&=\Biggl<\sum_{j,k}\sum_{s}A_{jk}\left[Hf_k+KC_k\right]Q^{s}_{k}\boldsymbol{e}_{s}\cdot\Biggl[\sum_{u}\sum_{\alpha,\beta}\mathscr{C}_{uj}Q^{\beta}_{u}\boldsymbol{e}_{\alpha}\boldsymbol{e}_{\beta}\boldsymbol{e}_{\alpha}+\sum_{v}\sum_{\alpha,\gamma}\mathscr{C}_{jv}Q^{\gamma}_{v}\boldsymbol{e}_{\alpha}\boldsymbol{e}_{\alpha}\boldsymbol{e}_{\gamma}\Biggr]\Biggr>\nonumber\\[5pt]
&=\Biggl<\sum_{k,u}\,\sum_{\alpha,\beta}\left(\sum_{j}A_{kj}\mathscr{C}_{ju}\right)\left[Hf_k+KC_k\right]{Q}^{\alpha}_{k}Q^{\beta}_{u}\boldsymbol{e}_{\beta}\boldsymbol{e}_{\alpha}\\[5pt]
&+\sum_{k,v}\,\sum_{\alpha,\gamma}\left(\sum_{j}A_{kj}\mathscr{C}_{jv}\right)\left[Hf_k+KC_k\right]{Q}^{\alpha}_{k}Q^{\gamma}_{v}\boldsymbol{e}_{\alpha}\boldsymbol{e}_{\gamma}\Biggr>\nonumber\\[5pt]
&=2\,\Biggl<\sum_{k}\dashuline{\left[Hf_k\boldsymbol{Q}_{k}+KC_k\boldsymbol{Q}_{k}\right]}\boldsymbol{Q}_{k}\Biggr>\nonumber
\end{align}
The underlined term in Eq.~(\ref{eq:force2}) represents the total force, $\boldsymbol{F}^{\text{c}}_{k}$, in the $k^{\text{th}}$ connector vector due to the spring and the dashpot, and may be written as
\begin{equation}\label{eq:fc_def}
\boldsymbol{F}^{\text{c}}_{k}=\boldsymbol{F}^{\text{s}}_{k}+KC_k\boldsymbol{Q}_{k}
\end{equation}
where $\boldsymbol{F}^{\text{s}}_{k}$ is as defined in Eq.~(\ref{eq:gen_spforce_def}), and we finally have
\begin{equation}\label{eq:force_term_final}
\left<\sum_{j,k}A_{jk}\left[Hf_k+KC_k\right]\boldsymbol{Q}_{k}\cdot\dfrac{\partial \widetilde{\boldsymbol{B}}}{\partial \boldsymbol{Q}_{j}}\right>=2\left<\sum_{k}\boldsymbol{F}^{\text{c}}_{k}\boldsymbol{Q}_{k}\right>=2\left<\sum_{k}\boldsymbol{Q}_{k}\boldsymbol{F}^{\text{c}}_{k}\right>
\end{equation}
Combining Equations~(\ref{eq:process2}), ~(\ref{eq:flow_term_final}), (\ref{eq:brown_term_final}), and \ref{eq:force_term_final}, we may write
\begin{equation}\label{eq:of_change_fin}
\begin{split}
\dfrac{d}{dt}\left<\widetilde{\boldsymbol{B}}\right>&=\left<\sum_{u,v}\mathscr{C}_{uv}\boldsymbol{Q}_{u}\boldsymbol{Q}_{v}\right>\cdot\boldsymbol{\kappa}^{T}+\boldsymbol{\kappa}\cdot\left<\sum_{u,v}\mathscr{C}_{uv}\boldsymbol{Q}_{u}\boldsymbol{Q}_{v}\right>+\dfrac{2k_BT}{\zeta}\left(N_{\text{b}}-1\right)\boldsymbol{\delta}\\[5pt]
&-\dfrac{2}{\zeta}\left<\sum_{k}\boldsymbol{Q}_{k}\boldsymbol{F}^{\text{c}}_{k}\right>
\end{split}
\end{equation}
and therefore, from Eq.~(\ref{eq:giesekus_exp_start_point}),
\begin{equation}
\begin{split}
\left<\sum_{u=1}^{N_{\text{b}}-1}\sum_{v=1}^{N_{\text{b}}-1}{\mathscr{C}_{uv}}\boldsymbol{Q}_{u}\boldsymbol{Q}_{v}\right>_{(1)}=\dfrac{2k_BT}{\zeta}\left(N_{\text{b}}-1\right)\boldsymbol{\delta}-\dfrac{2}{\zeta}\left<\sum_{k}\boldsymbol{Q}_{k}\boldsymbol{F}^{\text{c}}_{k}\right>
\end{split}
\end{equation}
with the resultant stress tensor expression given as
\begin{equation}\label{eq:kramers_equiv}
\begin{split}
\boldsymbol{\tau}_{\text{p}}&=\dfrac{n_{\text{p}}\zeta}{2}\left<\sum_{u=1}^{N_{\text{b}}-1}\sum_{v=1}^{N_{\text{b}}-1}{\mathscr{C}_{uv}}\boldsymbol{Q}_{u}\boldsymbol{Q}_{v}\right>_{(1)}=n_{\text{p}}k_BT\left(N_{\text{b}}-1\right)\boldsymbol{\delta}-n_{\text{p}}\left<\sum^{N_{\text{b}}-1}_{k=1}\boldsymbol{Q}_{k}\boldsymbol{F}^{\text{c}}_{k}\right>
\end{split}
\end{equation}
It is thus established that the stress tensor expression for free-draining bead-spring-dashpot chains is formally similar to that given by the Kramers expression, with the connector vector force suitably modified to account for the contribution from the dashpot. In order to obtain a closed-form expression that may be used for the calculation of stress tensor components from BD simulations, however, it is essential that the complete expression for $C_{k}$ be substituted into Eq.~(\ref{eq:fc_def}) and simplified.

Starting from Eq.~(\ref{eq:kramers_equiv}), we have
\begin{equation}\label{eq:st_begin}
\begin{split}
\boldsymbol{\tau}_{\text{p}}&=n_{\text{p}}k_BT\left(N_{\text{b}}-1\right)\boldsymbol{\delta}-n_{\text{p}}\left<\sum_{k}\boldsymbol{Q}_{k}\boldsymbol{F}^{\text{s}}_{k}\right>-n_{\text{p}}K\left<\sum_{k}C_k\boldsymbol{Q}_{k}\boldsymbol{Q}_{k}\right>
\end{split}
\end{equation}
Multiplying both sides of Eq.~(\ref{eq:decoup}) by $\boldsymbol{Q}_{k}$,
\begin{align}\label{eq:ck_qk}
C_k\boldsymbol{Q}_{k}&=\left(\dfrac{1}{1+\epsilon}\right)\sum_{l}\dashuline{\left(\dfrac{\boldsymbol{Q}_{k}}{Q_{k}}\right){\boldsymbol{\Lambda}}^{(k)}_{l}}\cdot\left(\boldsymbol{\kappa}\cdot\boldsymbol{Q}_l\right)-\left(\dfrac{k_BT}{\zeta}\right)\left(\dfrac{1}{1+\epsilon}\right)\sum_{l}\dotuline{\left(\dfrac{\boldsymbol{Q}_{k}}{Q_{k}}\right){\boldsymbol{J}}^{(k)}_{l}}\cdot\left(\dfrac{\partial \ln \psi}{\partial \boldsymbol{Q}_{l}}\right)\nonumber\\[5pt]
&+\left(\dfrac{1}{\zeta}\right)\left(\dfrac{1}{1+\epsilon}\right)\sum_{l}\left(\dfrac{\boldsymbol{Q}_{k}}{Q_{k}}\right){\boldsymbol{J}}^{(k)}_{l}\cdot\boldsymbol{F}^{\text{s}}_{l}
\end{align}
The dash- and dot-underlined terms in Eq.~(\ref{eq:ck_qk}) may be replaced by $\boldsymbol{\alpha}_{kl}$ and $\boldsymbol{\mu}_{kl}$, respectively, according to Eq.~(\ref{eq:uv_def}), to give
\begin{equation}\label{eq:ck_simp}
\begin{split}
C_k\boldsymbol{Q}_{k}&=\left(\dfrac{1}{1+\epsilon}\right)\sum_{l}\boldsymbol{\alpha}_{kl}\cdot\left(\boldsymbol{\kappa}\cdot\boldsymbol{Q}_l\right)-\left(\dfrac{k_BT}{\zeta}\right)\left(\dfrac{1}{1+\epsilon}\right)\sum_{l}\boldsymbol{\mu}_{kl}\cdot\left(\dfrac{\partial \ln \psi}{\partial \boldsymbol{Q}_{l}}\right)\\[5pt]
&+\left(\dfrac{1}{\zeta}\right)\left(\dfrac{1}{1+\epsilon}\right)\sum_{l}\boldsymbol{\mu}_{kl}\cdot\boldsymbol{F}^{\text{s}}_{l}
\end{split}
\end{equation}
Multiplying both sides of Eq.~(\ref{eq:ck_simp}) by $\boldsymbol{Q}_{k}$ again, 
\begin{equation}\label{eq:ck_qk_qk}
\begin{split}
C_k\boldsymbol{Q}_{k}\boldsymbol{Q}_{k}&=\left(\dfrac{1}{1+\epsilon}\right)\sum_{l}\boldsymbol{Q}_{k}\boldsymbol{\alpha}_{kl}\cdot\left(\boldsymbol{\kappa}\cdot\boldsymbol{Q}_l\right)-\left(\dfrac{k_BT}{\zeta}\right)\left(\dfrac{1}{1+\epsilon}\right)\sum_{l}\boldsymbol{Q}_{k}\boldsymbol{\mu}_{kl}\cdot\left(\dfrac{\partial \ln \psi}{\partial \boldsymbol{Q}_{l}}\right)\\[5pt]
&+\left(\dfrac{1}{\zeta}\right)\left(\dfrac{1}{1+\epsilon}\right)\sum_{l}\boldsymbol{Q}_{k}\boldsymbol{\mu}_{kl}\cdot\boldsymbol{F}^{\text{s}}_{l}
\end{split}
\end{equation}
Using Eqs.~(\ref{eq:alpha_def}) and (\ref{eq:chi_def}) to simplify the first term on the RHS of Eq.~(\ref{eq:ck_qk_qk}), summing over the index $k$, and taking an ensemble average on both sides of Eq.~(\ref{eq:ck_qk_qk}), we obtain
\begin{equation}\label{eq:ck_sum_av}
\begin{split}
&\left<\sum_{k}C_k\boldsymbol{Q}_{k}\boldsymbol{Q}_{k}\right>=\left(\dfrac{1}{1+\epsilon}\right)\left<\sum_{k,l}\boldsymbol{\kappa}:\dfrac{\chi^{(k)}_l\boldsymbol{Q}_l\boldsymbol{Q}_l\boldsymbol{Q}_k\boldsymbol{Q}_k}{\boldsymbol{Q}_l\boldsymbol{Q}_k}\right>\\[5pt]
&-\left(\dfrac{k_BT}{\zeta}\right)\left(\dfrac{1}{1+\epsilon}\right)\left<\sum_{k,l}\boldsymbol{Q}_{k}\boldsymbol{\mu}_{kl}\cdot\left(\dfrac{\partial \ln \psi}{\partial \boldsymbol{Q}_{l}}\right)\right>+\left(\dfrac{1}{\zeta}\right)\left(\dfrac{1}{1+\epsilon}\right)\left<\sum_{k,l}\boldsymbol{Q}_{k}\boldsymbol{\mu}_{kl}\cdot\boldsymbol{F}^{\text{s}}_{l}\right>
\end{split}
\end{equation}
The second and third terms on the RHS of Eq.~(\ref{eq:ck_sum_av}) are evaluated sequentially as shown below.
Starting with
\begin{equation}\label{eq:ck_process1}
\begin{split}
\Biggl<\sum_{k,l}\boldsymbol{Q}_{k}\boldsymbol{\mu}_{kl}\cdot\left(\dfrac{\partial \ln \psi}{\partial \boldsymbol{Q}_{l}}\right)\Biggr>&\equiv\Biggl<\sum_{k,l}\,\sum_{s,\alpha,\beta,\gamma}Q^{s}_{k}\mu^{\alpha\beta}_{kl}\dfrac{\partial\ln\psi}{\partial Q^{\gamma}_l}\left[\boldsymbol{e}_{s}\boldsymbol{e}_{\alpha}\boldsymbol{e}_{\beta}\cdot\boldsymbol{e}_{\gamma}\right]\Biggr>\\[5pt]
&=\sum_{k,l}\,\sum_{s,\alpha,\beta}\left[\int\,Q^{s}_{k}\mu^{\alpha\beta}_{kl}\left(\dfrac{1}{\psi}\right)\dfrac{\partial \psi}{\partial Q^{\beta}_l}\psi\,d\boldsymbol{Q}\right]\boldsymbol{e}_{s}\boldsymbol{e}_{\alpha}
\end{split}
\end{equation}
The integral in Eq.~(\ref{eq:ck_process1}) is solved as
\begin{equation}\label{eq:ck_process2}
\begin{split}
\int\,Q^{s}_{k}\,\mu^{\alpha\beta}_{kl}\dfrac{\partial \psi}{\partial Q^{\beta}_l}\,d\boldsymbol{Q}&=\int\Biggl\{\left[\dfrac{\partial}{\partial Q^{\beta}_{l}}\left(Q^{s}_{k}\,\mu^{\alpha\beta}_{kl}\psi\right)\right]-\left(Q^{s}_{k}\psi\right)\dfrac{\partial \mu_{kl}^{\alpha\beta}}{\partial Q^{\beta}_{l}}-\left(\mu_{kl}^{\alpha\beta}\psi\right)\dfrac{\partial Q_{k}^{s}}{\partial Q^{\beta}_{l}}\Biggr\}\,d\boldsymbol{Q}\\[10pt]
&=\dashuline{\int\left[\dfrac{\partial}{\partial Q^{\beta}_{l}}\left(Q^{s}_{k}\,\mu^{\alpha\beta}_{kl}\psi\right)\right]d\boldsymbol{Q}}-\int\left[Q^{s}_{k}\dfrac{\partial \mu_{kl}^{\alpha\beta}}{\partial Q^{\beta}_{l}}\right]\psi\,d\boldsymbol{Q}-\int\left[\mu_{kl}^{\alpha\beta}\delta_{kl}\delta^{s\beta}\right]\psi\,d\boldsymbol{Q}
\end{split}
\end{equation}
The underlined integral in Eq.~(\ref{eq:ck_process2}) vanishes due to the Gauss divergence theorem, and we get
\begin{equation}\label{eq:ck_process3}
\begin{split}
\int\,Q^{s}_{k}\,\mu^{\alpha\beta}_{kl}\dfrac{\partial \psi}{\partial Q^{\beta}_l}\,d\boldsymbol{Q}&=-\Biggl<Q^{s}_{k}\dfrac{\partial \mu_{kl}^{\alpha\beta}}{\partial Q^{\beta}_{l}}\Biggr>-\Biggl<\mu_{kl}^{\alpha\beta}\delta_{kl}\delta^{s\beta}\Biggr>
\end{split}
\end{equation}
Substituting Eq.~(\ref{eq:ck_process3}) into Eq.~(\ref{eq:ck_process1}), we obtain
\begin{equation}\label{eq:stress_brown_final}
\begin{split}
\Biggl<\sum_{k,l}\boldsymbol{Q}_{k}\boldsymbol{\mu}_{kl}\cdot\left(\dfrac{\partial \ln \psi}{\partial \boldsymbol{Q}_{l}}\right)\Biggr>&=-\Biggl<\sum_{k,l}\,\sum_{s,\alpha,\beta}Q^{s}_{k}\dfrac{\partial \mu_{kl}^{\alpha\beta}}{\partial Q^{\beta}_{l}}\boldsymbol{e}_{s}\boldsymbol{e}_{\alpha}\Biggr>-\Biggl<\sum_{k,l}\,\sum_{s,\alpha,\beta}\mu_{kl}^{\alpha\beta}\delta_{kl}\delta^{s\beta}\boldsymbol{e}_{s}\boldsymbol{e}_{\alpha}\Biggr>\\[5pt]
&=-\Biggl<\sum_{k,l}\boldsymbol{Q}_{k}\,\dfrac{\partial}{\partial \boldsymbol{Q}_{l}}\cdot\boldsymbol{\mu}^{T}_{kl}\Biggr>-\Biggl<\sum_{k}\boldsymbol{\mu}_{kk}^{T}\Biggr>
\end{split}
\end{equation}
With respect to the third term on the RHS of Eq.~(\ref{eq:ck_sum_av}), we note that
\begin{equation}\label{eq:tens_dot}
\boldsymbol{\mu}_{kl}\cdot\boldsymbol{F}^{\text{s}}_{l}=\boldsymbol{F}^{\text{s}}_{l}\cdot\boldsymbol{\mu}^{T}_{kl}
\end{equation}
which follows from the property of the tensor-dot product.
Combining Eqs.~(\ref{eq:ck_sum_av}), (\ref{eq:stress_brown_final}), and (\ref{eq:tens_dot}), we obtain
\begin{equation}\label{eq:ck_eq_av_final}
\begin{split}
&\left<\sum_{k}C_k\boldsymbol{Q}_{k}\boldsymbol{Q}_{k}\right>=\left(\dfrac{k_BT}{\zeta}\right)\left(\dfrac{1}{1+\epsilon}\right)\Biggl[\Biggl<\sum_{k,l}\boldsymbol{Q}_{k}\,\dfrac{\partial}{\partial \boldsymbol{Q}_{l}}\cdot\boldsymbol{\mu}^{T}_{kl}\Biggr>+\Biggl<\sum_{k}\boldsymbol{\mu}_{kk}^{T}\Biggr>\Biggr]\\[5pt]
&+\left(\dfrac{1}{\zeta}\right)\left(\dfrac{1}{1+\epsilon}\right)\left<\sum_{k,l}\boldsymbol{Q}_{k}\boldsymbol{F}^{\text{s}}_{l}\cdot\boldsymbol{\mu}^{T}_{kl}\right>+\left(\dfrac{1}{1+\epsilon}\right)\left<\sum_{k,l}\boldsymbol{\kappa}:\dfrac{\chi^{(k)}_l\boldsymbol{Q}_l\boldsymbol{Q}_l\boldsymbol{Q}_k\boldsymbol{Q}_k}{\boldsymbol{Q}_l\boldsymbol{Q}_k}\right>
\end{split}
\end{equation}
Finally, by substituting Eq.~(\ref{eq:ck_eq_av_final}) into Eq.~(\ref{eq:st_begin}), the stress tensor expression is obtained as
\begin{align}\label{eq:st_dim}
&\boldsymbol{\tau}_{\text{p}}=n_{\text{p}}k_BT\left(N_{\text{b}}-1\right)\boldsymbol{\delta}-n_{\text{p}}\left<\sum_{k}\boldsymbol{Q}_{k}\boldsymbol{F}^{\text{s}}_{k}\right>-\left(\dfrac{n_{\text{p}}k_BT}{\zeta}\right)\left(\dfrac{K}{1+\epsilon}\right)\Biggl[\Biggl<\sum_{k,l}\boldsymbol{Q}_{k}\,\dfrac{\partial}{\partial \boldsymbol{Q}_{l}}\cdot\boldsymbol{\mu}^{T}_{kl}\Biggr>\nonumber\\[5pt]
&+\Biggl<\sum_{k}\boldsymbol{\mu}_{kk}^{T}\Biggr>\Biggr]-\left(\dfrac{n_{\text{p}}}{\zeta}\right)\left(\dfrac{K}{1+\epsilon}\right)\left<\sum_{k,l}\boldsymbol{Q}_{k}\boldsymbol{F}^{\text{s}}_{l}\cdot\boldsymbol{\mu}^{T}_{kl}\right>\\[5pt]
&-\left(\dfrac{n_{\text{p}}\zeta}{2}\right)\left(\dfrac{\epsilon}{1+\epsilon}\right)\left<\sum_{k,l}\boldsymbol{\kappa}:\dfrac{\chi^{(k)}_l\boldsymbol{Q}_l\boldsymbol{Q}_l\boldsymbol{Q}_k\boldsymbol{Q}_k}{\boldsymbol{Q}_l\boldsymbol{Q}_k}\right>\nonumber
\end{align}
Upon scaling and simplification using the length- and timescales, $l_{H}$ and $\lambda_{H}$, the dimensionless form of the stress tensor is given by
\begin{align}\label{eq:stress_tensor_dimless_appendix}
\dfrac{\boldsymbol{\tau}_{\text{p}}}{n_{\text{p}}k_BT}&=\left(N_{\text{b}}-1\right)\boldsymbol{\delta}-\left<\sum_{k}\boldsymbol{Q}^{*}_k\boldsymbol{F}^{*\text{s}}_{k}\right>-\dfrac{1}{2}\left(\dfrac{\epsilon}{1+\epsilon}\right)\Biggl[\left<\sum_{k,l}\left(\boldsymbol{Q}^{*}_{k}\boldsymbol{F}^{*\text{s}}_{l}\right)\cdot\boldsymbol{\mu}_{kl}^{T}\right>+\left<\sum_{k}\boldsymbol{\mu}^{T}_{kk}\right>\nonumber\\[5pt]
&+\left<\sum_{k,l}\boldsymbol{Q}^{*}_{k}{\dfrac{\partial}{\partial \boldsymbol{Q}^{*}_l}\cdot\boldsymbol{\mu}_{kl}^{T}}\right>\Biggr]-\left(\dfrac{2\epsilon}{1+\epsilon}\right)\boldsymbol{\kappa}^{*}:\left<\sum_{k,l}\dfrac{\chi^{(k)}_{l}\boldsymbol{Q}^{*}_l\boldsymbol{Q}^{*}_l\boldsymbol{Q}^{*}_k\boldsymbol{Q}^{*}_k}{Q^{*}_lQ^{*}_k}\right>
\end{align}
which is reproduced as Eq.~(13) of the main text.

\section{\label{sec:var_red} Variance reduction}

\begin{figure*}[t]
\begin{tabular}{c c}
\includegraphics[width=75mm]{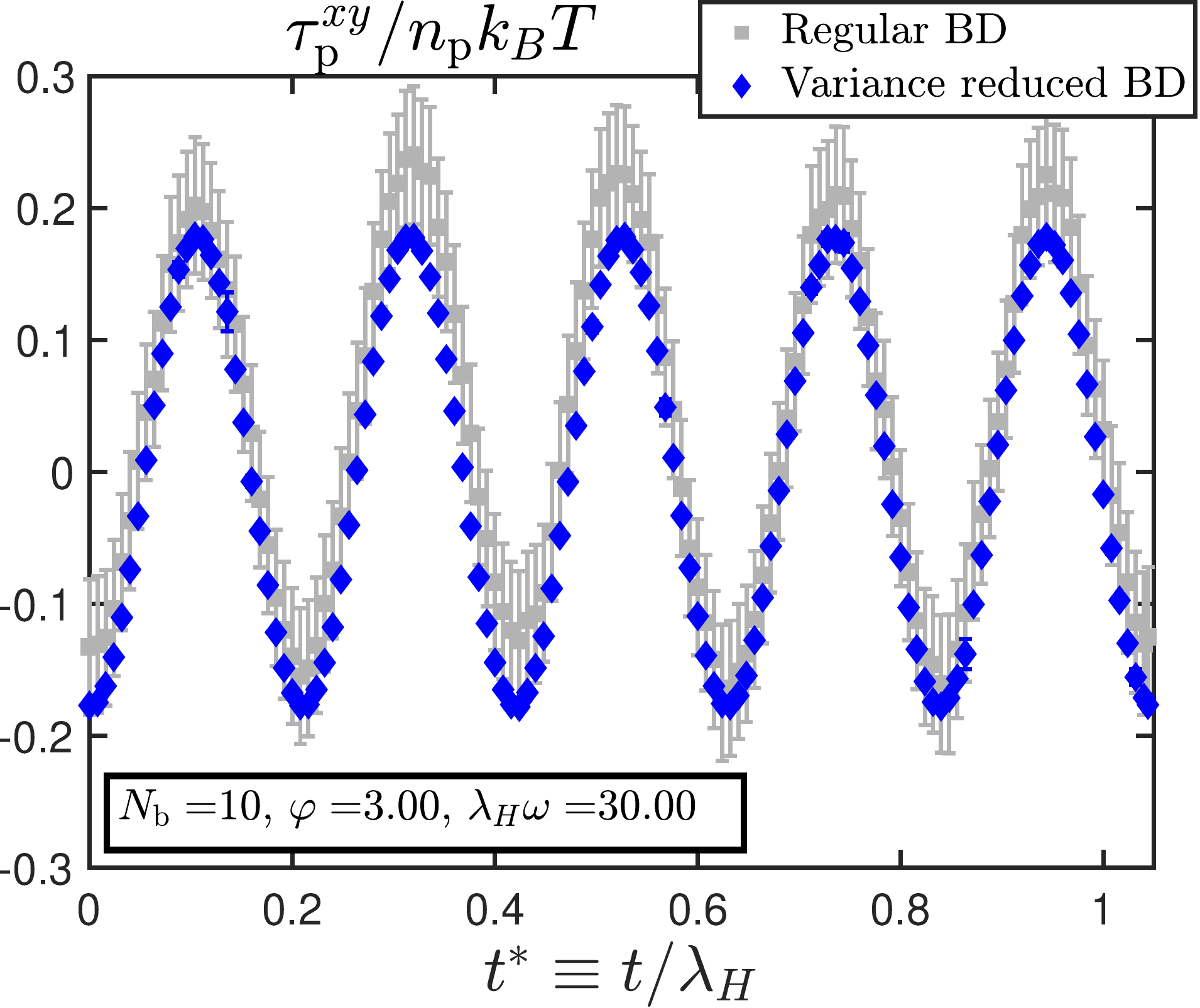}&
\includegraphics[width=75mm]{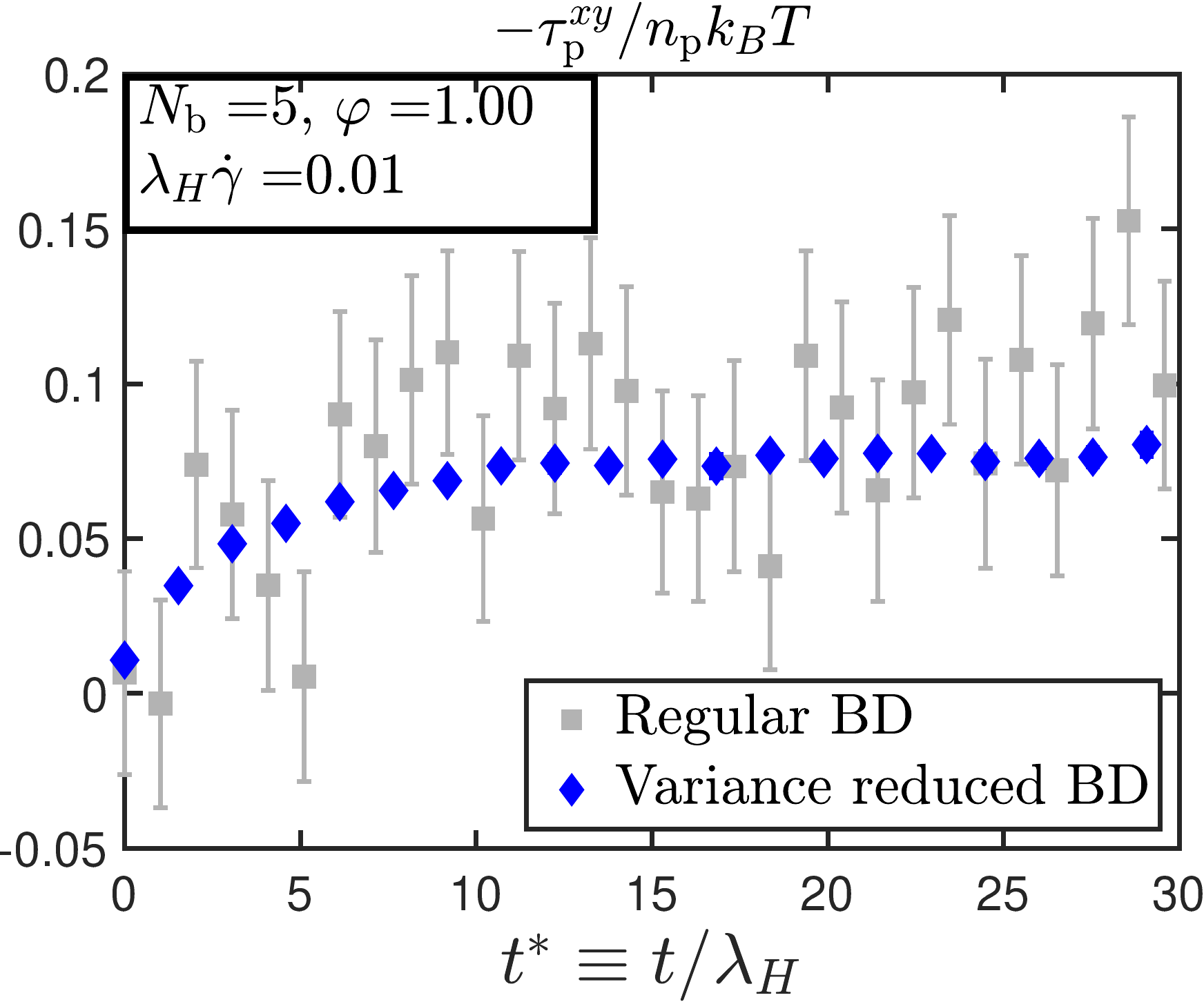} \\
 (a) &  (b)\\[5pt]
\end{tabular}
    
\caption{Illustration of the efficacy of variance reduction for (a) a ten-bead chain with internal friction in small amplitude oscillatory shear flow, and (b) a five bead chain with internal friction in simple shear flow.}
\label{fig:var-red-compare}
\end{figure*}

The variance reduction scheme proposed by~\citet{Wagner1997} is used in the present work. We provide below the expressions used in variance reduced BD simulations for the calculation of viscometric functions in flow. The procedure consists of simulating two trajectories in parallel, one of which accounts for the presence of the flow term, while the other is an equilibrium run without the flow term included. The governing equations for the two trajectories are as given below, reproduced from Eq. (5) in the paper,
\begin{equation}\label{eq:main_traj}
\begin{split}
d\widetilde{\bm{Q}}^{*}&=\Biggl[\bm{\mathcal{K}}^{*}\cdot\widetilde{\bm{Q}}^{*}-\left(\dfrac{\varphi}{1+2\varphi}\right){\bm{\mathcal{U}}}\cdot\left(\bm{\mathcal{K}}^{*}\cdot\widetilde{\bm{Q}}^{*}\right)-\dfrac{1}{4}\bm{\mathcal{D}}\cdot\widetilde{\bm{F}}^{*\text{s}}-\dfrac{1}{4}\left(\dfrac{\varphi}{1+2\varphi}\right)\widetilde{\bm{v}}^{*}\Biggr]dt^{*}+{\dfrac{1}{\sqrt{2}}}\bm{\mathcal{B}}\cdot d\bm{\widetilde{W}}^{*}
\end{split}
\end{equation}
\begin{equation}\label{eq:par_traj}
\begin{split}
d\widetilde{\bm{Q}}^{*}_{\text{eq}}&=\Biggl[-\dfrac{1}{4}\bm{\mathcal{D}}_{\text{eq}}\cdot\widetilde{\bm{F}}_{\text{eq}}^{*\text{s}}-\dfrac{1}{4}\left(\dfrac{\varphi}{1+2\varphi}\right)\widetilde{\bm{v}}_{\text{eq}}^{*}\Biggr]dt^{*}+{\dfrac{1}{\sqrt{2}}}\bm{\mathcal{B}}_{\text{eq}}\cdot d\bm{\widetilde{W}}^{*}
\end{split}
\end{equation}
The Wiener numbers that appear in Eqs.~(\ref{eq:main_traj}) and ~(\ref{eq:par_traj}) are identical. The initial configurations for both the trajectories are sampled from the same equilibrium distribution. 

In Fig.~\ref{fig:var-red-compare}, the effectiveness of variance reduction in simple and oscillatory shear flow has been illustrated by comparison against data obtained from control simulations without variance reduction, for the same ensemble size of $10^{4}$ trajectories. A step-size of $\Delta t^*=10^{-3}$ has been used for the oscillatory flow simulations, while $\Delta t^*=10^{-2}$ for simulations of simple shear flow.

\section{\label{sec:app_b} A recursive algorithm for the calculation of gradients of continued fractions}
The connector vectors and associated quantities appearing in this section are in their dimensionless form, with the asterisks omitted for the sake of notational simplicity. 

The discrete Euler version of the governing stochastic differential equation for bead-spring-dashpot chains with $N_{\text{b}}$ beads and $N\equiv\left(N_{\text{b}}-1\right)$ springs, and the stress tensor expression, are reproduced here from Sec.~II of the main paper
\begin{align}\label{eq:exp_eul_dimless}
\widetilde{\bm{Q}}\left(t_{n+1}\right)=&\widetilde{\bm{Q}}_{n}+\Biggl[\bm{\mathcal{K}}\left(t_{n}\right)\cdot\widetilde{\bm{Q}}\left(t_{n}\right)-\left(\dfrac{\varphi}{1+2\varphi}\right){\bm{\mathcal{U}}}\left(t_{n}\right)\cdot\left(\bm{\mathcal{K}}\left(t_{n}\right)\cdot\widetilde{\bm{Q}}\left(t_{n}\right)\right)-\dfrac{1}{4}\bm{\mathcal{D}}\left(t_{n}\right)\cdot\widetilde{\bm{F}}^{\text{s}}\left(t_{n}\right)\\[5pt]
&-\dashuline{\left(\dfrac{1}{4}\right)\left(\dfrac{\varphi}{1+2\varphi}\right)\widetilde{\bm{v}}\left(t_{n}\right)}\Biggr]\Delta t_{n}+ {\dfrac{1}{\sqrt{2}}}\Delta \bm{\widetilde{S}}_{n}\nonumber
\end{align}
\vskip-10pt
\begin{align}\label{eq:stress_tensor}
&\dfrac{\boldsymbol{\tau}_{\text{p}}}{n_{\text{p}}k_BT}=\left(N_{\text{b}}-1\right)\boldsymbol{\delta}-\Biggl[\left<\sum^{N}_{k=1}\bm{Q}_k\bm{F}^{\text{s}}_{k}\right>-\dfrac{1}{2}\left(\dfrac{\epsilon}{1+\epsilon}\right)\left<\sum^{N}_{k,l=1}\left(\bm{Q}_{k}\bm{F}^{\text{s}}_{l}\right)\cdot\boldsymbol{\mu}_{kl}^{T}\right>\Biggr]\\[5pt]
&-\dfrac{1}{2}\left(\dfrac{\epsilon}{1+\epsilon}\right)\Biggl[\left<\sum^{N}_{k=1}\boldsymbol{\mu}^{T}_{kk}\right>+\left<\sum^{N}_{k,l=1}\bm{Q}_{k}\dotuline{\dfrac{\partial}{\partial \bm{Q}_l}\cdot\boldsymbol{\mu}_{kl}^{T}}\right>\Biggr]-\left(\dfrac{2\epsilon}{1+\epsilon}\right)\boldsymbol{\kappa}:\left<\sum^{N}_{k,l=1}\dfrac{\chi^{(k)}_{l}\bm{Q}_l\bm{Q}_l\bm{Q}_k\bm{Q}_k}{Q_lQ_k}\right>\nonumber
\end{align}
where $\epsilon=2\varphi$, $\widetilde{\bm{v}}\equiv\left[\widehat{\bm{v}}_{1},\,\widehat{\bm{v}}_{2}\,...,\,\widehat{\bm{v}}_{N}\right]$, with $\widehat{\bm{v}}_{j}=\sum_{k=1}^{N}\left({\partial}/{\partial \bm{Q}_{k}}\right)\cdot\bm{V}_{jk}^{T}$.
Clearly, Eqs.~(\ref{eq:exp_eul_dimless}) and ~(\ref{eq:stress_tensor}) require the calculation of the divergences $\left({\partial}/{\partial \bm{Q}_k}\right)\cdot\bm{V}^{T}_{jk}$and $({\partial}/{\partial \bm{Q}_l})\cdot\boldsymbol{\mu}_{kl}^{T}$ for all values of $\{j,k,l\}\in [1,N]$. It is straightforward to evaluate these divergences numerically, using the central difference approximation scheme for the calculation of gradients. In this section, an analytical route for the calculation of these gradients is presented.

It may be seen that both $\bm{V}^{T}_{jk}$ and $\boldsymbol{\mu}_{kl}^{T}$ possess essentially the same structure, i.e; they may be written as a sum of $n_{\text{t}}$ tensors, as
\begin{equation}
\bm{V}^{T}_{jk}=\boldsymbol{\pi}_{1}+\boldsymbol{\pi}_{2}+\dots+\boldsymbol{\pi}_{n_{\text{t}}}
\end{equation} 
where each term on the RHS has the following general structure
\begin{equation}
\boldsymbol{\pi}_{q}=h\left(M_q,P_q\right)\left[\dfrac{\bm{Q}_i\bm{Q}_m}{Q_iQ_m}\right]
\end{equation}
where $q\in\left[1,n_{\text{t}}\right]$, and $i,m\in\left[\left(q-1\right),\left(q+1\right)\right]$. In general, the quantities $M_k$ and $P_k$ are defined recursively as follows
\begin{equation}\label{eq:recdef_m_p}
\begin{split}
M_k&=p\left(\dfrac{L^2_{k-1}}{1-M_{k-1}}\right); \quad \text {with} \quad M_{1}=0\\[5pt]
P_k&=p\left(\dfrac{L^{2}_{k}}{1-P_{k+1}}\right); \quad \text {with} \quad P_{N}=0
\end{split}
\end{equation}
where
\begin{equation}\label{eq:p_l_def}
p=\left(\dfrac{K}{\zeta+2K}\right)^2;\quad\, L_{k}\equiv\cos\theta_k=\dfrac{\bm{Q}_k\cdot\bm{Q}_{k+1}}{Q_kQ_{k+1}}
\end{equation}
For any scalar $h$, and any tensor $\bm{H}$, the divergence of their product obeys the following identity,
\begin{align}
\boldsymbol{\nabla}\cdot\left(h\bm{H}\right)=\left(\boldsymbol{\nabla}h\right)\cdot\bm{H}+h\boldsymbol{\nabla}\cdot\bm{H}
\end{align}
The evaluation of $\left({\partial}/{\partial \bm{Q}_k}\right)\cdot\bm{V}^{T}_{jk}$ or $({\partial}/{\partial \bm{Q}_l})\cdot\boldsymbol{\mu}_{kl}^{T}$ would require knowledge of the gradient of the scalar prefactor, ${\partial h}/{\partial \bm{Q}_k}$. The difficulty in analytically evaluating this gradient term may be illustrated by considering a specific form of $h(M_q,P_q)$, say, $h\equiv [1-M_6]^{-1}$, which is encountered in the calculation of $\boldsymbol{\Gamma}^{(7)}_{6}$, required for the construction of $\boldsymbol{\Lambda}^{(7)}_6$, and subsequently that of the elements of the block matrix, $\bm{\mathcal{U}}$, as evident from Eqs.~(A17), (A19) and (A23) of the main paper. Suppose it is desired to calculate the gradient of $h$ with respect to $\bm{Q}_3$. 
We may begin by representing $h$ as
\begin{align}\label{eq:cont_example}
h=\dfrac{1}{1-M_{6}}&=\dfrac{1}{1-\dfrac{pL_5^2}{1-\dfrac{pL_4^2}{1-\dfrac{pL_3^2}{1-\dfrac{pL_2^2}{1-pL^2_1}}}}}
\end{align}
where only $L_2$ and $L_3$ are functions of $\bm{Q}_3$. Clearly, it is not trivial to apply the quotient-rule to evaluate the gradient of $h$ with respect to $\bm{Q}_3$. Fortunately, continued fractions of the type indicated in Eq.~(\ref{eq:cont_example}), and finite continued products of such fractions, may be expressed as ratios of polynomials~\cite{malila2014,Cretney2014}. 

Suppose we define
\begin{align}\label{eq:i_def}
I_k&=\left(1-M_1\right)\left(1-M_2\right)..\left(1-M_{k-1}\right)\left(1-M_{k}\right)
\end{align}
It can be shown that
\begin{align}\label{eq:i_rec_def}
I_{k+2}&=1-pL_1^2-pL_2^2-p\Biggl[\sum_{i=3}^{k+1}L^{2}_{i}I_{i-1}\Biggr]
\end{align}

Similarly, defining
\begin{align}\label{eq:d_def}
D_k=\left(1-P_k\right)\left(1-P_{k+1}\right)..\left(1-P_{N-1}\right)\left(1-P_{N}\right)
\end{align}
where $N$ is the number of springs in the chain, it can be shown that
\begin{align}\label{eq:d_rec_def}
D_{k}&=1-pL_{N-1}^2-pL_{N-2}^2-p\Biggl[\sum_{i=k}^{N-3}L^{2}_{i}D_{i+2}\Biggr]
\end{align}

Note that while $I_0$ is not defined according to Eq.~(\ref{eq:i_def}), we set $I_0=I_1=1$ for programming convenience. Similarly, we set $D_{N+1}=D_{N}=1$ for the same reason.

Using Eqs.~(\ref{eq:i_def})\textendash(\ref{eq:d_rec_def}), it can be shown that
\begin{align}\label{eq:fwd_rational}
\left[\prod_{i=l}^{k-1}\left(\dfrac{1}{1-M_i}\right)\right]&=\dfrac{I_{l-1}}{I_{k-1}},
\end{align}
\begin{align}\label{eq:bkwd_rational}
\left[\prod_{i=k+1}^{l}\left(\dfrac{1}{1-P_i}\right)\right]&=\dfrac{D_{l+1}}{D_{k+1}},
\end{align}
\begin{align}\label{eq:inv_mk}
\dfrac{1}{1-M_{k}}=\dfrac{I_{k-1}}{I_{k}},
\end{align}
\begin{align}\label{eq:inv_pk}
\dfrac{1}{1-P_{k}}=\dfrac{D_{k+1}}{D_{k}},
\end{align}
and
\begin{align}\label{eq:inv_pk_mi}
\dfrac{1}{1-M_{i}-P_{k}}=\dfrac{1}{\left(\dfrac{I_{i}}{I_{i-1}}\right)+\left(\dfrac{D_{k}}{D_{k+1}}\right)-1}
\end{align}
Using the polynomial representations introduced above, we may concisely write
\begin{align}\label{eq:ratio_example}
\dfrac{\partial h}{\partial \bm{Q}_3}=\dfrac{\partial }{\partial \bm{Q}_3}\left[\dfrac{1}{1-M_{6}}\right]=\dfrac{\partial }{\partial \bm{Q}_3}\left(\dfrac{I_5}{I_6}\right)
\end{align}
The quotient rule may now be applied to the ratio of polynomials given on the RHS of Eq.~(\ref{eq:ratio_example}). The next task is to obtain general expressions for the gradients, $\partial I_k/\partial \bm{Q}_j$, and $\partial D_k/\partial \bm{Q}_j$.

\subsection{\label{sec:fwd_poly}Forward continued product}
We have seen how a recurrence relation for $I_{k}$ can be obtained. We now provide an expression for $I_k$ as a polynomial in $p$. Consider $I_k$ where $1\leq k\leq N$ is any integer. 
The degree, $n$, of the polynomial in $p$ that expresses $I_k$ is given by $n=\floor*{\dfrac{k}{2}}$ where $\floor*{i}$ represents the greatest integer lesser than or equal to $i$. We can then write
\begin{align}\label{eq:i_polynomial}
I_{k}&=1-p\sum_{i=1}^{k-1}\,L_{i}^2f\left(1,k+1,i\right)+p^{2}\sum_{i=1}^{k-3}L_{i}^2\,f\left(2,k-1,i\right)-p^{3}\sum_{i=1}^{k-5}L_i^{2}\,f\left(3,k-3,i\right)\cdots\nonumber\\[5pt]
&+\left(-1\right)^{n}p^{n}\sum_{i=1}^{k-(2n-1)}L^{2}_{i}\,f\left(n,\left[k-(2n-3)\right],i\right)
\end{align}
where the function $f\left(m,l,j\right)$ is defined recursively as follows
\begin{equation}\label{eq:recdef_f}
{f}\left(m,l,j\right)=\sum_{s=j+2}^{l} L^{2}_s\,{f}\left(m-1,l+2,s\right)
\end{equation}
with 
\begin{equation}\label{eq:rec_stop_f}
{f}\left(1,l,j\right)=1\quad\quad \forall \quad l,j
\end{equation}
While Eq.~(\ref{eq:i_polynomial}) provides an expression for $I_k$ as a polynomial in $p$, for the calculation of $\partial I_k/\partial \bm{Q}_j$, it is desirable to obtain an expansion for $I_k$ in terms of $L^2_{\nu}$. As evident from Eq.~(\ref{eq:p_l_def}), only $L_{j}$ and $L_{j-1}$ depend on $\bm{Q}_j$. The calculation of $\partial I_k/\partial \bm{Q}_j$ would therefore be simplified if an expansion in terms of $L^2_{\nu}$ is available. This is realized in the following expression, 
\begin{align}
I_k=1+\sum_{\nu=1}^{k-1}L^2_{\nu}\widetilde{q}^{(k)}_{\nu}
\end{align}
with
\begin{align}\label{eq:coeff_def_fwd_poly_composite}
\widetilde{q}^{(k)}_{\nu}=\sum_{\mu=1}^{n}\left(-1\right)^{\mu}p^{\mu}\widetilde{f}\left(\mu,[k-(2\mu-3)],-1;\,\widetilde{\bm{\ell}}\right)
\end{align}
where $1\leq\nu<k$; $\bm{\widetilde{\ell}}$ may in general be a set of numbers but $\bm{\widetilde{\ell}}=\{\nu\}$ in Eq.~(\ref{eq:coeff_def_fwd_poly_composite}), and the function $\widetilde{f}\left(m,l,j;\,\widetilde{\bm{\ell}}\right)$ is defined recursively as
\[\widetilde{f}\left(m,l,j;\,\widetilde{\bm{\ell}}\right)=\sum_{\substack{s=j+2 \\[2pt] s\notin\widetilde{\bm{\lambda}}\\[2pt] s>\min\left(\widetilde{\bm{\ell}}\right)}}^{l} L^{2}_s\,\widetilde{f}\left(m-1,l+2,s;\,\widetilde{\bm{\ell}}\right)\]
%\boxed{\widetilde{f}\left(m,l,j;\,\widetilde{\bm{\ell}}\right)=\sum_{\substack{s=j+2 \\[2pt] s\notin\widetilde{\bm{\lambda}}\\[2pt] s>\min\left(\widetilde{\bm{\ell}}\right)}}^{l} L^{2}_s\,\widetilde{f}\left(m-1,l+2,s;\,\widetilde{\bm{\ell}}\right)} \]
with 
\begin{equation}\label{eq:rec_stop_ftilde}
\widetilde{f}\left(1,l,j;\,\widetilde{\bm{\ell}}\right)=1\quad\quad \forall \quad l,j,\widetilde{\bm{\ell}}
\end{equation}

The set $\bm{\widetilde{\lambda}}$ is related to $\bm{\widetilde{\ell}}$ as follows. For the general case of ${\bm{\widetilde{\ell}}}\equiv\left\{a_1,a_2,a_3,..\right\}$, where the $a_i$ represent arbitrary integers, $\widetilde{\bm{\lambda}}$ is constructed as 
\begin{equation*}
\begin{split}
\bm{\widetilde{\lambda}}\equiv\left\{\left(a_1-1\right),a_1,\left(a_1+1\right),\left(a_2-1\right),a_2,\left(a_2+1\right),...\right\}\\[5pt]
\end{split}
\end{equation*}
For the present case of $\bm{\widetilde{\ell}}=\{\nu\}$, we have $\bm{\widetilde{\lambda}}=\{\nu-1,\nu,\nu+1\}$. 
The conditional summation over the index $s$ appearing in the boxed equation above may be understood as follows. For given values of $\{m,l,j;\,\widetilde{\bm{\ell}}\}$, the index $s$ runs from a lower limit of $j+2$ to an upper limit of $l$, with two constraints. Firstly, $s$ must not belong to the set $\bm{\widetilde{\lambda}}$, and secondly, the value of $s$ must be larger than the smallest entry in the set $\widetilde{\bm{\ell}}$ . A detailed illustration of the use of the recursive relations, $f\left(m,l,j\right)$ and $\widetilde{f}\left(m,l,j;\,\widetilde{\bm{\ell}}\right)$, has been provided below, in Sec.~\ref{sec:ex_impl_fwd}.
%%%%%%%%%%%%%%%%%%%%%%%%%%%%%%%%%%%%%%%%
\begin{figure}[t]
    \centerline{
   {\includegraphics[width=120mm]{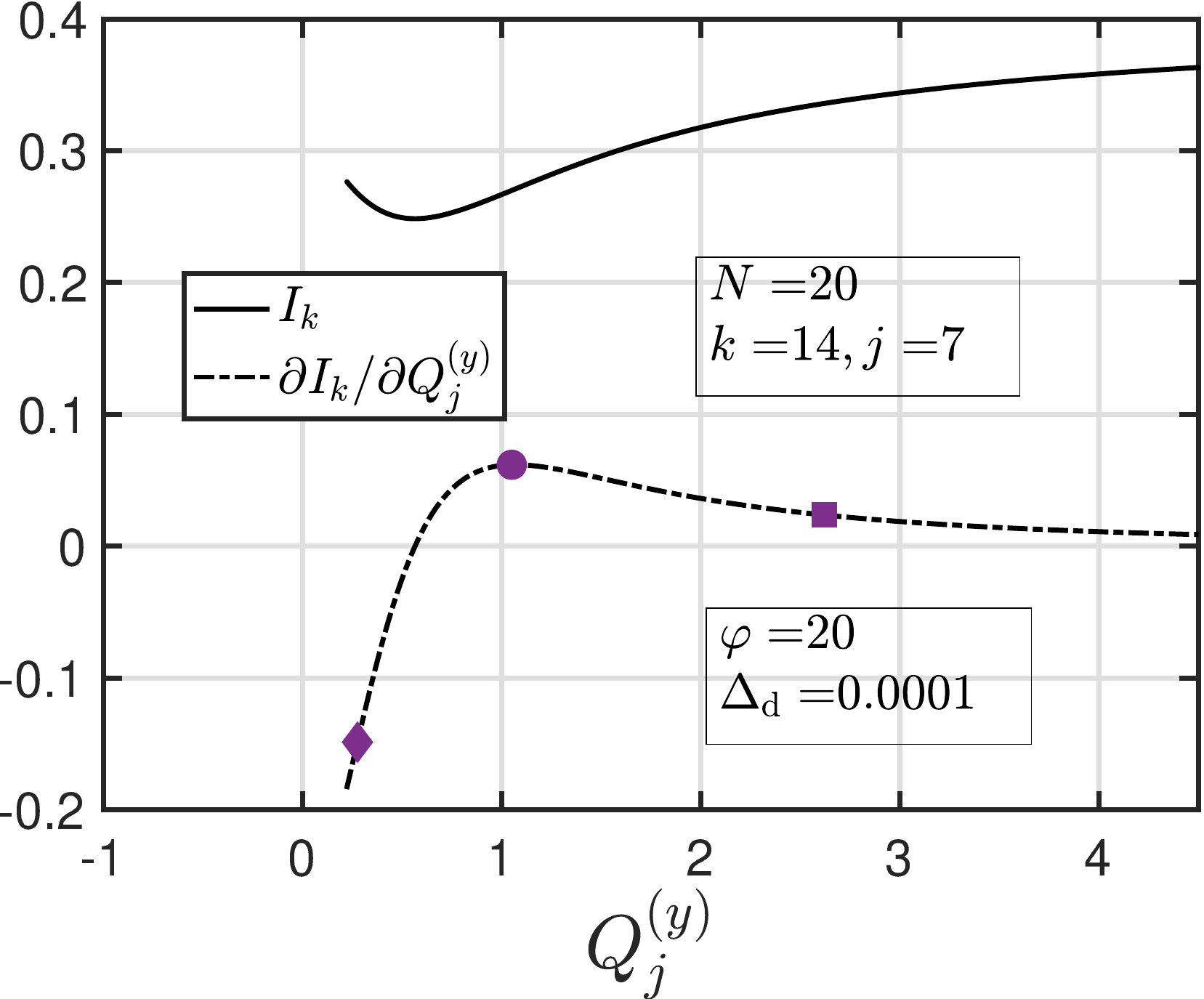}}     
    }
%\begin{figure}[t]
%\centering
%\includegraphics[width=0.7\linewidth]{fig1}  
\caption{Plot of $I_k$ [solid line, from Eq.~(\ref{eq:i_polynomial})] and its gradient in the $y-$direction [broken line, from Eq.~(\ref{eq:i_k_part_diff})]. Symbols indicate derivatives calculated using central-difference scheme, with a spatial discretization width of $\Delta_{\text{d}}$.}
\label{fig:i_k_derv_check}
\end{figure}
%%%%%%%%%%%%%%%%%%%%%%%%%%%%%%%%%%%%%%%%

The next task is to find a general expression for $\dfrac{\partial \widetilde{q}^{(k)}_{\nu}}{\partial \bm{Q}_j}$. Note that
\begin{align}
\dfrac{\partial \widetilde{q}^{(k)}_{\nu}}{\partial \bm{Q}_j}=0; \quad j<\nu+2
\end{align}
and the following equation holds when $j\geq \nu+2$,
\begin{equation}\label{eq:qtilde_derv_fwd}
\begin{split}
\dfrac{\partial \widetilde{q}^{(k)}_{\nu}}{\partial \bm{Q}_j}&=\widetilde{\theta}_{j\nu}\left[\sum_{\mu=2}^{n}\left(-1\right)^{\mu}p^{\mu}\widetilde{f}(\mu-1,\left[k-(2\mu-5)\right],-1;\,\{\nu,j-1\})\right]\dashuline{\dfrac{\partial L^2_{j-1}}{\partial \bm{Q}_j}}\\[10pt]
&+\widehat{\theta}_{jk}\left[\sum_{\mu=2}^{n}\left(-1\right)^{\mu}p^{\mu}\widetilde{f}(\mu-1,\left[k-(2\mu-5)\right],-1;\,\{\nu,j\})\right]\uline{\dfrac{\partial L^2_{j}}{\partial \bm{Q}_j}}; 
\end{split}
\end{equation}
where the underlined terms may be evaluated using Eq.~(\ref{eq:lsq_min}) and Eq.~(\ref{eq:lsq_same}) given in Sec.~\ref{sec:list_ident}, and the indicator functions, $\widetilde{\theta}_{j\nu}$ and $\widehat{\theta}_{jk}$ are defined as
 \begin{align}\label{eq:theta_ind}
\widetilde{\theta}_{j\nu}= \left\{
\begin{array}{ll}
        0; &  j\leq\nu+2\\[15pt]
        1;  & j>\nu+2
\end{array} 
\right. 
\end{align}
and
 \begin{align}\label{eq:thetahat_ind}
\widehat{\theta}_{jk}= \left\{
\begin{array}{ll}
        0; &  j=k\\[15pt]
        1;  & j\neq k
\end{array} 
\right. 
\end{align}
which enables us to write
 \begin{align}\label{eq:i_k_part_diff}
\dfrac{\partial I_k}{\partial \bm{Q}_j}= \left\{
\begin{array}{ll}
        0; &  j>k\\[15pt]
       \sum_{i=1}^{j-2}L^2_{i}\dfrac{\partial \widetilde{q}^{(k)}_{i}}{\partial \bm{Q}_j}+\widetilde{q}^{(k)}_{j-1}{\dfrac{\partial L^2_{j-1}}{\partial \bm{Q}_j}}+\widetilde{q}^{(k)}_{j}{\dfrac{\partial L^2_{j}}{\partial \bm{Q}_j}}; & j\leq k
\end{array} 
\right. 
\end{align}
In Fig.~\ref{fig:i_k_derv_check}, the derivative of the polynomial $I_k$, with respect to the $y-$th component of $\bm{Q}_j$, is plotted for arbitrarily chosen values of $k=14,\,j=7$, and the chain length, $N=20$. There is an excellent agreement between the derivative calculated numerically, and that obtained analytically using Eq.~(\ref{eq:i_k_part_diff}).

%%%%%%%%%%%%
\begin{figure}[t]
    \centerline{
   {\includegraphics[width=120mm]{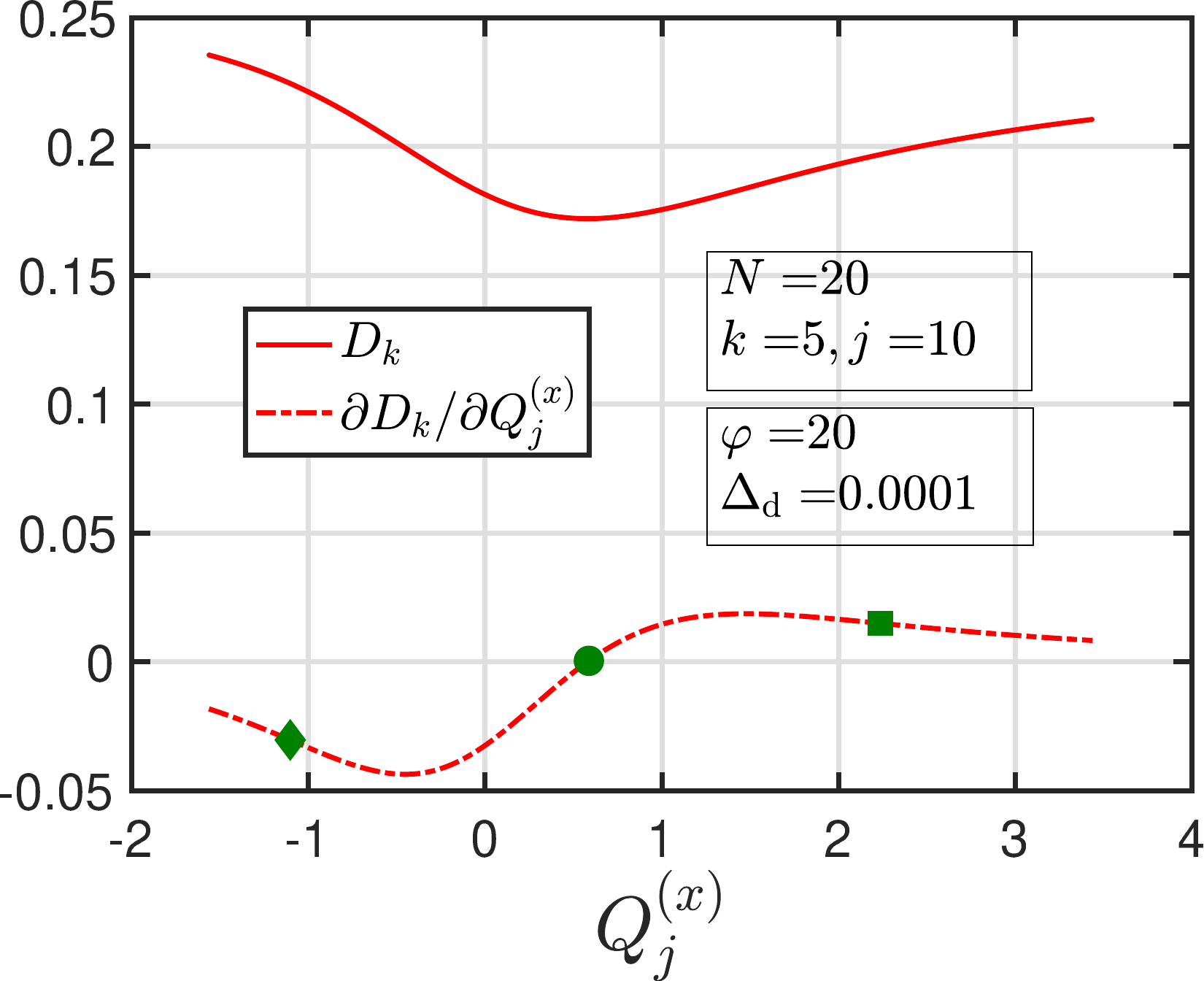}}     
    }
 %  \vskip-10pt 
%\begin{figure}[t]
%\centering
%\includegraphics[width=0.7\linewidth]{fig2}  
\caption{Plot of $D_k$ [solid line, from Eq.~(\ref{eq:d_polynomial})] and its gradient in the $x-$direction [broken line, from Eq.~(\ref{eq:d_k_part_diff})]. Symbols indicate derivatives calculated using central-difference scheme, with a spatial discretization width of $\Delta_{\text{d}}$.}
\label{fig:d_k_derv_check}
\end{figure}
%%%%%%%%%%%%%
%
\subsection{\label{sec:bkwd_poly}Backward continued product}
The backward continued product defined in Eq.~(\ref{eq:d_def}) may be expressed as a polynomial in $p$. Consider $D_k$ where $1\leq\,k\leq N$ is any integer. The degree, $n$, of the polynomial in $p$ that expresses $D_k$ is given by $n=\ceil*{\dfrac{N-k}{2}}$ where $\ceil*{i}$ represents the smallest integer greater than or equal to $i$. We write

\begin{align}\label{eq:d_polynomial}
D_{k}&=1-p\sum_{i=k}^{N-1}L_i^2g\left(1,k-2,i\right)+p^2\sum_{i=k+2}^{N-1}L^2_{i}g\left(2,k,i\right)-p^{3}\sum_{i=k+4}^{N-1}L^2_{i}g\left(3,k+2,i\right)\cdots\nonumber\\[5pt]
&+\left(-1\right)^{n}p^{n}\sum_{i=k+2(n-1)}^{N-1}L^{2}_ig\left(n,\left[k+2(n-2)\right],i\right)
\end{align}
where the function ${g}\left(m,l,\gamma\right)$ is written recursively as
\[
  {g}\left(m,l,\gamma\right)=\sum_{\substack{s=l}}^{\gamma-2} L^{2}_s\,{g}\left(m-1,l-2,s\right)
\]
with 
\begin{equation}
{g}\left(1,l,\gamma\right)=1\quad\quad \forall \quad l,\gamma
\end{equation}
A polynomial expansion for $D_k$ in terms of $L_{\nu}$ may be written as 
\begin{align}
D_k=1+\sum_{\nu=k}^{N-1}L^2_{\nu}\widetilde{w}^{(k)}_{\nu}
\end{align}
where
\begin{align}\label{eq:coeff_def_bkwd_poly_composite}
\widetilde{w}^{(k)}_{\nu}=\sum_{\mu=1}^{n}\left(-1\right)^{\mu}p^{\mu}\widetilde{g}(\mu,\left[k+2(\mu-2)\right],(N+1);\,\bm{\widetilde{\ell}})
\end{align}
where $k\leq\nu<N$, and the function $\widetilde{g}\left(m,l,j;\,\bm{\widetilde{\ell}}\right)$ is written recursively as follows
\[
  \widetilde{g}\left(m,l,j;\,\widetilde{\bm{\ell}}\right)=\sum_{\substack{s=l \\[2pt] s\notin\widetilde{\bm{\lambda}}\\[2pt] s>\min\left(\widetilde{\bm{\ell}}\right)}}^{j-2} L^{2}_s\,\widetilde{g}\left(m-1,l-2,s;\,\bm{\widetilde{\ell}}\right)
\]
with 
\begin{equation}
\widetilde{g}\left(1,l,j;\,\widetilde{\bm{\ell}}\right)=1\quad\quad \forall \quad l,j,\widetilde{\bm{\ell}}
\end{equation}
In the expansion given by Eq.~(\ref{eq:coeff_def_bkwd_poly_composite}), $\bm{\widetilde{\ell}}=\{\nu\}$, and $\bm{\widetilde{\lambda}}=\{\left(\nu-1\right),\nu,\left(\nu+1\right)\}$. 

The next task is to find a general expression for $\dfrac{\partial \widetilde{w}^{(k)}_{\nu}}{\partial \bm{Q}_j}$. Note that
\begin{align}
\dfrac{\partial \widetilde{w}^{(k)}_{\nu}}{\partial \bm{Q}_j}=0; \quad j<\nu+2
\end{align}
and the following equation holds when $j\geq\nu+2$,
\begin{equation}
\begin{split}
\dfrac{\partial \widetilde{w}^{(k)}_{\nu}}{\partial \bm{Q}_j}&=\widetilde{\theta}_{j\nu}\left[\sum_{\mu=2}^{n}\left(-1\right)^{\mu}p^{\mu}\widetilde{g}(\mu-1,\left[k+2(\mu-3)\right],(N+1);\,\{\nu,j-1\})\right]\dfrac{\partial L^2_{j-1}}{\partial \bm{Q}_j}\\[10pt]
&+\widehat{\theta}_{jN}\left[\sum_{\mu=2}^{n}\left(-1\right)^{\mu}p^{\mu}\widetilde{g}(\mu-1,\left[k+2(\mu-3)\right],(N+1);\,\{\nu,j\})\right]\dfrac{\partial L^2_{j}}{\partial \bm{Q}_j}; 
\end{split}
\end{equation}
which enables us to write
 \begin{align}\label{eq:d_k_part_diff}
\dfrac{\partial D_k}{\partial \bm{Q}_j}= \left\{
\begin{array}{ll}
        0; &  j<k\\[15pt]
       \sum_{i=k}^{j-2}L^2_{i}\dfrac{\partial \widetilde{w}^{(k)}_{i}}{\partial \bm{Q}_j}+\widetilde{w}^{(k)}_{j-1}\dfrac{\partial L^2_{j-1}}{\partial \bm{Q}_j}+\widetilde{w}^{(k)}_{j}\dfrac{\partial L^2_{j}}{\partial \bm{Q}_j}; & j\geq k
\end{array} 
\right. 
\end{align}
In Fig.~\ref{fig:d_k_derv_check}, the derivative of the polynomial $D_k$, with respect to the $x-$th component of $\bm{Q}_j$, is plotted for arbitrarily chosen values of $k=5,\,j=10$, and the chain length, $N=20$. There is an excellent agreement between the derivative calculated numerically, and that obtained analytically using Eq.~(\ref{eq:d_k_part_diff}).

%%%%%%%%%%%%%%%%%%%%%%%%%%%%%%%%%%%%%
\subsection{\label{sec:list_ident}List of tensor identities}

The following identities, useful in the numerical calculation of divergence, are stated without proof: 
\begin{equation}
\dfrac{\partial Q_i^2}{\partial \bm{Q}_k}=2\bm{Q}_{i}\delta_{ki}
\end{equation}
\begin{align}\label{eq:derv_bmqk_qk}
\dfrac{\partial}{\partial \bm{Q}_k}\left(\dfrac{\bm{Q}_{i}}{Q_i}\right)=\dfrac{1}{Q_k}\left[\boldsymbol{\delta}-\dfrac{\bm{Q}_k\bm{Q}_k}{Q_k^2}\right]\delta_{ki}
\end{align}
\begin{equation}
%\begin{split}
\dfrac{\partial}{\partial \bm{Q}_k}\cdot\left[\dfrac{\bm{Q}_i\bm{Q}_j}{Q_iQ_j}\right] = \left(\dfrac{1}{Q_iQ_j}\right)\left[\bm{Q}_{i}-\left(\dfrac{\bm{Q}_j\bm{Q}_j}{Q^2_j}\right)\cdot\bm{Q}_{i}\right]\delta_{kj}
+ 2\left(\dfrac{\bm{Q}_{j}}{Q_jQ_i}\right)\delta_{ki}
%\end{split}
\end{equation}
\begin{equation}\label{eq:gen_exp_derv_l_i_q_j}
\begin{split}
\dfrac{\partial L_{i}}{\partial \bm{Q}_k}&=\Biggr\{\left(\dfrac{1}{Q_{i}}\right)\Biggl[\left(\dfrac{\bm{Q}_{i+1}}{Q_{i+1}}\right)-L_{i}\left(\dfrac{\bm{Q}_{i}}{Q_i}\right)\Biggr]\Biggl\}\delta_{ki}\\[10pt]
&+\Biggr\{\left(\dfrac{1}{Q_{i+1}}\right)\Biggl[\left(\dfrac{\bm{Q}_{i}}{Q_i}\right)-L_{i}\left(\dfrac{\bm{Q}_{i+1}}{Q_{i+1}}\right)\Biggr]\Biggl\}\delta_{k,i+1}
\end{split}
\end{equation}
Setting $i=(j-1),\,k=j$ in Eq.~(\ref{eq:gen_exp_derv_l_i_q_j}), 
\begin{align}
\dfrac{\partial L_{j-1}}{\partial \bm{Q}_j}&=\dfrac{1}{Q_j}\left[\left(\dfrac{\bm{Q}_{j-1}}{\bm{Q}_{j-1}}\right)-L_{j-1}\left(\dfrac{\bm{Q}_j}{Q_j}\right)\right]
\end{align}
Setting $i=k=j$ in Eq.~(\ref{eq:gen_exp_derv_l_i_q_j}), 
\begin{align}
\dfrac{\partial L_{j}}{\partial \bm{Q}_j}&=\dfrac{1}{Q_j}\left[\left(\dfrac{\bm{Q}_{j+1}}{\bm{Q}_{j+1}}\right)-L_j\left(\dfrac{\bm{Q}_j}{Q_j}\right)\right]
\end{align}
\begin{equation}\label{eq:gen_exp_derv_l_i_sq_q_j}
\begin{split}
\dfrac{\partial L^2_{i}}{\partial \bm{Q}_k}&=\Biggr\{\left(\dfrac{2L_i}{Q_{i}}\right)\Biggl[\left(\dfrac{\bm{Q}_{i+1}}{Q_{i+1}}\right)-L_{i}\left(\dfrac{\bm{Q}_{i}}{Q_i}\right)\Biggr]\Biggl\}\delta_{ki}+\Biggr\{\left(\dfrac{2L_i}{Q_{i+1}}\right)\Biggl[\left(\dfrac{\bm{Q}_{i}}{Q_i}\right)-L_{i}\left(\dfrac{\bm{Q}_{i+1}}{Q_{i+1}}\right)\Biggr]\Biggl\}\delta_{k,i+1}
\end{split}
\end{equation}
Setting $i=(j-1),\,k=j$ in Eq.~(\ref{eq:gen_exp_derv_l_i_sq_q_j}) and simplifying,
\begin{align}\label{eq:lsq_min}
\dfrac{\partial L^2_{j-1}}{\partial \bm{Q}_j}&=\dfrac{2L_{j-1}}{Q_{j-1}Q^2_{j}}\left[Q_{j}\bm{Q}_{j-1}-L_{j-1}Q_{j-1}\bm{Q}_{j}\right]
\end{align}
Setting $i=k=j$ in Eq.~(\ref{eq:gen_exp_derv_l_i_sq_q_j}) and simplifying,
\begin{align}\label{eq:lsq_same}
\dfrac{\partial L^2_{j}}{\partial \bm{Q}_j}&=\dfrac{2L_j}{Q_{j+1}Q^2_j}\left[Q_{j}\bm{Q}_{j+1}-L_{j}Q_{j+1}\bm{Q}_{j}\right]
\end{align}
\begin{equation}\label{eq:mixed_derv}
\begin{split}
\dfrac{\partial}{\partial \bm{Q}_k}\left[L_{i}L_{j}\right]&=\left(\dfrac{L_i}{Q_j}\right)\left[\left(\dfrac{\bm{Q}_{j+1}}{Q_{j+1}}\right)-L_{j}\left(\dfrac{\bm{Q}_{j}}{Q_{j}}\right)\right]\delta_{kj}+\left(\dfrac{L_i}{Q_{j+1}}\right)\left[\left(\dfrac{\bm{Q}_{j}}{Q_{j}}\right)-L_{j}\left(\dfrac{\bm{Q}_{j+1}}{Q_{j+1}}\right)\right]\delta_{k,j+1}\\[10pt]
&+\left(\dfrac{L_j}{Q_i}\right)\left[\left(\dfrac{\bm{Q}_{i+1}}{Q_{i+1}}\right)-L_{i}\left(\dfrac{\bm{Q}_{i}}{Q_{i}}\right)\right]\delta_{ki}+\left(\dfrac{L_j}{Q_{i+1}}\right)\left[\left(\dfrac{\bm{Q}_{i}}{Q_{i}}\right)-L_{i}\left(\dfrac{\bm{Q}_{i+1}}{Q_{i+1}}\right)\right]\delta_{k,i+1}
\end{split}
\end{equation}
Setting $i=(k-1),\,j=k$ in Eq.~(\ref{eq:mixed_derv}) and simplifying,
\begin{equation}
\dfrac{\partial}{\partial \bm{Q}_k}\left[L_{k-1}L_{k}\right] = \dfrac{1}{Q_{k-1}Q^2_kQ_{k+1}}\Biggl\{Q_{k-1}Q_{k}L_{k-1}\bm{Q}_{k+1}-2Q_{k-1}Q_{k+1}L_{k-1}L_{k}\bm{Q}_k+Q_kQ_{k+1}L_{k}\bm{Q}_{k-1}\Biggr\}
\end{equation}
%\vspace{-20pt}

\subsection{\label{sec:ex_impl_fwd}Illustrative example for algorithmic approach to gradient calculation of forward continued product}
The application of a recursive-function-based route for the calculation of $I_k$ and $\dfrac{\partial I_k}{\partial \bm{Q}_j}$ will be made clear in this section using an illustrative example, for $k=8,\,j=4$ in a chain with $N=10$ springs. Note that the exact value of $N$ is immaterial for calculations of derivatives involving the forward continued product. The degree, $n$, of the polynomial in $p$ that expresses $I_{8}$ is given by
$n=\floor*{\dfrac{8}{2}}=4$. Using Eq.~(\ref{eq:i_polynomial}), we have
\begin{align}\label{eq:i8_def}
I_8&=1-p\sum_{i=1}^{8-1}\,L_{i}^2f\left(1,[8+1],i\right)+p^{2}\sum_{i=1}^{8-3}L_{i}^2\,f\left(2,[8-1],i\right)-p^{3}\sum_{i=1}^{8-5}L_i^{2}\,f\left(3,[8-3],i\right)\nonumber\\
&+\left(-1\right)^{4}p^{4}\sum_{i=1}^{8-7}L^{2}_{i}\,f\left(4,\left[8-5\right],i\right)\nonumber\\
&=1-p\sum_{i=1}^{7}\,L_{i}^2\dotuline{f\left(1,9,i\right)}+p^{2}\sum_{i=1}^{5}L_{i}^2\,f\left(2,7,i\right)-p^{3}\sum_{i=1}^{3}L_i^{2}\,f\left(3,5,i\right)+p^{4}\sum_{i=1}^{1}L^{2}_{i}\,f\left(4,3,i\right)
\end{align}
From Eq.~(\ref{eq:rec_stop_f}), the above underlined term is just unity, and Eq.~(\ref{eq:i8_def}) may be simplified as
\begin{align}\label{eq:i8_sec_step}
I_8&=1-p\sum_{i=1}^{7}\,L_{i}^2+p^{2}\Bigl[L_{1}^2\,f\left(2,7,1\right)+L_{2}^2\,f\left(2,7,2\right)+L_{3}^2\,f\left(2,7,3\right)+L_{4}^2\,f\left(2,7,4\right)\\[5pt]
&+L_{5}^2\,f\left(2,7,5\right)\Bigr]-p^{3}\left[L_1^{2}\,\dashuline{f\left(3,5,1\right)}+L_2^{2}\,f\left(3,5,2\right)+L_3^{2}\,f\left(3,5,3\right)\right]+p^{4}L^{2}_{1}\,{f\left(4,3,1\right)}\nonumber
\end{align}
The underlined term in Eq.~(\ref{eq:i8_sec_step}) will be evaluated as an example. Using Eq.~(\ref{eq:recdef_f})
\begin{align}\label{eq:examp_f_eval}
f\left(3,5,1\right)&=\sum_{s=1+2}^{5} L^{2}_s\,{f}\left([3-1],[5+2],s\right)=\sum_{s=3}^{5} L^{2}_s\,{f}\left(2,7,s\right)\nonumber\\[5pt]
&=L^{2}_3\,{f}\left(2,7,3\right)+L^{2}_4\,{f}\left(2,7,4\right)+L^{2}_5\,{f}\left(2,7,5\right)\nonumber\\[5pt]
&=L^{2}_3\,\left[\sum_{s=3+2}^{7} L^{2}_s\,{f}\left([2-1],[7+2],s\right)\right]+L^{2}_4\,\left[\sum_{s=4+2}^{7} L^{2}_s\,{f}\left([2-1],[7+2],s\right)\right]\nonumber\\[5pt]
&+L^{2}_5\,\left[\sum_{s=5+2}^{7} L^{2}_s\,{f}\left([2-1],[7+2],s\right)\right]\\[5pt]
&=L^2_3\left[L^2_5+L^2_6+L^2_7\right]+L^2_4\left[L^2_6+L^2_7\right]+L^2_5L^2_7\nonumber
\end{align}
Following a similar procedure, the complete expression for $I_8$ may be obtained as
\begin{align}\label{eq:complete_i8}
I_8&=1-p\left(L_1^2+L_2^2+L_3^2+L_4^2+L_5^2+L_6^2+L_7^2\right)
+p^2\Bigl[L_1^2\left(L_3^2+L_4^2+L_5^2+L_6^2+L_7^2\right)+\nonumber\\[5pt]
&+L^2_2\left(L_4^2+L_5^2+L_6^2+L_7^2\right)+L^2_3\left(L_5^2+L_6^2+L_7^2\right)+L^2_4\left(L_6^2+L_7^2\right)+L^2_5L^2_7\Bigr]\\[5pt]
&-p^3L^2_1\Bigl[L^2_3\left(L_5^2+L_6^2+L_7^2\right)+L^2_4\left(L_6^2+L_7^2\right)+L^2_5L^2_7\Bigr]-p^3L^2_2\Bigl[L^2_4\left(L_6^2+L_7^2\right)+L^2_5L^2_7\Bigr]\nonumber\\[5pt]
&-p^3L^2_3L^2_5L^2_7+p^{4}L^2_1L^2_3L^2_5L^2_7\nonumber
\end{align}
It is now desired to take the gradient of $I_8$ with respect to $\bm{Q}_4$. From Eq.~(\ref{eq:p_l_def}), it is clear that only $L_{3}$ and $L_{4}$ are functions of $\bm{Q}_4$. By grouping together the relevant terms on the RHS of Eq.~(\ref{eq:complete_i8}), the expression for $\dfrac{\partial I_8}{\partial \bm{Q}_4}$ may then be written as
\begin{align}\label{eq:bforce_derv_i8}
\dfrac{\partial I_8}{\partial \bm{Q}_4}&=\Biggl\{-p+p^2\left(L_1^2+L^2_5+L^2_6+L^2_7\right)-p^3L_5^2L_7^2-p^3L^2_1\left(L^2_5+L^2_6+L^2_7\right)+p^4L^2_1L^2_5L^2_7\Biggr\}\dashuline{\dfrac{\partial L^2_{3}}{\partial \bm{Q}_4}}\nonumber\\[5pt]
&+\Biggl\{-p+p^2\left(L_1^2+L^2_2+L^2_6+L^2_7\right)-p^3\left(L_1^2+L_2^2\right)\left(L_6^2+L_7^2\right)\Biggr\}\uline{\dfrac{\partial L^2_{4}}{\partial \bm{Q}_4}}
\end{align}
where the underlined terms may be evaluated using Eq.~(\ref{eq:lsq_min}) and Eq.~(\ref{eq:lsq_same}) given in Sec.~\ref{sec:list_ident}. Equation~(\ref{eq:bforce_derv_i8}) has been obtained using a bruteforce approach, by individually examining terms on the RHS of Eq.~(\ref{eq:complete_i8}) and retaining the ones that do not vanish when a gradient with respect to $\bm{Q}_4$ is taken. An algorithmic approach for obtaining an expression for $\dfrac{\partial I_8}{\partial \bm{Q}_4}$  is illustrated next. Starting with Eq.~(\ref{eq:i_k_part_diff}), 
\begin{equation}\label{eq:first_step_ex}
\dfrac{\partial I_8}{\partial \bm{Q}_4} = \sum_{i=1}^{2}L^2_{i}\dfrac{\partial \widetilde{q}^{(8)}_{i}}{\partial \bm{Q}_4}+\widetilde{q}^{(8)}_{3}{\dfrac{\partial L^2_{3}}{\partial \bm{Q}_4}}+\widetilde{q}^{(8)}_{4}{\dfrac{\partial L^2_{4}}{\partial \bm{Q}_4}}
= L^2_{1}\dfrac{\partial \widetilde{q}^{(8)}_{1}}{\partial \bm{Q}_4}+L^2_{2}\dfrac{\partial \widetilde{q}^{(8)}_{2}}{\partial \bm{Q}_4}+\widetilde{q}^{(8)}_{3}{\dfrac{\partial L^2_{3}}{\partial \bm{Q}_4}}+\widetilde{q}^{(8)}_{4}{\dfrac{\partial L^2_{4}}{\partial \bm{Q}_4}}
\end{equation}
The steps for the construction of $\dfrac{\partial \widetilde{q}^{(8)}_{1}}{\partial \bm{Q}_4}$, using Eq.~(\ref{eq:qtilde_derv_fwd}), may be written as 
\begin{equation}\label{eq:sec_step_ex}
\begin{split}
\dfrac{\partial \widetilde{q}^{(8)}_{1}}{\partial \bm{Q}_4}&=\widetilde{\theta}_{41}\left[\sum_{\mu=2}^{4}\left(-1\right)^{\mu}p^{\mu}\widetilde{f}(\mu-1,\left[8-(2\mu-5)\right],-1;\,\{1,3\})\right]{\dfrac{\partial L^2_{3}}{\partial \bm{Q}_4}}\\[10pt]
&+\widehat{\theta}_{48}\left[\sum_{\mu=2}^{4}\left(-1\right)^{\mu}p^{\mu}\widetilde{f}(\mu-1,\left[8-(2\mu-5)\right],-1;\,\{1,4\})\right]{\dfrac{\partial L^2_{4}}{\partial \bm{Q}_4}};
\end{split}
\end{equation}
recognizing that $\nu=1$, $\widetilde{\bm{\ell}}=\{\nu,j-1\}=\{1,3\}$ for the first term on the RHS of Eq.~(\ref{eq:sec_step_ex}), and $\widetilde{\bm{\ell}}=\{\nu,j\}=\{1,4\}$ for the second term on the RHS.
Upon simplifying Eq.~(\ref{eq:sec_step_ex}) using the indicator functions defined in Eqs.~(\ref{eq:theta_ind}) and (\ref{eq:thetahat_ind}), we obtain
\begin{equation}\label{eq:third_step_ex}
\begin{split}
\dfrac{\partial \widetilde{q}^{(8)}_{1}}{\partial \bm{Q}_4}&=\dotuline{\left[\sum_{\mu=2}^{4}\left(-1\right)^{\mu}p^{\mu}\widetilde{f}(\mu-1,\left[8-(2\mu-5)\right],-1;\,\{1,3\})\right]}{\dfrac{\partial L^2_{3}}{\partial \bm{Q}_4}}\\[10pt]
&+\left[\sum_{\mu=2}^{4}\left(-1\right)^{\mu}p^{\mu}\widetilde{f}(\mu-1,\left[8-(2\mu-5)\right],-1;\,\{1,4\})\right]{\dfrac{\partial L^2_{4}}{\partial \bm{Q}_4}};
\end{split}
\end{equation}
The contents within the square braces, underlined as shown above, will be evaluated explicitly next.
\begin{align}\label{eq:fourth_step_ex}
\sum_{\mu=2}^{4}\left(-1\right)^{\mu}p^{\mu}\widetilde{f}(\mu-1,\left[8-(2\mu-5)\right],-1;\,\{1,3\})&=p^2\widetilde{f}(1,9,-1;\,\{1,3\})-p^3\widetilde{f}(2,7,-1;\,\{1,3\})\nonumber\\[5pt]
&+p^4\widetilde{f}(3,5,-1;\,\{1,3\})
\end{align}
Now, 
\begin{equation}
\label{eq:fifth_step_ex}
\widetilde{f}(1,9,-1;\,\{1,3\})=1
\end{equation}
which follows from Eq.~(\ref{eq:rec_stop_ftilde}). The $\widetilde{f}$ appearing in the second term on the RHS of Eq.~(\ref{eq:fourth_step_ex}) is evaluated as
\[
 \widetilde{f}(2,7,-1;\,\{1,3\})=\sum_{\substack{s=-1+2 \\[2pt] s\notin\widetilde{\bm{\lambda}}\\[2pt] s>\min\left(\{1,3\}\right)}}^{7} L^{2}_s\,\widetilde{f}\left(1,9,s;\,\{1,3\}\right)
\]
where $\widetilde{\bm{\ell}}=\{1,3\}$, and $\widetilde{\bm{\lambda}}\equiv\{(1-1),1,(1+1),(3-1),3,(3+1)\}=\{0,1,2,3,4\}$, with duplicate entries in the set $\widetilde{\bm{\lambda}}$ discarded. 
We therefore obtain
\[
 \widetilde{f}(2,7,-1;\,\{1,3\})=\sum_{\substack{s=1 \\[2pt] s\notin\{0,1,2,3,4\}\\[2pt] s>1}}^{7} L^{2}_s\,\widetilde{f}\left(1,9,s;\,\{1,3\}\right)
 \]
%\[
% \widetilde{f}(2,7,-1;\,\{1,3\})=\sum_{\substack{s=1 \\[2pt] s\notin\{0,1,2,3,4\}\\[2pt] s>1}}^{7} L^{2}_s\,\widetilde{f}\left(1,9,s;\,\{1,3\}\right)=L^{2}_5\,\uline{\widetilde{f}\left(1,9,5;\,\{1,3\}\right)}+L^{2}_6\,\uline{\widetilde{f}\left(1,9,6;\,\{1,3\}\right)}
% +L^{2}_7\,\uline{\widetilde{f}\left(1,9,7;\,\{1,3\}\right)}
%\]
\begin{equation*}
\begin{split}
\qquad\qquad\qquad\qquad\qquad\qquad\quad&=L^{2}_5\,\uline{\widetilde{f}\left(1,9,5;\,\{1,3\}\right)}+L^{2}_6\,\uline{\widetilde{f}\left(1,9,6;\,\{1,3\}\right)}\\[5pt]
 &+L^{2}_7\,\uline{\widetilde{f}\left(1,9,7;\,\{1,3\}\right)}
\end{split}
\end{equation*}
Eq.~(\ref{eq:rec_stop_ftilde}) implies that each of the underlined terms in the above equation is unity, allowing us to write
\begin{equation}\label{eq:sixth_step_ex}
\widetilde{f}(2,7,-1;\,\{1,3\})=L^2_5+L^2_6+L^2_7
\end{equation}
Similarly,
\[
 \widetilde{f}(3,5,-1;\,\{1,3\})=\sum_{\substack{s=1 \\[2pt] s\notin\{0,1,2,3,4\}\\[2pt] s>1}}^{5} L^{2}_s\,\widetilde{f}\left(2,7,s;\,\{1,3\}\right)= L^{2}_5\,\widetilde{f}\left(2,7,5;\,\{1,3\}\right)
\]
We then have
\[
 \widetilde{f}(2,7,5;\,\{1,3\})=\sum_{\substack{s=7 \\[2pt] s\notin\{0,1,2,3,4\}\\[2pt] s>1}}^{7} L^{2}_s\,\widetilde{f}\left(1,9,s;\,\{1,3\}\right)= L^{2}_7\
\]
and may write
\begin{equation}\label{eq:seventh_step_ex}
 \widetilde{f}(3,5,-1;\,\{1,3\})=L^2_{5}L^2_{7}
\end{equation}
Using Eqs.~(\ref{eq:fifth_step_ex}), (\ref{eq:sixth_step_ex}), and (\ref{eq:seventh_step_ex}), Eq.~(\ref{eq:fourth_step_ex}) may be rewritten as
\begin{equation}\label{eq:eigth_step_ex}
\begin{split}
\sum_{\mu=2}^{4}\left(-1\right)^{\mu}p^{\mu}\widetilde{f}(\mu-1,\left[8-(2\mu-5)\right],-1;\,\{1,3\})&=p^2-p^3\left(L^2_5+L^2_6+L^2_7\right)+p^4L^2_{5}L^2_{7}
\end{split}
\end{equation}
Following along similar lines, the contents of the square braces in the second term on the RHS of Eq.~(\ref{eq:sec_step_ex}) may be evaluated to be
\begin{equation}\label{eq:ninth_step_ex}
\begin{split}
&\sum_{\mu=2}^{4}\left(-1\right)^{\mu}p^{\mu}\widetilde{f}(\mu-1,\left[8-(2\mu-5)\right],-1;\,\{1,4\})\\
&=p^2\widetilde{f}(1,9,-1;\,\{1,4\})-p^3\widetilde{f}(2,7,-1;\,\{1,4\})+p^4\widetilde{f}(3,5,-1;\,\{1,4\})\\
&=p^2-p^3\left(L^2_6+L^2_7\right)
\end{split}
\end{equation}
From Eqs.~(\ref{eq:eigth_step_ex}) and (\ref{eq:ninth_step_ex}), Eq.~(\ref{eq:third_step_ex}) may be rewritten as 
\begin{equation}\label{eq:derv_1_put}
\begin{split}
\dfrac{\partial \widetilde{q}^{(8)}_{1}}{\partial \bm{Q}_4}&={\left[p^2-p^3\left(L^2_5+L^2_6+L^2_7\right)+p^4L^2_{5}L^2_{7}\right]}{\dfrac{\partial L^2_{3}}{\partial \bm{Q}_4}}+\left[p^2-p^3\left(L^2_6+L^2_7\right)\right]{\dfrac{\partial L^2_{4}}{\partial \bm{Q}_4}}
\end{split}
\end{equation}
Following along similar lines, we obtain
\begin{equation}\label{eq:derv_2_put}
\begin{split}
\dfrac{\partial \widetilde{q}^{(8)}_{2}}{\partial \bm{Q}_4}&=\left[p^2-p^3\left(L^2_6+L^2_7\right)\right]{\dfrac{\partial L^2_{4}}{\partial \bm{Q}_4}}
\end{split}
\end{equation}
The steps for the construction of $\widetilde{q}^{(8)}_{4}$ are given next. Starting from Eq.~(\ref{eq:coeff_def_fwd_poly_composite}), 
\begin{equation}\label{eq:tenth_step_ex}
\begin{split}
\widetilde{q}^{(8)}_{4}&=\sum_{\mu=1}^{4}\left(-1\right)^{\mu}p^{\mu}\widetilde{f}(\mu,\left[8-(2\mu-3)\right],-1;\,\{4\})\\
&=-p\widetilde{f}(1,9,-1;\,\{4\})+p^2\widetilde{f}(2,7,-1;\,\{4\})-p^3\widetilde{f}(3,5,-1;\,\{4\})+p^4\widetilde{f}(4,3,-1;\,\{4\})
\end{split}
\end{equation}
Now,
\begin{equation}\label{eq:eleventh_step_ex}
\widetilde{f}(1,9,-1;\, \{4\})=1
\end{equation}
which follows from Eq.~(\ref{eq:rec_stop_ftilde}). The $\widetilde{f}$ appearing in the second term on the RHS of Eq.~(\ref{eq:tenth_step_ex}) is evaluated as
\[
 \widetilde{f}(2,7,-1;\,\{4\})=\sum_{\substack{s=-1+2 \\[2pt] s\notin\widetilde{\bm{\lambda}}\\[2pt] s>\min\left(\{4\}\right)}}^{7} L^{2}_s\,\widetilde{f}\left(1,9,s;\,\{4\}\right)
\]
\vspace{-10pt}
where $\widetilde{\bm{\lambda}}\equiv\{(4-1),4,(4+1)\}=\{3,4,5\}$. We therefore obtain
\[
 \widetilde{f}(2,7,-1;\,\{4\})=\sum_{\substack{s=1 \\[2pt] s\notin\{3,4,5\}\\[2pt] s>4}}^{7} L^{2}_s\,\widetilde{f}\left(1,9,s;\,\{4\}\right)= L^{2}_6\,\uline{\widetilde{f}\left(1,9,6;\,\{4\}\right)}+ L^{2}_7\,\uline{\widetilde{f}\left(1,9,7;\,\{4\}\right)}
\]
Eq.~(\ref{eq:rec_stop_ftilde}) implies that each of the underlined terms in the above equation is unity, allowing us to write
\begin{equation}\label{eq:twelfth_step_ex}
\widetilde{f}(2,7,-1;\,\{4\})=L^2_6+L^2_7
\end{equation}
Processing the terms,
\[
 \widetilde{f}(3,5,-1;\,\{4\})=\sum_{\substack{s=1 \\[2pt] s\notin\{3,4,5\}\\[2pt] s>4}}^{5} L^{2}_s\,\widetilde{f}\left(2,7,s;\,\{4\}\right)= 0 \, ;
\] 
\[
 \widetilde{f}(4,3,-1;\,\{4\})=\sum_{\substack{s=1 \\[2pt] s\notin\{3,4,5\}\\[2pt] s>4}}^{3} L^{2}_s\,\widetilde{f}\left(3,5,s;\,\{4\}\right)= 0
\] 
Using Eqs.~(\ref{eq:eleventh_step_ex}) and (\ref{eq:twelfth_step_ex}), Eq.~(\ref{eq:tenth_step_ex}) may be rewritten as
\begin{equation}\label{eq:q4_ans}
\widetilde{q}^{(8)}_{4}=-p+p^2\left(L^2_6+L^2_7\right)
\end{equation}
Following along similar lines, we obtain
\begin{equation}\label{eq:q3_ans}
\widetilde{q}^{(8)}_{3}=-p+p^2\left(L^2_5+L^2_6+L^2_7\right)-p^3L^2_5L^2_7
\end{equation}
Using Eqs.~(\ref{eq:derv_1_put}), (\ref{eq:derv_2_put}), (\ref{eq:q4_ans}) and (\ref{eq:q3_ans}), Eq.~(\ref{eq:first_step_ex}) may be rewritten as
\begin{equation*}
\begin{split}
\dfrac{\partial I_8}{\partial \bm{Q}_4} &=L^2_1\Biggl\{{\left[p^2-p^3\left(L^2_5+L^2_6+L^2_7\right)+p^4L^2_{5}L^2_{7}\right]}{\dfrac{\partial L^2_{3}}{\partial \bm{Q}_4}}+\left[p^2-p^3\left(L^2_6+L^2_7\right)\right]{\dfrac{\partial L^2_{4}}{\partial \bm{Q}_4}}\Biggr\}\\[5pt]
&+L^2_2\Biggl\{\left[p^2-p^3\left(L^2_6+L^2_7\right)\right]{\dfrac{\partial L^2_{4}}{\partial \bm{Q}_4}}\Biggr\}\nonumber\\[5pt]
&+\Biggl\{-p+p^2\left(L^2_5+L^2_6+L^2_7\right)-p^3L^2_5L^2_7\Biggr\}{\dfrac{\partial L^2_{3}}{\partial \bm{Q}_4}} + \Biggl\{-p+p^2\left(L^2_6+L^2_7\right)\Biggr\}{\dfrac{\partial L^2_{4}}{\partial \bm{Q}_4}}\\[5pt]
&=\Biggl\{-p+p^2\left(L^2_5+L^2_6+L^2_7\right)-p^3L^2_5L^2_7+L_1^2\Bigl[p^2-p^3\left(L^2_5+L^2_6+L^2_7\right)+p^4L^2_{5}L^2_{7}\Bigr]\Biggr\}{\dfrac{\partial L^2_{3}}{\partial \bm{Q}_4}}\\[5pt]
&+\Biggl\{-p+p^2\left(L^2_6+L^2_7\right)+L^2_1\Bigl[p^2-p^3\left(L^2_6+L^2_7\right)\Bigr]+L^2_2\Bigl[p^2-p^3\left(L^2_6+L^2_7\right)\Bigr]\Biggr\}{\dfrac{\partial L^2_{4}}{\partial \bm{Q}_4}}
\end{split}
\end{equation*}
which, upon simplification, yields
\begin{align}\label{eq:algo_derv_i8}
\dfrac{\partial I_8}{\partial \bm{Q}_4} &=\Biggl\{-p+p^2\left(L_1^2+L^2_5+L^2_6+L^2_7\right)-p^3L^2_5L^2_7-p^3L_1^2\left(L^2_5+L^2_6+L^2_7\right)+p^4L_1^2L^2_{5}L^2_{7}\Biggr\}{\dfrac{\partial L^2_{3}}{\partial \bm{Q}_4}}\nonumber\\[5pt]
&+\Biggl\{-p+p^2\left(L^2_1+L^2_2+L^2_6+L^2_7\right)-p^3\left(L^2_1+L^2_2\right)\left(L^2_6+L^2_7\right)\Biggr\}{\dfrac{\partial L^2_{4}}{\partial \bm{Q}_4}}
\end{align}
It is thus observed that Eq.~(\ref{eq:algo_derv_i8}), which has been obtained using the recursive-function-based route for the algorithmic calculation of the gradient, is identical to the expression for the gradient written using a bruteforce approach, given by Eq.~(\ref{eq:bforce_derv_i8}). 

\begin{figure}[t]
    \centerline{
   {\includegraphics[width=115mm]{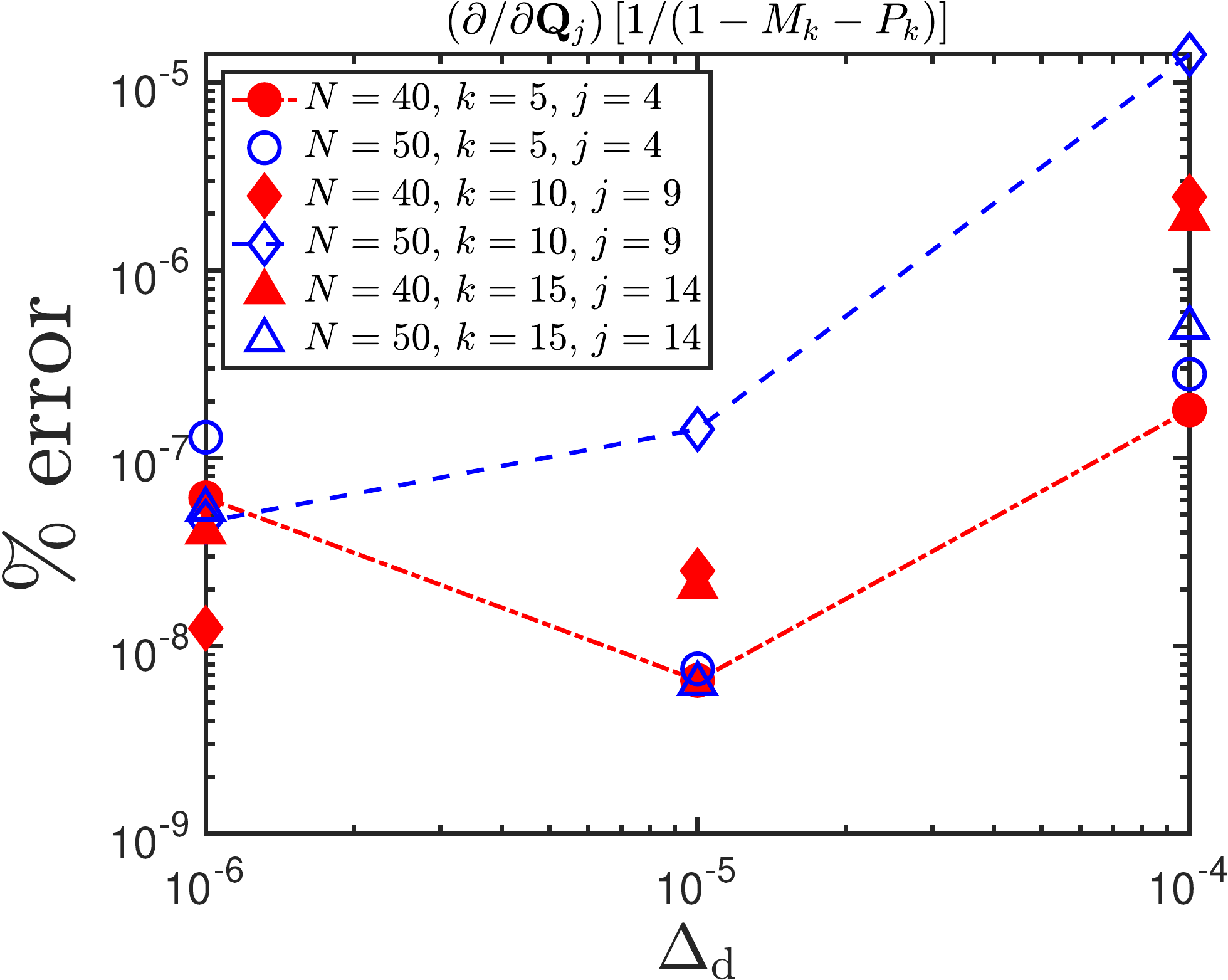}}     
    }
\caption{Variation of error in the calculation of gradient [Eq.~(\ref{eq:erdef_grad})], as a function of the spatial discretization width, for two different chain lengths. An internal friction parameter of $\varphi=200$ is used for all the data points.}
\label{fig:error_scaling_disc_width}
\vskip-10pt
\end{figure}

\subsection{\label{sec:num_div} Calculation of divergence terms in SDE and stress tensor expression}

As the first step, it is desired to examine the effect of the spatial discretization width, $\Delta_{\text{d}}$, on the accuracy of the numerical calculation of the gradient. As an example, the gradient of $\left[1/(1-M_k-P_k)\right]$, evaluated with respect to the connector vector $\bm{Q}_j$ for different values of $j,k,\,\text{and}\,N$, using the central-difference approximation [Eq.~(19) of the main paper], is compared against the solution obtained using the recursive algorithm detailed in Secs.~\ref{sec:fwd_poly}-~\ref{sec:list_ident}.

The error in the gradient evaluated using the central-difference approximation is calculated as
\begin{equation}\label{eq:erdef_grad}
\%\,\text{error}=\dfrac{|\bm{d}_{\text{num}}-\bm{d}_{\text{recursive}}|}{|\bm{d}_{\text{recursive}}|}\times 100
\end{equation}
where $\bm{d}\equiv \left({\partial}/{\partial \bm{Q}_j}\right)\left[1/(1-M_k-P_k)\right]$, and $|\bm{d}|=\sqrt{d^2_{x}+d^2_{y}+d^2_{z}}$.

In Fig.~\ref{fig:error_scaling_disc_width}, the variation of this error is plotted as a function of the discretization width, $\Delta_{\text{d}}$, for several test cases. For the data set denoted by empty diamond symbols, the error is seen to decrease nearly monotonically with the decrease in the spatial discretization width. However, for several other data sets, the error varies non-monotonically as the spatial discretization width is changed. Since the minima in the error, where it exists, is observed to occur in the neighbourhood of $\Delta_{\text{d}}=10^{-5}$, this value of the discretization width has been used in all our calculations. It is noted that the time required for the numerical calculation of the gradient is practically independent of the discretization width.

\begin{figure*}[t]
\begin{tabular}{c c}
\includegraphics[width=80mm]{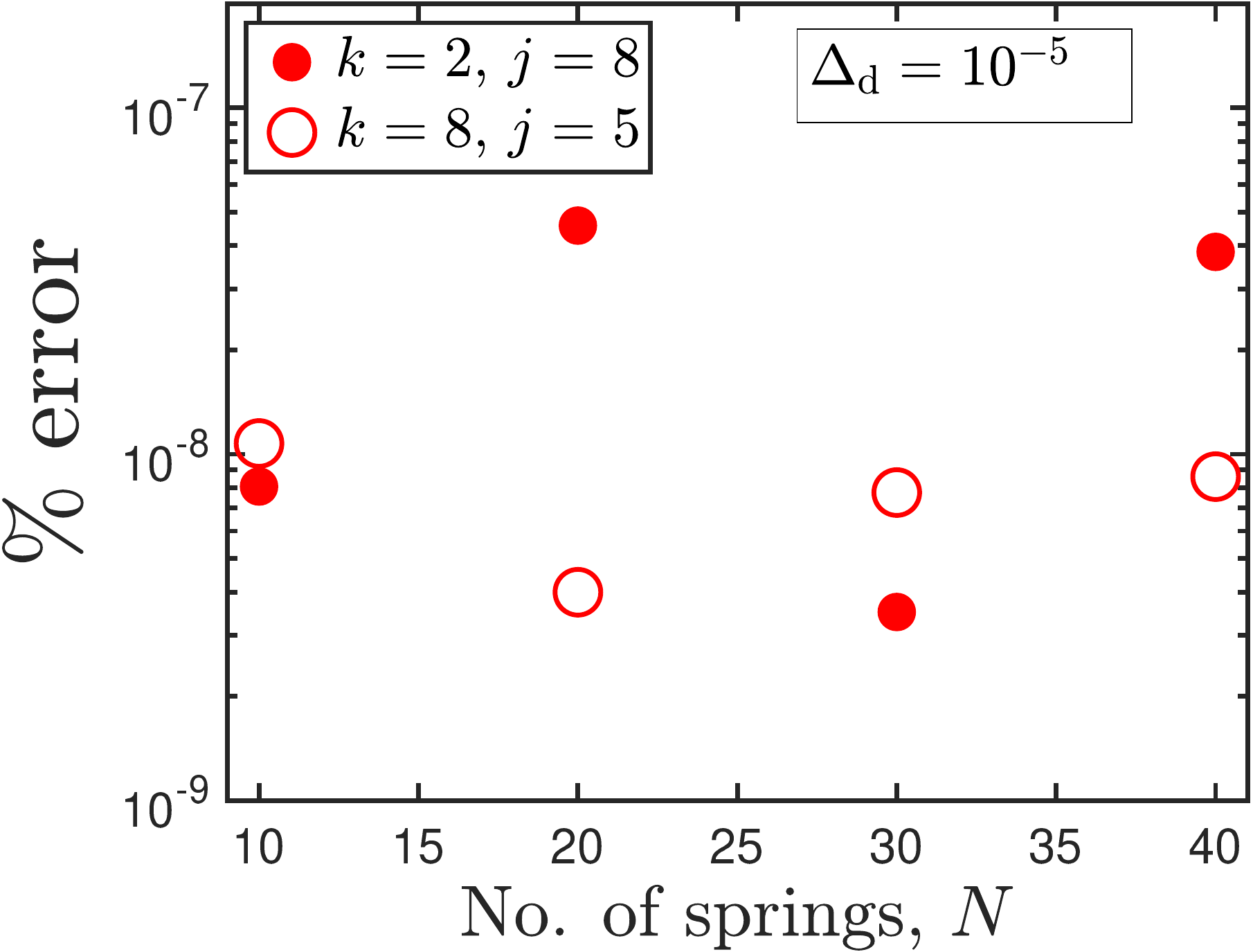}&
\includegraphics[width=80mm]{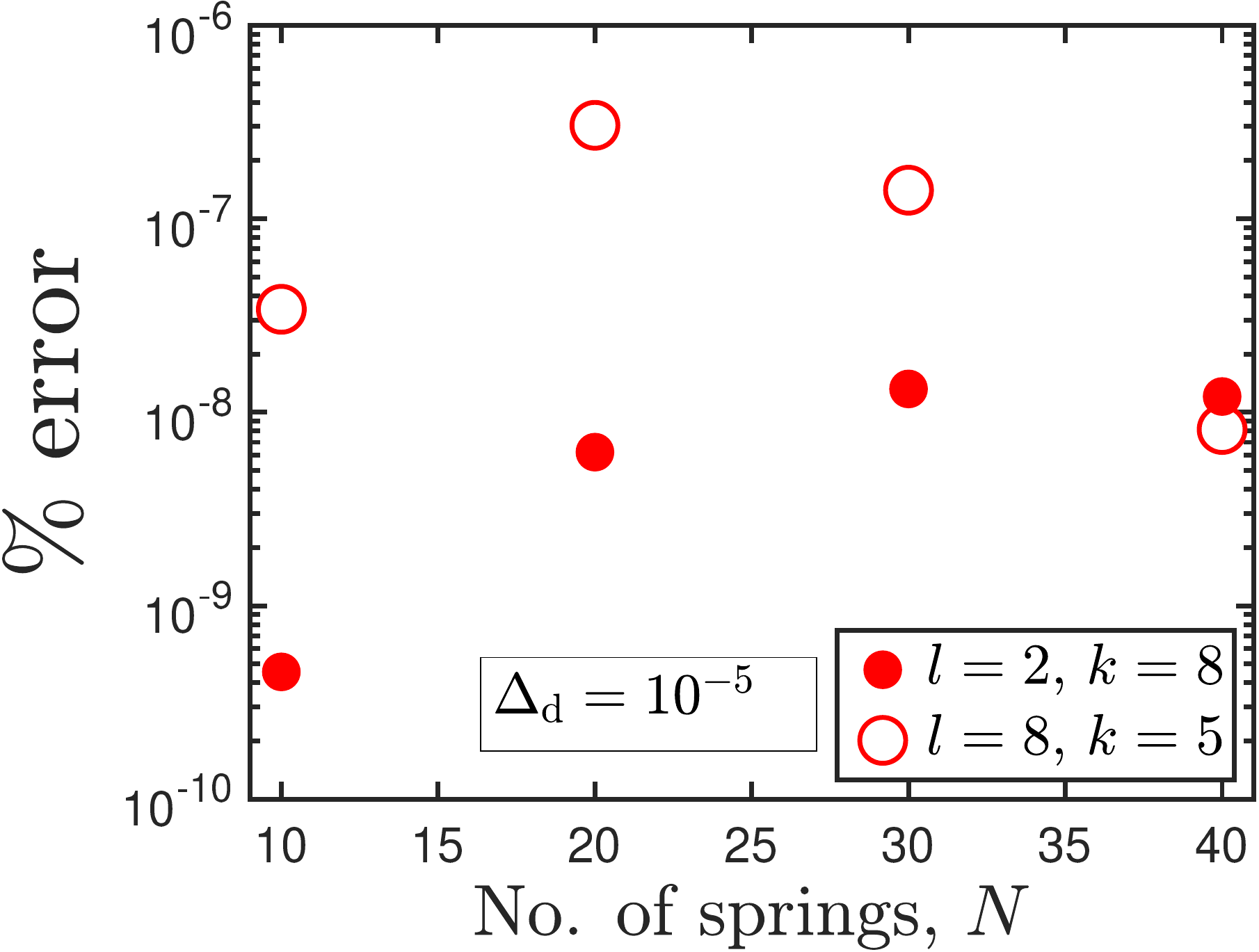} \\
 (a) &  (b)\\[5pt]
\end{tabular}
\caption{Error in the calculation of (a) $\left({\partial}/{\partial \bm{Q}_k}\right)\cdot\bm{V}^{T}_{jk}$  and (b) $\left({\partial}/{\partial \bm{Q}_l}\right)\cdot\boldsymbol{\mu}^{T}_{kl}$, as a function of chain length. An internal friction parameter of $\varphi=200$ is used for all the data points.}
\label{fig:error_scaling_chain_length}
\end{figure*}

\begin{figure*}[t]
\begin{tabular}{c c}
\includegraphics[width=80mm]{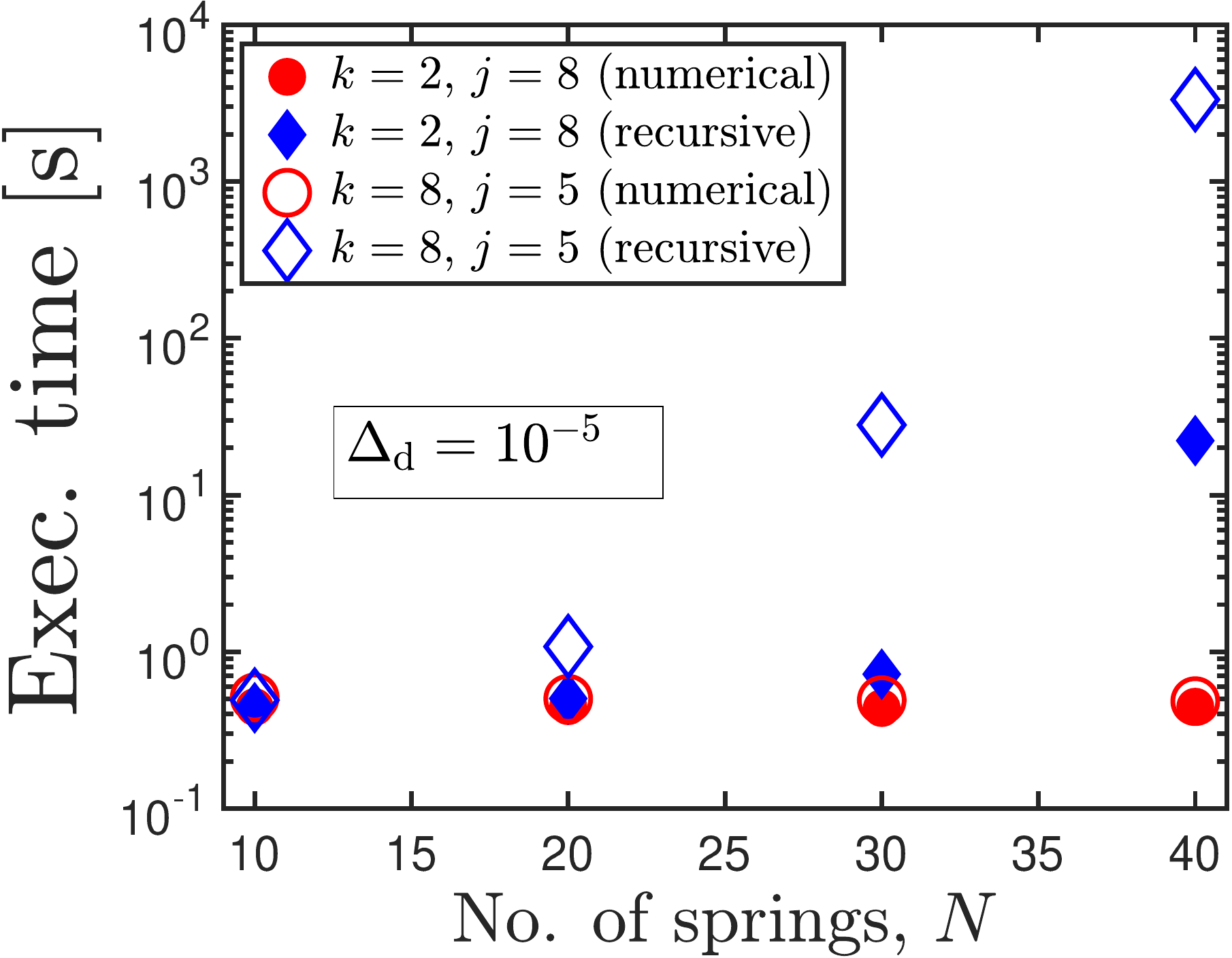}&
\includegraphics[width=80mm]{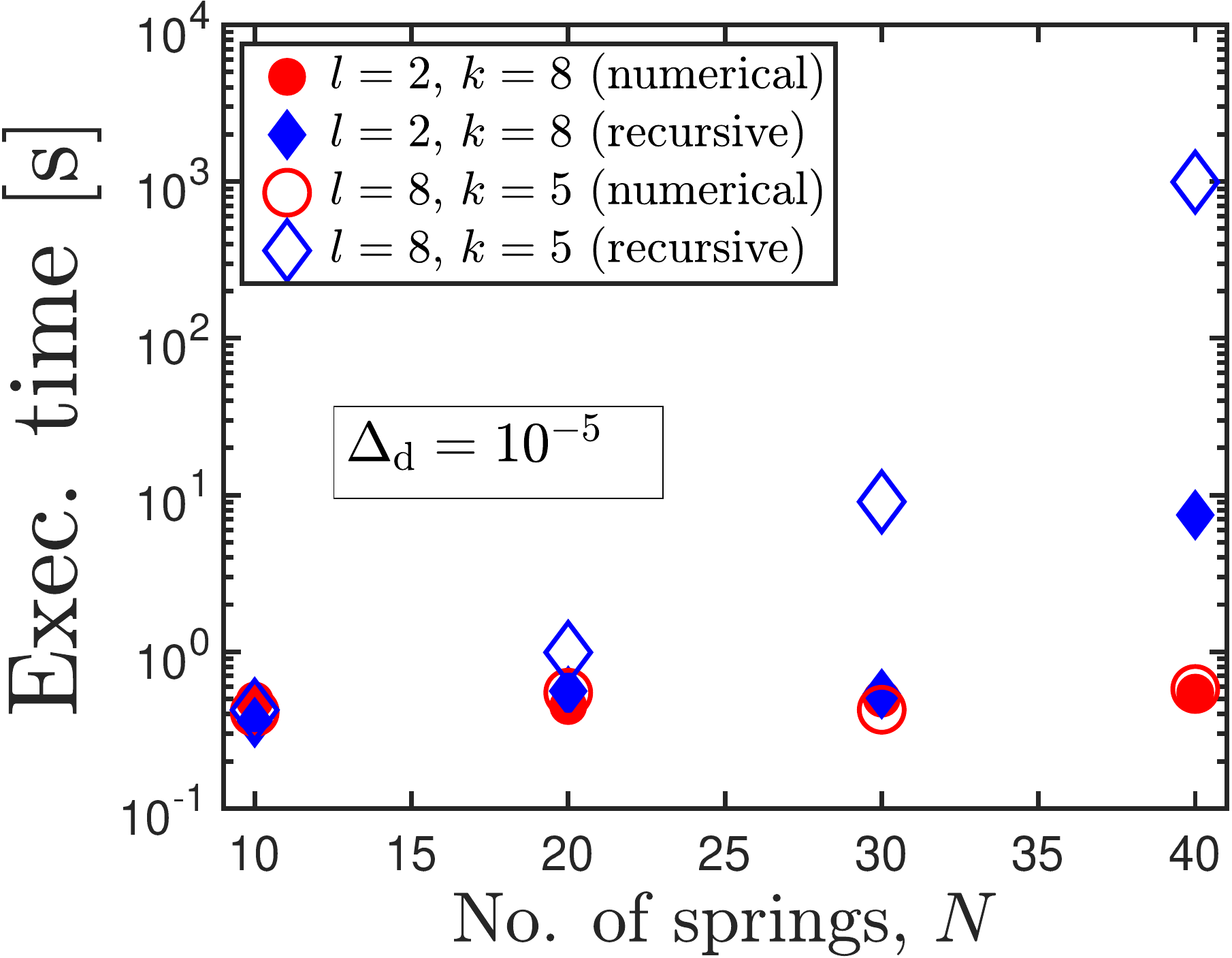} \\
 (a) &  (b)\\[5pt]
\end{tabular}
\caption{Execution time, in seconds, for the calculation of (a) $\left({\partial}/{\partial \mathbf{Q}_k}\right)\cdot\bm{V}^{T}_{jk}$  and (b) $\left({\partial}/{\partial \mathbf{Q}_l}\right)\cdot\boldsymbol{\mu}^{T}_{kl}$, using two different methods, as a function of chain length. An internal friction parameter of $\varphi=200$ is used for all the data points. All the runs were executed on on MonARCH, the HPC hosted at Monash University, on the same type of processor [16 core Xeon-E5-2667-v3 @ 3.20GHz servers with 100550MB usable memory].}
\label{fig:time_scaling_chain_length}
\end{figure*}

In Fig.~\ref{fig:error_scaling_chain_length}, the error in the calculation of divergence terms, which appear in the governing stochastic differential equation and the stress tensor expression, is plotted as a function of the chain length. The error is calculated as
\begin{equation}\label{eq:erdef_div}
\%\,\text{error}=\dfrac{|\bm{z}_{\text{num}}-\bm{z}_{\text{recursive}}|}{|\bm{z}_{\text{recursive}}|}\times 100
\end{equation}
where $\bm{z}\equiv \left({\partial}/{\partial \bm{Q}_k}\right)\cdot\bm{V}^{T}_{jk}\,\,\text{or}\,\,\left({\partial}/{\partial \bm{Q}_l}\right)\cdot\boldsymbol{\mu}^{T}_{kl}$, and various values of $j,k,\,\text{and}\,l$ have been considered. The error in all the cases is seen to be $\sim10^{-7}\%$.

In Fig.~\ref{fig:time_scaling_chain_length}, the execution time needed for calculating the divergence is plotted as a function of chain length. At lower values of the chain length, the execution times using the two approaches are comparable. With the increase in chain length, however, the time needed for recursive calculation is vastly greater than that for the numerical route. Furthermore, while the execution time using the direct route is nearly independent of the chain length, the time needed for the recursive route increases precipitously at higher chain lengths, due to the larger number of polynomial evaluations. In view of its faster execution execution time, and excellent accuracy ($\sim10^{-7}\%$), the numerical method for divergence calculation has been used in all our simulations.


%\bibliographystyle{aipnum4-1}
%\bibliography{supp_inf}